\documentclass[11pt,headings=big, Tightenlines,numbers=noenddot,DIV=14,a4paper]{article}%

\usepackage[margin=1 in]{geometry}
\usepackage{comment}
\usepackage{graphicx}  
\usepackage{dcolumn} 
\usepackage{bm,relsize}   
\usepackage{amsfonts,amsmath,amssymb,amsthm,mathtools}
\usepackage{physics}
\usepackage{slashed}
\usepackage{placeins}
\usepackage{adjustbox}
\usepackage[normalem]{ulem}
\usepackage{lscape}
\usepackage{multirow}
\usepackage{lipsum, color}
\usepackage{tabularray}
\usepackage{rotating}
\usepackage{chngpage}
\usepackage[usenames,dvipsnames,svgnames, table]{xcolor}
\usepackage[linktoc=page,bookmarks=false,colorlinks=true,linkbordercolor=RoyalBlue,citebordercolor=ForestGreen,urlbordercolor=CornflowerBlue]{hyperref}

\definecolor{zima_blue}{HTML}{1393C1}

\hypersetup{
    colorlinks=true,
    linkcolor=zima_blue,
    filecolor=zima_blue,      
    urlcolor=zima_blue,
    citecolor=zima_blue,
}

\usepackage[sort&compress,numbers,merge]{natbib}
\usepackage{breakcites}
\usepackage{array}
\usepackage{makecell}
\usepackage{tabularray}
\usepackage{multirow}

\newcommand{\be}{\begin{equation}} 
\newcommand{\ee}{\end{equation}}
\newcommand{\bea}{\begin{equation}\begin{aligned}} 
\newcommand{\eea}{\end{aligned}\end{equation}}

\newcommand{\mycomment}[1]{}

\usepackage{tikz-feynman}

\usepackage{empheq}
\usepackage[most]{tcolorbox}
\newtcbox{\mymath}[1][]{%
    nobeforeafter, math upper, tcbox raise base,
    enhanced, colframe=blue!20!black,
    colback=blue!15, boxrule=1pt,
    #1}

\usepackage[
    starfontserif 
    ]{starfont}
\usepackage{amsmath}
\usepackage{relsize}

\DeclareSymbolFont{starfontsym}{OT1}{sts}{m}{n}
\DeclareMathSymbol{\mathSun}{\mathord}{starfontsym}{115}
\DeclareMathSymbol{\mathMercury}{\mathord}{starfontsym}{102}
\DeclareMathSymbol{\mathVenus}{\mathord}{starfontsym}{103}
\DeclareMathSymbol{\mathTerra}{\mathord}{starfontsym}{76}
\DeclareMathSymbol{\mathvarTerra}{\mathord}{starfontsym}{108}
\DeclareMathSymbol{\mathMoon}{\mathord}{starfontsym}{100}
\DeclareMathSymbol{\mathvarMoon}{\mathord}{starfontsym}{97}
\DeclareMathSymbol{\mathMars}{\mathord}{starfontsym}{104}
\DeclareMathSymbol{\mathJupiter}{\mathord}{starfontsym}{106}
\DeclareMathSymbol{\mathSaturn}{\mathord}{starfontsym}{83}
\DeclareMathSymbol{\mathUranus}{\mathord}{starfontsym}{70}
\DeclareMathSymbol{\mathvarUranus}{\mathord}{starfontsym}{65}
\DeclareMathSymbol{\mathNeptune}{\mathord}{starfontsym}{71}
\DeclareMathSymbol{\mathPluto}{\mathord}{starfontsym}{74}
\DeclareMathSymbol{\mathvarPluto}{\mathord}{starfontsym}{72}

\begin{document}

\begin{flushright}
\footnotesize
MITP-26-024
\end{flushright}

\begin{flushright}
\footnotesize
\end{flushright}
\color{black}
\begin{center}

{\LARGE \bf Gravitational Waves from Black Hole Reheating \\[0.2cm] \it The Scalar-Induced Component}

\medskip
\bigskip\color{black}\vspace{0.5cm}

\renewcommand{\thefootnote}{\fnsymbol{footnote}}

{\large
Yann Gouttenoire$^{a,b}$\footnote{\href{mailto:yann.gouttenoire@gmail.com}{yann.gouttenoire@gmail.com}},
Nicholas Leister$^{b}$\footnote{\href{mailto:nleister@uni-mainz.de}{nleister@uni-mainz.de}},
Pedro Schwaller$^{b}$\footnote{\href{mailto:pedro.schwaller@uni-mainz.de}{pedro.schwaller@uni-mainz.de}}
}

\renewcommand{\thefootnote}{\arabic{footnote}}
\setcounter{footnote}{0}

{\it \small $^a$Institut d’Astrophysique de Paris (IAP), CNRS, Sorbonne Universit\'e, FR-75014, France}\\
{\it \small $^b$PRISMA$^{++}$ Cluster of Excellence $\&$ Mainz Institute for Theoretical Physics, Johannes Gutenberg University, 55099 Mainz, Germany}

\end{center}

\bigskip

\centerline{\bf Abstract}
\begin{quote}

The reheating of the universe by the evaporation of light primordial black holes (PBHs) can leave a stochastic gravitational-wave (GW) background in the early Universe. In the
monochromatic limit, their simultaneous evaporation produces an abrupt matter--to-radiation transition, triggering the so-called \textit{Poltergeist} GW signal, usually predicted to be dominant and observable. We revisit this result by including the irreducible mass spread implied by gravitational collapse in General Relativity, whose infrared tail scales as $d f_{\rm PBH}/d\ln M_{\rm PBH}\propto M_{\rm PBH}^{3.78}$. We show that this minimal width smooths reheating enough to suppress the Poltergeist background by orders of magnitude, down to the level of the scalar-induced GW signal produced during a generic early matter era, such as one driven by the decay of a heavy relic. We provide a complete decomposition of the scalar-induced spectrum into eight production channels and find that none, except the one from PBH formation, reaches either the $\Delta N_{\rm eff}$ bound or the projected sensitivity of future GW observatories. This reopens regions of ultra-light PBH parameter space previously thought to be excluded by these constraints.
\end{quote}
\newpage
\noindent\makebox[\linewidth]{\rule{\textwidth}{1pt}} 
\setcounter{tocdepth}{2}
\tableofcontents
\noindent\makebox[\linewidth]{\rule{\textwidth}{1pt}} 
\newpage

\section{Introduction and summary of results}

Primordial black holes (PBHs), first proposed in Refs.~\cite{Zeldovich:1967lct,Hawking:1971ei,Carr:1974nx,Carr:1975qj} and reviewed in Refs.~\cite{Sasaki:2018dmp,Carr:2020gox,Carr:2020xqk,Green:2020jor,Escriva:2022duf,Carr:2026hot}, provide a powerful probe of the early Universe. PBHs lighter than roughly $10^9\,{\rm g}$ evaporate before Big Bang nucleosynthesis (BBN)~\cite{Kohri:1999ex,Carr:2009jm,Acharya:2020jbv,Keith:2020jww,Boccia:2024nly,Holst:2024ubt,Wu:2025ovd,Wang:2025pum}, making them a minimal mechanism for reheating the Universe~\cite{Carr:1976zz,Lidsey:2001nj,Hidalgo:2011fj,Dalianis:2021dbs,RiajulHaque:2023cqe,Gross:2024wkl}.

Ultra-light PBHs can form from enhanced ultra-slow roll during inflation~\cite{Carr:1993aq,Ivanov:1994pa,Clesse:2015wea,Ballesteros:2017fsr,Pattison:2017mbe,Espinosa:2017sgp,Kannike:2017bxn,Inomata:2017okj,Franciolini:2018vbk,Byrnes:2018txb,Ezquiaga:2019ftu,Ballesteros:2020qam,Fumagalli:2020adf,Pi:2021dft,Kristiano:2022maq,Karam:2022nym,Franciolini:2023agm}, gauge-field production in axion inflation~\cite{Linde:2012bt,Domcke:2020zez}, spherical domains produced during inflation~\cite{Garriga:2015fdk,Deng:2016vzb,Deng:2017uwc,Deng:2020mds,Kusenko:2020pcg,Maeso:2021xvl,Escriva:2023uko,He:2023yvl,Huang:2023mwy,Kitajima:2020kig,Kasai:2023ofh,Kasai:2023qic}, domain-wall annihilation~\cite{Vachaspati:2017hjw,Ferrer:2018uiu,Gelmini:2023ngs,Gelmini:2022nim,Gouttenoire:2023gbn,Gouttenoire:2023ftk,Ferreira:2024eru,Gouttenoire:2025ofv,Lu:2024ngi,Ge:2019ihf,Ge:2023rrq,Dunsky:2024zdo}, cosmic-string collapse~\cite{Hawking:1987bn,Caldwell:1995fu,Jenkins:2020ctp,Blanco-Pillado:2021klh,Brandenberger:2021zvn}, first-order phase transitions through late-blooming~\cite{Kodama:1982sf,Hsu:1990fg,Liu:2021svg,Hashino:2021qoq,Kawana:2022olo,Lewicki:2023ioy,Gouttenoire:2023naa,Baldes:2023rqv,Gouttenoire:2023bqy,Salvio:2023ynn,Gouttenoire:2023pxh,Jinno:2023vnr,Flores:2024lng,Lewicki:2024ghw,Lewicki:2024sfw,Cai:2024nln,Ai:2024cka,Arteaga:2024vde,Banerjee:2024cwv,Hashino:2025fse,Murai:2025hse,Zou:2025sow,Franciolini:2025ztf,Wang:2025hwc,Kierkla:2025vwp,Wang:2026zvz,Ai:2026zrs}, matter compression by bubble walls~\cite{Crawford:1982yz,Gross:2021qgx,Baker:2021sno,Kawana:2021tde,Huang:2022him}, bubble collisions~\cite{Hawking:1982ga,Moss:1994iq,Ashoorioon:2020hln,Jung:2021mku}, scalar condensates~\cite{Dolgov:1992pu,Kasai:2022vhq,Cotner:2016cvr,Martin:2019nuw}, dissipative dark sectors~\cite{Chang:2018bgx,Flores:2020drq,Domenech:2023afs,Chakraborty:2022mwu,Ralegankar:2024zjd,Milligan:2025zbu}, during an early matter era~\cite{Harada:2016mhb,Harada:2017fjm,Carr:2017edp,Kokubu:2018fxy,Harada:2022xjp,Harada:2022xjp,deJong:2023gsx,Conzinu:2023fui,Saito:2024hlj,Ebrahimian:2025syf,Ye:2025wif}, or from primordial magnetic fields~\cite{Maiti:2026lvx}.

Such PBH scenarios predict signatures accessible to cosmological observations. Hawking evaporation can produce dark matter~\cite{Fujita:2014hha,Lennon:2017tqq,Morrison:2018xla,Hooper:2019gtx,Baldes:2020nuv,Masina:2020xhk,Masina:2021zpu,Bernal:2020bjf,Bernal:2020ili,Gondolo:2020uqv,Cheek:2021cfe,Cheek:2022dbx,Cheek:2022mmy,Haque:2023awl,Friedlander:2023jzw,Arcadi:2024tib,Franciolini:2026fdv,Thoss:2026slt}, the baryon asymmetry~\cite{Zeldovich:1976vw,Carr:1976zz,Toussaint:1978br,Turner:1979zj,Turner:1979bt,Barrow:1990he,Majumdar:1995yr,Upadhyay:1999vk,Dolgov:2000ht,Baumann:2007yr,Fujita:2014hha,Hook:2014mla,Hamada:2016jnq,Barman:2022pdo,Barman:2024slw,IguazJuan:2025vmd,Klipfel:2026nzx}, hot relics~\cite{Anantua:2008am,Dolgov:2011cq,Hooper:2020evu,Calza:2021czr,Cheek:2021odj,Cheek:2022dbx,Franciolini:2022htd,Shallue:2024hqe,Eby:2024mhd,Sanchis:2025awq}, hot spots~\cite{Das:2021wei,He:2022wwy,He:2024wvt,Hamaide:2023ayu,Gunn:2024xaq,Altomonte:2025hpt,Vanvlasselaer:2026vkh}, or high-energy cosmic rays~\cite{Baker:2025zxm,Baker:2025cff,Airoldi:2025opo,Klipfel:2025jql}. Dark matter and the baryon asymmetry can also be produced by superradiance~\cite{Ferraz:2020zgi,Bernal:2022oha,Calza:2023rjt,Ghoshal:2023fno,Manno:2025dhw,Jia:2025vqn,Neves:2025kxp}. Another well-studied probe of ultra-light PBHs is scalar-induced gravitational waves (SIGWs), where tensor modes $h$ are sourced at second order in the Newtonian potential, $h\propto \Phi^2$~\cite{Matarrese:1993zf,Matarrese:1997ay,Ananda:2006af,Baumann:2007yr,Espinosa:2018eve,Kohri:2018awv,Inomata:2019zqy,Inomata:2019ivs,Pearce:2023kxp,Pearce:2025ywc}. The first contribution is the SIGW signal produced at PBH formation~\cite{Saito:2008jc}, which has received considerable attention in the context of the PTA signal~\cite{Vaskonen:2020lbd,Kohri:2020qqd,Inomata:2020xad,DeLuca:2020agl,Domenech:2020ers,Franciolini:2023pbf,Gouttenoire:2025jxe}, but which for $M_{\rm PBH}<10^9~\rm g$ lies above $\rm kHz$ frequencies, cf.~Eq.~\eqref{eq:f0_form}, near the upper edge of observable frequencies. The second contribution comes from SIGWs produced at PBH evaporation, which have garnered significant interest~\cite{Inomata:2025wiv,Inomata:2020lmk,Papanikolaou:2020qtd,Domenech:2020ssp,Domenech:2021wkk,Bhaumik:2022pil,Gross:2024wkl,Gross:2025hia,Balaji:2024hpu,Domenech:2024cjn,Domenech:2025ffb,Papanikolaou:2025ddc}, see reviews~\cite{Domenech:2023fuz,Domenech:2024kmh}. The dominant signal is the Poltergeist mechanism~\cite{Inomata:2020lmk,Inomata:2025wiv}, sourced at reheating by the rapid oscillation of Newtonian potential $\Phi$ that have been enhanced during the preceding PBH-dominated epoch. The predicted SIGW signal has been shown to lie in frequency bands accessible to next-generation gravitational wave observatories, see e.g.~\cite{Bhaumik:2022pil}. The Poltergeist signal has also been studied in scenarios beyond PBH reheating~\cite{Inomata:2019ivs,Inomata:2025wiv,Croon:2019rqu,White:2021hwi,Kawasaki:2023rfx,Harigaya:2023mhl,Yu:2025jgx}. 

A central but often overlooked input is the PBH mass distribution. Broad mass functions are already known to reduce the amplitude of SIGW signals~\cite{Inomata:2020lmk}. In this work, we emphasize that an extended mass distribution is not an optional model choice but a generic consequence of gravitational collapse in general relativity. Near the collapse threshold, the PBH mass at formation $M_{\rm \mathsmaller{PBH}}$ is related to the initial density contrast $\delta$ through the critical-collapse scaling discovered by \textit{Choptuik}~\cite{Choptuik:1992jv} and further developed in Refs.~\cite{Evans:1994pj,Maison:1995cc,Gundlach:2007gc}:
\begin{equation} {\it Choptuik's ~law:}\qquad \qquad\label{eq:choptuik_main_intro} M_{\rm \mathsmaller{PBH}} \propto (\delta-\delta_c)^{\gamma_{\rm M}} , \end{equation} 
where $\delta_c$ is a threshold and $\gamma_M\simeq 0.36$ is a critical exponent. Inserting this scaling into the standard statistical prescriptions for the PBH abundance, either Press--Schechter formalism~\cite{Press:1973iz} or peak theory~\cite{Bond:1990iw}, see Fig.~\ref{fig:peakvsPS}, yields the PBH mass distribution~\cite{Niemeyer:1997mt,Yokoyama:1998xd,Carr:2018rid,Karam:2022nym} \begin{equation} \label{eq:pbhdist} \frac{df_{\rm \mathsmaller{PBH}}}{d\log M_{\rm \mathsmaller{PBH}}} \propto M_{\rm \mathsmaller{PBH}}^{1 + \frac{1}{\gamma_{\rm M}}} \exp\left[-c_1\left(\frac{M_{\rm \mathsmaller{PBH}}}{\langle M_{\rm \mathsmaller{PBH}} \rangle}\right)^{c_2}\right], \end{equation} where $c_1$, $c_2$ parametrize the distribution shape~\cite{Karam:2022nym}. The infrared slope $df_{\rm \mathsmaller{PBH}}/d\log M_{\rm \mathsmaller{PBH}}\propto M_{\rm \mathsmaller{PBH}}^{3.78}$ is universal, an irreducible prediction of general relativity that cannot be removed by tuning the source.

\begin{figure}
\vspace{-1cm}
    \centering
    \includegraphics[width=0.4\linewidth]{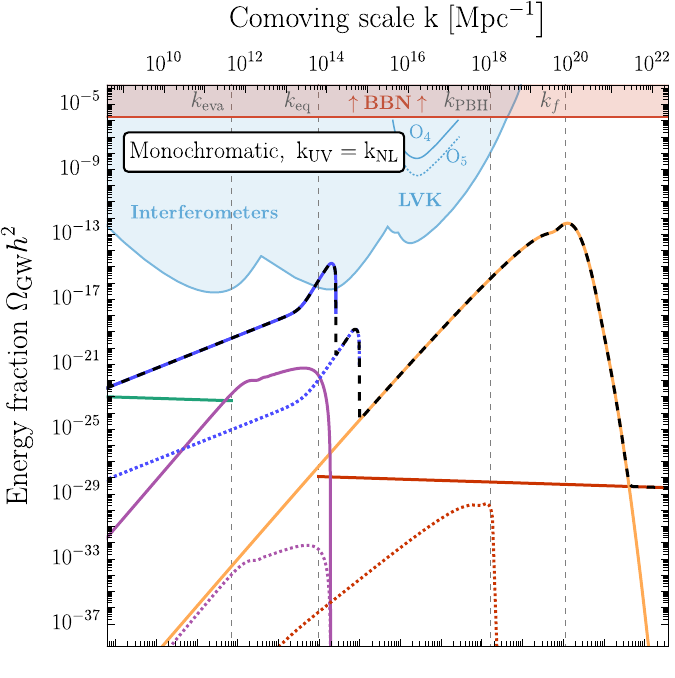}\hspace{3mm}
    \includegraphics[width=0.4\linewidth]{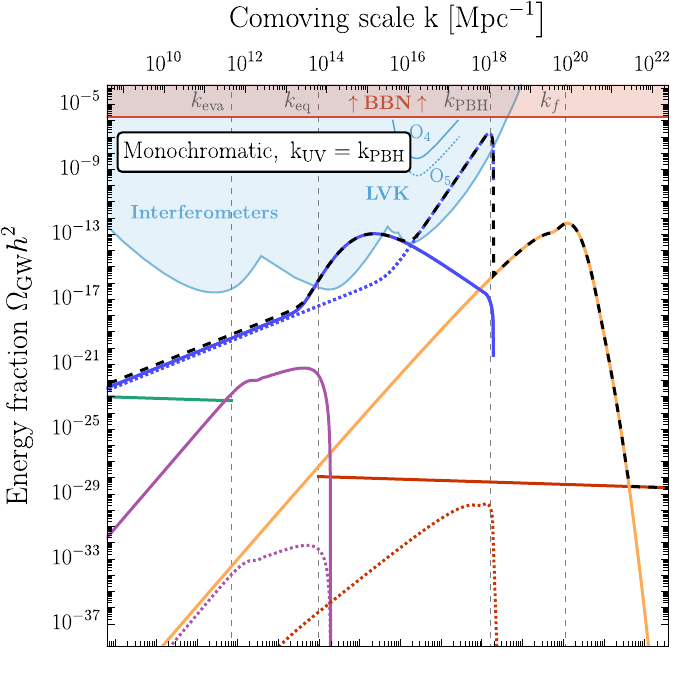}\\[-0.8em]
    \includegraphics[width=0.4\linewidth]{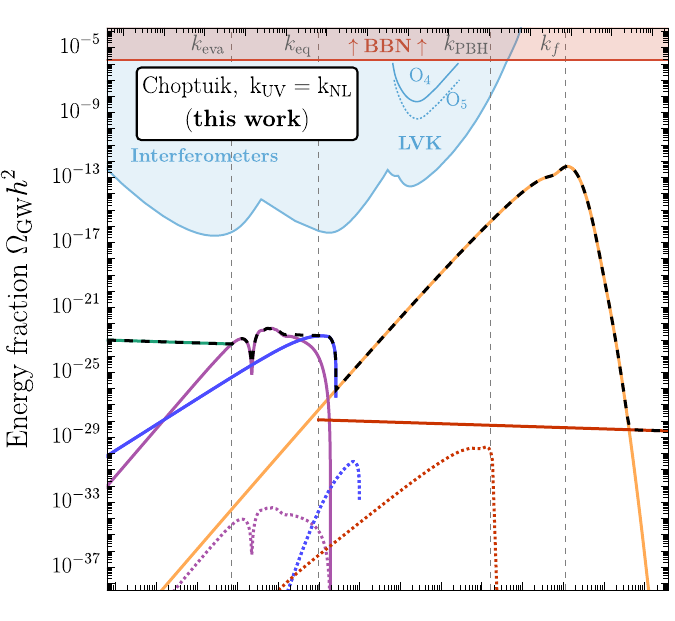}\hspace{3mm}
    \includegraphics[width=0.4\linewidth]{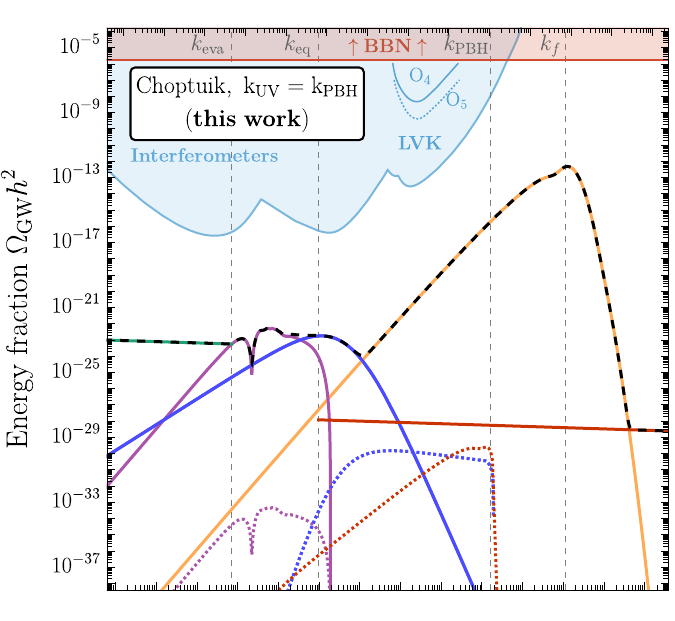}\\[-0.8em]
    \includegraphics[width=0.4\linewidth]{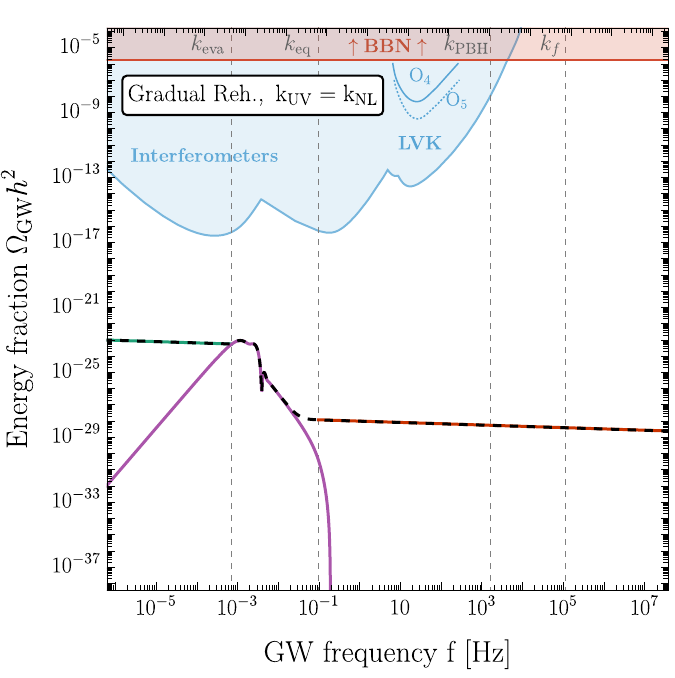}\hspace{3mm}
    \includegraphics[width=0.4\linewidth]{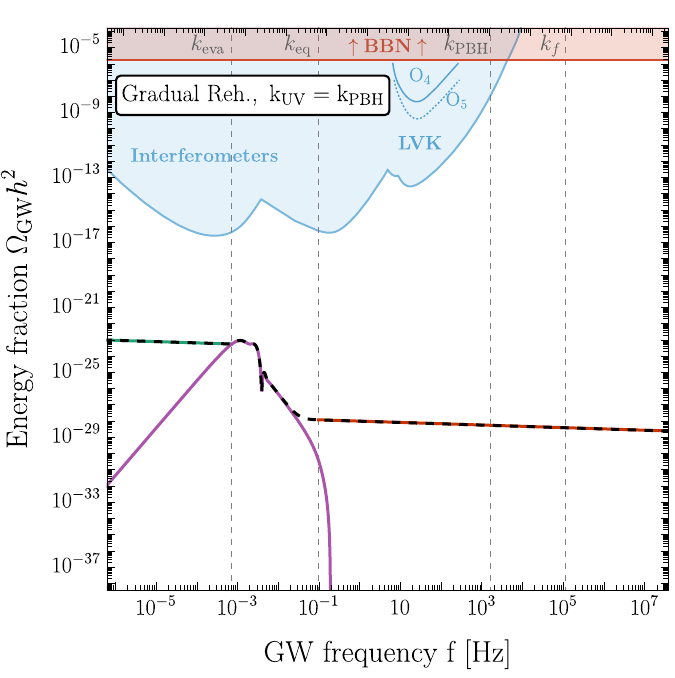}
    \includegraphics[width=0.8\linewidth]{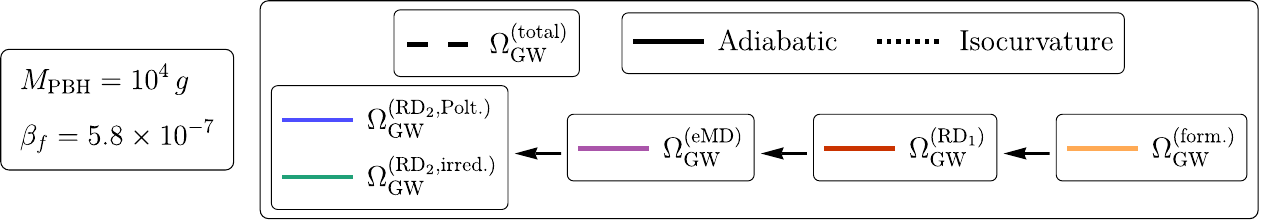}
    \caption{Summary of all present-day SIGW components at the benchmark $(M_{\rm \mathsmaller{PBH}},\beta_f)=(10^4\,\mathrm{g},\,6\times 10^{-7})$, i.e.\ $\langle\eta_{\rm eva}\rangle/\eta_{\rm eq}\simeq 225$ (matching Ref.~\cite{Inomata:2020lmk}), compared with the combined sensitivity envelope of planned GW interferometers (light blue, see Fig.~\ref{fig:GW_reach_longversion}). \textbf{{\it Middle row:} our new results for PBHs with the Choptuik mass distribution in Eq.~\eqref{eq:pbhdist}.} \textit{Top:} monochromatic PBHs. \textit{Bottom:} gradual reheating: PBHs are replaced by unstable heavy particles with constant decay rate. \textit{Left:} non-linear cutoff $k_{\rm \mathsmaller{UV}}=k_{\rm \mathsmaller{NL}}$. \textit{Right:} fluid cutoff $k_{\rm \mathsmaller{UV}}=k_{\rm \mathsmaller{PBH}}$, see Tab.~\ref{tab:summary_cut-off}. The Choptuik smoothing brings the Poltergeist signal, $\Omega_{\rm GW}^{\rm (RD_2, Polt.)}$, down to a level comparable with the eMD contribution, $\Omega_{\rm GW}^{\rm (eMD)}$. The formation component, $\Omega_{\rm GW}^{\rm (form.)}$, is evaluated at curvature-peak width $\Delta=0.5$.}
    \label{fig:GW_all_sources}
\end{figure}

\begin{figure}
    \centering
    \includegraphics[width=0.48\linewidth]{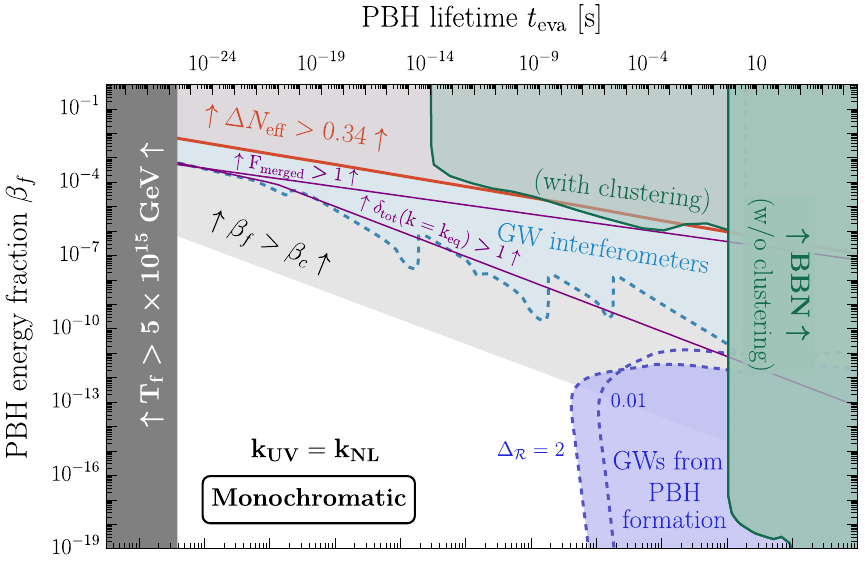}
    \includegraphics[width=0.48\linewidth]{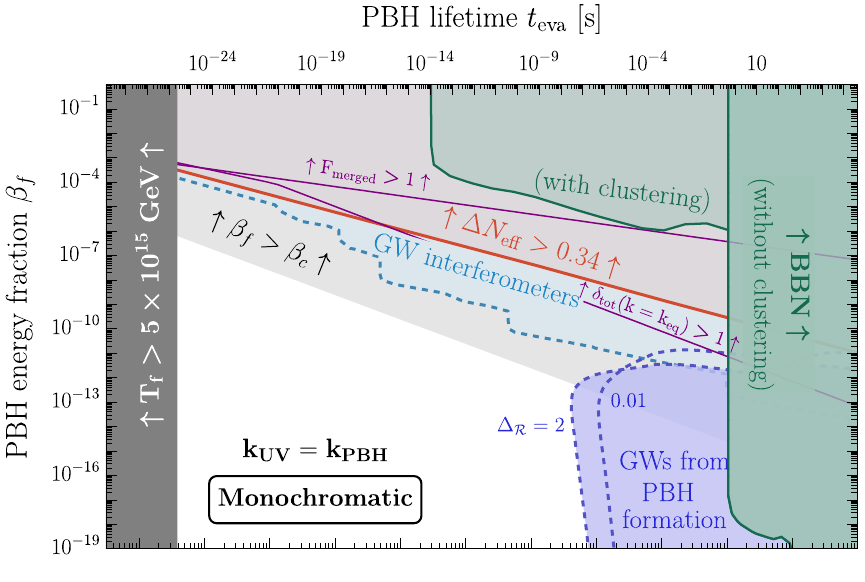}
    \includegraphics[width=0.48\linewidth]{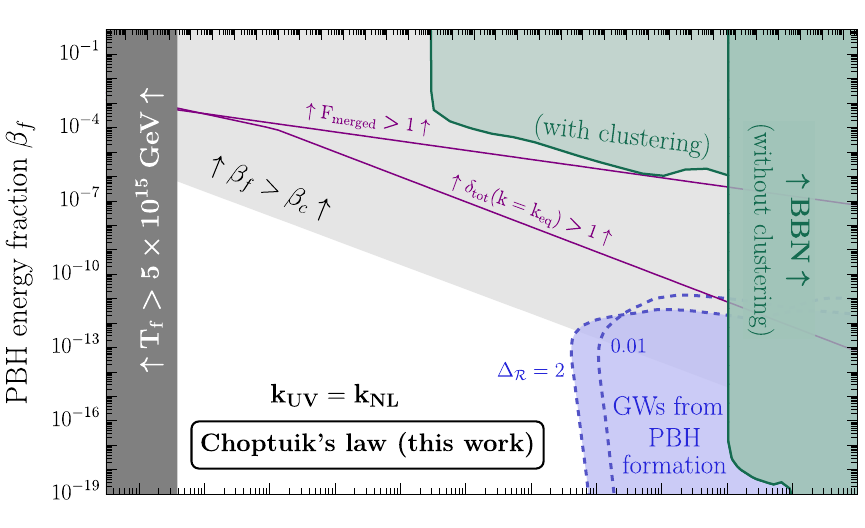}
    \includegraphics[width=0.48\linewidth]{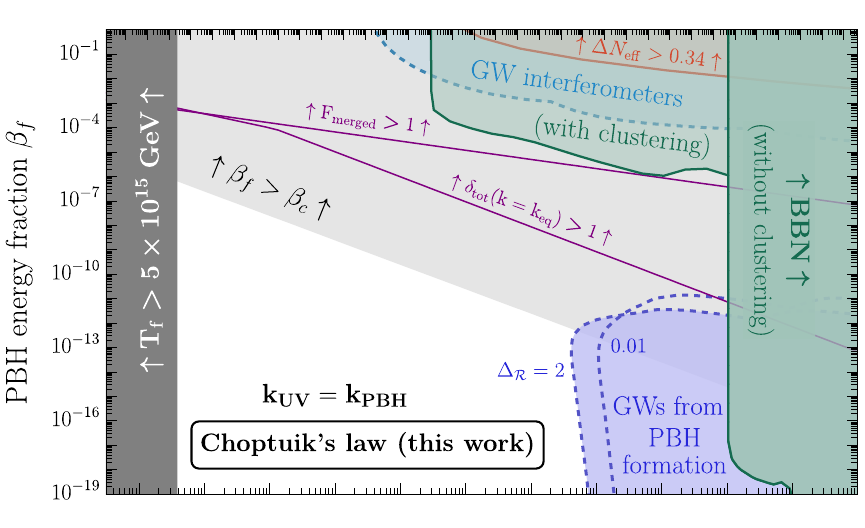}
    \includegraphics[width=0.48\linewidth]{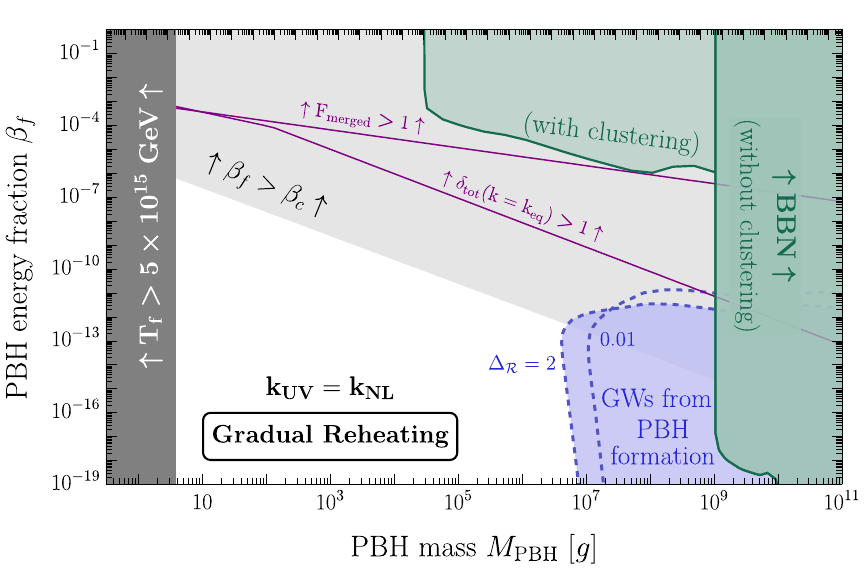}
    \includegraphics[width=0.48\linewidth]{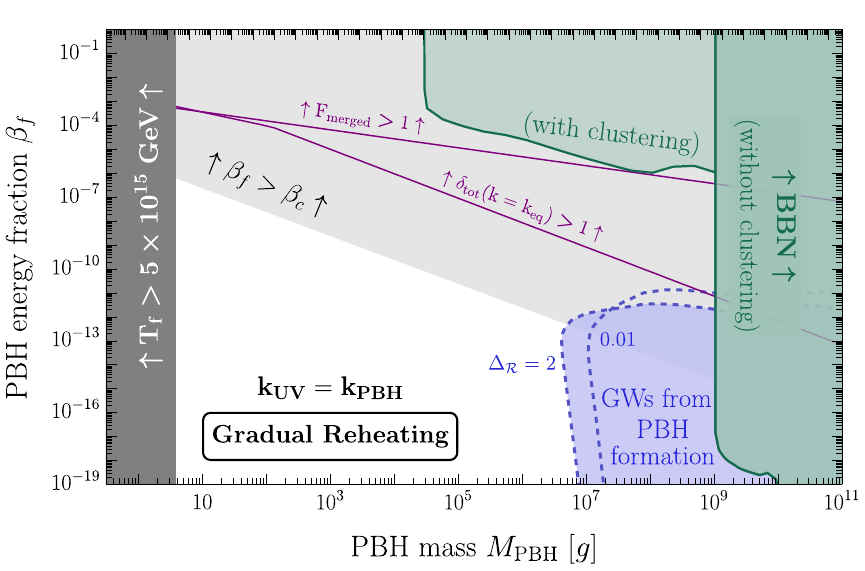}
\caption{%
\label{fig:beta_vs_MPBH_standard}
Exclusion and sensitivity contours in the $(M_{\rm \mathsmaller{PBH}}, \beta_f)$ plane of the ultra-light PBH parameter space. \textbf{Top:} idealised monochromatic mass function. \textbf{Middle:} PBH mass function with the universal IR tail predicted by Choptuik's critical scaling law.  In this case, the smoothed reheating transition suppresses the Poltergeist contribution $\Omega_{\rm GW}^{\rm (RD_2,Polt.)}$, weakens the experiments reach, and reopens regions previously thought to be excluded.  Sensitivity curves correspond to the most optimistic combination of planned GW interferometers (SKA, Theia, $\mu$Ares, LISA, AEDGE, BBO, DECIGO, ET, and CE) assuming total astrophysical-foreground subtraction (black line in Fig.~\ref{fig:GW_reach_longversion}). \textbf{Bottom:} Gradual reheating which assumes that PBHs decay with a constant decay rate as if they were unstable heavy particles.  The similarity between the middle and bottom panels shows that Choptuik broadening makes PBH domination, for $k_{\rm UV}=k_{\rm NL}$, qualitatively equivalent to a conventional eMD in terms of SIGW production. The green region shows the BBN constraint including PBH clustering, derived in Ref.~\cite{Holst:2024ubt}, which complements the conventional bound of Ref.~\cite{Carr:2020gox} where clustering is neglected.} 
\end{figure}

\paragraph{Key results.} In this work, we include this universal infrared tail in the full SIGW computation. To remain conservative, we adopt a sharp ultraviolet cutoff, corresponding to $c_2\to+\infty$ (see black dotted line in Fig.~\ref{fig:psi_PBH}), which maximizes the resulting signal. Our main findings are:
\begin{enumerate}\setlength\itemsep{2pt}
\item \textbf{Steepened suppression factor.} The subhorizon suppression factor $\mathcal{S}_\Phi$ of the Newtonian potential $\Phi$ across the PBH-to-radiation transition steepens from the monochromatic scaling $\mathcal{S}_\Phi\propto k^{-1/3}$ to $\mathcal{S}_\Phi\propto k^{-4/3}$, see Figs.~\ref{fig:pbhepoch_longpaper}
and \ref{fig:alpha_vs_gamma_M}. This reduces the Poltergeist signal by orders of magnitude compared to the monochromatic prediction, as shown by the top and middle panels of Fig.~\ref{fig:GW_all_sources}.
\item \textbf{PBH reheating behaves as a generic matter phase.} Once the critical-collapse mass spread is included, the SIGW signal from PBH reheating becomes comparable to the signal generated during a generic early matter-dominated phase~\cite{Inomata:2019zqy}, such as one driven by the decay of a heavy relic. The large parametric enhancement of the monochromatic Poltergeist signal is therefore lost, with the separate SIGW contribution from PBH formation being the only exception, see the middle and bottom panels of Fig.~\ref{fig:GW_all_sources}.
\item \textbf{BBN and detector reach are relaxed.} The suppression of the Poltergeist signal relaxes both the BBN bound on $\Delta N_{\rm eff}$ and the projected reach of future gravitational-wave observatories. This reopens regions of the $(M_{\rm \mathsmaller{PBH}},\beta_f)$ plane previously thought to be excluded or detectable by these probes, see Fig.~\ref{fig:beta_vs_MPBH_standard}. Existing studies quoting larger Poltergeist amplitudes~\cite{Inomata:2020lmk,Papanikolaou:2020qtd,Domenech:2020ssp,Domenech:2021wkk,Bhaumik:2022pil,Gross:2024wkl,Gross:2025hia,Balaji:2024hpu,Domenech:2024cjn,Domenech:2025ffb,Inomata:2025wiv,Papanikolaou:2025ddc,Domenech:2023fuz,Domenech:2024kmh} effectively rely on idealized mass distributions that omit the universal infrared tail imposed by critical collapse.
\item \textbf{Eight SIGW channels.} We complete the analysis by computing all scalar-induced contributions to $\Omega_{\rm GW}$, summarized in Tab.~\ref{tab:overview}. These contributions can be comparable to the reduced Poltergeist component and can even dominate in parts of parameter space, although they remain below the projected sensitivity of future experiments. Complementary high-frequency channels from PBH mergers and direct Hawking graviton emission are summarized here, with details left to the companion paper~\cite{YannNicoPedro}.
\end{enumerate}

\begin{table}[h]
\centering
\small
\renewcommand{\arraystretch}{1.4}
\setlength{\tabcolsep}{4pt}
\begin{tabular}{
>{\centering\arraybackslash}m{2.4cm}
>{\centering\arraybackslash}m{2.9cm}
>{\centering\arraybackslash}m{3.1cm}
>{\centering\arraybackslash}m{2.9cm}
>{\centering\arraybackslash}m{3.1cm}
}
 &\makecell{\bf{PBH}\\ \bf{formation}\\($\rm form$)\\~}
 &\makecell{\bf{Before}\\ \bf{PBH-domination}\\($\rm RD_1$)\\~}
 &\makecell{\bf{PBH domination}\\(eMD)}
 &\makecell{\bf{Reheating}\\($\rm RD_2$)}\\[2mm]
&\includegraphics[width=\linewidth]{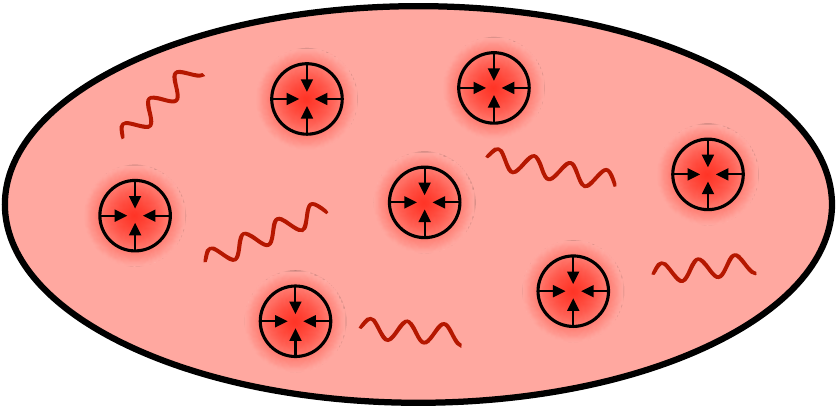}
&\includegraphics[width=\linewidth]{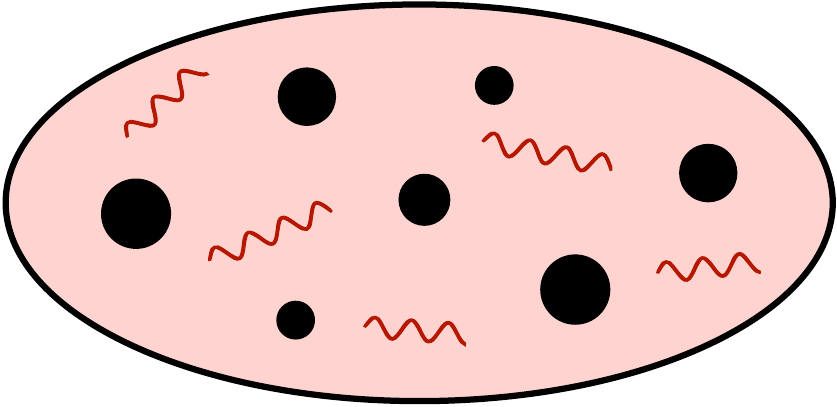}
&\includegraphics[width=\linewidth]{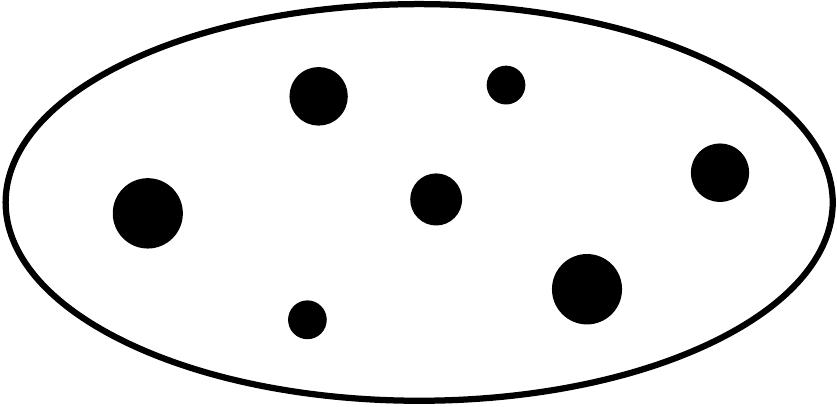}
&\includegraphics[width=\linewidth]{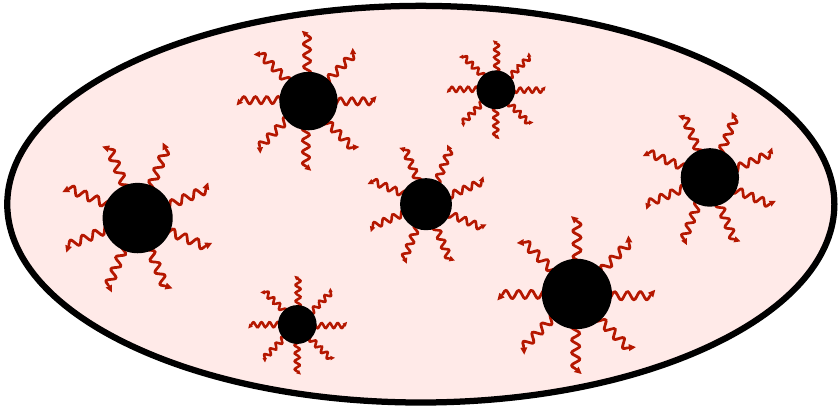} \\[2mm]\hline
\makecell{\rule{0pt}{3.2ex}\bf{Newtonian}\\\bf{potential} \\($\Phi$)\\[1mm]}
& \makecell{$\mathcal{P}_{\mathcal{R}}$\\(Sec.~\ref{sec:transfer_function})}
& \makecell{$T_\Phi^{\rm (RD_1)}$\\(Sec.~\ref{chap:eRD})}
& \makecell{$T_\Phi^{\rm (eMD)}$\\(Sec.~\ref{sec:transfer_function})}
& \multirow{2}{*}[-4.5mm]{\makecell{$\mathcal{S}_\Phi$\\(Sec.~\ref{sec:pert_equation_synchronous})}}  \\
\makecell{\bf{PBH}\\\bf{isocurvature} \\($S$)\\[1mm]}
& \makecell{$\mathcal{P}_{S}$\\(Sec.~\ref{sec:transfer_function})}
& \makecell{$T_S^{\rm (RD_1)}$\\(Sec.~\ref{chap:eRD})}
& \makecell{$T_S^{\rm (eMD)}$\\(Sec.~\ref{sec:transfer_function})}
&  \\[2mm]\hline
\multirow{3}{*}{\makecell{~\\\bf{Tensor $[\Phi]$}\\~}}
& \multirow{3}{*}{\makecell{$\Omega_{\rm GW}^{\rm (form)}[\Phi]$\\(Sec.~\ref{chap:SIGW_formation})}}
& \multirow{3}{*}{\makecell{$\Omega_{\rm GW}^{\rm (RD_1,irred.)}[\Phi]$\\
(Sec.~\ref{chap:eRD})}}
& \multirow{3}{*}{\makecell{$\Omega_{\rm GW}^{\rm (eMD)}[\Phi]$\\(Sec.~\ref{chap:eMD})}}
& \makecell{\rule{0pt}{4ex}$\Omega_{\rm GW}^{\rm (RD_2,irred.)}[\Phi]$\\(Sec.~\ref{chap:eRD})}\\
&
&
&
&
\makecell{\rule{0pt}{3.2ex}$\Omega_{\rm GW}^{\rm (RD_2,Polt.)}[\Phi]$\\(Sec.~\ref{chap:poltergeist})}\\
\makecell{~\\\bf{Tensor $[S]$}\\~}
&
& \makecell{$\Omega_{\rm GW}^{\rm (RD_1,univ.)}[S]$\\(Sec.~\ref{chap:eRD})}
& \makecell{$\Omega_{\rm GW}^{\rm (eMD)}[S]$\\(Sec.~\ref{chap:eMD})}
& \makecell{\rule{0pt}{3.2ex}$\Omega_{\rm GW}^{\rm (RD_2,Polt.)}[S]$\\(Sec.~\ref{chap:poltergeist})} \\\hline
\end{tabular}
\caption{
Roadmap of the scalar perturbations (upper block) and induced GW components (lower block) computed in this work across the four phases of an ultra-light PBH cosmology: PBH formation, $\rm RD_1$, $\rm eMD$, and $\rm RD_2$, for adiabatic ($\Phi$) and isocurvature ($S$) seeds. Sections of derivation are given in parentheses. The same GW components are also listed in Tab.~\ref{tab:GW_sources}, organised by physical origin.}
\label{tab:overview}
\end{table}

\newpage 
\paragraph{Benchmark spectra and parameter scan.} 
In Fig.~\ref{fig:GW_all_sources}, we show the present-day SIGW spectrum for the benchmark $M_{\rm \mathsmaller{PBH}}=10^4\,{\rm g}$ and $\beta_f=6\times10^{-7}$, corresponding to $\langle\eta_{\rm eva}\rangle/\eta_{\rm eq}\simeq225$, or $N_{\rm MD}\simeq9.1$, to compare directly with Ref.~\cite{Inomata:2020lmk}. The plot includes the eight SIGW contributions listed in Tab.~\ref{tab:overview}. The poltergeist contribution, $\Omega_{\rm GW}^{\rm (RD_2,Polt.)}$, is the most sensitive to the rate of the matter-to-radiation transition, see also Fig.~\ref{fig:GWSig_UV}. The contribution sourced during PBH domination, $\Omega_{\rm GW}^{\rm (eMD)}$, is only weakly sensitive to this rate, see Fig.~\ref{fig:GWSig_RD_comp_MD}, while the remaining contributions are essentially unaffected.

We compare two ultraviolet prescriptions. The fiducial choice,
$k_{\rm \mathsmaller{UV}}=k_{\rm \mathsmaller{NL}}$, cuts the source when the
PBH density field becomes non-linear and gives the controlled perturbative
prediction. The alternative,
$k_{\rm \mathsmaller{UV}}=k_{\rm \mathsmaller{PBH}}$, extends the integral to
the mean PBH separation and should be interpreted as a conservative upper
envelope, see Sec.~\ref{sec:UV_cut_off} for more details. Once the universal IR tail in the PBH mass distribution is included, the Poltergeist is
strongly suppressed and becomes comparable to the other SIGW channels rather
than parametrically dominant. 

The corresponding phenomenology in the
$(M_{\rm \mathsmaller{PBH}},\beta_{\rm f})$ plane is shown in
Fig.~\ref{fig:beta_vs_MPBH_standard}. We restrict to the basic PBH mass window
set by the lower bound on the Horizon mass from the maximal inflation scale (see Sec.~\ref{sec:lightest_PBH}) and
by the requirement of evaporation before BBN:
$1\,{\rm g}\lesssim M_{\rm \mathsmaller{PBH}}\lesssim10^9\,{\rm g}$.\footnote{
In the pre-BBN evaporation window ($M_{\rm \mathsmaller{PBH}}\lesssim10^9\rm g$), the PBH-induced isocurvature mode peaks at
$k_{\rm PBH}\gtrsim10^{14}\,{\rm Mpc}^{-1}$ (see Eq.~\eqref{eq:kPBH_D}), well above the proton diffusion scale
$k_p\sim10^{10}\,{\rm Mpc}^{-1}$. Baryon-isocurvature perturbations sourced by
PBH evaporation are therefore washed out by proton diffusion and evade the
recently derived bounds on small-scale baryon inhomogeneities
\cite{Bagherian:2025puf}.
} We compute the
spectra using the universal Choptuik infrared scaling with a sharp high mass
cutoff shown by the black dotted line in Fig.~\ref{fig:psi_PBH}. This maximises
the signal: a smoother high mass tail would broaden evaporation further and
additionally suppress the SIGW amplitude. Our spectra are therefore conservative upper
envelopes. For projected GW sensitivities we use the most optimistic combined
envelope of future probes shown in
Fig.~\ref{fig:GW_reach_longversion}~\cite{LISACosmologyWorkingGroup:2022jok,NANOGrav:2023hvm,Punturo:2010zz,Reitze:2019iox,Janssen:2014dka,Crowder:2005nr,AEDGE:2019nxb,Badurina:2019hst,KAGRA:2021kbb,LIGOScientific:2014pky,Sesana:2019vho,Garcia-Bellido:2021zgu,Blas:2021mqw}. 
Only narrow, highly tuned regions remain potentially observable. In this work, we only focus on the SIGWs components. In \cite{YannNicoPedro}, we show that the high-frequency components, from Hawking gravitons and binary mergers, summarised in Sec.~\ref{sec:HF_GW}, are subdominant, see also~\cite{Cheek:2022mmy}. 

Fig.~\ref{fig:beta_vs_MPBH_standard} also displays the main consistency
limits. The purple contour $k_{\rm \mathsmaller{NL}}<k_{\rm eq}$ marks where
perturbations which enter the horizon at the beginning of the PBH domination becomes non-linear at the time of evaporation.
This matters because both the Poltergeist signal is peaked at $k_{\rm eq}$. In principle, beyond this contour the linear SIGW prediction carries large theoretical uncertainty.

The remaining purple contour, $F_{\rm merged}>1$, delimits the region in which PBH mergers become important. Following~\cite{Peters:1964zz,Ding:2023smy}, the binary inspiral time is
\begin{equation}
t_{\mathrm{mer}}
=
\frac{3}{85}
\frac{a_x^4}{2 M_{\rm \mathsmaller{PBH}}^3}
(1-e^2)^{7/2}
\frac{c^5}{G^3},
\qquad
a_x
=
\alpha
\left(
\frac{3 M_{\rm\mathsmaller{PBH}}}
{4 \pi \rho_{\rm \mathsmaller{PBH}}\left(t_{\rm eq}\right)}
\right)^{1/3}
X^{4/3},
\qquad
X\equiv \frac{x^3}{\bar{x}^3},
\end{equation}
where $e$ is the initial eccentricity, $a_x$ the semi-major axis, and $x$ and $\bar{x}$ the initial comoving separation of the pair and the mean PBH distance at matter-radiation equality. The dimensionless coefficient $\alpha \simeq 0.1$, fixed by the numerical decoupling of a PBH pair from the Hubble flow~\cite{Franciolini:2021xbq}, relates $a_x$ to $x$, $\bar{x}$, and $M_{\rm PBH}$ through the relation above. The contour $F_{\rm merged}=1$ is then set by $t_{\rm mer}=t_{\rm eva}$ at the longest inspiral time, reached for $e=0$ and $X=1$. Beyond it, repeated mergers generate heavier PBHs and the mass function evolves in time~\cite{Holst:2024ubt}, broadening it further and providing an additional suppression of the SIGW signal. Our prediction should therefore be interpreted as a conservative upper envelope above that purple line.

Successive mergers also affect the BBN bound. The solid green contour (without clustering) in
Fig.~\ref{fig:beta_vs_MPBH_standard} shows the standard bound neglecting mergers, while the contour (with clustering) includes heavier merger remnants whose delayed
evaporation can occur close to, or after, the onset of BBN~\cite{Holst:2024ubt}.

\clearpage

\begin{table}[t]
   \centering
    \renewcommand{\arraystretch}{1.15}
    \setlength{\tabcolsep}{6pt}
    \begin{tabular}{|c|l|c|}
        \hline
        \textbf{Symbol} & \textbf{Meaning} & \textbf{Reference Eqn.} \\
        \hline\hline

        $x^\mu=(\eta,\vec{x})$ & space--time coordinates & \eqref{eq:FLRW} \\
        \hline
        $a(\eta)$ & scale factor & \eqref{eq:FLRW} \\
        \hline
        $t,\eta$ & cosmic and conformal time & \eqref{eq:cosmic_time} \\
        \hline
        $H,\mathcal H$ & cosmic and conformal Hubble parameters & \eqref{eq:Hubble_prime} \\
        \hline
        $\omega$ & equation-of-state parameter & \eqref{eq:Hubble_prime} \\
        \hline
        $\overline{\rho},\overline p$ & background energy and pressure density & \eqref{eq:Hubble_prime} \\
        \hline

        $\rho_{\rm \mathsmaller{PBH}},\rho_{\rm r}$ & PBH and radiation energy densities & \eqref{eq:rhocomoving} \\
        \hline
        $\Gamma$ & PBH evaporation rate in conformal time & \eqref{eq:Gamma_def} \\
        \hline
        $M_{{\rm PBH},f}$ & PBH mass at formation, later called  $M_{{\rm PBH}}$  & \eqref{eq:PBH_mass_decay} \\
        \hline
        $t_{\rm eva},\,\eta_{\rm eva}$ & cosmic and conformal PBH lifetime & \eqref{eq:t_eva_def}, \eqref{eq:Gamma_lifetime} \\
        \hline
        $\beta_f,\beta_c$ & PBH energy fraction at formation and its critical value& \eqref{eq:beta_f_c} \\
        \hline
        $N_{\rm MD}$ & duration of the PBH-induced matter era in expansion e-folds & \eqref{eq:NMD_main} \\
        \hline
        $M_{\rm H},\,\gamma_{\rm H}$ & horizon mass and PBH-to-horizon mass ratio & \eqref{eq:M_f_def}, \eqref{eq:def_gammaH}, \eqref{eq:def_gammaH_rm} \\
        \hline
        $k_f,\,a_f$ & comoving Hubble scale and scale factor at PBH formation & \eqref{eq:M_f_def} \\
        \hline
        $k_{\rm eq}$ & comoving Hubble scale at PBH--radiation equality & \eqref{eq:keq_main} \\
        \hline
        $k_{\rm \mathsmaller{PBH}}$ & comoving inverse mean PBH separation (fluid UV cut-off) & \eqref{eq:kUV_def_main} \\
        \hline
        $k_{\rm eva}$ & comoving Hubble scale at evaporation & \eqref{eq:keva_def_main}, \eqref{eq:Phi_split_RD2} \\
        \hline
        $\mathcal P_{\mathcal R}$ & curvature power spectrum & \eqref{eq:log_normal} \\
        \hline
        $\mathcal A_{\mathcal R},\,\Delta,\,k_f$ & amplitude, width and peak scale of $\mathcal P_{\mathcal R}$ & \eqref{eq:log_normal} \\
        \hline
        $M_f$ & horizon mass associated with the peak scale $k_f$ & \eqref{eq:Mstar_def_maintext} \\
        \hline
        $\delta_m,\delta_c$ & smoothed density contrast and collapse threshold & \eqref{eq:choptuik_main} \\
        \hline
        $\gamma_{\rm M},\,\mathcal K$ & critical exponent and normalization in Choptuik scaling & \eqref{eq:choptuik_main} \\
        \hline
        $r_m,\,\kappa$ & compaction scale and geometric conversion factor $k_f r_m$ & \eqref{eq:MH_rm_def_maintext}, \eqref{eq:kappa_def_maintext} \\
        \hline
        $\Phi,\Psi$ & scalar potentials in Newtonian gauge & \eqref{eq:metric_main} \\
        \hline
        $h_{ij},\,h_\lambda$ & tensor perturbation and its helicity modes & \eqref{eq:metric_main}, \eqref{eq:h_lambda_k} \\
        \hline
        $S$ & PBH isocurvature perturbation & \eqref{eq:Transfer_fct_def} \\
        \hline
        $T_\Phi,T_S$ & scalar transfer functions & \eqref{eq:Transfer_fct_def}, \eqref{eq:transfer_adia_iso_main} \\
        \hline
        $\mathcal P_\Phi,\mathcal P_S$ & power spectra of the adiabatic and isocurvature seeds & \eqref{eq:Def_Power_Spectra_Phi_S} \\
        \hline

        $k$ & comoving wavenumber & \eqref{eq:Fourier_transform_Phi} \\
        \hline
        $x=k\eta$ & dimensionless time variable & \eqref{eq:tensor_from_greens_function} \\
        \hline
        $G_h$ & tensor Green's function & \eqref{eq:green_function_diff_eq} \\
        \hline
        $I_X$ & time-integrated SIGW kernel & \eqref{eq:kernel_standard} \\
        \hline
        $\mathcal P_h$ & tensor power spectrum & \eqref{eq:P_h_def} \\
        \hline
        $\Omega_{\rm GW}$ & GW energy density per logarithmic interval & \eqref{eq:SIGW_basic_formula} \\
        \hline
        $\mathcal D(T_\star,T_0)$ & redshift factor from production to today & \eqref{eq:_Omega_GW_0_DW_app}, \eqref{eq:mathcal_D_def} \\
        \hline

        $k_{\rm \mathsmaller{UV}}$ & UV cutoff for the scalar source & \eqref{eq:Phi_Polt_main} \\
        \hline
        $k_{\rm \mathsmaller{NL}}$ & nonlinear cutoff scale & \eqref{eq:kNL_adi}, \eqref{eq:kNL_iso} \\
        \hline
        $k_D$ & Silk damping scale of the radiation fluid & \eqref{eq:kD_def} \\
        \hline
    \end{tabular}
    \caption{Symbols used throughout this work.}
    \label{tab:notation}
\end{table}

\paragraph{Outline.} The remainder of this work systematically analyses GW signatures from realistic PBH populations in the early Universe. Sec.~\ref{sec:PBHgas} presents the background dynamics of PBH-dominated scenarios. Sec.~\ref{sec:PBH_mass_distrib} introduces the PBH gas formed from super-horizon collapse and derives the resulting extended mass distribution using Press--Schechter statistics and peak theory. Sec.~\ref{sec:perturbations} derives the transfer functions and suppression factors of the cosmological perturbations that source the GWs. Sec.~\ref{chap:basics_SIGW_main} defines the framework for SIGW production and computes the signal from PBH evaporation in Sec.~\ref{chap:poltergeist}. Sec.~\ref{sec:UV_cut_off} addresses the sensitivity of the SIGW signal to the UV cut-off of the scalar power spectrum. We then turn to complementary sources of gravitational waves, namely SIGWs sourced during the early matter-dominated phase (Sec.~\ref{chap:eMD}), {during} the early radiation era (Sec.~\ref{chap:eRD}), {from} direct enhancement of the curvature spectrum at PBH formation (Sec.~\ref{chap:SIGW_formation}), and high-frequency GWs from PBH mergers and Hawking emission (Sec.~\ref{sec:HF_GW}). Finally, we conclude. A list of the symbols used throughout this work is provided in  Tab.~\ref{tab:notation}.

\clearpage

\section{Background cosmology of evaporating PBHs}
We briefly review the background cosmology relevant for primordial black
holes (PBHs).
On sufficiently large scales, the Universe is well described by a spatially flat, homogeneous, and isotropic Friedmann--Lemaître--Robertson--Walker (FLRW) spacetime. In conformal time \(\eta\), the line element takes the form
\begin{equation}
\label{eq:FLRW}
\mathrm{d}s^2
= a^2(\eta)\left(-\mathrm{d}\eta^2 + \delta_{ij}\,\mathrm{d}x^i \mathrm{d}x^j\right),
\end{equation}
where \(a(\eta)\) is the scale factor. The conformal time \(\eta\) is related to the cosmic time \(t\) by
\begin{equation}
\label{eq:cosmic_time}
\mathrm{d}\eta \equiv \frac{\mathrm{d}t}{a(t)}.
\end{equation}
Derivatives with respect to $\eta$ are denoted by a prime.
The expansion rate is characterized by the Hubble parameter $H$ and the
conformal Hubble parameter $\mathcal{H}$, defined respectively by
\begin{equation}
H \equiv \frac{\dot a}{a},
\qquad
\mathcal{H} \equiv \frac{a'}{a} = aH.
\end{equation}
The comoving Hubble horizon is then defined as the inverse conformal
Hubble parameter, $\mathcal{H}^{-1}$, which sets the characteristic scale
of causal connectivity in comoving coordinates.
For a homogeneous and isotropic Universe filled with a perfect fluid of
equation-of-state parameter $\omega\equiv \overline{p}/\overline{\rho}$,
the Einstein equations reduce to the standard Friedmann and continuity
relations,
\begin{align}
\label{eq:Hubble_prime}
\mathcal{H}^2 = \frac{8\pi G}{3}\overline{\rho}a^2, \qquad
\mathcal{H}' = -\frac{1+3\omega}{2}\mathcal{H}^2,\qquad
\overline{\rho}' + 3(1+\omega)\mathcal{H}\overline{\rho} = 0.
\end{align}

\subsection{From formation to evaporation}
\subsubsection*{Black Hole Properties and Evaporation}
\label{sec:PBHgas}
Once formed, primordial black holes act as pressureless matter on large
scales and can be described as a non-relativistic fluid with average
energy density $\rho_{\rm \mathsmaller{PBH}}$. As such, they redshift like ordinary
matter. However, unlike stable cold dark matter, PBHs lose mass through Hawking
evaporation \cite{Hawking:1974sw}, eventually disappearing altogether.
Expressed in conformal time, the associated mass loss is described by the
logarithmic rate \footnote{Here, we adopt Hawking's
standard evaporation formula, disregarding potential quantum gravity
effects such as the memory burden mechanism
\cite{Dvali:2018xpy,Dvali:2020wft,Dvali:2024hsb}, which has been proposed
as a means to avoid information loss. This effect can stabilize the
black hole once it has lost approximately half of its initial mass,
thereby significantly extending its lifetime. As a consequence, the
range of PBH parameters consistent with evaporation before BBN would be
substantially modified.}
\begin{equation}
\label{eq:Gamma_def}
\Gamma(\eta;M_{\rm \mathsmaller{PBH}})
\equiv
-\,\frac{{\rm d}\ln M_{\rm \mathsmaller{PBH}}}{{\rm d}\eta}
=
a(\eta)\,
\frac{A\,M_{\rm pl}^4}{M_{\rm \mathsmaller{PBH}}^3(\eta)},
\qquad
A=\frac{\pi\,\mathcal{G}}{480}\,g_{{\rm H}\star}\simeq 2.7,
\end{equation}
where the greybody factor $\mathcal{G}\simeq 3.8$ captures the partial
back-scattering of Hawking radiation by gravitational and centrifugal
potentials, as computed in
\cite{Page:1976df,Page:1976ki,Page:1977um,MacGibbon:1990zk,MacGibbon:1991tj}
and revisited in \cite{Baldes:2020nuv,Cheek:2021odj}. The quantity
$g_{{\rm H}\star}(T_{\rm \mathsmaller{PBH}})$ accounts for the number of spin-weighted
degrees of freedom accessible at the Hawking--Bekenstein temperature
\cite{Bekenstein:1974ax,Hawking:1976de}
\begin{equation}
T_{\rm \mathsmaller{PBH}}
\equiv
\frac{M_{\rm pl}^2}{M_{\rm \mathsmaller{PBH}}}
\simeq
100~{\rm GeV}
\left( \frac{10^{11}~\rm g}{M_{\rm \mathsmaller{PBH}}} \right).
\end{equation}
In the regime of interest in which $M_{\rm \mathsmaller{PBH}}\lesssim 10^{11}~\rm g$
and for the particle spectrum predicted in the Standard Model, we have
$g_{{\rm H}\star}(T_{\rm \mathsmaller{PBH}})\simeq 108$ \cite{Hooper:2019gtx}.
Moreover, we assume the absence of heavier BSM particles that would
increase this number close to final evaporation, when the PBH
temperature increases rapidly. Throughout this work we use the reduced Planck mass
$M_{\rm pl}\simeq 2.44 \times 10^{18}~{\rm GeV}$.
For simplicity, PBHs are assumed to be non-rotating.
This approximation is well justified for PBHs formed via superhorizon
collapse, for which the resulting spin distribution peaks well below
the percent level, $\mathcal{A}_{\mathcal{R}}<10^{-2}$
\cite{DeLuca:2019buf}. Integrating the mass-loss equation yields the time evolution of the PBH mass \footnote{This is an idealized scenario in which mass growth from accretion is neglected. Recent studies~\cite{Haque:2026vvp,Kallifatides:2026sik} indicate that accretion can lead to substantial PBH mass growth for sufficiently large collapse fractions $\gamma_H$. The two analyses, however, find different critical values, $\gamma_{H,c}=0.395$ and $0.495$, at which the absorption rate enters a runaway regime that formally leads to divergent mass growth. As a result, the precise onset of absorption dominance remains uncertain. In Sec.~\ref{sec:PBH_mass_distrib} we find $\gamma_H\simeq [0.2, 4]$ depending on the width of the primordial power spectrum and on the time at which we evaluate $\gamma_{\rm H}$, see Tab.~\ref{tab:kappa_gammaH}. Unlike the analyses of Refs.~\cite{Haque:2026vvp,Kallifatides:2026sik}, which treat the radiation bath as an external reservoir with fixed temperature, realistic scenarios—particularly those involving ultra-light PBHs that temporarily dominate the cosmic energy budget—do not admit an infinite heat source: the radiation component becomes subdominant due to differential redshifting and partial absorption. We therefore expect absorption effects to modify the initial PBH masses by an $\mathcal{O}(1)$ factor, without qualitatively altering the overall picture, which may be captured by the mapping $M_{{\rm PBH},f}\rightarrow M_{{\rm PBH},f}+M_{{\rm PBH,abs}}$.
}
\begin{equation}
\label{eq:PBH_mass_decay}
M_{\rm \mathsmaller{PBH}}(t)
=
M_{\rm PBH,f}
\left(1-\frac{t}{t_{\rm eva}}\right)^{1/3},
\end{equation}
where $M_{\rm PBH,f}$ denotes the PBH mass at formation and $t_{\rm eva}$
is the total PBH lifetime,
\begin{equation}
\label{eq:t_eva_def}
t_{\rm eva}
=
\frac{M_{\rm PBH,f}^{3}}{3 A M_{\rm pl}^{4}}
\simeq
0.41~{\rm s}
\left( \frac{108}{g_{{\rm H}\star}}\right)
\left(\frac{M_{\rm PBH,f}}{10^9~{\rm g}}\right)^3 .
\end{equation}
Using the conformal-time definition in Eq.~\eqref{eq:Gamma_def}, the
evaporation rate can equivalently be expressed as
\begin{equation}
\label{eq:Gamma_lifetime}
\Gamma(\eta;M_{\rm \mathsmaller{PBH}})
=
\frac{a(\eta)}{3\,[t_{\rm eva}-t(\eta)]},
\end{equation}
where $t(\eta)$ is the cosmic time corresponding to conformal time
$\eta$, and $\eta_{\rm eva}$ is defined by $t(\eta_{\rm eva})=t_{\rm eva}$.
For notational simplicity, we will denote the formation mass
$M_{\rm PBH,f}$ by $M$ in the following.

\begin{figure}
\centering\includegraphics[width=0.65\linewidth]{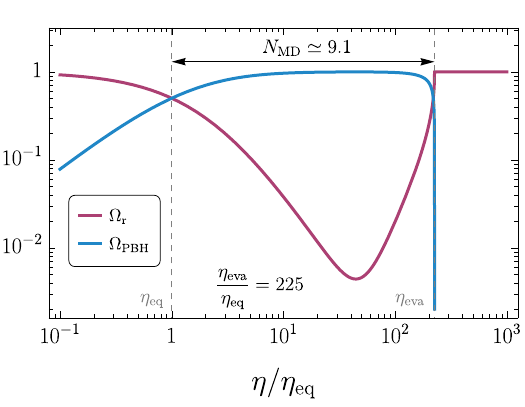}
    \caption{Evolution of the PBH and radiation energy fractions, $\Omega_{\rm \mathsmaller{PBH}}$ and $\Omega_{\rm r}$. PBHs redshift more slowly than radiation, leading to the onset of PBH domination at $\eta_{\rm eq}$, defined by $\Omega_{\rm \mathsmaller{PBH}}=\Omega_{\rm r}$. This is followed by an early matter-dominated era, which ends at $\eta_{\rm eva}$ when PBHs evaporate and reheat the Universe.}
    \label{fig:eMD}
\end{figure}

\subsubsection*{Cosmological History}
We focus on ultra-light PBHs that evaporate prior to Big-Bang Nucleosynthesis and assume a monochromatic mass function characterized by a single mass scale $M_{\rm \mathsmaller{PBH}}$, which we identify with the average PBH mass at formation. PBHs form during radiation domination when overdensities re-enter the horizon, with a mass set by the horizon mass at that time and controlled by an $\mathcal{O}(1)$ collapse efficiency factor $\gamma_{\rm H}$. Their subsequent evolution is governed by Hawking evaporation. Consistency with the maximal temperature attained after inflation imposes a lower bound on the PBH mass, while the requirement that evaporation occur before BBN selects the ultra-light window relevant for this work,
\begin{equation}
1~{\rm g} \;\lesssim\; M_{\rm \mathsmaller{PBH}} \;\lesssim\; 10^{9}~{\rm g}.
\end{equation}
If the initial PBH energy fraction at formation, $\beta_f$, exceeds a critical value $\beta_c$, PBHs come to dominate the energy density before evaporating, giving rise to an early matter-dominated epoch\footnote{For the sake of simplicity, we omit in this equation and the ones below the factors involving ratios of relativistic degrees of freedom, which are reported in App.~\ref{App:PBHgas}.}
\begin{equation}
\label{eq:beta_f_c}
    \beta_f>\beta_c\simeq2.5\times 10^{-14}\gamma_{\rm H}^{-1/2}\left(\frac{10^8{\rm g}}{M_{{\rm PBH}}}\right)\qquad \Rightarrow \qquad {\rm \textbf{eMD}}
\end{equation}
This phase begins at conformal time $\eta_{\rm eq}$, when the PBH and radiation energy densities become equal, and ends at $\eta_{\rm eva}$, when PBHs evaporate. We show the evolution of relative matter and density energy fractions in Fig.~\ref{fig:eMD} and characterize the duration of the eMD between equality and evaporation by the number of e-folds,
\begin{equation}
\label{eq:NMD_main}
N_{\rm MD} \;\equiv\; \ln\!\left(\frac{a(\eta_{\rm eva})}{a(\eta_{\rm eq})}\right),
\end{equation}
which depends primarily on $\beta_f$ and $M_{{\rm PBH}}$.

PBH evaporation reheats the Universe and injects entropy into the thermal bath, diluting any pre-existing relic abundances. The associated entropy injection can be summarized by a dilution factor
\begin{equation}
D\equiv \frac{S(T_{\rm eva})}{S(T_{\rm eva}^-)}\simeq \frac{\beta_f}{\beta_{\rm c}}\simeq \exp\left(\frac{3}{4}N_{\rm MD}\right).
\end{equation}
In the opposite regime, $\beta_f<\beta_c$, PBHs never dominate the energy density, the eMD phase is absent, and one recovers $D=1$, corresponding to standard adiabatic expansion.

\begin{figure}
\centering{\includegraphics[width=0.75\linewidth]{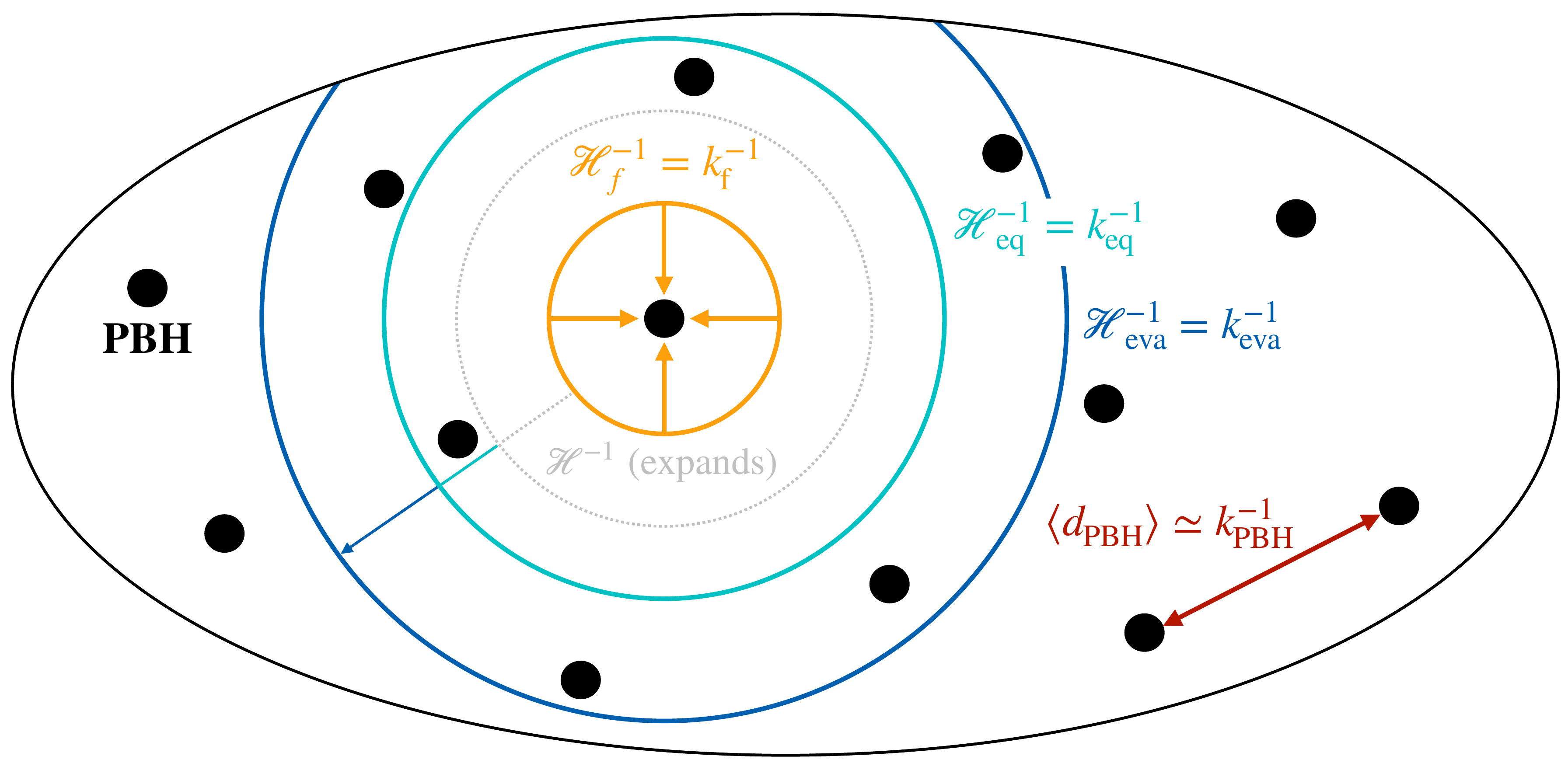}}
    \caption{Schematic illustration of the characteristic comoving scales during a PBH-dominated epoch in the early Universe. The comoving volume is populated by an ensemble of randomly distributed PBHs, shown as black dots. Indicated are the PBH separation cutoff $k_{\rm \mathsmaller{PBH}}^{-1}$, together with the formation scale $k_f^{-1}$, the equality scale $k_{\rm eq}^{-1}$, and the evaporation scale $k_{\rm eva}^{-1}$, corresponding to the Hubble horizon at PBH formation, early matter–radiation equality, and reheating from Hawking evaporation, respectively. $k_f^{-1}$ is therefore associated with the horizon crossing of a sufficiently large overdensity for PBH formation and hence the comoving size of the collapsing region (denoted by inward arrows). As the comoving Hubble radius grows (denoted by outward arrows), these characteristic scales increase accordingly.}
    \label{fig:scales}
\end{figure}
\subsubsection*{Characteristic comoving scales}
Here, we summarize the physical interpretation of the characteristic scales associated with the formation, evolution, and evaporation of PBHs. The goal is to provide an intuitive overview of their meaning and mutual relations, while retaining only the final expressions required for phenomenological applications. All technical derivations including degree-of-freedom factors can be found in App.~\ref{App:PBHgas}.
\begin{itemize}
\item[] \textbf{Formation scale} $\mathbf{k_f}$ The formation scale corresponds to the comoving size of the Hubble horizon, $k_f=\mathcal{H}_f$, at the time when a sufficiently large overdensity re-enters the horizon and collapses to form a PBH. It therefore also represents the comoving size of the collapsing region itself. This scale is directly tied to the PBH mass through the horizon mass $M_{{\rm H},f}$ and the scale factor $a_f\equiv a(\eta_f)$ at formation,
\begin{equation}
\label{eq:M_f_def}
M_{{\rm PBH},f} \equiv \gamma_{\rm H} M_{{\rm H},f},\qquad \textrm{with}\quad M_{\rm H,f}=a_f\frac{4\pi M_{\rm pl}^2 }{\mathcal{H}_f}.
\end{equation}
\item[] \textbf{PBH scale} $\mathbf{k_{PBH}}$
This scale is set by the mean comoving separation between PBHs at formation. On scales smaller than $k_{\rm \mathsmaller{PBH}}^{-1}$ the discrete nature of the PBH distribution becomes important, and a fluid description of the PBH ensemble breaks down. Up to mild dependence on relativistic degrees of freedom, the corresponding comoving wavenumber is given by
\begin{equation}
\label{eq:kUV_def_main}
k_{\rm \mathsmaller{PBH}} \simeq 4.9 \times 10^{14}~{\rm Mpc}^{-1}
\left( \frac{10^8~{\rm g}}{M_{\rm \mathsmaller{PBH}}} \right)^{5/6}.
\end{equation}
\item[] \textbf{Equality scale} $\mathbf{k_{eq}}$
If the initial PBH energy fraction at formation $\beta_f$ exceeds a critical value, PBHs come to dominate the energy density of the Universe prior to their evaporation, leading to an early matter-dominated phase. The scale $k_{\rm eq}^{-1}$ corresponds to the Hubble horizon at the onset of this PBH-dominated era. Up to mild corrections ($c_1=\mathcal{O}(1)$ from relativistic dofs), one finds relative to the PBH scale
\begin{equation}
\label{eq:keq_main}
\frac{k_{\rm eq}}{k_{\rm \mathsmaller{PBH}}} \simeq \sqrt{2}\gamma_{\rm H}^{1/3}\beta_f^{2/3}.
\end{equation}

\item[] \textbf{Evaporation scale} $\mathbf{k_{eva}}$
The evaporation scale is defined as the comoving Hubble radius at the time when PBHs complete their Hawking evaporation and reheat the Universe. The evaporation scale is set by the PBH mass as
\begin{equation}
\label{eq:keva_def_main}
k_{\rm eva} \simeq 4.2 \times 10^{5}~{\rm Mpc}^{-1}
\left( \frac{10^8~{\rm g}}{M_{\rm \mathsmaller{PBH}}} \right)^{3/2}.
\end{equation}
\end{itemize}
Since the comoving Hubble radius grows with time, the relative ordering of these scales obeys the following hierarchy:
\begin{equation}
k_{\rm eva} \ll k_{\rm eq} \ll k_{\rm \mathsmaller{PBH}} \ll k_f.
\end{equation}
Fig.~\ref{fig:scales} provides a schematic overview of the physical
origin of the characteristic comoving scales introduced above, and
illustrates their relative hierarchy in cosmic time.

\subsection{PBH mass distribution}
\label{sec:PBH_mass_distrib}
\begin{figure}[t]
\centering
\includegraphics[width=0.49\columnwidth]{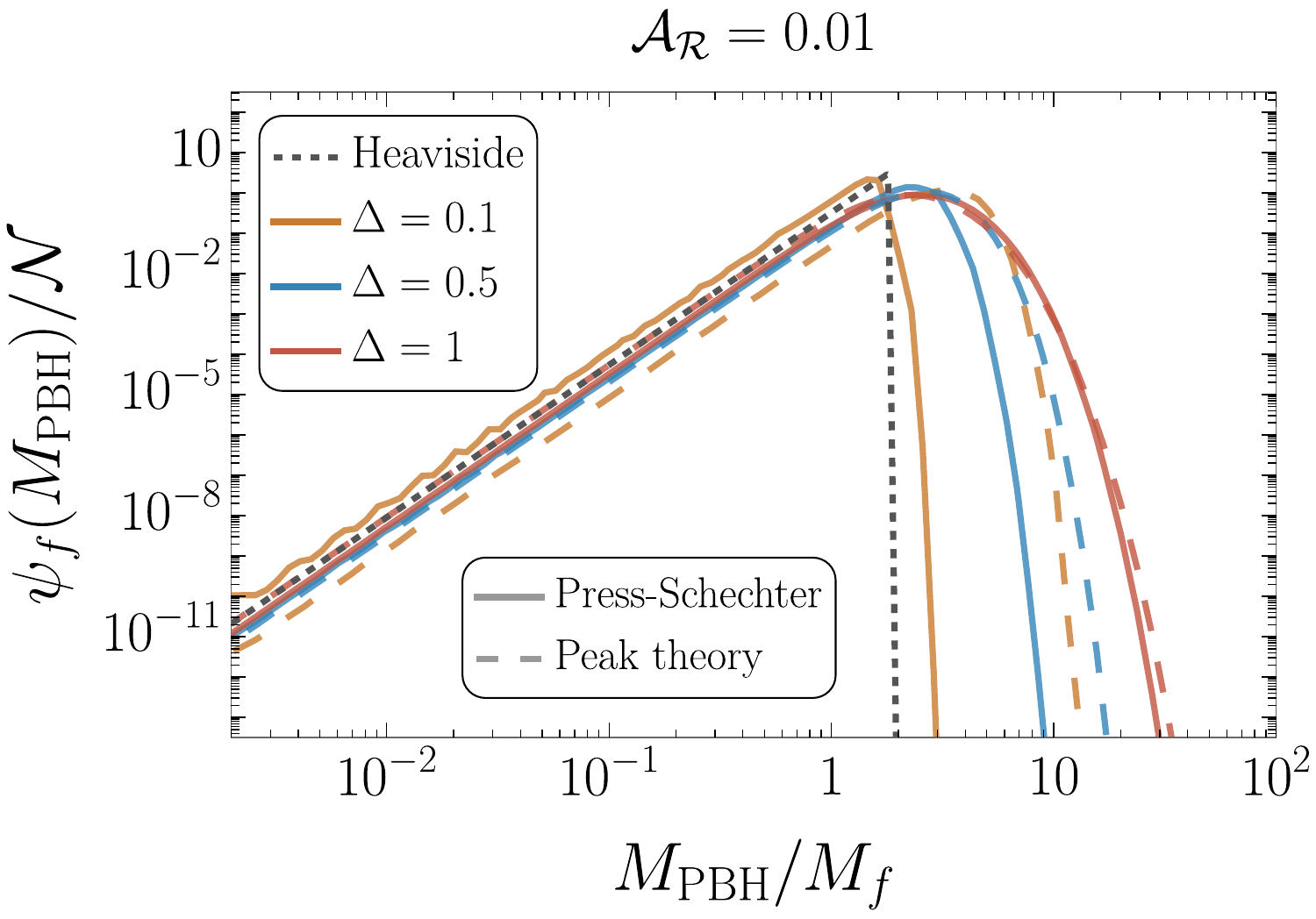}\includegraphics[width=0.49\columnwidth]{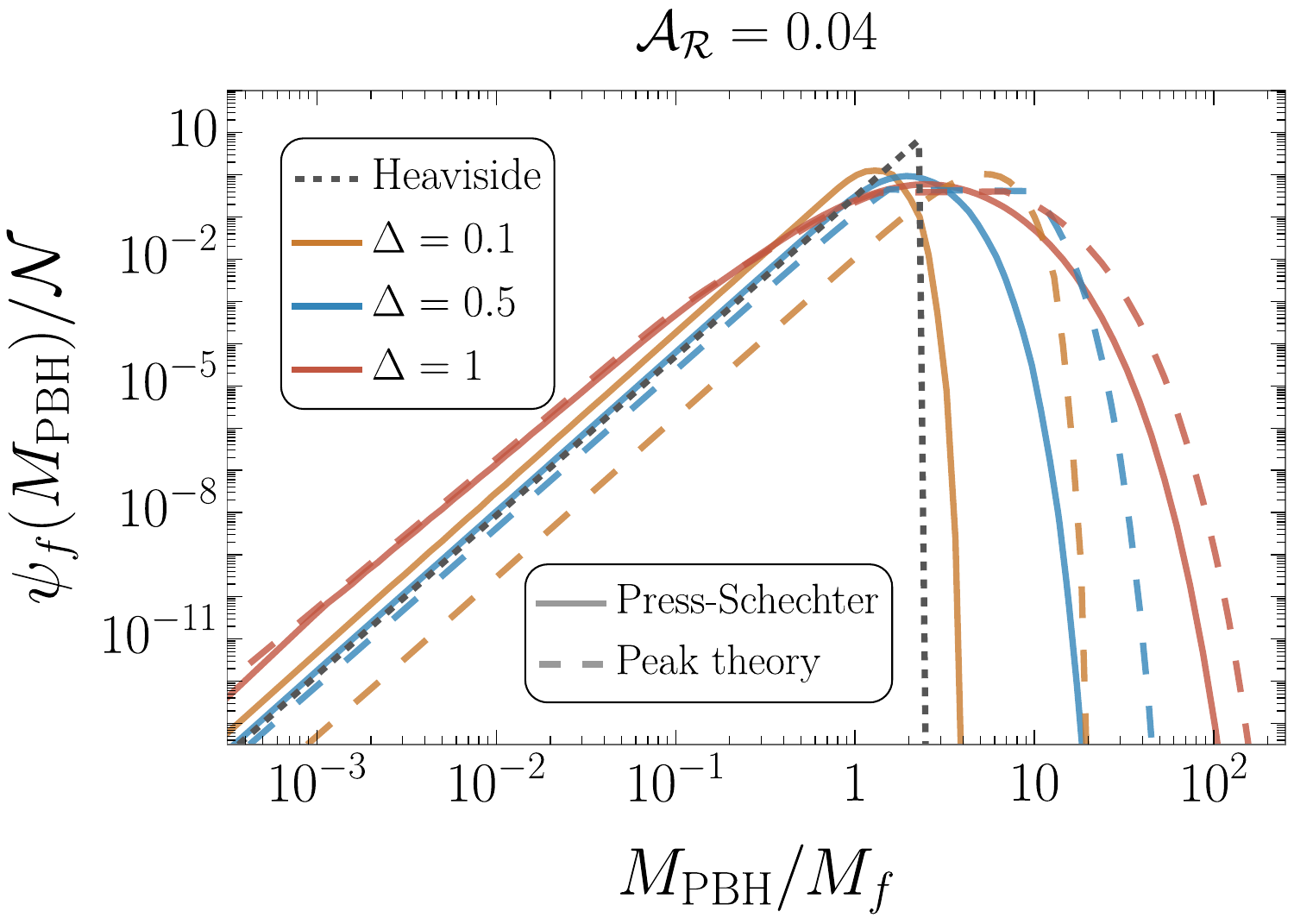}
\caption{PBH mass distribution $\psi_f(M_{\rm \mathsmaller{PBH}})$ for different widths $\Delta$ and amplitudes $\mathcal{A}_{\mathcal{R}}$ of the log-normal curvature power spectrum in Eq.~\eqref{eq:log_normal}. To remain agnostic about the model-dependent UV part, we conservatively model the mass distribution as a power law $\psi_f(M_{\rm \mathsmaller{PBH}})\propto M_{\rm \mathsmaller{PBH}}^{1+1/\gamma_{\rm M}}\propto M_{\rm \mathsmaller{PBH}}^{3.78}$ with a sharp cutoff at $M_{\rm cut}$, see Eq.~\eqref{eq:psi_f_maintext}. The reference mass $M_f$ is the horizon mass at $k_f=\mathcal{H}$, while the horizon mass entering Choptuik scaling is $M_H(r_m)=\kappa^2 M_f$, with $\kappa$ for the log-normal spectra given in Eq.~\eqref{eq:kappa_lognormal_fit_app} of App.~\ref{app:PBH_mass_distrib}.}
\label{fig:psi_PBH}
\end{figure}

This subsection summarizes the ingredients of the PBH mass distribution underlying our analysis. A complete derivation, the explicit fitting formula for the geometric factor $\kappa$ defined below, and a detailed comparison of analytic approaches are presented in App.~\ref{app:PBH_mass_distrib}.
\begin{figure}
\centering
\includegraphics[width=0.49\linewidth]{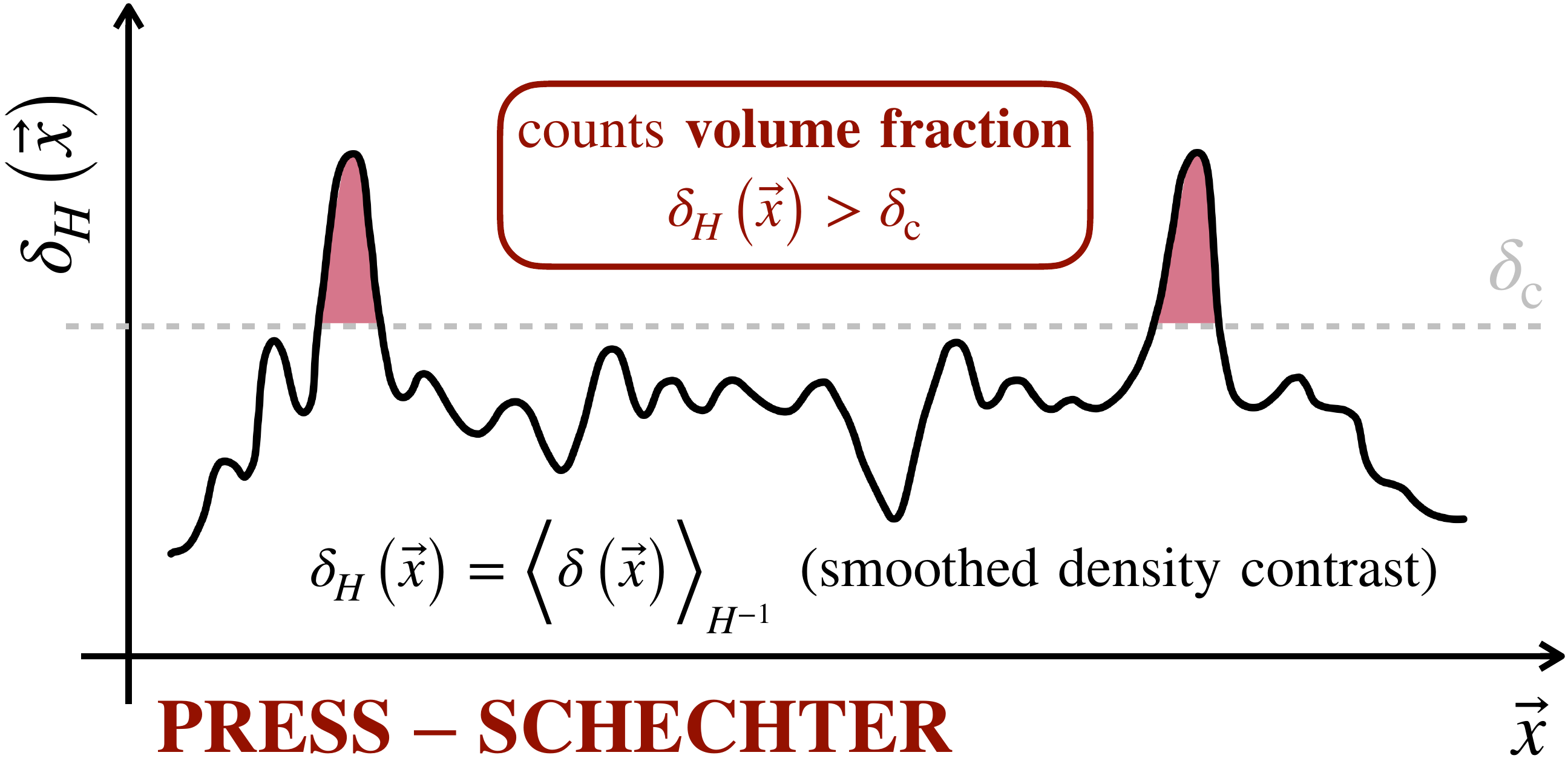}\hspace{3mm}\includegraphics[width=0.49\linewidth]{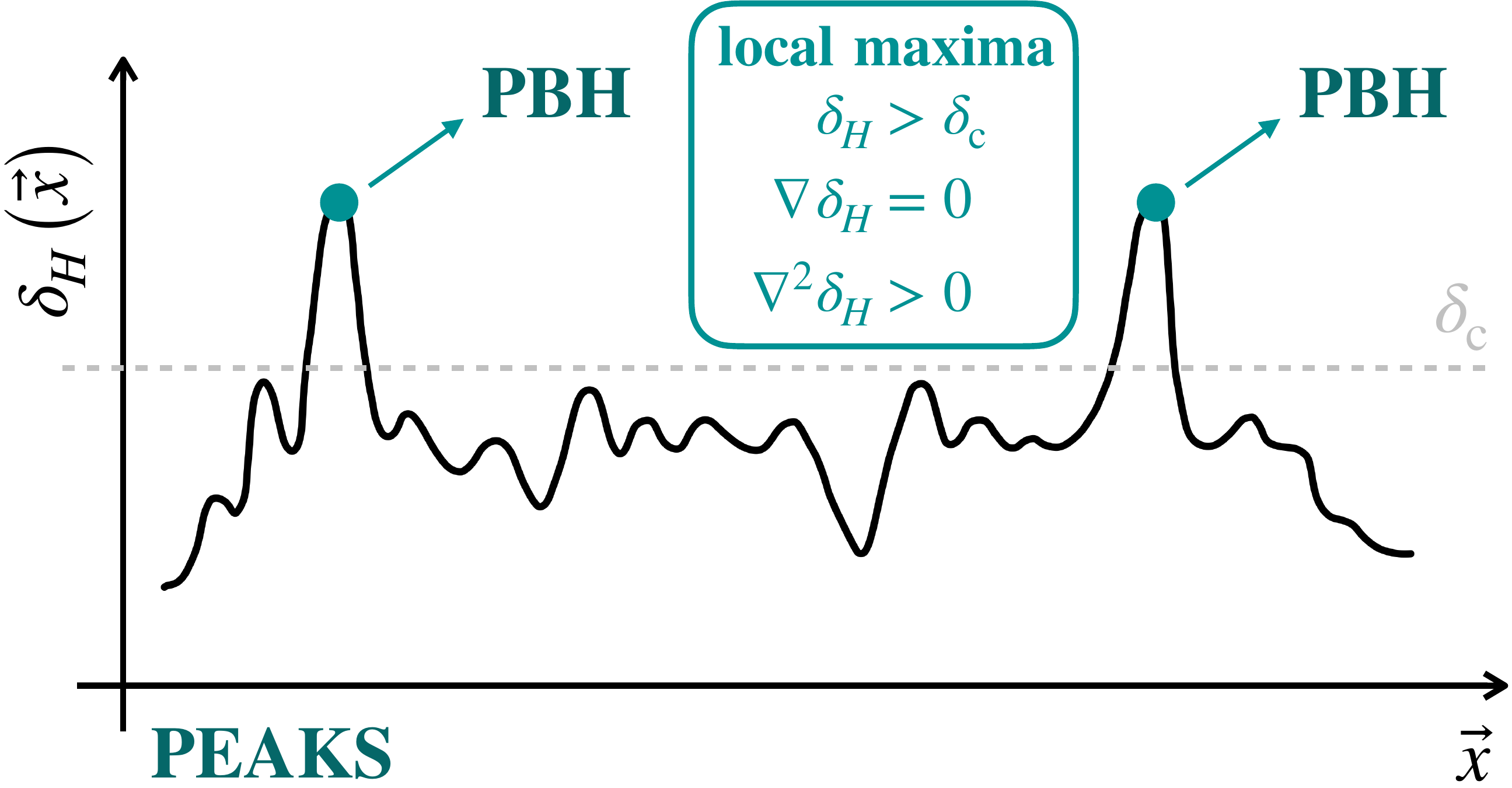}
\caption{Schematic comparison of the Press--Schechter and peak-theory prescriptions for primordial black hole formation. In both panels, $\delta_R$ denotes the density contrast smoothed on a fixed scale $R=\mathcal{H}^{-1}$. \textbf{Left:} Press--Schechter estimates the PBH abundance from the volume fraction of regions in which $\delta_R$ exceeds the collapse threshold $\delta_c$, treating all such regions on the same footing. \textbf{Right:} peak theory identifies PBH formation sites with local maxima of $\delta_R$ above threshold.}
\label{fig:peakvsPS}
\end{figure}
PBHs are assumed to form during radiation domination from the collapse of large overdensities as they re-enter the Hubble horizon~\cite{Carr:1975qj}. Their abundance is characterized by the energy fraction $\beta$, defined as the fraction of the total energy density converted into PBHs. To track the distribution over masses we use the differential quantity
\begin{equation}
\frac{d\beta(M_{\rm \mathsmaller{PBH}})}{d\ln M_{\rm \mathsmaller{PBH}}},
\end{equation}
which gives how the PBH energy density is distributed over masses at formation. The normalized PBH mass function is then
\begin{equation}
\label{eq:psi_def}
\psi_f(M_{\rm \mathsmaller{PBH}})\equiv \frac{1}{\beta_f}\frac{d\beta(M_{\rm \mathsmaller{PBH}},T_f)}{d\ln M_{\rm \mathsmaller{PBH}}},\qquad \int d\ln M_{\rm \mathsmaller{PBH}}\,\psi_f(M_{\rm \mathsmaller{PBH}})=1,
\end{equation}
where $\beta_f$ is the total PBH energy fraction at formation.
The PBH mass spectrum is set by the statistics of the primordial curvature perturbations, encoded in $\mathcal{P}_{\mathcal{R}}(k)$. Throughout this work we adopt the log-normal parametrization
\begin{equation}
\label{eq:log_normal}
\mathcal{P}_{\mathcal{R}}(k)=\frac{\mathcal{A}_{\mathcal{R}}}{\sqrt{2\pi}\,\Delta}\exp\!\left[-\frac{\ln^2(k/k_f)}{2\Delta^2}\right],
\end{equation}
peaked at the scale $k_f$ with width $\Delta$, and we assume Gaussian statistics. The horizon mass evaluated when this peak scale crosses the horizon is
\begin{equation}
\label{eq:Mstar_def_maintext}
M_f\equiv M_H(\mathcal{H}=k_f) .
\end{equation}
Near the collapse threshold, the PBH mass is not fixed only by the horizon mass. Numerical simulations of gravitational collapse show that, for a one-parameter family of initial data labeled by an amplitude $p$, there exists a critical value $p_c$ separating dispersion from black-hole formation~\cite{Choptuik:1992jv,Evans:1994pj,Niemeyer:1997mt,Niemeyer:1999ak}. For $p>p_c$ and close to threshold, the solution first approaches a universal self-similar critical solution. This critical solution has one unstable mode. The time spent near it is therefore controlled by the distance to threshold, and the final black-hole mass obeys the universal scaling law~\cite{Choptuik:1992jv,Evans:1994pj,Maison:1995cc,Gundlach:2007gc}
\begin{equation}
\label{eq:choptuik_main}
{\it Choptuik's ~law:}\qquad \qquad  M_{\rm \mathsmaller{PBH}}
=
\mathcal{K}\,M_H\,
(\delta_m-\delta_c)^{\gamma_{\rm M}} .
\end{equation}
Here $\delta_m$ is the smoothed density contrast at horizon crossing, $\delta_c$ is the collapse threshold, $\mathcal{K}\sim 4$ is a coefficient, and $\gamma_{\rm M}$ is the critical exponent. The scaling in Eq.~\eqref{eq:choptuik_main} relies on three assumptions: the collapsing region is approximately spherically symmetric, the perturbation is close to threshold, $\delta_m-\delta_c\ll\delta_c$, and the equation of state is fixed during collapse. The exponent $\gamma_{\rm M}$ is universal for a given matter content, while $\delta_c$ and $\mathcal{K}$ depend on the perturbation profile. For radiation domination one finds $\gamma_{\rm M}\simeq0.36$~\cite{Niemeyer:1997mt,Niemeyer:1999ak,Yokoyama:1998xd}, while typical values of the threshold lie in the range $\delta_c\in[0.40,0.67]$~\cite{Musco:2018rwt,Escriva:2019phb}.
 An important subtlety is that the horizon mass entering Eq.~\eqref{eq:choptuik_main} should be associated with the real-space scale controlling collapse, namely the radius $r_m$ at which the compaction function is maximal, rather than directly with the Fourier scale $k_f^{-1}$~\cite{Shibata:1999zs,Musco:2018rwt}. We therefore define
\begin{equation}
\label{eq:MH_rm_def_maintext}
M_H\equiv M_H(r_m)\equiv M_H(\mathcal{H}=1/r_m).
\end{equation}
where $r_m$ is the radius at which the compaction function is maximal. The compaction function $\mathcal{C}=2\delta M/R$ measures the excess mass inside a spherical region normalized by its areal radius, and it is this scale rather than the wavelength $k_f^{-1}$ that controls the collapse~\cite{Shibata:1999zs,Harada:2015yda,Musco:2012au,Musco:2018rwt,Young:2019yug,Escriva:2019nsa,Escriva:2021aeh,Musco:2020jjb}.
The ratio between the two reference scales is parametrized by~\cite{Musco:2020jjb,Franciolini:2022tfm,Ferrante:2022mui,Frosina:2023nxu}
\begin{equation}
\label{eq:kappa_def_maintext}
\kappa\equiv k_f r_m,
\end{equation}
which for the log-normal spectrum of Eq.~\eqref{eq:log_normal} depends only on the width $\Delta$. An explicit fitting formula is given in Eq.~\eqref{eq:kappa_lognormal_fit_app} of App.~\ref{app:PBH_mass_distrib}, and representative values for the widths used in this work are reported in Tab.~\ref{tab:kappa_gammaH}. Since PBHs form during radiation domination, $M_H\propto\mathcal{H}^{-2}$, and therefore
\begin{equation}
\label{eq:MH_Mstar_kappa_maintext}
M_H(r_m)=M_H(\mathcal{H}=k_f/\kappa)=\kappa^2 M_f .
\end{equation}
With these geometric factors in hand, we turn to the statistical computation of the mass fraction. Two analytic frameworks are commonly used to compute the PBH abundance from $\mathcal{P}_{\mathcal{R}}(k)$, namely the Press--Schechter (PS) formalism~\cite{Press:1973iz,Bond:1990iw} and peak theory~\cite{Bardeen:1985tr}, schematically compared in Fig.~\ref{fig:peakvsPS}. In the PS framework, the PBH abundance is estimated from the volume fraction of regions in which the smoothed density contrast exceeds $\delta_c$, while peak theory instead counts local maxima above threshold. Despite these different counting prescriptions, both approaches yield the same infrared tail of the logarithmic mass function~\cite{Niemeyer:1999ak,Green:1999xm,Young:2019yug,Gow:2020bzo},
\begin{equation}
\label{eq:IR_tail_maintext}
\psi_f(M_{\rm \mathsmaller{PBH}})\propto M_{\rm \mathsmaller{PBH}}^{1+1/\gamma_{\rm M}} .
\end{equation}
This scaling refers to $d\beta/d\ln M_{\rm \mathsmaller{PBH}}$, whereas a mass function defined per unit mass, $d\beta/dM_{\rm \mathsmaller{PBH}}$, scales instead as $M_{\rm \mathsmaller{PBH}}^{1/\gamma_{\rm M}}$. The ultraviolet part of the distribution, including the precise position and shape of the high-mass cutoff, is model dependent~\cite{Niemeyer:1997mt,Yokoyama:1998xd,Carr:2018rid,Karam:2022nym}.
This behavior is illustrated in Fig.~\ref{fig:psi_PBH}, which shows $\psi_f(M_{\rm \mathsmaller{PBH}})$ for several widths $\Delta$ and amplitudes $\mathcal{A}_{\mathcal{R}}$ of the log-normal spectrum. To retain only the universal IR behavior and discard the model-dependent UV tail, we adopt the conservative sharp-cutoff form
\begin{equation}
\label{eq:psi_f_maintext}
{\it Choptuik's~ mass~ dist:}\quad \quad \psi_{f}^{{\rm sharp}}(M_{\rm \mathsmaller{PBH}})\!=\!\frac{1+\gamma_{\rm M}}{\gamma_{\rm M}}\!\left(\frac{M_{\rm \mathsmaller{PBH}}}{M_{\rm cut}}\right)^{1+1/\gamma_{\rm M}}\!\!\!\!\Theta(M_{\rm cut}-M_{\rm \mathsmaller{PBH}}) ,
\end{equation}
with $M_{\rm cut}$ the upper cutoff of the distribution.
From this mass function we extract a characteristic PBH mass. The PBH energy density contained in an interval $d\ln M_{\rm \mathsmaller{PBH}}$ is $d\rho_{\rm PBH}=\rho_{\rm PBH}\,\psi_f(M_{\rm \mathsmaller{PBH}})\,d\ln M_{\rm \mathsmaller{PBH}}$, so the corresponding contribution to the number density is $dn_{\rm PBH}=d\rho_{\rm PBH}/M_{\rm \mathsmaller{PBH}}$. The average PBH mass, defined as the mass density divided by the number density, then reads
\begin{equation}
\langle M_{\rm \mathsmaller{PBH}}\rangle\equiv \frac{\rho_{\rm PBH}}{n_{\rm PBH}}=\left[\int d\ln M_{\rm \mathsmaller{PBH}}\,\psi_f(M_{\rm \mathsmaller{PBH}})\,M_{\rm \mathsmaller{PBH}}^{-1}\right]^{-1}.
\end{equation}
It is convenient to express this average in units of $M_f$, defining
\begin{equation}
\label{eq:def_gammaH}
\gamma_{\rm H}^{(k_f)}\equiv \frac{\langle M_{\rm \mathsmaller{PBH}}\rangle}{M_f}\simeq
\begin{cases}
1.2, & \text{Press--Schechter},\\[4pt]
2.7, & \text{peak theory},
\end{cases}
\end{equation}
where the numerical values are computed from the PBH mass distribution shown in Fig.~\ref{fig:psi_PBH} for $\Delta=0.1$ and $\mathcal{A}_{\mathcal{R}}=0.01$. The larger peak-theory value reflects the heavier high-mass tail produced when counting rare density maxima. Equivalently, if the average mass is normalized to the horizon mass at the compaction scale, $M_H(r_m)$, we have
\begin{equation}
\label{eq:def_gammaH_rm}
\gamma_{\rm H}^{(r_m)}\equiv \frac{\langle M_{\rm \mathsmaller{PBH}}\rangle}{M_H(r_m)}=\frac{\gamma_{\rm H}^{(k_f)}}{\kappa^2}\simeq
\begin{cases}
0.16, & \text{Press--Schechter},\\[4pt]
0.37, & \text{peak theory},
\end{cases}
\end{equation}
in agreement with the commonly quoted estimate $\gamma_{\rm H}^{(r_m)}\sim 0.2$~\cite{Carr:2020gox}. The full dependence of $\kappa$, $\gamma_{\rm H}^{(k_f)}$, and $\gamma_{\rm H}^{(r_m)}$ on $\Delta$ and $\mathcal{A}_{\mathcal{R}}$ is collected in Tab.~\ref{tab:kappa_gammaH} of App.~\ref{app:PBH_mass_distrib}. In what follows we treat 
\begin{equation}
    \gamma_{\rm H}\sim \gamma_{\rm H}^{(k_f)} \sim \gamma_{\rm H}^{(r_m)},
\end{equation} 
as an $\mathcal{O}(1)$ parameter and identify $M_{\rm \mathsmaller{PBH}}\equiv\langle M_{\rm \mathsmaller{PBH}}\rangle$ as the characteristic PBH mass.

\newpage
\section{Perturbations in a PBH-dominated universe}
\label{sec:perturbations}
\subsection{The averaged two-fluid background}

Before turning to cosmological perturbations and the generation of
gravitational waves from scalar metric fluctuations, we briefly recall
the evolution of the homogeneous background.
The background dynamics are governed by a PBH-induced matter component
and a radiation bath, with energy densities $\rho_{\rm \mathsmaller{PBH}}$ and
$\rho_{\rm r}$, respectively, and scale factor $a$.
Following Ref.~\cite{Inomata:2020lmk}, the standard fluid equations are
supplemented by a source term that accounts for PBH evaporation.
It is convenient to express the background evolution in terms of
comoving energy densities.
The resulting system of equations reads
\begin{equation}
\label{eq:rhocomoving}
\left(\rho_{\rm \mathsmaller{PBH}} a^3\right)' = -\,\Gamma\,\rho_{\rm \mathsmaller{PBH}} a^3, \qquad
\left(\rho_{\rm r} a^4\right)' = \;\;\Gamma\,\rho_{\rm \mathsmaller{PBH}} a^4, \qquad
\mathcal{H} 
= a\,\sqrt{\frac{\rho_{\rm \mathsmaller{PBH}}+\rho_{\rm r}}{3M_{\rm pl}^2}} ,
\end{equation}
where primes denote derivatives with respect to conformal time $\eta$,
and $\Gamma$ is the PBH evaporation rate defined in
Eq.~\eqref{eq:Gamma_def}.
Solving the background evolution equations~\eqref{eq:rhocomoving} 
yields the
following expressions for the energy densities:
\begin{align}
\label{eq:rhom_general}
\rho_{\rm \mathsmaller{PBH}}(\eta;M_{\rm \mathsmaller{PBH}})
&=
\rho_{\rm \mathsmaller{PBH},0}\left(\frac{a_0}{a}\right)^3
\exp\!\left[-\int_0^\eta \!{\rm d}\tilde{\eta}\,\Gamma(\tilde{\eta})\right]=
\rho_{\rm \mathsmaller{PBH},0}\left(\frac{a_0}{a}\right)^3
\left(1-\frac{t(\eta)}{t_{\rm eva}(M_{\rm \mathsmaller{PBH}})}\right)^{1/3},
\\[6pt]
\label{eq:rhor_general}
\rho_{\rm r}(\eta;M_{\rm \mathsmaller{PBH}})
&=
\left(\frac{a_0}{a}\right)^4
\left[
\rho_{\rm r,0}
+
\rho_{\rm \mathsmaller{PBH},0}
\int_0^\eta \!{\rm d}\tilde{\eta}\,
\frac{a^2(\tilde{\eta})}{a_0}\,
\frac{1}{3\,t_{\rm eva}(M_{\rm \mathsmaller{PBH}})}
\left(1-\frac{t(\tilde{\eta})}{t_{\rm eva}(M_{\rm \mathsmaller{PBH}})}\right)^{-2/3}
\right].
\end{align}
Relaxing the idealized assumption of a monochromatic PBH mass spectrum and instead allowing for an extended mass distribution $\psi(M_{\rm \mathsmaller{PBH}})$ implies that different PBHs evaporate at different times, thereby prolonging the reheating transition. To obtain a conservative lower bound on the duration of this modified reheating period, we model the mass function using Choptuik's distribution with a sharp cutoff, $\psi_{\rm f,sharp}(M_{\rm \mathsmaller{PBH}})$, defined in Eq.~\ref{eq:psi_f_maintext}. The use of a sharp cutoff further allows us to remain largely model agnostic while capturing the minimal physical broadening implied by critical collapse.
We therefore define averaged background quantities on causally connected regions of the Universe, taken to be Hubble-sized patches. Concretely, for any function $\mathcal{F}(M_{\rm \mathsmaller{PBH}})$ we perform the replacement
\begin{equation}
\mathcal{F}(M_{\rm \mathsmaller{PBH}}) \rightarrow \langle \mathcal{F}(M_{\rm \mathsmaller{PBH}}) \rangle_{\mathcal{H}^{-1}},
\end{equation}
where the average is taken over the PBH mass distribution within a given Hubble volume. We further assume that, at the times of interest, the number of PBHs contained in a Hubble patch is sufficiently large for this ensemble average to be well defined and given by\footnote{Variance of statistical fluctuations due to the finite number of PBHs within a Hubble volume scale as $1/N$ and vanish in the large-$N$ limit. Residual effects of such fluctuations are captured by isocurvature perturbations.}
\begin{equation}
  \langle\mathcal{F}(M_{\rm \mathsmaller{PBH}})\rangle_{\mathcal{H}^{-1}}\simeq\int \mathcal{F}(M_{\rm \mathsmaller{PBH}})\,\psi(M_{\rm \mathsmaller{PBH}}){\rm d}\log M_{\rm \mathsmaller{PBH}}.
\end{equation}
We define the average time at which the universe is reheated as
\begin{equation}
\langle\eta_{\rm eva}\rangle=\int \eta_{\rm eva}(M_{\rm \mathsmaller{PBH}})\,\psi(M_{\rm \mathsmaller{PBH}}){\rm d}\log M_{\rm \mathsmaller{PBH}}.
\end{equation}\begin{figure}
    \centering
    \includegraphics[width=0.7\linewidth]{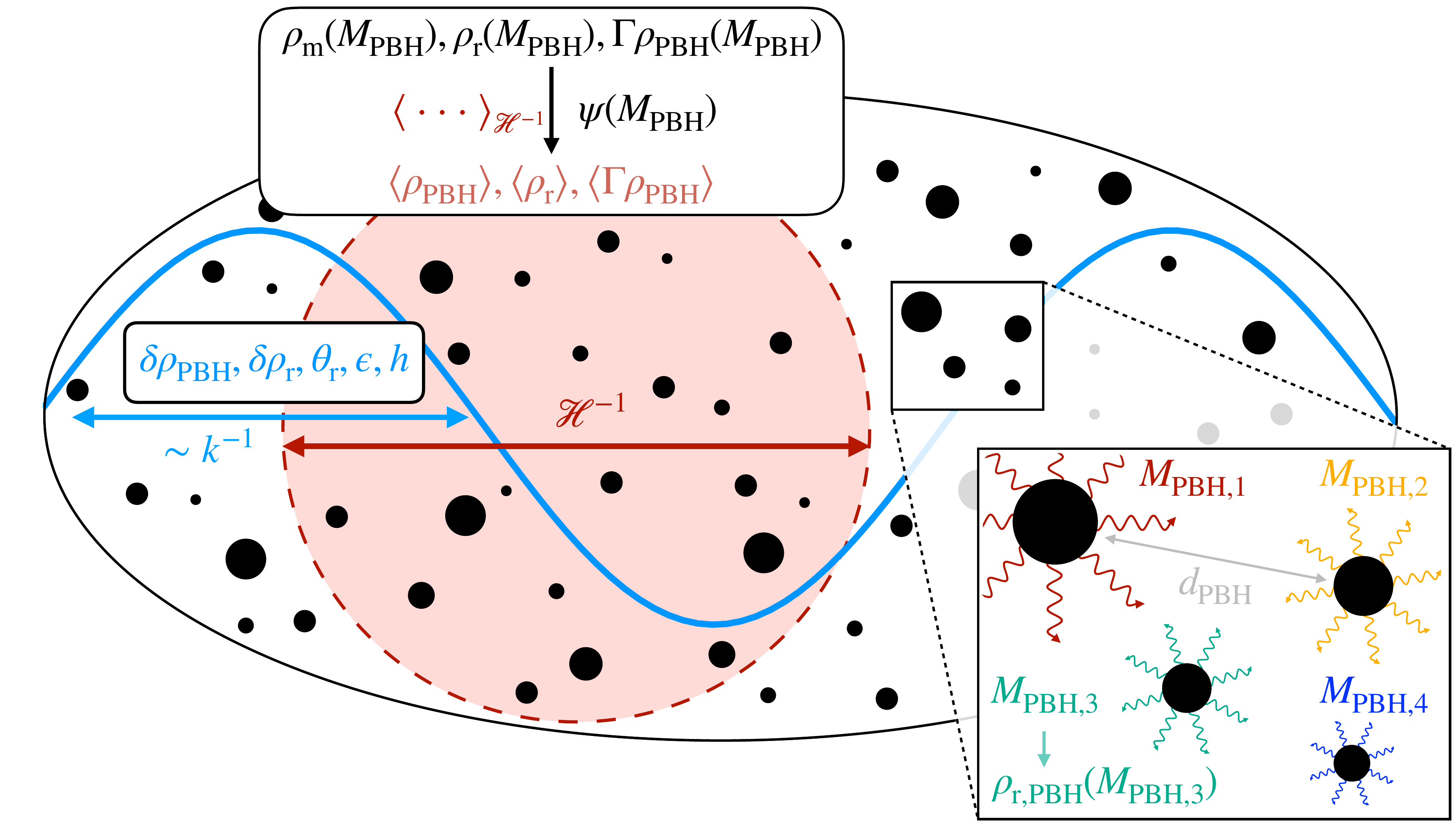}
    \caption{Schematic representation of the averaged background quantities
$(\rho_{\rm \mathsmaller{PBH}}, \rho_{\rm r}, \Gamma \rho_{\rm \mathsmaller{PBH}})$ and the associated
long-wavelength perturbations
$(\delta\rho_{\rm \mathsmaller{PBH}}, \delta\rho_{\rm r}, \theta_{\rm r}, \epsilon, h)$.
The figure depicts the comoving universe in the presence of a
long-wavelength metric perturbation (blue), whose scale is much larger than the mean separation between PBHs.
Such perturbations therefore evolve on an ensemble-averaged background, represented by the red Hubble sized patch containing many PBHs with different masses. On sufficiently small scales, of order the PBH separation $d_{\rm \mathsmaller{PBH}}$, local regions are instead dominated by individual PBHs, whose properties determine the local energy densities $\rho_{\rm r}(M_{\rm \mathsmaller{PBH}})$ and $\rho_{\rm \mathsmaller{PBH}}(M_{\rm \mathsmaller{PBH}})$. The ensemble is characterized by the PBH mass distribution
$\psi(M_{\rm \mathsmaller{PBH}})$, over which single-PBH quantities are averaged, yielding $\left(\rho_{\rm \mathsmaller{PBH}}, \rho_{\rm r}, \Gamma \rho_{\rm \mathsmaller{PBH}}\right)
\xrightarrow{\psi(M_{\rm \mathsmaller{PBH}})}
\left(\langle \rho_{\rm \mathsmaller{PBH}} \rangle,
\langle \rho_{\rm r} \rangle,
\langle \Gamma \rho_{\rm \mathsmaller{PBH}} \rangle\right)$.
    \label{fig:PT_Schematic}}
\end{figure}
Accordingly, the space-averaged matter and radiation energy densities are obtained by 
\begin{align}
\label{eq:rho_m_bkg}
    \langle \rho_{\rm \mathsmaller{PBH}} \rangle(\eta) = \int \rho_{\rm \mathsmaller{PBH}}(\eta;M_{\rm \mathsmaller{PBH}}) \psi(M_{\rm \mathsmaller{PBH}}) {\rm d}\log M_{\rm \mathsmaller{PBH}},\\
    \label{eq:rho_r_bkg}
     \langle \rho_{\rm r} \rangle(\eta) = \int \rho_{\rm r}(\eta;M_{\rm \mathsmaller{PBH}}) \psi(M_{\rm \mathsmaller{PBH}}) {\rm d}\log M_{\rm \mathsmaller{PBH}},
\end{align}
where $\rho_{\rm \mathsmaller{PBH}}(\eta;M_{\rm \mathsmaller{PBH}})$ and $\rho_{\rm r}(\eta;M_{\rm \mathsmaller{PBH}})$ are given by Eqs.~\eqref{eq:rhom_general} and \eqref{eq:rhor_general} with $a(\eta)$ given by Eq.~\eqref{eq:scale_factor_full} below.
Those background quantities, $\left<\rho_m\right>$ and $\left<\rho_r\right>$, will be needed  when solving the linear perturbation evolution equations in Sec.~\ref{sec:pert_equation_synchronous}. Another useful quantity is 
\begin{equation}
\label{eq:Gamma_rho_m_bkg}
     \langle \Gamma\rho_{\rm \mathsmaller{PBH}} \rangle(\eta) = \int \Gamma(\eta;M_{\rm \mathsmaller{PBH}})\rho_{\rm \mathsmaller{PBH}}(\eta;M_{\rm \mathsmaller{PBH}}) \psi(M_{\rm \mathsmaller{PBH}}) {\rm d}\log M_{\rm \mathsmaller{PBH}},
\end{equation}
where $\Gamma(\eta;M_{\rm \mathsmaller{PBH}})$ is given by Eq.~\eqref{eq:Gamma_def}.

In order to disentangle the coupled differential equations governing the scale factor $a(\eta)$ and the background quantities, we compute $a(\eta)$ under the simplifying assumption of instantaneous PBH reheating \footnote{We have verified numerically that this approximation affects our results only at the percent level.} which occurs at $\langle\eta_{\rm eva}\rangle$.
For $\eta < \left<\eta_{\rm eva}\right>$, the universe is a matter-radiation mixture and $a(\eta)$ is given by Eq.~\eqref{eq:y_eta_matter_radiation} of App.~\ref{app:Transfer_function} (see also Ref.~\cite{Baumann:2022mni}).
For $\eta > \left<\eta_{\rm eva}\right>$, the universe is radiation dominated so $a(\eta)=c_1\eta+c_2$. Fixing the constant $c_1$ and $c_2$ by imposing the continuity of the scale factor and its derivative at $\eta=\left<\eta_{\rm eva}\right>$ leads to
\begin{equation}
\label{eq:scale_factor_full}
a(\eta)=a_{\mathrm{eq}}\left\{\begin{array}{ll}
\left(\frac{\eta}{\eta_\star}\right)^2+2\left(\frac{\eta}{\eta_\star}\right) & \left(\eta \leq \left<\eta_{\mathrm{eva}}\right>\right), \\
\dfrac{2 \eta\left(\left<\eta_{\mathrm{eva}}\right>+\eta_\star\right)-{\left<\eta_{\mathrm{eva}}\right>^2}}{\eta_\star^2} & \left(\eta>\left<\eta_{\mathrm{eva}}\right>\right),
\end{array}\right.
\end{equation}
with $a_{\rm eq}=a_f\rho_{{\rm r},f}/\rho_{{\rm \mathsmaller{PBH}},f}$ and $\eta_\star=\eta_{\rm eq}/(\sqrt{2}-1)$. Consequently, the Hubble factor reads\footnote{It is also customary~\cite{Domenech:2020ssp} to introduce the shifted time variable $\bar{\eta}\equiv \eta-\left<\eta_{\rm eva}\right>/2$ such that the scale factor after evaporation, $\eta>\left<\eta_{\mathrm{eva}}\right>$,  becomes $a/a_{\rm eq}=\left(2\bar{\eta}(\left<\eta_{\rm eva}\right>+\eta_\star)+\left<\eta_{\rm eva}\right>\eta_\star\right)/\eta_\star^2\xrightarrow[\eta_{\rm eq}\to 0]{} 2\bar{\eta} \left<\eta_{\rm eva}\right>/\eta_\star^2 $ and the Hubble rate simplifies to $\mathcal{H}= 2(\left<\eta_{\mathrm{eva}}\right>+\eta_\star)/[2\bar{\eta}(\left<\eta_{\mathrm{eva}}\right>+\eta_\star)+\left<\eta_{\mathrm{eva}}\right>\eta_{\star}] \xrightarrow[\eta_{\rm eq}\to 0]{} \bar{\eta}^{-1} $. \label{footnote:etabar}}
\begin{equation}
\label{eq:Hubble_full}
\mathcal{H}(\eta)=\left\{\begin{array}{ll}
\dfrac{2(\eta+\eta_\star)}{\eta(\eta+2\eta_{\star})} & \left(\eta \leq \left<\eta_{\mathrm{eva}}\right>\right), \\[8pt]
\dfrac{2(\left<\eta_{\mathrm{eva}}\right>+\eta_\star)}{2\eta(\left<\eta_{\mathrm{eva}}\right>+\eta_\star)-\left<\eta_{\mathrm{eva}}\right>^2} & \left(\eta>\left<\eta_{\mathrm{eva}}\right>\right).
\end{array}\right.
\end{equation}
Note that we have 
\begin{align}
    &k_{\rm eq} \equiv \mathcal{H}(\eta_{\rm eq})=\frac{4-2\sqrt{2}}{\eta_{\rm eq}}\simeq \frac{1.17}{\eta_{\rm eq}}, \\
    &k_{\rm eva} \equiv \mathcal{H}(\eta_{\rm eva})=\dfrac{2(\left<\eta_{\mathrm{eva}}\right>+\eta_\star)}{\left<\eta_{\mathrm{eva}}\right>(\left<\eta_{\mathrm{eva}}\right>+2\eta_{\star})} \xrightarrow[\left<\eta_{\mathrm{eva}}\right>/\eta_{\rm eq} \gg 1]{} \frac{2}{\left<\eta_{\mathrm{eva}}\right>} .
\end{align}
\begin{figure}[t]
    \centering
    \includegraphics[width=0.7\linewidth]{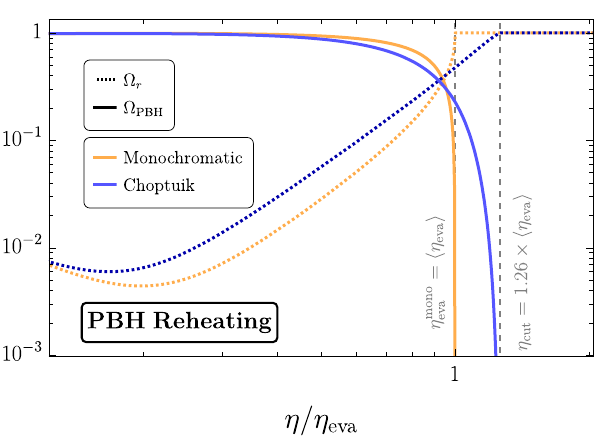}
    \caption{Background energy-density fractions during PBH reheating for monochromatic and extended mass functions following the Choptuik scaling. We show the evolution of the PBH and radiation components, $\Omega_{\rm \mathsmaller{PBH}}$ and $\Omega_r$, as functions of conformal time normalized to the evaporation time, $\eta/\eta_{\rm eva}$. $\langle \eta_{\rm eva}\rangle$ and $\eta_{\rm eva}^{\rm mono}$ coincide, while the cutoff evaporation time $\eta_{\rm cut}$ associated with the cutoff mass of the extended distribution is larger by a factor $\eta_{\rm cut} \simeq1.26\langle \eta_{\rm eva}\rangle$.}
    \label{fig:extended_Mono_reh}
\end{figure}

\subsection{Transfer function before evaporation}
\label{sec:transfer_function}

We now study the linear perturbations in the presence of a PBH gas on scales much smaller than the formation scale, $k\ll k_f$. In the Newtonian gauge, the perturbed FLRW metric reads
\begin{equation}
\label{eq:metric_main}
{\rm d}s^2 = a(\eta)^2\left[-(1 + 2\Phi)\,{\rm d}\eta^2 +\left( (1 - 2\Psi)\delta_{ij}+h_{ij}\right) {\rm d}x^i {\rm d}x^j\right],
\end{equation}
where $\Phi$ and $\Psi$ denote the Newtonian potentials, and $\eta$ is the conformal time. Newtonian gauge is particularly convenient as $\Phi$ coincides with the Newtonian potential in the non-relativistic limit. In the absence of anisotropic stress, these potentials coincide $\Phi = \Psi$~\cite{Baumann:2022mni}. We work in Fourier space 
\begin{equation}
   \Phi(\mathbf{k},\eta) = \int d^3\mathbf{x}\, e^{-i\mathbf{k}\cdot \mathbf{x}} \Phi(\mathbf{x},\eta).
\end{equation}
Using the perturbed Einstein and conservation equations in the Newton gauge and neglecting anisotropic stress ($\Phi=\Psi$), one finds the following evolution equation for $\Phi$ 
\begin{equation}
\label{eq:Phi_eom_short}
\Phi''+3\mathcal{H}(1+c_s^2)\Phi'
+\left[3(c_s^2-\omega)\mathcal{H}^2+c_s^2k^2\right]\Phi
=\frac{3}{2}\mathcal{H}^2\,\delta_{p,{\rm nad}},
\end{equation}
where $\delta_{p,{\rm nad}}\equiv \delta p_{\rm nad}/\overline{\rho}$ and $\delta p_{\rm nad} \equiv \delta p - c_s^2 \delta \rho$ denote the non-adiabatic density contrast and pressure perturbation, respectively.\footnote{
In perturbation theory, energy densities and pressure contributions of the different components are split into a homogeneous background part and a small fluctuation,
$\rho_i(t,\mathbf{x})=\bar\rho_i(t)+\delta\rho_i(t,\mathbf{x})$ and $p_i(t,\mathbf{x})=\bar p_i(t)+\delta p_i(t,\mathbf{x})$.
These matter perturbations enter the Einstein equations through the perturbed energy-momentum tensor and are discussed in more detail in App.~\ref{app:two_fluid_perturbations}.
}
In case of a mixture of perfect fluids, PBH + radiation, the non-adiabatic density contrast reads
\begin{equation}
\label{eq:omega_csSqr_deltapnad}
    \delta_{p,{\rm nad}} =\frac{\left<\Omega_{\rm \mathsmaller{PBH}}\right>\left<\Omega_{\rm r}\right>(1+\omega_{\rm \mathsmaller{PBH}})(1+\omega_{\rm r})}{1+\omega}(c_{\rm \mathsmaller{PBH}}^2-c_{\rm r}^2)S\,,
\end{equation}
where we have introduced the relative background energy fractions
\begin{equation}
    \langle\Omega_{\rm \mathsmaller{PBH}}\rangle\equiv \frac{\langle\rho_{\rm \mathsmaller{PBH}}\rangle}{\langle\rho_{\rm \mathsmaller{PBH}}\rangle+\langle\rho_{\rm r}\rangle}\,, \qquad {\rm and} \qquad   \langle\Omega_{\rm r}\rangle\equiv \frac{\langle\rho_{\rm r}\rangle}{\langle\rho_{\rm \mathsmaller{PBH}}\rangle+\langle\rho_{\rm r}\rangle}\,.
\end{equation}
as well as the effective equation-of-state parameter $\omega\equiv\bar p/\bar\rho$ 
and the adiabatic sound speed $c_s^2\equiv \bar p'/\bar\rho'$.The latter take the simple forms $\omega_{\rm r}=c_{\rm r}^2=1/3$ and $\omega_{\rm \mathsmaller{PBH}}=c_{\rm \mathsmaller{PBH}}^2=0$ during pure radiation and matter eras, respectively. 
The quantity $S$ is the PBH--radiation isocurvature perturbation
\begin{equation}
S \;\equiv\; \delta_{\rm \mathsmaller{PBH}}-\frac{3}{4}\delta_{\rm r}.
\end{equation}
The latter obeys the following evolution equation in Fourier space
\begin{align}
S''+\mathcal{H}S'-\frac{c_s^2 k^2}{1+\omega}\,S
&=\frac{2(c_{\rm r}^2-c_{\rm \mathsmaller{PBH}}^2)}{3(1+\omega)}\frac{k^4}{\mathcal{H}^2}\Phi\,. 
\label{eq:S_eom_short}
\end{align}
We refer to App.~\ref{app:Transfer_function} for a derivation of Eqs.~\eqref{eq:Phi_eom_short} and \eqref{eq:S_eom_short}. Such derivations can also be found in Refs.~\cite{Domenech:2024wao,Zeng:2025tno}, or in the textbook~\cite{Peter:2013avv}.
Numerically solving the system of Eqs.~\eqref{eq:Phi_eom_short} and \eqref{eq:S_eom_short} through the radiation–matter transition shows that $\Phi$ freezes to a constant in the matter era,
\begin{figure}[t]
\centering
\includegraphics[width=0.7\textwidth]{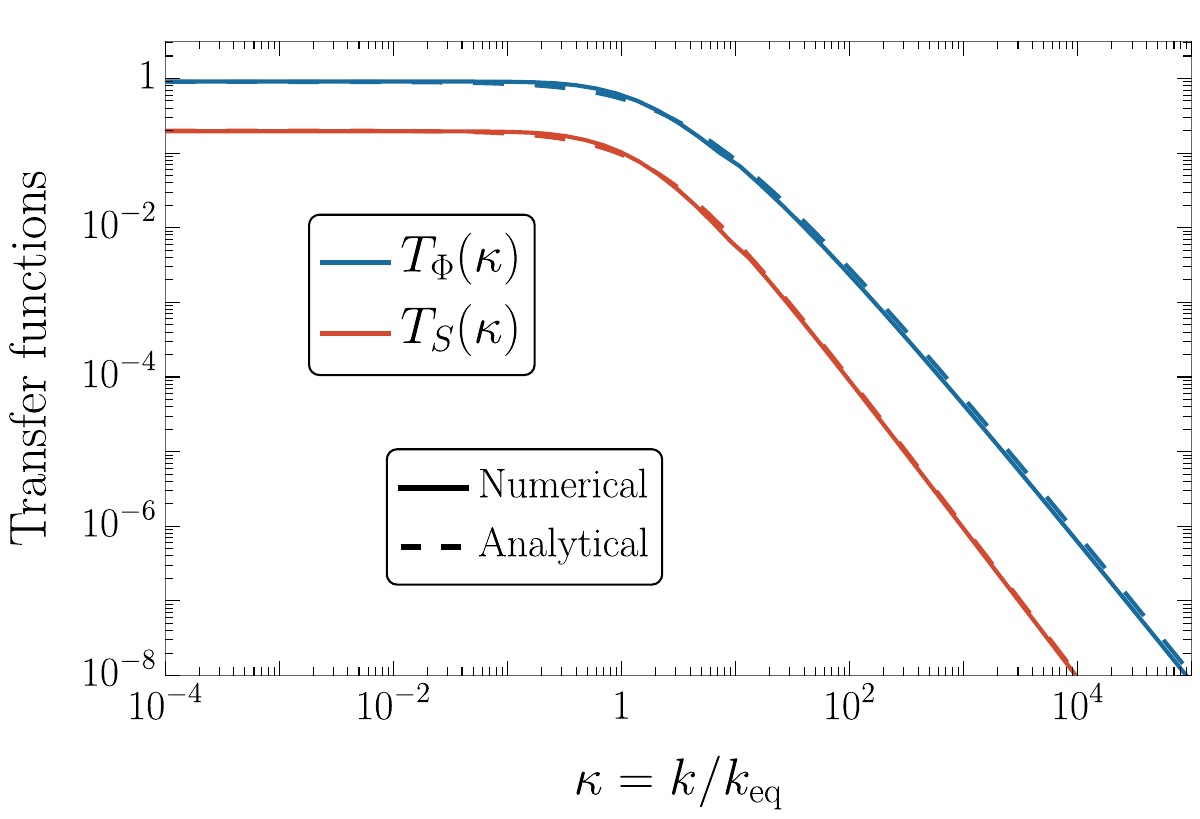}
\caption{\label{fig:Scalar_pert_Transfer_fct_main} Transfer function of the Newtonian potential $\Phi$ assuming adiabatic (\textbf{blue}) and isocurvature  (\textbf{red}) initial conditions. The solid lines are obtained from numerically integrating Eqs.~\eqref{eq:Phi_eom_short} and \eqref{eq:S_eom_short} with the dashed lines correspond to 
Eq.~\eqref{eq:Transfer_fct_def}.}
\end{figure}
\begin{equation}
\label{eq:Transfer_fct_def}
\Phi(k,y\!\gg\!1)=T_\Phi^{\rm (eMD)}(\kappa)\Phi(k,0)+T_S^{\rm (eMD)}(\kappa)S(k,0), 
\qquad \kappa\equiv k/k_{\rm eq},\qquad y\equiv a/a_{\rm eq},
\end{equation}
where the superscript ${\text{``eMD''}}$ indicates that the transfer functions account for effects from the PBH-dominated era but not for effects from the PBH evaporation which are discussed in the next section, Sec.~\ref{sec:pert_equation_synchronous}. 
Fig.~\ref{fig:Scalar_pert_Transfer_fct_main} shows that the numerical solution for the transfer functions can be very well approximated by the semi-analytical formulae (see App.~\ref{app:Transfer_function} and Refs.~\cite{Kodama:1986fg,Hu:1995en} for derivations)\footnote{The analytical expressions in Eq.~\eqref{eq:transfer_adia_iso_main} cease to be valid for $k \ll k_{\rm eva}$, as the corresponding modes enter the horizon during the radiation-dominated era following PBH evaporation. This regime is nevertheless captured by the numerical computation, which explains the discrepancy between the blue solid and dashed curves at $k \lesssim k_{\rm eva}$ in Fig.~\ref{fig:GWSig_UV}.
}
\begin{equation}
\label{eq:transfer_adia_iso_main}
T_\Phi^{\rm (eMD)}(\kappa)\simeq \left(\frac{10}{9}+\frac{0.09\kappa^2}{\ln\left(1+0.13\kappa\right)} \right)^{-1},\qquad 
 T_S^{\rm (eMD)}(\kappa) \simeq \left(5^{1/c}+\kappa^{2/c}\right)^{-c},\qquad c=2.
\end{equation}
The initial values $\Phi(k,0)$ and $S(\mathbf{k},0)$ in Eq.~\eqref{eq:Transfer_fct_def} are treated as Gaussian random fields\footnote{
For a statistically homogeneous field $S$, the power spectrum, bispectrum, and trispectrum are defined through the Fourier-space correlators
\begin{align}
\langle S(\mathbf{k}_1) S(\mathbf{k}_2) \rangle
&=
(2\pi)^3 \delta^{(3)}(\mathbf{k}_1+\mathbf{k}_2)\, P_S(k_1), \\
\langle S(\mathbf{k}_1) S(\mathbf{k}_2) S(\mathbf{k}_3) \rangle_c
&=
(2\pi)^3 \delta^{(3)}(\mathbf{k}_1+\mathbf{k}_2+\mathbf{k}_3)\,
B_S(k_1,k_2,k_3), \\
\langle S(\mathbf{k}_1) S(\mathbf{k}_2) S(\mathbf{k}_3) S(\mathbf{k}_4) \rangle_c
&=
(2\pi)^3 \delta^{(3)}(\mathbf{k}_1+\mathbf{k}_2+\mathbf{k}_3+\mathbf{k}_4)\,
T_S(k_1,k_2,k_3,k_4),
\end{align}
where the subscript $c$ denotes the connected part. For Gaussian fields, all connected correlators beyond the two-point function vanish identically. However, the PBH-induced isocurvature perturbations $S$ are not Gaussian but originate from the Poisson statistics of the discrete PBH number density. As a consequence, connected correlators exist at all orders. In Fourier space, the Poisson process produces scale-independent (shot-noise) contributions,
\begin{equation}
P_S(k)=\frac{1}{\bar n_{\rm \mathsmaller{PBH}}}, \qquad
B_S(k_1,k_2,k_3)=\frac{1}{\bar n_{\rm \mathsmaller{PBH}}^2}, \qquad
T_S(k_1,k_2,k_3,k_4)=\frac{1}{\bar n_{\rm \mathsmaller{PBH}}^3},
\end{equation}
and similarly for higher-point functions. Here $P_S \equiv (2\pi^2/k^3)\mathcal{P}_S$ denotes the dimensionful power spectrum. Expressed in terms of dimensionless spectra and evaluated at wavenumbers $k \ll k_{\rm \mathsmaller{PBH}}\sim \bar n_{\rm \mathsmaller{PBH}}^{1/3}$, corresponding to the inverse mean PBH separation, the bispectrum and trispectrum scale as $(k/k_{\rm \mathsmaller{PBH}})^6$ and $(k/k_{\rm \mathsmaller{PBH}})^9$, respectively. These contributions become relevant only for modes approaching the PBH spacing, where perturbative treatments are expected to break down, and are therefore neglected in the present work.}
and statistically homogeneous with power spectra $\mathcal{P}_{\Phi}(k)$ and $\mathcal{P}_{S}(k)$ defined by
\begin{align}
\label{eq:Def_Power_Spectra_Phi_S}
    \langle \Phi(\mathbf{k},0)\Phi(\mathbf{k}',0)\rangle=  \frac{2\pi^2}{k^3}\mathcal{P}_{\Phi}(k)\delta(\mathbf{k}-\mathbf{k}^\prime),\qquad \langle S(\mathbf{k},0)S(\mathbf{k}',0)\rangle=  \frac{2\pi^2}{k^3}\mathcal{P}_{S}(k)\delta(\mathbf{k}-\mathbf{k}^\prime).
\end{align}
The adiabatic curvature perturbations originate from quantum fluctuations of the inflaton field and are conveniently described in terms of the gauge-invariant comoving curvature perturbation $\mathcal{R}$~\cite{Baumann:2022mni}. Its relation to the Newtonian potential $\Phi$ reads
\begin{equation}
\label{eq:R_vs_Phi}
    -\mathcal{R}
    =
    \Phi + \frac{2}{3}\frac{\mathcal{H}^{-1}\Phi'+\Phi}{1+\omega}
    \xrightarrow[k\eta \ll 1]{}
    \frac{5+3\omega}{3(1+\omega)}\,\Phi,
\end{equation}
where the superhorizon limit follows from $\Phi'\simeq 0$ in the absence of non-adiabatic pressure perturbations. This implies
\begin{equation}
\label{eq:P_Phi_main}
\mathcal{P}_\Phi(k)
=
\left( \frac{3(1+\omega(0))}{5+3\omega(0)}\right)^{2}
\mathcal{P}_{\mathcal{R}}(k)
=
\left( \frac{2}{3}\right)^{2}
\mathcal{A}_{\mathcal{R}}
\left(\frac{k}{k_{\rm CMB}}\right)^{n_s-1},
\end{equation}
 where the normalization is set at $\eta\to 0$, deep in the radiation-dominated era with $\omega(0)=1/3$, prior to PBH domination.\footnote{A generalization to an arbitrary pre-PBH equation of state (e.g.\ kination) can be found in Ref.~\cite{Domenech:2024wao}.} Modes entering the horizon during a PBH-dominated phase ($k<k_{\rm eq}$) instead inherit the standard prefactor $(3/5)^2$. This difference is consistently accounted for by the transfer function in Eq.~\eqref{eq:transfer_adia_iso_main} through a factor $10/9$.
We adopt the usual power-law parametrization for the primordial spectrum,
$\mathcal{P}_{\mathcal{R}}(k)=\mathcal{A}_{\mathcal{R}}(k/k_{\rm CMB})^{n_s-1}$,
with $\mathcal{A}_{\mathcal{R}}\simeq 2.1\times 10^{-9}$, $n_s\simeq 0.96$, and pivot scale $k_{\rm CMB}\simeq 0.05\,{\rm Mpc}^{-1}$, consistent with \textit{Planck} observations~\cite{Planck:2018vyg}. While this corresponds to a nearly scale-invariant spectrum on CMB scales, significant departures may arise at much smaller scales $k\gg k_{\rm CMB}$.
In particular, the discrete (Poisson) nature of PBHs induces an isocurvature component at formation~\cite{Ali-Haimoud:2018dau,Carr:2018rid,Papanikolaou:2020qtd,Gerlach:2025vco}, with power spectrum
\begin{equation}
\label{eq:P_S_main}
\mathcal{P}_{S}(k)
=
\frac{2}{3\pi}
\left(\frac{k}{k_{\rm \mathsmaller{PBH}}}\right)^{3},
\qquad
k_{\rm \mathsmaller{PBH}}
=
a_f\left(\frac{4\pi}{3}n_{\rm \mathsmaller{PBH}}(t_f)\right)^{1/3},
\end{equation}
where $k_{\rm \mathsmaller{PBH}}^{-1}$ corresponds to the typical comoving separation between neighboring PBHs at formation.

\subsection{Suppression factor after evaporation}
\label{sec:pert_equation_synchronous}

\begin{figure}[t]
    \centering
\includegraphics[width=0.48\linewidth]{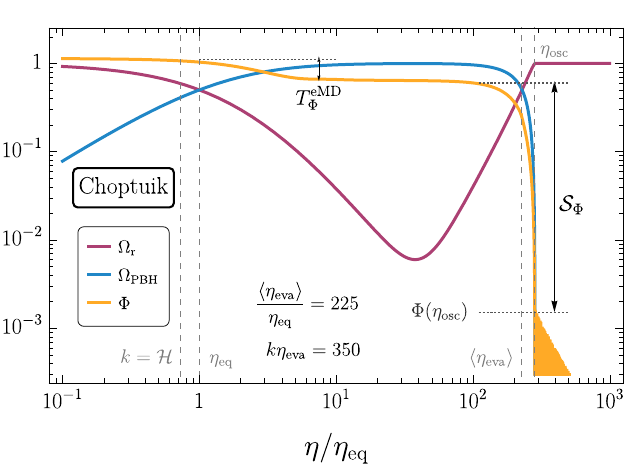}
\includegraphics[width=0.48\linewidth]{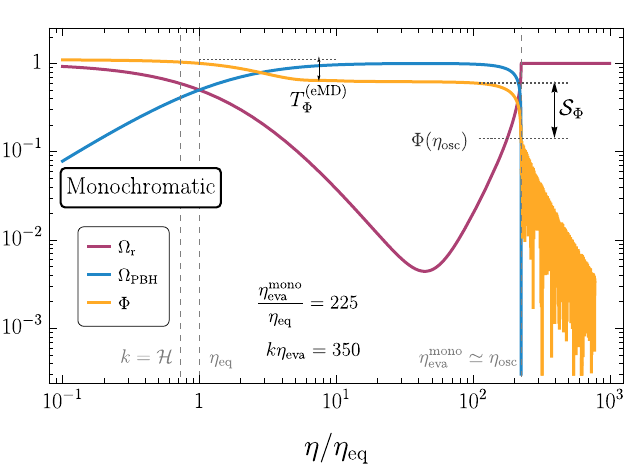}
    \caption{Shown is the evolution of energy densities ($\Omega_{\rm \mathsmaller{PBH}}$, light blue) and ($\Omega_r$, dark blue) during PBH domination. The grey curve shows $\Phi_k$ (normalized to $10/9$ in RD), with the RD-to-MD transition governed by $\mathcal{T}$, and the MD-to-RD transition from PBH evaporation by $\mathcal{S}$.  We choose $\langle\eta_{\rm eva}\rangle/\eta_{\rm eq}=225$ here and in all future plots for better comparability with \cite{Inomata:2020lmk}. \textbf{Left:} Mass distribution following the Choptuik scaling. \textbf{Right:} Monochromatic mass distribution.}
\label{fig:pbhepoch_longpaper}
\end{figure}

In the presence of linear metric perturbations, the PBH decay rate as measured by comoving observers, $\Gamma_{\rm obs}(\vec{x})$, may acquire a dependence on the position $\vec{x}$. Indeed, comoving observers with $x^i=\text{const}$ measure a proper conformal time $d\eta_c^2 = (1+2\Phi)\,d\eta^2
$. Hence, they observe PBH to decay with a rate $\Gamma_{\rm obs}=d\log{M_{\rm \mathsmaller{PBH}}}/d\eta_c =\Gamma(1-\Phi)$ where $\Gamma=d\log{M_{\rm \mathsmaller{PBH}}}/d\eta$. Additionally, if the PBH fluid has a non-vanishing velocity with respect to the observer, then the decay rate needs to be Lorentz transformed as well.
However, the effect of PBH decay on linear perturbations can be simplified by working in the synchronous gauge, where the metric takes the form~\cite{Audren:2014bca,Poulin:2016nat,Inomata:2020lmk}
\begin{equation}
    {\rm d}s^2 = a^2\left({\rm d}\eta^2 - H_{ij} {\rm d}x^i {\rm d}x^j\right).
\end{equation}
Here, $H_{ij} = \delta_{ij} + \hat{k}_i \hat{k}_j h + \left(\hat{k}_i \hat{k}_j - \frac{1}{3} \delta_{ij}\right) 6\varepsilon+\frac{1}{2}h_{ij}$, with $\hat{k}_i \equiv k_i/|k|$. The quantities $h$ and $\varepsilon$ are the trace and traceless component of the scalar mode in the synchronous gauge.\footnote{$h$ and $\varepsilon$ corresponds $h$ and $\eta$ in~\cite{Ma:1995ey,Audren:2014bca,Poulin:2016nat} and  to $\gamma$ and $\varepsilon$ in~\cite{Inomata:2020lmk}.} For simplicity, we neglect sources of anisotropic stress.
Linearized Einstein equations in the synchronous gauge read~\cite{Ma:1995ey,Audren:2014bca,Poulin:2016nat,Inomata:2020lmk}:
\begin{align}
\label{eq:syncrhonous_gauge_eq_first}
    \delta_{\rm \mathsmaller{PBH}}^{\prime} &= -\frac{h^{\prime}}{2} \\
    \delta_{\mathrm{r}}^{\prime} &= -\frac{4}{3} \left(\theta_{\mathrm{r}} + \frac{h^{\prime}}{2}\right) +  \frac{\left<\Gamma\rho_{\rm \mathsmaller{PBH}}\right>}{\left<\rho_{\mathrm{r}}\right>} 
    \label{eq:syncrhonous_gauge_eq_second}
    \left(\delta_{\rm \mathsmaller{PBH}} - \delta_{\mathrm{r}}\right), \\
    \theta_{\mathrm{r}}^{\prime} &= \frac{k^2}{4} \delta_{\mathrm{r}} - \frac{\left<\Gamma\rho_{\rm \mathsmaller{PBH}}\right>}{\left<\rho_{\mathrm{r}}\right>} \theta_{\mathrm{r}},    \label{eq:syncrhonous_gauge_eq_third}\\
    k^2 \varepsilon-\frac{1}{2} \frac{a^{\prime}}{a} h^{\prime} & =-\frac{3}{2} \mathcal{H}^2\left(\frac{\left<\rho_{\rm \mathsmaller{PBH}}\right>}{\left<\rho_{\mathrm{tot}}\right>} \delta_{\rm \mathsmaller{PBH}}+\frac{\left<\rho_{\mathrm{r}}\right>}{\left<\rho_{\mathrm{tot}}\right>} \delta_{\mathrm{r}}\right),    \label{eq:syncrhonous_gauge_eq_four} \\
    \label{eq:syncrhonous_gauge_eq_last}
k^2 \varepsilon^{\prime} & =2 \mathcal{H}^2 \frac{\left<\rho_{\mathrm{r}}\right>}{\left<\rho_{\mathrm{tot}}\right>} \theta_{\mathrm{r}},
\end{align}
where $\theta \equiv \partial_i v^i$ is the velocity divergence.
The initial conditions for the perturbations  are~\cite{Inomata:2020lmk}:
\begin{align}
& \delta_{\mathrm{r}}=-\frac{2}{3} C(k \eta)^2,\quad \delta_{\rm \mathsmaller{PBH}}=\frac{3}{4} \delta_{\mathrm{r}}, \quad \theta_{\mathrm{r}}=-\frac{1}{18} C\left(k^4 \eta^3\right), \\
& h=C(k \eta)^2, \quad \varepsilon=2 C-\frac{1}{18} C(k \eta)^2,
\end{align}
where $C= \mathcal{R}/2$.
These hold for RD, which is why we start the numerical evolution long before the onset of PBH domination. To translate back to Newtonian gauge, we use~\cite{Ma:1995ey,Poulin:2016nat,Inomata:2020lmk}: 
\begin{equation}
\label{eq:Phi_synchronous}
    \Phi(k,\eta) = \varepsilon - \mathcal{H} \frac{(6\varepsilon + h)'}{2k^2}.
\end{equation}
We refer to 
App.~\ref{app:Phi_PBH_evaporation} for more details on the derivation of linear perturbations in synchronous gauge and for a formulation in Newtonian gauge as well.
Solving for the system of Eqs.~\eqref{eq:Phi_eom_short} and \eqref{eq:S_eom_short} followed by Eqs.~(\ref{eq:syncrhonous_gauge_eq_first}-\ref{eq:syncrhonous_gauge_eq_last}), with $\left<\rho_{\rm \mathsmaller{PBH}}\right>$, $\left<\rho_r\right>$, and $\Gamma\left<\rho_{\rm \mathsmaller{PBH}}\right>$ determined by Eqs.~\eqref{eq:rho_m_bkg}, \eqref{eq:rho_r_bkg}, and \eqref{eq:Gamma_rho_m_bkg}, we obtain the evolution of the scalar potential $\Phi$ in Eq.~\eqref{eq:Phi_synchronous} during both the transition from radiation to matter when PBHs dominate at $\eta_{\rm eq}$ and from matter back to radiation when PBHs evaporate at $\langle\eta_{\rm eva}\rangle$. We show the results and differences for monochromatic and extended mass functions in Fig.~\ref{fig:pbhepoch_longpaper}.

The amplitude of the scalar field $\Phi_{\rm osc}$, right before it starts oscillating, just after PBH evaporation, can be decomposed into two distinct contributions,

\begin{figure}[t!]
\centering
\raisebox{0cm}
{\makebox{\includegraphics[width=0.7\textwidth, scale=1]{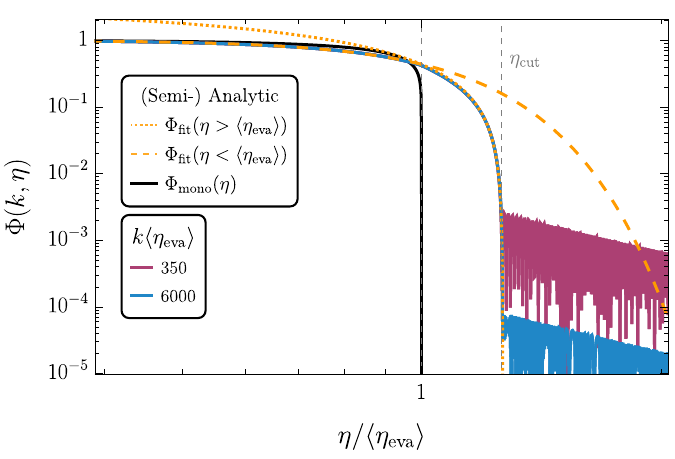}}}\hspace{1cm}
\caption{Scalar evolution during PBH reheating. The figure displays the evolution of the Newtonian potential for different momentum modes (blue, green) as well as the best fit (orange) describing the universal evolution prior to the onset of oscillation. For comparison, the analytic approximation and neglecting decoupling for the Newtonian potential in the case of monochromatic mass distribution (black) is shown.}
\label{fig:phi_fit+suppre}
\end{figure}
\begin{equation}
\label{eq:Phi_osc}
\Phi_{\rm osc}(k)\equiv \Phi(k,\eta_{\rm osc})=\left[T_\Phi^{\rm (eMD)}(k)\Phi(k,0)+T_S^{\rm (eMD)}(k)S(k,0)\right]\,\mathcal{S}_\Phi(k)\,,
\end{equation}
where $\Phi(k,0)$ and $S(k,0)$ are the primordial seeds.
The transfer functions $T_\Phi^{\rm (eMD)}(k)$ and $T_S^{\rm (eMD)}(k)$, given in Eq.~\eqref{eq:Transfer_fct_def}, capture the decay of modes that entered the horizon during the radiation-dominated (RD) era, prior to the onset of PBH domination (eMD). Once the eMD phase begins, these modes effectively freeze until PBH evaporation. The quantity $\mathcal{S}_\Phi(k)$ is the \textit{suppression factor} which describes the decay of $\Phi$ during PBH evaporation phase. 
The physical origin of the decay encapsulated in $\mathcal{S}_\Phi(k)$ is clear from examining the Poisson equation in the subhorizon limit \cite{Inomata:2020lmk}:
\begin{equation}
\label{eq:poisson_subhorizon}
k^2 \Phi \simeq \frac{3}{2} \mathcal{H}^2\left(\frac{\left<\rho_{\rm \mathsmaller{PBH}}\right>}{\left<\rho_{\mathrm{tot}}\right>} \delta_{\mathrm{m}}+\frac{\left<\rho_{\mathrm{\rm r}}\right>}{\left<\rho_{\mathrm{tot}}\right>} \delta_{\mathrm{r}}\right),
\end{equation}
where $\delta_{\rm \mathsmaller{PBH}}$ and $\delta_{\rm r}$ denote the PBH and radiation density contrasts.
During the eMD, one has $\delta_{\rm \mathsmaller{PBH}}\propto a$ while $\delta_{\rm r}$ remains approximately constant (and oscillatory), so the PBH contribution dominates and $\Phi$ remains on an almost constant plateau. As Hawking evaporation reduces $\rho_{\rm m}$, the two source terms become comparable, driving $\Phi$ to follow the oscillatory behaviour of $\delta_{\rm r}$.

During evaporation, and before it starts oscillating at a time $\eta_{\rm osc}(k)$, the potential in Eq.~\eqref{eq:poisson_subhorizon} follows the PBH
fraction $\Omega_m\equiv \rho_{\rm \mathsmaller{PBH}}/\rho_{\rm tot}$,
\begin{equation}
\label{eq:Phi_ana_main}
\Phi(k,\eta)\ \propto\ \langle\Omega_{\rm m}(\eta)\rangle,\qquad \textrm{for}\quad \eta\lesssim \eta_{\rm osc}(k),
\end{equation}
so that the suppression factor can be estimated as
\begin{equation}
\label{eq:supp_combined_main}
\mathcal{S}_\Phi(k)
\simeq
\frac{\Phi(k,\eta_{\rm osc})}{\Phi(k,\eta\ll\eta_{\rm osc})}
\simeq
\langle\Omega_{\rm m}(\eta_{\rm osc})\rangle,
\end{equation}
where $\langle\Omega_{\rm m}\rangle$ is the PBH energy fraction averaged over the extended PBH
mass function obeying the Choptuik scaling
\begin{equation}
\label{eq:rho_avg_0_main}
\langle\Omega_{\rm m}\rangle(\eta)
=
\int\!\frac{{\rm d}M}{M}\,\psi(M)\,\Omega_{\rm m}(\eta;M)
=
\frac{1+\gamma_{\rm M}}{\gamma_{\rm M}}
\int_{M(\eta)}^{M_{\rm cut}}\!\frac{{\rm d}M}{M}
\left(\frac{M}{M_{\rm cut}}\right)^{1+1/\gamma_{\rm M}}
\Bigl(1 - \frac{t(\eta)}{t_{\rm eva}(M)}\Bigr)^{1/3}.
\end{equation}
We exploit $t_{\rm eva}\propto M^3$ and $t\propto\eta^3$ to change variables to
$\eta_{\rm eva}\propto M$, and evaluate the resulting integral in closed form
with \textsc{Mathematica} (see App.~\ref{app:suppression_fac} for more details)
\begin{equation}
\label{eq:fully_analytic_sol_rhom}
\langle \Omega_{\rm m} \rangle = -(-1)^{2/3}(1+\gamma_{\rm M})\left.\frac{\eta}{\eta_{\rm eva}}
\left(\frac{\eta_{\rm eva}}{\eta_{\rm cut}}\right)^{\! 1+\frac{1}{\gamma_{\rm M}}}
\!\!{}_{2}F_1\left( -\frac{1}{3}, \frac{1}{3\gamma_{\rm M}}; 1 + \frac{1}{3\gamma_{\rm M}};\frac{\eta_{\text{eva}}^3}{\eta^3} \right)
\right|_{\eta_{\rm eva}=\eta}^{\eta_{\rm eva}=\eta_{\rm cut}},
\end{equation}
where ${}_2F_1$ denotes the hypergeometric function and 
\begin{equation}
\label{eq:eta_cut_def}
    \eta_{\rm cut}\equiv \frac{1+2\gamma_{\rm M}}{1+\gamma_{\rm M}}\left<\eta_{\rm eva}\right>,
\end{equation} 
with $\eta_{\rm cut}\simeq 1.26\left<\eta_{\rm eva}\right>$ for $\gamma_{\rm M}\simeq 0.36$.

\begin{figure}[t!]
\centering
\raisebox{0cm}
{\makebox{\includegraphics[width=0.7\textwidth, scale=1]{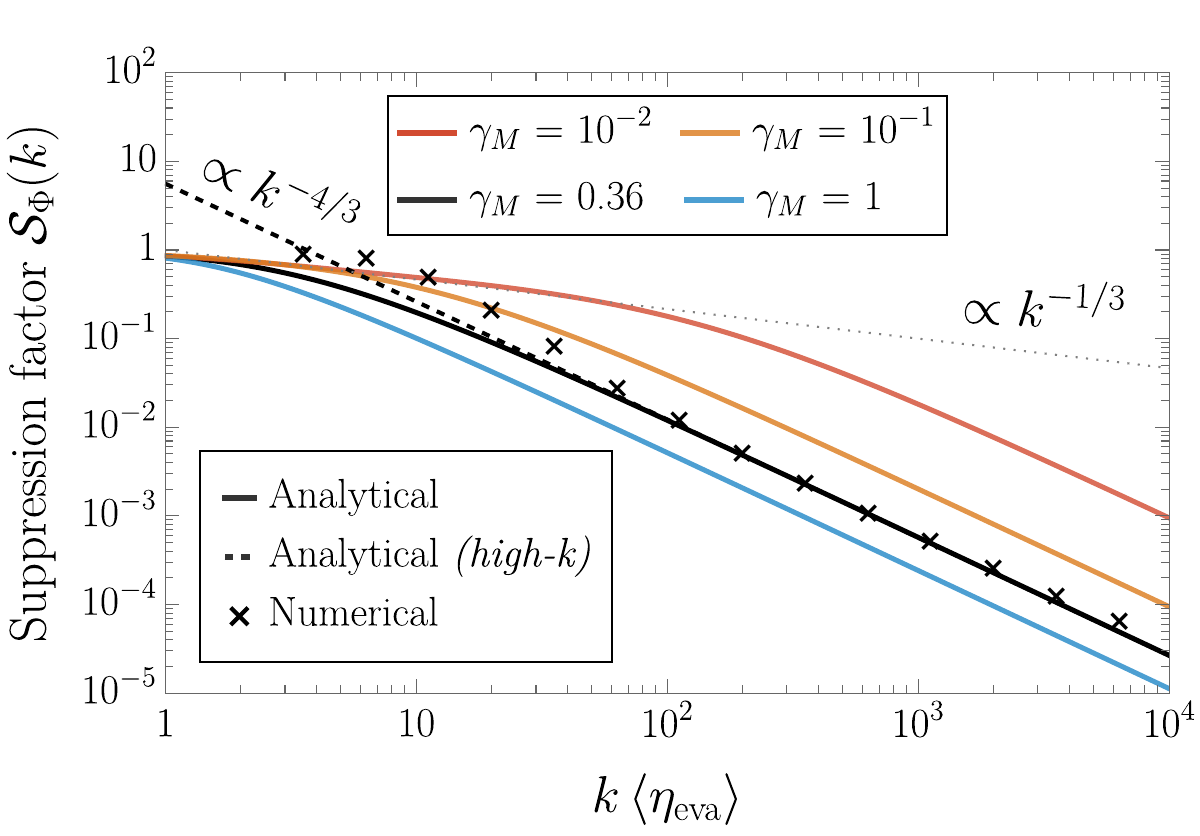}}}
\caption{Suppression factor $\mathcal{S}_\Phi(k)$ of the Newtonian potential $\Phi$ for different values of the 
critical exponent $\gamma_{\rm M}$ entering the Choptuik's PBH mass distribution $\psi(M)$ in 
Eq.~\eqref{eq:psi_f_maintext}.  
In the sub-horizon limit $k\langle\eta_{\rm eva}\rangle \gg 1$, the analytical estimate in 
Eq.~\eqref{eq:supp_combined_main} --- with $\eta_{\rm osc}$ determined from 
Eq.~\eqref{eq:dec_limit_first_derivative_main} --- (\textbf{solid lines}) and the analytical estimate in 
Eq.~\eqref{eq:suppression_factor_final} (\textbf{dashed line}) show excellent agreement with the 
numerical computation (\textbf{black crosses}) obtained from Eq.~\eqref{eq:Phi_synchronous}.  
For the value $\gamma_{\rm M}\simeq 0.36$ predicted by General Relativity 
\cite{Choptuik:1992jv,Abrahams:1993wa,Evans:1994pj,Maison:1995cc,Neilsen:1998qc,Koike:1999eg,
Gundlach:2007gc,Ianniccari:2024ltb}, we find the asymptotic behaviour $\mathcal{S}_\Phi(k)\propto k^{-4/3}$.  
In the limit $\gamma_{\rm M}\to 0$, corresponding to a monochromatic PBH mass function, we recover 
the scaling $\mathcal{S}_\Phi(k)\propto k^{-1/3}$  
in agreement with Ref.~\cite{Inomata:2020lmk}.  
However, for $k\langle\eta_{\rm eva}\rangle \gtrsim \gamma_{\rm M}^{-1}$ the suppression transitions 
back to the asymptotic behaviour $\mathcal{S}_\Phi(k)\propto k^{-4/3}$.
}
\label{fig:alpha_vs_gamma_M}
\end{figure}

The remaining ingredient is the oscillation time $\eta_{\rm osc}(k)$, defined as the
moment when $\Phi$ starts oscillating, also called decoupling time in Ref.~\cite{Inomata:2020lmk}. The latter can be determined as the time when the derivative $\Phi'$ equates the gradient term $k\Phi$ in the evolution equation,
\begin{equation}
\label{eq:dec_limit_first_derivative_main}
\left|\frac{\Phi'}{\Phi}\right|
\simeq \sqrt{\frac{2}{3}}\,k
\qquad\text{at}\qquad \eta=\eta_{\rm osc}(k).
\end{equation}
In App.~\ref{app:suppression_fac}, we show that the criterion in Eq.~\eqref{eq:dec_limit_first_derivative_main} gives better results than the one proposed in Ref.~\cite{Inomata:2020lmk} that relies on the second derivative of $\Phi$, see in particular Fig.~\ref{fig:alpha_vs_gamma_M_app} for a comparison of the two approaches.
Using the asymptotic form of
$\langle\Omega_{\rm m}\rangle$ near the evaporation endpoint $\eta\to \eta_{\rm cut}^-$, we find
$\Phi\propto(\eta_{\rm cut}-\eta)^{4/3}$ and solve
Eq.~\eqref{eq:dec_limit_first_derivative} for $\eta_{\rm osc}$, obtaining
$\eta_{\rm cut}-\eta_{\rm osc}\propto k^{-1}$.  Substituting this into
Eq.~\eqref{eq:suppression_factor_ana} leads to the high–$k$ limit
\begin{equation}
\label{eq:suppression_factor_final_main}
{\it Choptuik:}\qquad \qquad   \mathcal{S}_\Phi(k)
    \simeq 
    \kappa_0 \left(\frac{k_{\rm eva}}{k}\right)^{4/3},
    \qquad
    \kappa_0 \equiv
    \frac{3^{2/3}(1+\gamma_M)^{7/3}}{2^{4/3}\gamma_M(1+2\gamma_M)^{4/3}},
\end{equation}
with $\kappa_0\simeq 2.3$ for the value of the Choptuik exponent predicted by GR,
$\gamma_{\rm M}\simeq 0.36$. This
$\mathcal{S}_\Phi(k)\propto k^{-4/3}$ scaling for a Choptuik-mass-distribution is one of the main results of this section. 

In Fig.~\ref{fig:alpha_vs_gamma_M}, we compare the high-$k$ limit in 
Eq.~\eqref{eq:suppression_factor_final_main} with the analytical estimate in 
Eq.~\eqref{eq:supp_combined_main}, as well as with the 
fully numerical determination of $\mathcal{S}_\Phi(k)$.  
The latter is obtained by measuring the decrease of $\Phi$ from its plateau value 
down to its value at the onset of the first oscillation.  
Here $\Phi$ is computed from Eq.~\eqref{eq:Phi_synchronous} after numerically 
integrating Eqs.~(\ref{eq:syncrhonous_gauge_eq_first}--\ref{eq:syncrhonous_gauge_eq_last}), 
with $\langle\rho_m\rangle$, $\langle\rho_r\rangle$, and 
$\langle\Gamma\rho_m\rangle$ determined from 
Eqs.~\eqref{eq:rho_m_bkg}--\eqref{eq:Gamma_rho_m_bkg}, averaged over $\psi(M_{\rm \mathsmaller{PBH}})$ in Eq.~\eqref{eq:psi_f_maintext} with $\gamma_{\rm M}\simeq 0.36$.  
We have verified that the numerical extraction of $\mathcal{S}_\Phi(k)$ is independent of the 
duration of the matter era, $\eta_{\rm eva}/\eta_{\rm eq}$.  
This is expected, since the dependence of $\Phi$ on $\eta_{\rm eva}/\eta_{\rm eq}$ appears 
only through the transfer functions $T_\Phi$ and $T_S$ in Eq.~\eqref{eq:Phi_osc}, 
as illustrated in Fig.~\ref{fig:Scalar_pert_Transfer_fct_main}.  
We find that the numerical solution for $\Phi(\eta)$ before it starts oscillating is well approximated by 
the following fit (see Fig.~\ref{fig:phi_fit+suppre}, left panel):
\begin{equation}
\label{eq:Phi_fit}
F_\Phi(\eta)\equiv \frac{\Phi(k,\eta)}{\Phi(k,\eta\ll\eta_{\rm eva})} \simeq
\begin{cases}
\exp\!\left[-0.84\left(\dfrac{\eta}{\langle\eta_{\rm eva}\rangle}\right)^{3.43}\right], & 
\bigl(\eta < \langle\eta_{\rm eva}\rangle\bigr), \\[0.4cm]
3.5\left(\dfrac{\eta_{\rm cut}-\eta}{\eta_{\rm cut}}\right)^{4/3}, &
\bigl(\langle\eta_{\rm eva}\rangle\leq \eta \leq \eta_{\rm osc}(k)\bigr).
\end{cases}
\end{equation} 
which captures both the early plateau and the late–time decay, and reproduces
the numerically extracted suppression factor to good accuracy. Note that $\mathcal{S}_{\Phi}(k)\equiv F_\Phi(k,\eta_{\rm osc})$, see Eq.~\eqref{eq:supp_combined_main}.

Finally, the same formalism reproduces the known monochromatic limit.
Taking $\gamma_{\rm M}\to 0$ in the analytic expression for
$\langle\Omega_{\rm m}\rangle$ yields
\begin{equation}
\label{eq:supp_mono_main}
{\it Mono.:}\qquad \qquad  \mathcal{S}_\Phi(k)\xrightarrow[\gamma_{\rm M}\to 0]{}
\left(\frac{\sqrt{12}}{k\langle\eta_{\rm eva}\rangle}\right)^{1/3},
\end{equation}
in agreement with Ref.~\cite{Inomata:2020lmk}.
However, as shown in Fig.~\ref{fig:alpha_vs_gamma_M}, the monochromatic scaling $\mathcal{S}_\Phi(k)\propto k^{-1/3}$ transitions to
the Choptuik-mass-function scaling $\mathcal{S}_\Phi(k)\propto k^{-4/3}$ behaviour once
\begin{equation}
\label{eq:transition_mono_ext_main}
k\langle\eta_{\rm eva}\rangle
~\gtrsim~
k_{\rm trans}\langle\eta_{\rm eva}\rangle
\equiv
\frac{\sqrt{3}}{2^{1/3}\gamma_{\rm M}}.
\end{equation}
We refer the reader to App.~\ref{app:suppression_fac} for a deeper analytical understanding of the transition wavenumber in Eq.~\eqref{eq:transition_mono_ext_main}. Note that $\gamma_M$ enters only through the prefactor $\mathcal{S}_\Phi$, encoded in $\kappa_0$ in Eq.~\eqref{eq:suppression_factor_final_main}. By contrast, the asymptotic scaling $\mathcal{S}_\Phi\propto k^{-4/3}$ is independent of the precise value of $\gamma_M$. The latter only sets the transition scale, given in Eq.~\eqref{eq:transition_mono_ext_main}, below which the monochromatic scaling $\mathcal{S}_\Phi\propto k^{-1/3}$ is recovered, see Fig.~\ref{fig:alpha_vs_gamma_M}.

\clearpage
\section{The SIGW formalism}
\label{chap:basics_SIGW_main}
\subsection{From scalar sources to GWs}

At first order in cosmological perturbation theory, scalar, vector, and tensor perturbations decouple. However, at second order, scalar perturbations act as sources for tensor perturbations, thereby generating GWs~\cite{Ananda:2006af,Baumann:2007zm,Kohri:2018awv,Espinosa:2018eve}. These scalar-induced gravitational waves (SIGWs) arise inevitably at quadratic order in scalar perturbations, even in the absence of primordial tensor modes. Solving the equation of motion (EOM) for the tensor modes $h_{ij}$, in the Newtonian gauge defined by Eq.~\eqref{eq:metric_main}, yields
\begin{equation}
\label{eq:tensor_eom}
h''_{ij}(\mathbf{x}, \eta) + 2\mathcal{H} h'_{ij}(\mathbf{x}, \eta) - \Delta h_{ij}(\mathbf{x}, \eta) = \mathcal{P}_{ij}^{ab} \mathcal{S}_{ab}(\mathbf{x}, \eta),
\end{equation}
where $\mathcal{H} = aH$ is the conformal Hubble parameter, $\mathcal{P}_{ij}^{ab}$ denotes the transverse-traceless projector with $a,b,i,j\in \{1,2,3\}$, and $\mathcal{S}_{ab}$ represents the source term quadratic in scalar perturbations ($\Phi$ and its derivatives)~\cite{Kohri:2018awv,Domenech:2021ztg}:
\begin{equation}
\label{eq:source_term_0}
    \mathcal{S}_{ab}(\mathbf{x},\eta)= 4\left[\partial_a\Phi\partial_b\Phi + \frac{2}{3(1+\omega)}\partial_a(\Phi+\Phi'/\mathcal{H})\partial_b(\Phi+\Phi'/\mathcal{H})\right],
\end{equation}
where an underlying perfect fluid with equation-of-state parameter $\omega$ is assumed. The scalar and tensor fluctuations can be decomposed into Fourier modes
\begin{equation}
\label{eq:Fourier_transform_Phi}
\Phi(\mathbf{x},\eta) = \int \frac{d\mathbf{k}^3}{(2\pi)^3} e^{i\mathbf{k}\cdot \mathbf{x}}\Phi(\mathbf{k},\eta) ,
\qquad
h_{ij}(\mathbf{x},\eta)=\int\frac{d \mathbf{k}^3}{(2\pi)^3}\,e^{i\mathbf{k}\cdot \mathbf{x}}\,h_{ij}(\mathbf{k},\eta).
\end{equation}
The tensor modes can be additionally decomposed into a sum over the two transverse polarizations labeled by $\lambda = \{\times, +\}$
\begin{equation}
    h_{ij}(\mathbf{k},\eta)=\sum_\lambda e^\lambda_{ij}(\mathbf{k})\,h_{\lambda}(\mathbf{k}, \eta),
\end{equation}
with the polarisation tensors $e_{ij}^{\lambda}$ forming an orthonormal basis $e_{ij}^{\lambda}e_{ij}^{\tilde{\lambda}}=\delta_{\lambda\tilde{\lambda}}$, and satisfying $\mathcal{P}_{ij}^{ab}e_{ij}^{\lambda}=e_{ab}^{\lambda}$. 
As we discussed in Sec.~\ref{sec:transfer_function}, the initial condition for the scalar potential can arise from both adiabatic and isocurvature perturbations, which we denote by $\Phi(k,0)$ and $S(k,0)$, respectively. Since the evolution equations are linear, the system can be decomposed into a linear combination of an initially adiabatic system and an initially isocurvature system
\begin{equation}
\label{eq:generic_transfer_fct}
\Phi(\mathbf{k},\eta) = T_\Phi(k,\eta)\Phi(k,0)+T_S(k,\eta)S(k,0),
\end{equation}
where the transfer functions  $T_\Phi$ and $T_{S}$ describe the evolution of $\Phi(\mathbf{k},\eta)$ assuming adiabatic and isocurvature initial conditions, respectively. 
We suppose that the two seeds are uncorrelated
\begin{equation}
\label{eq:S_Phi_unroorelated}
    \left<\Phi(k,0)\, S(k,0) \right>=0.
\end{equation}
The Fourier transform of $\mathcal{S}_{ab}(\mathbf{x},\eta)$ in Eq.~\eqref{eq:source_term_0} projected on the $\lambda$ basis reads~\cite{Domenech:2021ztg}
\begin{equation}
\label{eq:source_f_X}
   \mathcal{S}_\lambda(\mathbf{k},\eta) =4 \int \frac{d^3 q}{(2 \pi)^3} e_\lambda^{i j}(k) q_i q_j X_{\mathbf{q}} X_{\mathbf{k}-\mathbf{q}} f_X\left(\eta, \frac{q}{k},\frac{|\mathbf{k}-\mathbf{q}|}{k}\right),\qquad  X=\{\Phi,S\},
\end{equation}
with the source function being given by\footnote{\label{footnote:convention_KT}The definition of the source function $f_X$ in Eq.~\eqref{eq:source_term} differs from the one in Ref.~\cite{Kohri:2018awv}, denoted $f_X^{\rm KT}$, by a factor 
\begin{equation}
    f_X^{\rm KT}~=~2 \left(\frac{3 + 3 \omega(0)}{5 + 3 \omega(0)}\right)^2f_X~=~\frac{8}{9}f_X.
\end{equation} 
The extra factor of \(2\) in their \(f_\Phi^{\rm KT}\) yields a factor of \(4\) in \(\Omega_{\rm GW}\), which is exactly compensated by a factor of \(1/4\) arising from their convention for \(\mathcal{P}_h\) and its relation to \(\Omega_{\rm GW}\), compared with our definition in Eq.~\eqref{eq:tensorpowerspectrum_general}. The factor \(\left(\frac{3+3\omega(0)}{5+3\omega(0)}\right)^2\) with $\omega(0)=1/3$ instead comes from the different normalization choice, since Ref.~\cite{Kohri:2018awv} is written in terms of \(\mathcal P_{\mathcal R}\) whereas we use \(\mathcal P_\Phi\), normalized to the early radiation era, preceding PBH domination, see Eq.~\eqref{eq:P_Phi_main}. }
\begin{equation}
\label{eq:source_term}
\begin{aligned}
f_X\left(\eta, \frac{q}{k},\frac{|\mathbf{k}-\mathbf{q}|}{k}\right)= & ~T_X(q \eta) T_X(|\mathbf{k}-\mathbf{q}| \eta) \\
& +\frac{2}{3(1+\omega)}\left(T_X(q \eta)+\frac{T_X^{\prime}(q \eta)}{\mathcal{H}}\right)\left(T_X(|\mathbf{k}-\mathbf{q}| \eta)+\frac{T_X^{\prime}(|\mathbf{k}-\mathbf{q}| \eta)}{\mathcal{H}}\right).
\end{aligned}
\end{equation} 
The differential equation for the tensor modes $h$ in Eq.~\eqref{eq:tensor_eom}, in Fourier space and projected on the $\lambda$ basis, can be solved by Green's method~\cite{Domenech:2019quo}
\begin{equation}
\label{eq:tensor_from_greens_function}
    h_{\lambda}(\mathbf{k}, \eta)=\int_0^{x= k \eta}{\rm d}\overline{x}\,G_h(x,\overline{x})\,\frac{\mathcal{S}_{\lambda}(\mathbf{k},\overline{x})}{k^2},
\end{equation}
where $x \equiv k\eta$ and $\bar{x} \equiv k\bar{\eta}$. The Green’s function $G_h$ is defined by the inhomogeneous differential equation
\begin{equation}
\label{eq:green_function_diff_eq}
G_h^{\prime \prime}(x,\bar{x}) + 2\frac{\mathcal{H}}{k}\,G_h^{\prime}(x,\bar{x}) + G_h(x,\bar{x})
= \delta(x-\bar{x}) \, .
\end{equation}
and for $\mathcal{H}= 2/\left[(1+3\omega )\eta\right]$ with constant $\omega$ reads~\cite{Domenech:2021ztg,Domenech:2025ffb}\footnote{\label{footnote:green_fct}We use the same convention for $G_h(x,\tilde{x})$ as in Refs.~\cite{Domenech:2019quo,Domenech:2025ffb} (defined in Eq.~\eqref{eq:hgreen}). This is related to $G_h(\eta,\tilde{\eta})$ in Ref.~\cite{Domenech:2021ztg} by $G_h(x,\bar{x}) = k\,G_h(\eta,\bar{\eta})$ and to $G_k(\eta,\bar{\eta})$ in Ref.~\cite{Kohri:2018awv,Inomata:2019ivs,Inomata:2019zqy}  by $G_h(x,\bar{x}) = \frac{a(\bar{\eta})}{a({\eta})}k\,G_h(\eta,\bar{\eta})$.\label{footnote:green_fct_notation}}
\begin{equation}
\label{eq:hgreen}
G_h(x,\bar{x}) = \frac{\pi}{2}\,\frac{\bar{x}^{\,b+3/2}}{x^{\,b+1/2}}
\left[
J_{b+1/2}(\bar{x})\,Y_{b+1/2}(x)
-
J_{b+1/2}(x)\,Y_{b+1/2}(\bar{x})
\right]
\Theta(x-\bar{x}) \, ,
\end{equation}
where $b \equiv (1-3\omega)/(1+3\omega)$, and $J_\alpha$ and $Y_\alpha$ denote the cylindrical Bessel functions of the first and second kind, respectively.
Inserting Eq.~\eqref{eq:source_f_X} in Eq.~\eqref{eq:tensor_from_greens_function}, we obtain
\begin{equation}
\label{eq:h_lambda_k}
h_\lambda(\mathbf{k},\eta)
=
4\sum_{X=\{\Phi,S\}}
\int \frac{d^3 q}{(2\pi)^3}\,
e_\lambda^{ij}(\mathbf{k})\,
\frac{q_i q_j}{k^2}\,
X_{\mathbf q}\,X_{\mathbf{k}-\mathbf q}\,
I_X(x,u,v),
\end{equation}
 with the kernel function $I_X\left(x, u, v\right)$ given by 
\begin{equation}
\label{eq:kernel_standard}
    I_X\left(x, u, v\right)\equiv \int_0^x \mathrm{d} \overline{x} \, G_h(x, \overline{x}) \, f_X\left(\overline{x}, u, v\right),
\end{equation}
where $v\equiv q/k$ and $u \equiv |\mathbf{k}-\mathbf{q}|/k$ parametrise the geometry of the triangle formed by $\mathbf{k}$, $\mathbf{q}$ and $\mathbf{k}-\mathbf{q}$.
We define the GW power spectrum, summed over polarizations, as 
\begin{equation}
\label{eq:P_h_def}
\mathcal{P}_{h}(k, \eta) \equiv \frac{k^3}{2\pi^2}\sum_{\lambda} \int d^3\mathbf{x} \, e^{i\mathbf{k} \cdot \mathbf{x}} \left< h_{\lambda}(\eta,0)\, h_{\lambda}(\eta,\mathbf{x})\right> =
\frac{k^3}{2\pi^2}
\sum_{\lambda}
\left\langle
\left|h_{\lambda}(\mathbf{k},\eta)\right|^2
\right\rangle_{\rm r},
\end{equation}
where the prefactor $k^3/2\pi^2$ ensures that $\mathcal{P}_{h}(k,\eta)\, d\ln k$ is defined per unit of logarithmic interval $[k,\,k+dk]$. The subscript $``\rm r"$ denotes that the disconnected contribution proportional to $(2\pi)^3\delta^{(3)}(0)$ has been removed.
Substituting the expression for \(h_\lambda\), one gets
\begin{equation}
\begin{aligned}
\left\langle
|h_\lambda(\mathbf{k},\eta)|^2
\right\rangle_{\rm r}
&=
16
\sum_{X=\{\Phi,S\}}
\int \frac{d^3 q}{(2\pi)^3}
\int \frac{d^3 p}{(2\pi)^3}\,
e_\lambda^{ij}(\mathbf{k})e_\lambda^{mn *}(\mathbf{k})
\frac{q_i q_j}{k^2}\frac{p_m p_n}{k^2}
\\
&\hspace{1.7cm}\times
I_X(x,u,v)\,I_X(x,\bar{u},\bar{v})\,
\Big\langle
X_{\mathbf q}X_{\mathbf{k}-\mathbf q}
X_{\mathbf p}X_{-\mathbf{k}-\mathbf p}
\Big\rangle_{\rm r},
\end{aligned}
\end{equation}
where $\bar v \equiv p/k$ and $\bar u \equiv |\mathbf{k}+\mathbf p|/k$.
Assuming Gaussian scalar perturbations and using Wick's theorem, one finds
\begin{equation}
\langle X_{\mathbf q}X_{\mathbf{k}-\mathbf q}
X_{\mathbf p}X_{-\mathbf{k}-\mathbf p}\rangle_{\rm r}
=
2\,(2\pi)^3\delta^{(3)}(\mathbf q+\mathbf p)\,
\frac{(2\pi^2)^2}{q^3|\mathbf{k}-\mathbf q|^3}
\mathcal{P}_X(q)\mathcal{P}_X(|\mathbf{k}-\mathbf q|).
\end{equation}
The delta function enforces $\mathbf{p}=-\mathbf{q}$, so that $\bar{v}=v$ and 
$\bar{u}=u$ upon integration over $\mathbf{p}$, and both kernel 
functions reduce to $I_X(x,u,v)$.
Using the polarization completeness relation
\begin{equation}
\sum_\lambda e^\lambda_{ij}(\hat{\mathbf k})\,e^{\lambda *}_{mn}(\hat{\mathbf k})
=
\frac12
\left[
P_{im}P_{jn}+P_{in}P_{jm}-P_{ij}P_{mn}
\right],
\end{equation}
where $
P_{ij}\equiv \delta_{ij}-\hat k_i \hat k_j
$
is the projector orthogonal to \(\hat{\mathbf k}\), we obtain
\begin{equation}
\sum_\lambda
\left[
e_\lambda^{ij}(\mathbf{k}) q_i q_j
\right]^2
=
\frac{k^4}{32}
\left[
4v^2-(1+v^2-u^2)^2
\right]^2.
\end{equation}
Changing variables to $u=q/k$ and $v=|\mathbf{k}-\mathbf{q}|/k$, such that $\mathrm{d}^3 q = 2\pi k^3 u v\,\mathrm{d}u\,\mathrm{d}v$, the tensor power spectrum reads~\cite{Domenech:2021ztg}
\begin{equation}
\label{eq:tensorpowerspectrum_general}
\overline{\mathcal{P}_h(k,\eta)}
=
8 \int_0^{\infty} \mathrm{d}v
\int_{|1-v|}^{1+v} \mathrm{d}u\,
\mathcal{K}(u,v)
\sum_{X\in\{\Phi,S\}}
\overline{I_X^2(x,k,u,v)}\,
\mathcal{P}_X(uk)\,\mathcal{P}_X(vk),
\end{equation}
with
\begin{equation}
\mathcal{K}(u,v)
=
\left(
\frac{4v^2-(1+v^2-u^2)^2}{4uv}
\right)^2.
\end{equation}
The overline denotes time averaging.
It is sometimes convenient to introduce $t=u+v-1$ and $s=u-v$, for which the integration domain factorizes as $t\ge 0$, $|s|\le 1$ and $\mathrm{d}u\,\mathrm{d}v=\tfrac{1}{2}\mathrm{d}t\,\mathrm{d}s$. In these variables,
\begin{equation}
\label{eq:tensorpowerspectrum_general_s_t}
\overline{\mathcal{P}_h(k,\eta)}
=
4 \int_0^\infty \mathrm{d}t \int_{-1}^{1} \mathrm{d}s\,
\mathcal{K}(s,t)
\sum_{X\in\{\Phi,S\}}
\overline{I_X^2}\,
\mathcal{P}_X(ku)\,\mathcal{P}_X(kv),
\end{equation}
with
\begin{equation}
\mathcal{K}(s,t)
=
\left[
\frac{t(2+t)(1-s^2)}{(1+t)^2-s^2}
\right]^2.
\end{equation}
The GW energy density is given by Weinberg's formula~\cite{Weinberg:1972kfs,Maggiore:2007ulw} 
\begin{equation}
\rho_{\rm GW}(\eta) \;=\; \frac{1}{32\pi G\,a^2(\eta)} 
\left\langle h_{ij}'(\mathbf{x}, \eta)\, h_{ij}'(\mathbf{x}, \eta) \right\rangle ,
\end{equation}
where $\left<\cdot \right>$ denotes time and space average over wavelengths small compared to the horizon. Then the GW energy density per logarithmic interval is
\begin{equation}
\frac{d\rho_{\rm GW}}{d\ln k} \;=\; 
\frac{k^2}{32\pi G\,a^2(\eta)} \, \overline{\mathcal{P}_{h}(k, \eta)} ,
\end{equation}
where $\overline{\mathcal{P}_{h}(k, \eta)}$ is the time-average of the tensor power spectrum given in Eq.~\eqref{eq:tensorpowerspectrum_general}.
Finally, the dimensionless GW density parameter relative to the critical density is
\begin{equation}
\label{eq:SIGW_basic_formula}
\Omega_{\rm GW}(k,\eta) 
\;\equiv\; \frac{1}{\rho_{\rm tot}(\eta)} \frac{d\rho_{\rm GW}}{d\ln k}
\;=\; \frac{k^2}{12 \mathcal{H}^2}   \overline{\mathcal{P}_{h}(k, \eta)}.
\end{equation}
At late times $\eta_\star\gg\eta_{\rm eva}$, subhorizon GWs propagate freely and $\Omega_{\rm GW}(k,\eta) $ is frozen during radiation to a value which we denote $\Omega_{\rm GW,\star}$.

\subsection{Redshifting to today}
The present-day GW power spectrum reads
\begin{align}
\label{eq:_Omega_GW_0_DW_app}
\Omega_{\rm GW,0}h^2 ~=~ \mathcal{D}(T_{\star}, T_0)~\Omega_{\rm GW,\star} \,,
\end{align}
where $ \mathcal{D}(T_{\star}, T_0) $ is the redshift factor of the radiation energy density between temperature $T_{\star}$ and $T_0$,
\begin{equation}
\label{eq:mathcal_D_def}
    \mathcal{D}(T_{\star}, T_0) \equiv h^2\frac{ \rho_{\rm tot}(T_{\star})}{\rho_{\rm tot}(T_{0})}\left( \frac{a(T_{\star})}{a(T_0)} \right)^{4} = \Omega_{\rm rad,0}h^2\Delta(T_{\star}, T_0),
\end{equation}
where we have assumed adiabatic evolution $ T \propto g_{s,\star}^{-1/3} a^{-1} $ for the radiation energy density $ \rho_{\rm r}(T) = \pi^2 g_\star T^4 / 30 $, and introduced the quantity
\begin{equation}
\label{eq:Delta_def}
    \Delta(T_1, T_2) \equiv  \left( \frac{g_{\star}(T_1)}{g_{\star}(T_2)} \right)
    \left( \frac{g_{s,\star}(T_2)}{g_{s,\star}(T_1)} \right)^{4/3} = 0.39 \left( \frac{g_{\star}(T_{\star})}{106.75}\right)\left( \frac{106.75}{g_{s,\star}(T_{\star})}\right)^{4/3}.
\end{equation}
We use $ g_{\star}(T_0) \simeq 3.38 $, $ g_{s,\star}(T_0) \simeq 3.94 $, and the effective number of relativistic species $ N_{\rm eff} \simeq 3.045 $ \cite{Mangano:2005cc, deSalas:2016ztq}.
The current fraction of radiation energy is given by $ \Omega_{\rm rad,0}h^2= h^2 \rho_{\rm r}(T_0)/\rho_0 \simeq 4.18 \times 10^{-5} $, with the current temperature of the cosmic microwave background $ T_0 \simeq 2.73~\mathrm{K} $ \cite{ParticleDataGroup:2022pth}, and the current critical density is $\rho_{0} = 3 M_{\rm pl}^2 H_0^2$ with $H_0 \simeq 100 h~\rm km/s/Mpc$ and $h$ being the reduced Hubble constant. Hence we have
\begin{align}
\label{eq:def_mathcalD}
    \mathcal{D}(T_{\star}, T_0)\simeq 1.62 \times 10^{-5}\left( \frac{g_{\star}(T_{\star})}{106.75}\right)\left( \frac{106.75}{g_{s,\star}(T_{\star})}\right)^{4/3}.
\end{align}
 For an adiabatic evolution, the GW frequency today $f_0$ redshifts with respect to the frequency at emission $f_{\star}$ by a factor 
\begin{equation}
    \frac{f_0}{f_{\star}}=\frac{a(T_{\star})}{a(T_0)} = \left(\frac{g_{\star,s}(T_0)}{g_{\star,s}(T_{\star})}\right)^{1/3} \frac{T_0}{T_{\star}} .
\end{equation}

\subsection{A roadmap of GW sources}
\label{sec:SIGW_sources}

We consider a cosmological history in which the Universe passes through four
successive epochs: an epoch of PBH formation ($\rm form$) around a conformal time
$\eta \sim \eta_f$ (Eq.~\eqref{eq:M_f_def}), an initial radiation-dominated
era for $\eta<\eta_{\rm eq}$ (${\rm RD}_1$), an early matter-dominated era
($\rm eMD$) driven by a gas of PBHs during $\eta_{\rm eq}<\eta<\eta_{\rm eva}$,
and a final radiation-dominated era following PBH evaporation for
$\eta>\eta_{\rm eva}$ (${\rm RD}_2$),
\begin{equation}
\label{eq:different_eras}
{\rm form}\;\longrightarrow\; {\rm RD}_1 \;\longrightarrow\; {\rm eMD}
\;\longrightarrow\; {\rm RD}_2\,.
\end{equation}
Gravitational waves are sourced at second order in the Newtonian potential
$\Phi$. The time evolution of the Newtonian potential through each cosmological
epoch is encoded in the transfer function $T_X(k,\eta)$, such that
\begin{equation}
\label{eq:Phi_Polt_main}
\Phi(k,\eta)
=\sum_{X\in\{\Phi,S\}} T_X(k,\eta)\,X(k,0)\,\Theta\!\left(k_{\rm
\mathsmaller{UV}}^{\mathsmaller{(X)}}(\eta)-k\right),
\end{equation}
where
\begin{equation}
\label{eq:transfer_fc_reh}
T_X(k,\eta)=\begin{cases}
T_X^{\rm (RD_1)}(k,\eta), & \eta<\eta_{\rm eq},\\[4pt]
T_X^{\rm (eMD)}(k)\,F_\Phi(\eta), & \eta_{\rm eq}<\eta\leq\eta_{\rm osc}(k),\\[4pt]
T_X^{\rm (eMD)}(k)\,\mathcal{S}_\Phi(k)\,T_\Phi^{\rm (RD_2)}(k,\eta),
& \eta>\eta_{\rm osc}(k).
\end{cases}
\end{equation}
Each factor captures a distinct phase of the evolution:
\begin{itemize}
    \item $T_X^{\rm (RD_1)}$ describes the oscillation and decay of sub-horizon
    modes after horizon entry during $\rm RD_1$, prior to PBH domination at
    $\eta_{\rm eq}$ (Eq.~\eqref{eq:TS_eRD}).
    \item $T_X^{\rm (eMD)}$ accounts for the subsequent freezing of sub-horizon
    modes during the eMD era (Eq.~\eqref{eq:transfer_adia_iso_main}).
    \item $F_\Phi(\eta)$ describes the exponential decay of the Newtonian
    potential $\Phi$ toward the end of eMD (Eq.~\eqref{eq:Phi_fit}).
    \item $\mathcal{S}_\Phi(k) \equiv F_\Phi\!\left(\eta_{\rm osc}(k)\right)$
    encodes the suppression of $\Phi$ induced by PBH evaporation, evaluated at
    the conformal time $\eta_{\rm osc}(k) \sim \eta_{\rm eva}$ at which the
    Newtonian potential begins to oscillate
    (Eqs.~\eqref{eq:suppression_factor_final_main}
    and~\eqref{eq:dec_limit_first_derivative_main}). This is \textit{a central
    result of this work}: it encodes the effect of the irreducible broadening of
    the PBH mass function $\psi(M_{\rm \mathsmaller{PBH}})$
    (Eq.~\eqref{eq:psi_f_maintext}), generalising the monochromatic
    approximation of Eq.~\eqref{eq:supp_mono_main} adopted in the
    literature~\cite{Inomata:2019ivs,Inomata:2020lmk,Domenech:2020ssp,%
    Papanikolaou:2020qtd,Domenech:2021wkk,Bhaumik:2022pil,Domenech:2024wao}.
    \item $T_\Phi^{\rm (RD_2)}$ describes the oscillatory decay of the potential
    in the post-evaporation radiation-dominated era $\rm RD_2$
    (Eq.~\eqref{eq:TRD2_def}).
\end{itemize}
Finally, the Heaviside function imposes a UV cutoff
$k_{\rm \mathsmaller{UV}}^{\mathsmaller{(X)}}(\eta)$ that removes modes beyond
the regime of validity of the GW calculation --- specifically, modes that have
entered the nonlinear regime or that lie outside the range of applicability of
the PBH fluid description (see Sec.~\ref{sec:UV_cut_off} for details).

Given the cosmological history in Eqs.~\eqref{eq:different_eras}
and~\eqref{eq:Phi_Polt_main}, the integration over GW emission time $\bar{x}$
in the kernel function $I_X\left(x, u, v\right)$ defined in
Eq.~\eqref{eq:kernel_standard} can be split into four pieces,
\begin{equation}
\int_0^x \mathrm{d} \overline{x} =
\underbrace{\int_{\mathcal{O}(0.1)x_{\rm f}}^{\mathcal{O}(10)x_{\rm f}}
\mathrm{d} \overline{x}}_{\rm (form)}
+\underbrace{\int_{\mathcal{O}(10)x_{\rm f}}^{x_{\rm eq}}
\mathrm{d} \overline{x}}_{\rm (RD_1)}
+\underbrace{\int_{x_{\rm eq}}^{x_{\rm osc}}
\mathrm{d} \overline{x}}_{\rm (eMD)}
+\underbrace{\int_{x_{\rm osc}}^{x}
\mathrm{d} \overline{x}}_{\rm (RD_2)},
\end{equation}
where each term corresponds, in chronological order, to GW production during
PBH formation, during ${\rm RD}_1$, during $\rm eMD$, and during
the radiation-dominated era following PBH evaporation (${\rm RD}_2$). Neglecting cross-correlations between
components emitted at different epochs
and between adiabatic ($X=\Phi$) and isocurvature
($X=S$) perturbations (Eq.~\eqref{eq:S_Phi_unroorelated}), the total SIGW
spectrum decomposes as (see Tab.~\ref{tab:GW_sources})
\begin{equation}
\label{eq:GW_total_PBH_reheating}
\Omega_{\rm GW}\simeq
\Omega_{\rm GW}^{\rm (form)}[\Phi]
+\sum_{X\in\{\Phi,S\}}\left[\Omega_{\rm GW}^{\rm (RD_1)}[X]
+\Omega_{\rm GW}^{\rm (eMD)}[X]
+\Omega_{\rm GW}^{\rm (RD_2)}[X]\right]\,,
\end{equation}
where the bracket $[X]$ indicates the nature of the primordial source $X(k,0)$,
with power spectrum given by Eq.~\eqref{eq:log_normal} for
$\Omega_{\rm GW}^{\rm (form)}$ and by Eqs.~\eqref{eq:P_Phi_main}
and~\eqref{eq:P_S_main} for the remaining contributions. We now discuss each contribution in turn, starting from the latest epoch of GW production and moving backward in time.

\begin{table*}[t]
\centering
\renewcommand{\arraystretch}{1.85}
\begin{tabular}{|c|c|c|c|c|c|c|}
\hline
\rule{0pt}{5ex}
\textbf{\shortstack{GW emission\\[1pt] epoch}}
& \textbf{\shortstack{Seed\\[2.pt]}}
& \textbf{\shortstack{GW\\[1pt] component}}
& \textbf{\shortstack{Physical\\[1 pt] label}}
& \textbf{\shortstack{Scalar \\[1pt] range}}
& \textbf{\shortstack{Section\\[2.pt]}}
& \textbf{\shortstack{Figures\\[-0.2 pt]}}
\\[-0.2ex]
\hline\hline

$\rm PBH~formation$
& $\Phi_{\rm peak}$
& $\Omega_{\rm GW}^{\rm (form)}[\Phi]$
& formation
& full range
& Sec.~\ref{chap:SIGW_formation}
& Fig.~\ref{fig:formation}
\\
\cline{1-7}

\multirow{2}{*}{\shortstack{before PBH-\\[3pt] domination\\[3pt] (${\rm RD}_1$)}}
& $\Phi_{\rm CMB}$
& $\Omega_{\rm GW}^{\rm (RD_1,irred.)}[\Phi]$
& irreducible
& \multirow{2}{*}{$k>k_{\rm eq}$}
& \multirow{2}{*}{Sec.~\ref{chap:eRD}}
& Fig.~\ref{fig:GW_all_sources}
\\
\cline{2-4}\cline{7-7}
& $ S\simeq \delta_{\rm \mathsmaller{PBH}}$
& $\Omega_{\rm GW}^{\rm (RD_1,univ.)}[S]$
& universal
& 
&
& Fig.~\ref{fig:IsoeRD}
\\
\cline{1-7}

\multirow{2}{*}{\shortstack{during PBH-\\[3pt] domination\\[3pt] (${\rm eMD}$)}}
& $\Phi_{\rm CMB}$
& $\Omega_{\rm GW}^{\rm (eMD)}[\Phi]$
& \multirow{2}{*}{eMD}
& \multirow{2}{*}{$k>k_{\rm eva}$}
& \multirow{2}{*}{Sec.~\ref{chap:eMD}}
& \multirow{2}{*}{Fig.~\ref{fig:GWSig_RD_comp_MD}}
\\
\cline{2-3}
&  $S\simeq \delta_{\rm \mathsmaller{PBH}}$
& $\Omega_{\rm GW}^{\rm (eMD)}[S]$
& 
& 
&
&
\\
\hline

\multirow{3}{*}{\shortstack{after PBH-\\[3pt] domination\\[3pt] (${\rm RD_2}$)}}
& \multirow{2}{*}{$\Phi_{\rm CMB}$}
& $\Omega_{\rm GW}^{\rm (RD_2,irred.)}[\Phi]$
& irreducible
& $k<k_{\rm eva}$
& \multirow{3}{*}{Sec.~\ref{chap:poltergeist}}
& Fig.~\ref{fig:GW_all_sources}
\\
\cline{3-5}\cline{7-7}
&
& $\Omega_{\rm GW}^{\rm (RD_2,Polt.)}[\Phi]$
& \multirow{2}{*}{Poltergeist}
& \multirow{2}{*}{$k>k_{\rm eva}$}
&
& \multirow{2}{*}{Fig.~\ref{fig:GWSig_UV}}
\\
\cline{2-3}
&  $S\simeq \delta_{\rm \mathsmaller{PBH}}$
& $\Omega_{\rm GW}^{\rm (RD_2,Polt.)}[S]$
& 
& 
&
&
\\
\hline

\end{tabular}
\caption{Summary of the scalar-induced gravitational-wave sources classified by
epoch, perturbation type, physical origin, and corresponding figures.}
\label{tab:GW_sources}
\end{table*}

\begin{enumerate}

\item $\boldsymbol{\Omega_{\rm GW}^{\rm (RD_2)}}$
(Sec.~\ref{chap:poltergeist}):  
The GW signal generated after PBH evaporation splits into two physically
distinct components,
\begin{equation}
\Omega_{\rm GW}^{\rm (RD_2)}
=
\Omega_{\rm GW}^{\rm (RD_2,\rm Polt.)}
+
\Omega_{\rm GW}^{\rm (RD_2,\rm irred.)}\,,
\end{equation}
depending on whether scalar modes re-enter the horizon before
($k>k_{\rm reh}$) or after ($k<k_{\rm reh}$) evaporation.

For $k>k_{\rm reh}$, the scalar potential $\Phi$ remains frozen during eMD
and starts oscillating once radiation domination is restored. This sudden
activation sources strong acoustic oscillations in the radiation fluid,
leading to a parametrically enhanced GW signal scaling as
$(k\eta_{\rm eva})^8$. The enhancement is controlled by the sharpness of
the eMD-to-${\rm RD}_2$ transition and the amplitude of the scalar transfer
functions. This contribution is commonly referred to as
Poltergeist GWs~\cite{Inomata:2020lmk,Inomata:2025wiv,
Papanikolaou:2020qtd,Domenech:2020ssp,Domenech:2021wkk,
Papanikolaou:2022chm,Bhaumik:2022zdd,Domenech:2024wao}.

Both adiabatic and isocurvature perturbations contribute,
\begin{equation}
\Omega_{\rm GW}^{\rm (RD_2,\rm Polt.)}
=
\Omega_{\rm GW}^{\rm (RD_2,\rm Polt.)}[\Phi]
+
\Omega_{\rm GW}^{\rm (RD_2,\rm Polt.)}[S]\,.
\end{equation}
The adiabatic and isocurvature components have been computed in
Refs.~\cite{Inomata:2020lmk,Inomata:2025wiv,Domenech:2020ssp,Bhaumik:2022zdd}
and
Refs.~\cite{Domenech:2021wkk,Papanikolaou:2022chm,Bhaumik:2022zdd,
Domenech:2024wao}, respectively. We extend these results to extended PBH
mass distributions, including Choptuik critical scaling.

For $k<k_{\rm reh}$, the modes enter the horizon only after PBH evaporation
and are therefore insensitive to both the intermediate PBH era and PBH
isocurvature perturbations. The resulting signal reduces to the standard
radiation-era contribution,
\begin{equation}
\label{eq:Omega_RD2_irred}
\Omega_{\rm GW}^{\rm (RD_2,\rm irred.)}[S]=0,
\qquad
\Omega_{\rm GW}^{\rm (RD_2,\rm irred.)}
=
\Omega_{\rm GW}^{\rm (RD_2,\rm irred.)}[\Phi]\,,
\end{equation}
already present in the absence of PBHs~\cite{Ananda:2006af,
Baumann:2007zm,Kohri:2018awv}. We refer to this unavoidable component as
the \textit{irreducible} gravitational-wave background.

\item $\boldsymbol{\Omega_{\rm GW}^{\rm (eMD)}}$ (Sec.~\ref{chap:eMD}):  
SIGWs sourced during the PBH-dominated era, where the gravitational
potential remains approximately constant~\cite{Kohri:2018awv}. Both
adiabatic and isocurvature contributions are included,
\begin{equation}
\Omega_{\rm GW}^{\rm (eMD)}
=
\Omega_{\rm GW}^{\rm (eMD)}[\Phi]
+
\Omega_{\rm GW}^{\rm (eMD)}[S]\,.
\end{equation}
The adiabatic component has been studied for constant and time-dependent
decay rates~\cite{Inomata:2019ivs,Pearce:2023kxp,Pearce:2025ywc}, while the
isocurvature contribution was investigated in
Ref.~\cite{Papanikolaou:2022chm}. We revisit both contributions in the
context of extended PBH mass distributions.

\item $\boldsymbol{\Omega_{\rm GW}^{\rm (RD_1)}}$ (Sec.~\ref{chap:eRD}):  
During the first radiation-dominated era, SIGWs are sourced by both
adiabatic perturbations and PBH isocurvature which we write as
\begin{equation}
\Omega_{\rm GW}^{\rm (RD_1)}
=
\Omega_{\rm GW}^{\rm (RD_1,irred.)}
+
\Omega_{\rm GW}^{\rm (RD_1,univ.)},
\end{equation}
with
\begin{equation}
\Omega_{\rm GW}^{\rm (RD_1,irred.)}
\equiv
\Omega_{\rm GW}^{\rm (RD_1)}[\Phi],
\qquad
\Omega_{\rm GW}^{\rm (RD_1,univ.)}
\equiv
\Omega_{\rm GW}^{\rm (RD_1)}[S]\,.
\end{equation}
The adiabatic term is the standard radiation-era SIGW background, and thus
has the same origin as the irreducible component generated after evaporation,
$\Omega_{\rm GW}^{\rm (RD_2,\rm irred.)}$ in Eq.~\eqref{eq:Omega_RD2_irred}. The isocurvature term is instead
sourced by Poisson fluctuations in the PBH number density. We therefore call
it the \textit{universal} contribution: its spectral shape is fixed by the
statistics of a population of localized objects, largely independently of
their microscopic nature, whether PBHs, solitons, or axion clumps%
~\cite{Domenech:2021and,Lozanov:2023aez,Lozanov:2023knf,Lozanov:2023rcd,Domenech:2025ffb}.

\item $\boldsymbol{\Omega_{\rm GW}^{\rm (form)}}$
(Sec.~\ref{chap:SIGW_formation}):  
SIGWs generated at PBH formation, sourced by the enhanced primordial
curvature power spectrum responsible for the collapse into PBHs
\cite{Saito:2008jc,Pi:2020otn}. These perturbations source GWs at
second order upon horizon re-entry, directly linking the GW signal to
the small-scale primordial power spectrum.
\end{enumerate}
Additional GW sources associated with PBHs (direct Hawking emission,
gravitational clustering, and binary inspirals) are discussed in a
companion work~\cite{YannNicoPedro} and summarized in
Sec.~\ref{sec:HF_GW}.

\subsection{Ultraviolet cut-off $k_{\mathsmaller{\mathrm{UV}}}$}
\label{sec:UV_cut_off}

Our computation of the induced GW spectrum rests on two assumptions, namely that scalar perturbations remain in the linear regime and that the PBH population can be described as a pressureless perfect fluid with $c_s^2 = \omega = 0$. Both assumptions break down at sufficiently small scales, which motivates an ultraviolet cut-off $k_{\rm \mathsmaller{UV}}$ in Eq.~\eqref{eq:Phi_Polt_main} excluding modes for which the perturbative treatment loses its validity. We adopt two physically motivated choices, the non-linear scale $k_{\rm \mathsmaller{NL}}$ marking the breakdown of linear theory and the inter-PBH separation $k_{\rm \mathsmaller{PBH}}$ marking the breakdown of the fluid description of the PBH gas. The two are used as alternatives rather than combined: $k_{\rm \mathsmaller{NL}}$ provides a conservative lower estimate of the signal by restricting to modes for which linear theory is unambiguously valid, while $k_{\rm \mathsmaller{PBH}}$ provides a conservative upper bound by extrapolating the linear calculation up to the scale at which the fluid description itself fails. A third cut-off, associated with dissipative damping of the radiation fluid and denoted $k_D$, is briefly discussed below for completeness but is not included in the present analysis. The three scales are summarised in Tab.~\ref{tab:summary_cut-off}. 

\paragraph{Non-linear cut-off.} The most conservative choice retains only modes that remain in the linear regime at the epoch $\eta$ of GW emission, that is, $k_{\rm \mathsmaller{UV}}^{\mathsmaller{(X)}} = k_{\rm \mathsmaller{NL}}^{\mathsmaller{(X)}}$, with the non-linear scale defined implicitly by the condition that the PBH density contrast reaches unity. For adiabatic and isocurvature initial conditions this gives, respectively,
\begin{align}
\delta_{\rm \mathsmaller{PBH}}\!\left(k_{\rm \mathsmaller{NL}}^{(\Phi)},\eta\right) \equiv 1, \qquad &\Phi(k,\eta) = T_\Phi(k,\eta)\,\Phi(k,0), \label{eq:kNL_adi}\\
\delta_{\rm \mathsmaller{PBH}}\!\left(k_{\rm \mathsmaller{NL}}^{(S)},\eta\right) \equiv 1, \qquad &\Phi(k,\eta) = T_S(k,\eta)\,S(k,0). \label{eq:kNL_iso}
\end{align}
The non-linear scale inherits a time dependence through $\eta$, which matters most during eMD, when the density contrast grows linearly with the scale factor. As shown in App.~\ref{sec:Time-dependent_Cutoff}, evaluating the cut-off at the end of the eMD era is an excellent approximation, so that Eqs.~\eqref{eq:kNL_adi} and~\eqref{eq:kNL_iso} are understood to hold at $\eta \simeq \eta_{\rm eva}$. Combining them with the sub-horizon Poisson equation,
\begin{equation}
\label{eq:Poisson_NL}
k^2\Phi(k,\eta) = \frac{3}{2}\mathcal{H}^2\delta_{\rm tot},
\end{equation}
and the transfer-function decomposition of Eq.~\eqref{eq:Transfer_fct_def}, one obtains
\begin{equation}
\label{eq:cutoff_def}
k_{\rm \mathsmaller{NL}} \simeq \frac{\sqrt{3/2}\;k_{\rm eva}}{\left[T_\Phi^{\rm (eMD)}(k_{\rm \mathsmaller{NL}})\,\Phi(k_{\rm \mathsmaller{NL}},0) + T_S^{\rm (eMD)}(k_{\rm \mathsmaller{NL}})\,S(k_{\rm \mathsmaller{NL}},0)\right]^{1/2}},
\end{equation}
where $T_X^{\rm (eMD)}(k)$ from Eq.~\eqref{eq:transfer_adia_iso_main} describes the eMD evolution of adiabatic ($X=\Phi$) and isocurvature ($X=S$) perturbations. Since the two channels are statistically uncorrelated (Eq.~\eqref{eq:S_Phi_unroorelated}), their non-linear scales can be determined independently,
\begin{equation}
\label{eq:kNL_adi_iso}
k_{\rm \mathsmaller{NL}}^{\rm adia} \simeq \frac{\sqrt{3/2}\;k_{\rm eva}}{\left[T_\Phi^{\rm (eMD)}(k_{\rm \mathsmaller{NL}}^{\rm adia})\,\Phi(k_{\rm \mathsmaller{NL}}^{\rm adia},0)\right]^{1/2}}, \qquad k_{\rm \mathsmaller{NL}}^{\rm iso} \simeq \frac{\sqrt{3/2}\;k_{\rm eva}}{\left[T_S^{\rm (eMD)}(k_{\rm \mathsmaller{NL}}^{\rm iso})\,S(k_{\rm \mathsmaller{NL}}^{\rm iso},0)\right]^{1/2}}.
\end{equation}
Substituting $\Phi(k,0) = \mathcal{P}_\Phi^{1/2}(k)$ and $S(k,0) = \mathcal{P}_S^{1/2}(k)$ from Eqs.~\eqref{eq:P_Phi_main} and~\eqref{eq:P_S_main}, and evaluating the adiabatic spectrum at $k_{\rm eva}$, yields
\begin{align}
\label{eq:cutoffs_final}
k_{\rm \mathsmaller{NL}}^{\rm adia} &\simeq 235\,k_{\rm eva}\left(\frac{k_{\rm eva}}{k_{\rm CMB}}\right)^{(1-n_s)/4}\left[\frac{9/10}{T_\Phi^{\rm (eMD)}(k_{\rm \mathsmaller{NL}}^{\rm adia})}\right],\\
\label{eq:cutoffs_final_2}
k_{\rm \mathsmaller{NL}}^{\rm iso} &\simeq \left(\frac{675\pi}{8}\right)^{1/7}k_{\rm eva}^{4/7}\,k_{\rm \mathsmaller{PBH}}^{3/7}\left[\frac{1/5}{T_S^{\rm (eMD)}(k_{\rm \mathsmaller{NL}}^{\rm iso})}\right].
\end{align}
The non-linear scale sets the maximum wavenumber that can be consistently included within perturbation theory and therefore controls both the peak frequency and the amplitude of the GW spectrum. The signal is maximised when $\beta_{\rm f} \gg \beta_{\rm c}$, since then $k_{\rm eq} > k_{\rm \mathsmaller{NL}}$ and the dominant modes enter the horizon during the eMD phase rather than before it. In that regime, the transfer functions reduce to their super-horizon plateau values $T_\Phi^{\rm (eMD)} \simeq 9/10$ and $T_S^{\rm (eMD)} \simeq 1/5$.

The use of $k_{\rm \mathsmaller{NL}}$ as the ultraviolet cut-off, set by the strict condition $\delta_{\rm \mathsmaller{PBH}} = 1$, is a deliberately conservative choice in light of the lattice simulations of Ref.~\cite{Fernandez:2023ddy}. Those simulations find that GW production during an eMD epoch remains efficient up to the light-crossing scale of the largest halos forming prior to reheating, $k_{\rm cross} \simeq 8\,k_{\rm \mathsmaller{NL}}$, beyond which the signal is suppressed by the incoherence of emission across the halo volume. In the intermediate range $k_{\rm \mathsmaller{NL}} < k < k_{\rm cross}$, non-linear halo collapse contributes additional power on top of the linear prediction, so that truncating the integral at $k_{\rm \mathsmaller{NL}}$ underestimates the eMD-era contribution $\Omega_{\rm GW}^{\rm (eMD)}$. No equivalent lattice study is currently available for the post-reheating Poltergeist component $\Omega_{\rm GW}^{\rm (Polt.)}$, which is sourced during radiation domination by the rapid decay of scalar potentials at the end of the eMD phase, and we apply the same conservative prescription in that regime.

\paragraph{Fluid cut-off.} The alternative choice extrapolates the linear calculation beyond $k_{\rm \mathsmaller{NL}}^{\mathsmaller{(X)}}$ up to the scale at which the fluid description of the PBH gas itself breaks down. The fluid approximation is valid only on scales larger than the mean comoving inter-PBH separation at formation, which sets
\begin{equation}
\label{eq:kuv=kPBH}
k_{\rm \mathsmaller{UV}}^{\mathsmaller{(X)}} = k_{\rm \mathsmaller{PBH}}, \qquad k_{\rm \mathsmaller{PBH}} \equiv a_f\!\left(\frac{4\pi}{3}\,n_{\rm \mathsmaller{PBH}}(t_f)\right)^{1/3}.
\end{equation}
This extrapolation includes modes for which linear theory no longer applies and therefore overestimates the GW amplitude. The resulting spectrum should accordingly be interpreted as a conservative upper bound. Including it is nevertheless useful in the present context, since one of the central conclusions of this work is that the GW signal from PBH reheating is substantially smaller than previously estimated in the literature, and any overestimate only strengthens that conclusion.\footnote{Note that $k_{\rm \mathsmaller{PBH}}$ coincides with the initial isocurvature non-linearity scale $k_{\rm \mathsmaller{NL}}^{(S)}(0)$, defined by $\mathcal{P}_S\!\left(k_{\rm \mathsmaller{NL}}^{(S)}(0)\right) \equiv 1$, up to a numerical factor $(3\pi/2)^{1/3} \simeq 1.67$. Because PBH density perturbations grow as $\delta_{\rm \mathsmaller{PBH}} \propto a$ during matter domination, however, the non-linearity scale shifts to significantly smaller wavenumbers by the time of evaporation, $k_{\rm \mathsmaller{NL}}^{(S)}(\langle\eta_{\rm eva}\rangle) \ll k_{\rm \mathsmaller{NL}}^{(S)}(0)$, so that the two cut-offs can differ substantially.}

\paragraph{Dissipation cut-off.} The radiation fluid is not perfect. Photons, neutrinos, and other relativistic species have finite mean free paths $\ell_{\rm mfp}$, and on scales comparable to $\ell_{\rm mfp}$ they diffuse across overdense regions into underdense ones, erasing the underlying scalar inhomogeneity. In a radiation-dominated background this process is known as Silk damping~\cite{Silk:1967kq,Weinberg:1971mx,Hu:1995em,Jeong:2014gna} and is governed by shear viscosity, with heat conduction subdominant at early times. The net effect on linear scalar perturbations is an exponential suppression of the Newtonian potential on small scales~\cite{Domenech:2025bvr,Yu:2024xmz},
\begin{equation}
\label{eq:Phi_Silk}
\Phi(k,\eta) \simeq \Phi_{\rm rad}(k,\eta)\,\exp\!\left[-\frac{k^2}{k_D^2(\eta)}\right],
\end{equation}
where $\Phi_{\rm rad}$ is the perfect-fluid solution and the comoving Silk damping scale $k_D$ is the cumulative diffusion length built up over a Hubble time,
\begin{equation}
\label{eq:kD_def}
k_D^{-2}(\eta) = \int_0^\eta {\rm d}\tilde\eta\,\frac{2\,\eta_{\rm sv}(\tilde\eta)}{3a(\rho+p)} \sim \mathcal{H}^{-1}(\eta)\,\ell_{\rm mfp}(\eta),
\qquad
\ell_{\rm mfp}(\eta) \sim \frac{1}{a(\eta)\,n_j(\eta)\,\sigma_j(\eta)},
\end{equation}
with $\eta_{\rm sv}$ the shear viscosity of the plasma, $n_j$ and $\sigma_j$ the number density and scattering cross-section of the species $j$ dominating the viscosity, and all lengths comoving. The middle expression makes the physical content transparent: $k_D^{-1}$ is the geometric mean of the Hubble radius and the mean free path. Refs.~\cite{Domenech:2025bvr,Yu:2024xmz} have shown that this scale has important consequences for the standard SIGW background, and analogous modifications are expected for the Poltergeist mechanism, though a quantitative determination of $k_D$ in the PBH-reheating scenario lies beyond the scope of the present work.

A second dissipative effect, distinct from the suppression in Eq.~\eqref{eq:Phi_Silk}, acts during the post-reheating phase. Once $\Phi$ resumes oscillating as an acoustic wave in the radiation fluid, the same viscous transport drains energy from these oscillations at a rate $\tau_d^{-1}(k) \sim k^2\,\mathcal{H}(\eta)/k_D^2(\eta)$, so that the oscillating Newtonian potential acts as a coherent GW source only over a finite window of duration $\tau_d(k)$. The resulting suppression of the low-frequency tail of the GW spectrum is analogous to the mechanism identified in Ref.~\cite{Domenech:2025bvr} and is expected to be subdominant relative to the cut-off on the scalar source in Eq.~\eqref{eq:Phi_Silk}. We leave both effects for future work.

\begin{table}[h!]
\centering
\renewcommand{\arraystretch}{1.3}
\begin{tabular}{|
    >{\centering\arraybackslash}p{3cm}|
    >{\centering\arraybackslash}p{3cm}|
    >{\centering\arraybackslash}p{3cm}
    >{\centering\arraybackslash}p{3cm}|
}
\hline
\multicolumn{4}{|c|}{\multirow{1}{*}{\centering\makecell[c]{\textbf{Cut-off scales}} } }\\\hline\hline
\multicolumn{2}{|c|}{\multirow{1}{*}{\centering\makecell[c]{\textbf{Origin}} } }
& \multicolumn{2}{c|}{\multirow{1}{*}{\centering \makecell[c]{\centering \makecell[c]{\textbf{Expression}}}}}
 \\\hline
\multicolumn{2}{|c|}{\multirow{2}{*}{\centering\makecell[c]{breakdown of fluid picture} } }
& \multicolumn{2}{c|}{\multirow{2}{*}{\centering \makecell[c]{\centering \makecell[c]{$k_{\rm \mathsmaller{PBH}}$ [Eq.~\eqref{eq:kUV_exp_0}]}}}} \\
\multicolumn{2}{|c|}{}&&\\\hline
\multicolumn{1}{|c|}{\multirow{4}{*}{\centering\makecell[c]{Non-linearities\\RD} } } & 
\multicolumn{1}{c|}{\multirow{2}{*}{\centering\makecell[c]{$k_{\rm \mathsmaller{NL}}^{\rm iso}$} } }
& \multicolumn{2}{c|}{\multirow{2}{*}{\centering \makecell[c]{\centering \makecell[c]{
$ 
\left(\frac{675\pi}{8}\right)^{1/7}
k_{\rm eva}^{4/7}\,k_{\rm \mathsmaller{PBH}}^{3/7}
\left[\frac{1/5}{T_S^{\rm (eMD)}(k_{\rm \mathsmaller{NL}}^{\rm iso})}\right] $ \,  [Eq.~\eqref{eq:cutoffs_final_2}]}}}}
 \\
&&&\\\cline{2-4}
\multicolumn{1}{|c|}{ } & 
\multicolumn{1}{c|}{\multirow{2}{*}{\centering\makecell[c]{$k_{\rm \mathsmaller{NL}}^{\rm adia}$} } }
& \multicolumn{2}{c|}{\multirow{2}{*}{\centering \makecell[c]{\centering \makecell[c]{$235\,k_{\rm eva}
\left(\frac{k_{\rm eva}}{k_{\rm CMB}}\right)^{(1-n_s)/4}
\left[\frac{9/10}{T_\Phi^{\rm (eMD)}(k_{\rm \mathsmaller{NL}}^{\rm adia})}\right]$ [Eq.~\eqref{eq:cutoffs_final}]}}}}\\
&&&\\\hline
\multicolumn{1}{|c|}{\multirow{4}{*}{\centering\makecell[c]{Non-linearities\\eMD} } } & 
\multicolumn{1}{c|}{\multirow{2}{*}{\centering\makecell[c]{$k_{\rm UV,eMD}^{\rm iso}$} } }
& \multicolumn{2}{c|}{\multirow{2}{*}{\centering \makecell[c]{\centering \makecell[c]{
${\rm Min}\left[k_{\rm eq},k^{\rm iso}_{\rm \mathsmaller{NL}}\right]$ \,  [Eq.~\eqref{eq:cutoff_eMD}]}}}}
 \\
&&&\\\cline{2-4}
\multicolumn{1}{|c|}{ } & 
\multicolumn{1}{c|}{\multirow{2}{*}{\centering\makecell[c]{$k_{\rm UV,eMD}^{\rm adia}$} } }
& \multicolumn{2}{c|}{\multirow{2}{*}{\centering \makecell[c]{\centering \makecell[c]{${\rm Min}\left[k_{\rm eq},k_{\rm \mathsmaller{NL}}^{\rm adia}\right]$ [Eq.~\eqref{eq:cutoff_eMD}]}}}}\\
&&&\\\hline
\end{tabular}
\caption{\label{tab:summary_cut-off} Summary of the different cut-off scales used in this work, see discussion in Sec.~\ref{sec:UV_cut_off}.}
\end{table}

\clearpage
\section{SIGWs following PBH evaporation}
\label{chap:poltergeist}

In this section we compute the gravitational-wave signal emitted during the radiation era that follows PBH evaporation ($\rm RD_2$), when the frozen Newtonian potential resumes oscillating. The signal naturally splits into two contributions, distinguished by when the sourcing scalar modes crossed the Hubble horizon:
\begin{itemize}\setlength\itemsep{2pt}
    \item modes that entered during the preceding eMD phase, remained frozen until evaporation, and are released at reheating --- their GW contribution is the Poltergeist signal $\Omega_{\rm GW}^{\rm (RD_2,Polt.)}[\Phi]$.
    \item modes that enter only during $\rm RD_2$ itself and contribute as in a standard radiation era --- their GW contribution is the irreducible background $\Omega_{\rm GW}^{\rm (RD_2,irred.)}[\Phi]$.
\end{itemize}
The two populations are separated by the reheating wavenumber $k_{\rm reh}$, the comoving Hubble scale at evaporation:
\begin{equation}
\label{eq:Phi_split_RD2}
    \Phi(k,\eta) =
    \underbrace{\Phi_{\rm eMD}(k,\eta)\,\Theta(k-k_{\rm reh})}_{\displaystyle\;\to\;\Omega_{\rm GW}^{\rm (RD_2,Polt.)}[\Phi]}
    \;+\;
    \underbrace{\Phi_{\rm RD_2}(k,\eta)\,\Theta(k_{\rm reh}-k)}_{\displaystyle\;\to\;\Omega_{\rm GW}^{\rm (RD_2,irred.)}[\Phi]}.
\end{equation}
Inserted into the quadratic GW source, the two pieces of $\Phi$ generate the two adiabatic contributions indicated under the braces. Adding the isocurvature contribution sourced by the PBH-number-density fluctuations $S$, the total $\rm RD_2$ SIGW spectrum reads
\begin{equation}
\label{eq:Omega_GW_RD2_split}
    \Omega_{\rm GW}^{\rm (RD_2)}
    = \Omega_{\rm GW}^{\rm (RD_2,Polt.)}[\Phi]
    + \Omega_{\rm GW}^{\rm (RD_2,irred.)}[\Phi]
    + \Omega_{\rm GW}^{\rm (RD_2,Polt.)}[S],
\end{equation}
where $[\Phi]$ and $[S]$ label the adiabatic and isocurvature seed channels. The irreducible isocurvature contribution vanishes, since post-evaporation modes are insensitive to the PBH Poisson source, which has decayed by then.
Mixed terms arising from cross-correlations between the Poltergeist and irreducible mode sets are neglected: where their frequency supports overlap, such terms can at most enhance the total signal by an $\mathcal{O}(1)$ factor through constructive interference, or partially suppress it through destructive interference. We show the representative benchmark spectra in Fig.~\ref{fig:GWSig_UV}. The technical derivations are collected in App.~\ref{app:Poltergeist_formalism}.

\subsection{The Poltergeist kernel}

The dominant contribution to Eq.~\eqref{eq:GW_total_PBH_reheating},
denoted $\Omega_{\rm GW}^{\rm (RD_2)}$, is generated during the reheating
stage immediately following PBH evaporation.
At this stage, the sudden disappearance of the PBH-dominated matter
component triggers rapid oscillations of the Newtonian potential in the
subsequent RD era (noted $\rm RD_2$).
These oscillations correspond to acoustic waves in the radiation fluid
and act as an efficient source of SIGWs~\cite{Inomata:2019ivs}.
The sharper the transition between the PBH-dominated and RD epochs,
the stronger the resulting sound waves and the larger the induced GW signal.
In the case of PBH evaporation, the decay rate develops a pole,
$\Gamma \to \infty$ as $t \to t_{\rm eva}$, leading to an almost
instantaneous reheating.
This extreme non-adiabaticity gives rise to a pronounced GW signal,
commonly referred to as the Poltergeist mechanism~\cite{Inomata:2020lmk,
Domenech:2020ssp, Papanikolaou:2020qtd, Domenech:2021wkk, Bhaumik:2022pil,
Domenech:2024wao}, which typically dominates GW production from evaporating PBHs.
We now outline the calculation of the Poltergeist GW spectrum
$\Omega_{\rm GW}^{\rm (RD_2)}$. Full details are provided in
App.~\ref{app:Poltergeist_formalism}.

Immediately after PBH evaporation, the Newtonian potential is set by its
value at the onset of oscillations, $\Phi(k,\eta_{\rm osc})$, given in
Eq.~\eqref{eq:Phi_osc}, and subsequently evolves as
\begin{equation}
\label{eq:Phi_Polt_2}
\Phi(k,\eta>\eta_{\rm osc})
=
\sum_{X\in\{\Phi,S\}} T_X(k,\eta)\,X(k,0),
\end{equation}
where the composite transfer function factorises as (Eq.~\eqref{eq:transfer_fc_reh})
\begin{equation}
\label{eq:transfer_fc_reh_pol}
T_X(k,\eta)
=
T_X^{\rm (eMD)}(k)\,\mathcal{S}_\Phi(k)\,T_\Phi^{\rm (RD_2)}(k,\eta).
\end{equation}
Each factor encodes a distinct phase of the evolution:
$T_X^{\rm (eMD)}$ describes the early matter-dominated era
(Eq.~\eqref{eq:transfer_adia_iso_main}), $\mathcal{S}_\Phi$ the
suppression induced by PBH evaporation
(Eq.~\eqref{eq:suppression_factor_final_main}), and
$T_\Phi^{\rm (RD_2)}$ the subsequent oscillatory decay during
radiation domination~\cite{Domenech:2020ssp},
\begin{equation}
\label{eq:TRD2_def}
T_\Phi^{\rm (RD_2)}(k,\eta)
=
\frac{C_1\, j_1(c_s k \bar{\eta})+C_2\, y_1(c_s k \bar{\eta})}{c_s k \bar{\eta}},
\qquad
\bar{\eta}\equiv \eta-\frac{\eta_{\rm eva}}{2},
\qquad
c_s=\frac{1}{\sqrt{3}},
\end{equation}
where $j_n$ and $y_n$ are spherical Bessel functions of the first and
second kind, respectively.
Imposing continuity of $\Phi$ and $\Phi'$ across the transition fixes
the integration constants,
\begin{equation}
\label{eq:C1C2_app}
C_1
=
-\frac{1}{8}\left(c_s k \eta_{\rm eva}\right)^3
y_2\!\left(\frac{c_s k \eta_{\rm eva}}{2}\right),
\qquad
C_2
=
\frac{1}{8}\left(c_s k \eta_{\rm eva}\right)^3
j_2\!\left(\frac{c_s k \eta_{\rm eva}}{2}\right).
\end{equation}
The source function defined in Eq.~\eqref{eq:source_f_X}, during $\rm RD_2$, reads
\begin{align}
\label{eq:f_X_RD_2}
f_X^{\rm (RD_2)}(\eta,u,v)
&=
T_X(ku,\eta)\,T_X(kv,\eta)
+\tfrac{1}{2}
\!\left(T_X(ku,\eta)+\tfrac{T_X'(ku,\eta)}{\mathcal{H}}\right)
\!\left(T_X(kv,\eta)+\tfrac{T_X'(kv,\eta)}{\mathcal{H}}\right).
\end{align}
Near reheating, where oscillation amplitudes are largest and GW
production is most efficient, the $T_X'$ terms are enhanced by a
factor $(k/k_{\rm eva})^2$ relative to the remaining terms,
so the source simplifies to
\begin{align}
\label{eq:f_X_2_dominant}
f_X^{\rm (RD_2)}(\eta,u,v)
\simeq
\frac{T_X'(ku,\eta)\,T_X'(kv,\eta)}{2\mathcal{H}^2}.
\end{align}
This enhancement arises because modes that were frozen during the
PBH-dominated era begin to oscillate on timescales much shorter than
the Hubble time --- a defining feature of the Poltergeist mechanism.\\
\noindent
The contribution to the GW kernel, defined in Eq.~\eqref{eq:kernel_standard}, due to GW emission after the $\Phi$ starts oscillating is
\begin{equation}
\label{eq:kernel_standard_RD2}
    I^{\rm (RD_2)}_{X}(\bar{x},u,v)\equiv \int_{x_{\rm osc}}^x \mathrm{d} \overline{x} \, G_h(x, \overline{x}) \, f_X^{\rm (RD_2)}\left(\overline{x}, u, v\right).
\end{equation}
Since $\eta_{\rm osc}\in[\eta_{\rm eva},\eta_{\rm cut}]$, with $\eta_{\rm cut}\simeq 1.26\eta_{\rm eva}$ (see Eq.~\eqref{eq:eta_cut_def}), we can safely identify $x_{\rm osc}\sim x_{\rm eva}$ in the lower boundary of Eq.~\eqref{eq:kernel_standard_RD2}.
Substituting Eq.~\eqref{eq:f_X_2_dominant} into
Eq.~\eqref{eq:kernel_standard_RD2} yields the kernel function
\begin{align}
\label{eq:KernelIRDGen}
I^{\rm (RD_2)}_{X}(\bar{x},u,v)
\simeq
\frac{1}{2k^2}
\int_{x_{\rm eva}/2}^{\bar{x}} \mathrm{d}\tilde{\bar{x}}
\;\tilde{\bar{x}}^2\,G_h^{\rm (RD_2)}(\bar{x},\tilde{\bar{x}})\,
T_X'\!\left(uk,\tilde{\eta}\right)
T_X'\!\left(vk,\tilde{\eta}\right),
\qquad X\in\{\Phi,S\},
\end{align}
where $\bar{x}\equiv k\bar{\eta}=x-x_{\rm eva}/2$, $x\equiv k\eta$,
$x_{\rm eva}\equiv k\eta_{\rm eva}$, and
$\tilde{\eta}=(\tilde{\bar{x}}+x_{\rm eva}/2)/k$.
The tensor Green function $G_h^{\rm (RD_2)}(\bar{x},\tilde{\bar{x}})$ is given by
Eq.~\eqref{eq:G_RD2}.
In the {case of interest, $x_{\rm eva}\gg 1$, differentiating
Eq.~\eqref{eq:transfer_fc_reh} gives~\cite{Domenech:2020ssp}
\begin{align}
\label{eq:tprime_main}
T_X'\!\left(uk,\tilde{\eta}\right)
\simeq
-\,\mathcal{S}_\Phi(uk)\,T_X^{\rm (eMD)}(uk)\,c_s u k
\left(\frac{x_{\rm eva}}{2\tilde{\bar{x}}}\right)^{\!2}
\sin\!\left[c_s u\left(\tilde{\bar{x}}-x_{\rm eva}/2\right)\right].
\end{align}
Inserting Eqs.~\eqref{eq:tprime_main} and~\eqref{eq:hgreen} into
Eq.~\eqref{eq:KernelIRDGen} yields
\begin{equation}
\label{eq:IX_RD_final_app}
I^{\rm (RD_2)}_{X}(\bar{x}, u, v)
\simeq
\frac{c_s^2 u v}{32\bar{x}}\,x_{\mathrm{eva}}^4\,
\mathcal{S}_{\Phi}(uk)\,\mathcal{S}_{\Phi}(vk)\,
T_{X}^{\rm (eMD)}(uk)\,T_{X}^{\rm (eMD)}(vk)\,
\mathcal{I}_{\mathrm{osc}}(\bar{x}, u, v),
\end{equation}
where the oscillatory integral is
\begin{equation}
\label{eq:I_osc}
\mathcal{I}_{\mathrm{osc}}(\bar{x}\to\infty,u,v)
=
\int_0^{\infty} \frac{\mathrm{d} x_2}{x_2+x_3}\,
\sin(x_1-x_2)\,\sin(c_s u\, x_2)\,\sin(c_s v\, x_2),
\end{equation}
with $x_1\equiv \bar{x}-x_{\rm eva}/2$,
$x_2\equiv \tilde{\bar{x}}-x_{\rm eva}/2$,
and $x_3\equiv x_{\rm eva}/2$.
The GW spectrum sourced by the Poltergeist effect
during PBH reheating ($\rm RD_2$), as given by
Eq.~\eqref{eq:tensorpowerspectrum_general}, takes the form
\begin{equation}
\label{eq:Omega_GW_Polt_main}
\Omega_{\rm GW}^{\rm (RD_2)}(k,\eta_{\rm eva})
=
\Omega_{\rm GW}^{\rm (RD_2)}[\Phi]
+
\Omega_{\rm GW}^{\rm (RD_2)}[S],
\end{equation}
with
\begin{multline}
\label{eq:Omega_GW_Polt_X_main}
\Omega_{\rm GW}^{\rm (RD_2)}[X]
=
\frac{1}{864}
\left(\frac{k}{k_{\rm eva}}\right)^{\!8}
\int_0^\infty \mathrm{d}v
\int_{|1-v|}^{1+v} \mathrm{d}u\;
\left[4v^2-\left(1+v^2-u^2\right)^2\right]^2
\\
\times\;
\overline{\mathcal{I}_{\rm osc}^2}(u,v)\,
\mathcal{S}_\Phi^2(uk)\,\mathcal{S}_\Phi^2(vk)\,
T_X^{(\rm eMD)\,2}(uk)\,T_X^{(\rm eMD)\,2}(vk)\,
\mathcal{P}_X(uk)\,\mathcal{P}_X(vk),
\end{multline}
where $X\in\{\Phi,S\}$ labels adiabatic and isocurvature modes,
$k_{\rm eva}$ is the comoving horizon scale at PBH evaporation,
and the primordial power spectra $\mathcal{P}_X$ are defined in
Eqs.~\eqref{eq:P_Phi_main} and~\eqref{eq:P_S_main}. All intermediate steps leading to Eq.~\eqref{eq:Omega_GW_Polt_X_main},
including the explicit form of the kernel, are presented in
App.~\ref{app:Poltergeist_formalism}.
The present-day GW abundance is obtained by redshifting the spectrum
from the evaporation epoch,
\begin{equation}
\label{eq:Omega_0_Polt_today}
\Omega_{\rm GW,0}^{\rm (RD_2)}(k)
=
\mathcal{D}(T_{\rm eva},T_0)\,
\Omega_{\rm GW}^{\rm (RD_2)}(k,\eta_{\rm eva}),
\end{equation}
where the redshift factor $\mathcal{D}(T_{\rm eva},T_0)$ is defined in
Eq.~\eqref{eq:def_mathcalD}.
The remaining momentum integrals are evaluated both analytically and
numerically. The resulting GW spectra are presented in
Sec.~\ref{sec:analytic_results_Omega_GW}.

\subsection{Resonant and low-frequency contributions}
\label{sec:analytic_results_Omega_GW}

As discussed above, the Poltergeist mechanism sources gravitational waves
when scalar perturbations frozen during the PBH-dominated era begin to
oscillate after reheating. In the regime $k/k_{\rm eva}\gg 1$ relevant for
PBH evaporation, the resulting SIGW spectrum admits a transparent analytic
interpretation that captures its dominant features. Using
Eq.~\eqref{eq:tensor_from_greens_function}, the tensor modes at first order in  $\mathcal{O}(k_{\rm eva}/k)$ can be written as
\begin{equation}
h_{\lambda}(\mathbf{k}, \eta)
\simeq
\int_0^{\eta}\mathrm{d}\bar{\eta}\,G_h(\eta,\bar{\eta})
\int \frac{\mathrm{d}^3\mathbf{q}}{(2\pi)^3}\;
\mathcal{K}_\lambda(\mathbf{k},\mathbf{q})\,
\Phi_{\mathbf{q}}^\prime(\bar{\eta})\,
\Phi_{\mathbf{k}-\mathbf{q}}^\prime(\bar{\eta}),
\end{equation}
where $\mathcal{K}_\lambda(\mathbf{k},\mathbf{q})\equiv 2\,e_\lambda^{ij}(\mathbf{k})\,q_i\,q_j/\mathcal{H}^2$ is a kinematic kernel, and the tensor Green function
and scalar modes oscillate as (see Eqs.~\eqref{eq:tprime_app}
and~\eqref{eq:hgreen_app})
\begin{equation}
G_h(\eta,\bar{\eta})\propto \sin\!\left[k(\eta-\bar{\eta})\right],
\qquad
\Phi_{\mathbf{q}}(\bar{\eta})
\propto
\sin\!\left[c_s q\!\left(\bar{\eta}+\tfrac{\eta_{\rm eva}}{2}\right)\right].
\end{equation}
After reheating, the tensor Green function oscillates with frequency $k$,
while scalar perturbations propagate as acoustic waves with sound speed
$c_s = 1/\sqrt{3}$. Two kinematic configurations dominate GW
production. Detailed derivations are given in App.~\ref{app:Poltergeist_formalism}.

\begin{enumerate}

\item \emph{Resonant configuration.}
This contribution arises when the two scalar mode frequencies together
match the tensor frequency,
\begin{equation}
|\mathbf{k}| \simeq c_s\!\left(|\mathbf{q}|+|\mathbf{q}-\mathbf{k}|\right),
\qquad \text{or equivalently} \qquad
u+v \simeq c_s^{-1}.
\end{equation}
The scalar source then oscillates in phase with the tensor Green function,
so contributions accumulate coherently over time. In the long-time limit,
the time integral produces a Dirac delta that enforces the kinematic resonance
condition, yielding (see Eq.~\eqref{eq:I_osc_res})
\begin{align}
\label{eq:I_osc_res_main}
\overline{\mathcal{I}^2_{\rm osc,res}}(u,v)
&\simeq
\frac{1}{32}\,\mathrm{Ci}^2\!\left(\left|1-c_s(u+v)\frac{k}{k_{\rm eva}}\right|\right)
\;\xrightarrow{\;k\gg k_{\rm eva}\;}
\frac{\pi c_s}{32}\frac{k_{\rm eva}}{k}\,
\delta\!\left(c_s^{-1}-u-v\right),
\end{align}
where the cosine and sine integrals are defined as
\begin{equation}
\label{eq:CI_SI_main}
\mathrm{Ci}(x)\equiv -\int_x^\infty \frac{\cos t}{t}\,\mathrm{d}t,
\qquad
\mathrm{Si}(x)\equiv \int_0^x \frac{\sin t}{t}\,\mathrm{d}t.
\end{equation}

\item \emph{Low-frequency configuration.}
The second contribution originates from configurations in which both scalar
momenta are comparable and much larger than the GW wavenumber,
\begin{equation}
|\mathbf{q}|\sim|\mathbf{q}-\mathbf{k}|\gg|\mathbf{k}|,
\qquad \text{or equivalently} \qquad
u \sim v \gg 1.
\end{equation}
In this regime the relevant phase of the time integral is
$\Omega = k - c_s\!\left(|\mathbf{q}|-|\mathbf{k}-\mathbf{q}|\right)$.
Since $|\mathbf{q}|\simeq|\mathbf{k}-\mathbf{q}|$, the difference of scalar
frequencies satisfies $c_s\!\left(|\mathbf{q}|-|\mathbf{k}-\mathbf{q}|\right)
\sim \mathcal{O}(k)$, so the phase $\Omega$ varies slowly compared to the
individual scalar oscillations. For an oscillatory integral
$\int \mathrm{d}\eta\,e^{i\Omega\eta}\sim 1/\Omega$, the oscillation-averaged
square scales as $1/\Omega^2$. The kernel is therefore enhanced by the small
frequency mismatch, leading to (see Eq.~\eqref{eq:I_osc_LF_2})
\begin{align}
\label{eq:I_osc_LF_2_main}
\overline{\mathcal{I}^2_{\rm osc,LF}}(u,v)
&\simeq
\frac{1}{8}\!\left[
\mathrm{Ci}^2\!\left(\frac{k}{k_{\rm eva}}\right)
+\left(\frac{\pi}{2}-\mathrm{Si}\!\left(\frac{k}{k_{\rm eva}}\right)\right)^{\!2}
\right]
\simeq
\frac{1}{8}\frac{k_{\rm eva}^2}{k^2}
+\mathcal{O}\!\left(\frac{k_{\rm eva}}{k}\right)^{\!3}.
\end{align}
\end{enumerate}
The Poltergeist-induced GW spectrum in Eq.~\eqref{eq:Omega_GW_Polt_main} can
accordingly be written as the superposition of a resonant and a low-frequency
contribution for both adiabatic and isocurvature perturbations,
\begin{equation}
\label{eq:Omega_GW_0_Reh_ana_main}
\Omega_{\rm GW}^{\rm (RD_2)}(k,\eta_{\rm eva})
\simeq
\sum_{X\in\{\Phi,S\}}
\Bigl[
\Omega_{\rm GW,res}^{\rm (RD_2)}[X]
+
\Omega_{\rm GW,LF}^{\rm (RD_2)}[X]
\Bigr].
\end{equation}
The resonant contribution captures the dominant part of the signal and
determines its overall shape and normalisation, while the low-frequency
contribution affects only the infrared tail and is numerically subdominant.
Both can be treated analytically. The derivations and their regimes of validity
are discussed in App.~\ref{app:Poltergeist_formalism}.\\
\noindent
We now summarise the final interpolating expressions for the resonant
contribution that are used throughout the remainder of this work. These
expressions provide a compact and accurate description of the dominant GW
signal and interpolate smoothly between the relevant momentum regimes. For
adiabatic perturbations,
\begin{equation}
\label{eq:Omega_GW_adia}
\Omega_{\rm GW,res}^{\rm (RD_2)}[\Phi]
\simeq 
\mathcal{A}_{\mathcal{R}}^2
\left(\frac{k}{k_{\rm eva}}\right)^{\!5/3}\left(\frac{k}{k_{\rm CMB}}\right)^{\!2 n_s-2}\!\!\!\!
\frac{
7.7\times 10^{-3}
}{
\bigl[1+ 0.24 \,(k/k_{\rm eq})^{-4n(\omega)/9}\bigr]^9
},
\end{equation}
with $  n(1/3)\simeq-1.83$ (see Eq.~\eqref{eq:transfer_adia_PL}), while for isocurvature perturbations,
\begin{equation}
\label{eq:Omega_GW_iso}
\Omega_{\rm GW,res}^{\rm (RD_2)}[S]\simeq
\left(\frac{k}{k_{\rm \mathsmaller{PBH}}}\right)^6
\left(\frac{k}{k_{\rm eva}}\right)^{5/3}
\,
\frac{
1.5\times 10^{-6}
}{
\left[
1+0.38\,\left(k/k_{\rm eq}\right)
\right]^8
}.
\end{equation}
The derivation of these expressions, together with a discussion of their
regimes of validity, is presented in App.~\ref{app:Poltergeist_formalism}.
The present-day GW energy density is obtained from Eq.~\eqref{eq:Omega_0_Polt_today}.

\subsubsection*{Comparison with numerical results}

In Fig.~\ref{fig:GWSig_UV}, we compare our analytical estimates for the peak
amplitude of the SIGW spectrum with full numerical results for
$\Omega_{\rm GW,0}$, finding excellent agreement across the relevant parameter
space for both adiabatic and isocurvature sources.
The figure also shows a direct comparison between the SIGW signal from a
monochromatic PBH mass\footnote{We also confirm earlier studies on monochromatic PBHs~\cite{Inomata:2020lmk} as shown in Fig.~\ref{fig:GWSigMono_longversion}.} function and our new results for an extended mass
distribution following Choptuik's critical scaling with a sharp UV cut-off.
In both cases, numerical results (dashed) are shown alongside the corresponding
analytical approximations (solid). The extended Choptuik mass distribution
leads to a substantial suppression of the GW signal — by several orders of
magnitude relative to the monochromatic case — primarily due to enhanced
damping at large wavenumbers. This suppression affects both the adiabatic and
isocurvature contributions, with the latter exhibiting an even stronger
reduction. The reason is that the isocurvature power spectrum peaks at high
wavenumbers ($k \sim k_{\rm \mathsmaller{PBH}}$), where the suppression factor scales as
$\mathcal{S}_\Phi^4(k)\sim (k_{\rm \mathsmaller{PBH}}/k_{\rm eva})^{-16/3}$
(Eq.~\eqref{eq:suppression_factor_final_main}), compared to the milder
$(k_{\rm \mathsmaller{PBH}}/k_{\rm eva})^{-4/3}$ (Eq.~\eqref{eq:supp_mono_main}) for a
monochromatic mass function.

\subsection{Maximal GW signal}
\label{sec:NL_dynamics}

We now estimate the maximal post-reheating Poltergeist signal for the two UV
prescriptions introduced in Sec.~\ref{sec:UV_cut_off} and summarised in
Tab.~\ref{tab:summary_cut-off}. Their impact on the spectrum is shown in
Fig.~\ref{fig:GWSig_UV}.

\paragraph{NL cutoff.}
The conservative choice
$k_{\rm \mathsmaller{UV}}=k_{\rm \mathsmaller{NL}}$ cuts the convolution at the
scale where the scalar source becomes non-linear. For the adiabatic
contribution, the maximal signal follows by evaluating
Eq.~\eqref{eq:Omega_GW_adia} near
$k\simeq k_{\rm \mathsmaller{NL}}^{\rm adia}\lesssim k_{\rm eq}$. The
low-frequency branch of Eq.~\eqref{eq:ana_IRtal} is subdominant, and after
redshifting to today we obtain
\begin{equation}
\label{eq:Omega_peak_adia}
\Omega_{\rm GW,0}^{\rm (RD_2)}[\Phi]\!
\left(k_{\rm \mathsmaller{NL}}^{\rm adia}\right)h^2
\simeq
1.8\times 10^{-22}
\left(\frac{M_{\rm \mathsmaller{PBH}}}{10^4\,\mathrm{g}}\right)^{1/10}
\lesssim
1.4\times 10^{-21},
\end{equation}
where the upper value corresponds to
$M_{\rm \mathsmaller{PBH}}=10^9\,{\rm g}$, the largest mass compatible with
BBN. The isocurvature contribution is maximized at
$k\simeq k_{\rm \mathsmaller{NL}}^{\rm iso}\lesssim k_{\rm eq}$, a hierarchy
realized for large initial PBH fractions, and gives
\begin{equation}
\label{eq:Omega_peak_iso}
\Omega_{\rm GW,0}^{\rm (RD_2)}[S]\!
\left(k_{\rm \mathsmaller{NL}}^{\rm iso}\right)h^2
\simeq
5.8\times 10^{-26}
\left(\frac{M_{\rm \mathsmaller{PBH}}}{10^4\,\mathrm{g}}\right)^{-38/21}
\leq
1.0\times 10^{-18},
\end{equation}
with the upper value reached at
$M_{\rm \mathsmaller{PBH}}=1\,{\rm g}$, the smallest mass allowed by the
maximal reheating temperature, Eq.~\eqref{eq:M_min_app}. Thus, with the
non-linear cutoff, both adiabatic and isocurvature Poltergeist
signals remain below the projected sensitivity of planned GW experiments, see Fig.~\ref{fig:beta_vs_MPBH_standard}-left-middle.

\paragraph{Fluid cutoff.}
The aggressive choice
$k_{\rm \mathsmaller{UV}}=k_{\rm \mathsmaller{PBH}}$ extends the convolution to
the mean PBH separation. Since $k_{\rm \mathsmaller{PBH}}/k_{\rm eq}\propto \beta_f^{-2/3}$ (Eq.~\eqref{eq:keq_main}), this scale lies above the spectral
maximum for any realistic $\beta_f$, so that the signal is cut only after its peak, the approximation is exact for small $\beta_f$ and receives only mild corrections for $\beta_f=\mathcal{O}(1)$.
Evaluating Eqs.~\eqref{eq:Omega_GW_adia} and~\eqref{eq:Omega_GW_iso}, the adiabatic and isocurvature spectra peak at
$k_{\rm peak}^{\rm adia}\simeq1.3\,k_{\rm eq}$ and
$k_{\rm peak}^{\rm iso}\simeq60.5\,k_{\rm eq}$, respectively, giving
\begin{align}
\label{eq:Omega_peak_UV_adia}
\Omega_{\rm GW,0}^{\rm (RD_2)}[\Phi]\!
\left(k_{\rm peak}^{\rm adia}\right)h^2
&\simeq
8.2\times10^{-26}
\left(\frac{k_{\rm eq}}{k_{\rm eva}}\right)^{5/3}
\simeq
1.7\times10^{-20}
\left(\frac{M_{\rm \mathsmaller{PBH}}}{10^4\,{\rm g}}\right)^{10/9}
\left(\frac{\beta_f}{10^{-5}}\right)^{10/9},
\\
\label{eq:Omega_peak_UV_iso}
\Omega_{\rm GW,0}^{\rm (RD_2)}[S]\!
\left(k_{\rm peak}^{\rm iso}\right)h^2
&\simeq
1.0\times10^{-18}\,
\frac{k_{\rm eq}^{23/3}}
{k_{\rm eva}^{5/3}\,k_{\rm \mathsmaller{PBH}}^6}
\simeq
2.7\times10^{-27}
\left(\frac{M_{\rm \mathsmaller{PBH}}}{10^4\,{\rm g}}\right)^{10/9}
\left(\frac{\beta_f}{10^{-5}}\right)^{46/9}.
\end{align}
This prescription gives an optimistic upper envelope within the coarse-grained
PBH-fluid description, but still does not yield an observable signal in most of the parameter space. We check that the Poltergeist signal is the only contribution leading to the $\Delta N_{\rm eff}$ and future GW observatories constraints shown in Fig.~\ref{fig:beta_vs_MPBH_standard}-right-middle.

Fig.~\ref{fig:GWSig_UV} (bottom panel) confirms that, for the extended Choptuik mass distribution, the peak amplitude is essentially insensitive to the choice of cutoff as long as $k_{\rm \mathsmaller{UV}}\gg k_{\rm eq}$: the dominant contribution comes from modes near $k_{\rm eq}$, well below either UV scale. In contrast, the monochromatic isocurvature spectrum peaks at $k_{\rm \mathsmaller{UV}}$ itself and is therefore strongly cutoff-dependent.

\subsubsection*{The irreducible contribution from inflationary SIGWs}
After evaporation, the Universe returns to RD, and scalar perturbations that enter the horizon during this second RD era, $\mathrm{RD}_2$, inevitably source SIGWs. As discussed in App.~\ref{App:lowerbound}, this contribution can be estimated using the standard radiation-dominated result for a power-law primordial curvature spectrum. Since SIGWs are dominantly generated around horizon entry during radiation domination, only modes entering after the onset of $\mathrm{RD}_2$ contribute to this irreducible signal. We therefore approximate the present-day contribution by
\begin{equation}
    \Omega_{{\rm GW},0}^{\rm (RD_2)}(k)
    \simeq
    \mathcal{D}(T,T_0)\,
    Q(n_s) A_\zeta^2
    \left(\frac{k}{k_{\rm CMB}}\right)^{2(n_s-1)}
    \Theta(k_{\rm eva}-k),
\end{equation}
where $\mathcal{D}(T,T_0)$ accounts for the standard redshifting to today. This $\mathrm{RD}_2$ signal is not diluted by the preceding $\rm eMD$ and provides an irreducible SIGW background associated with the primordial scalar power spectrum.

\begin{figure}[h!]
\centering
\includegraphics[width=0.48\textwidth]{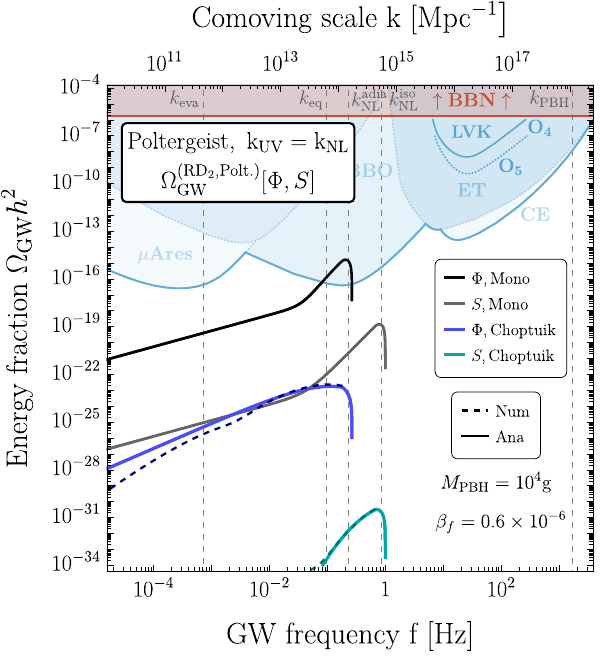}
\includegraphics[width=0.48\textwidth]{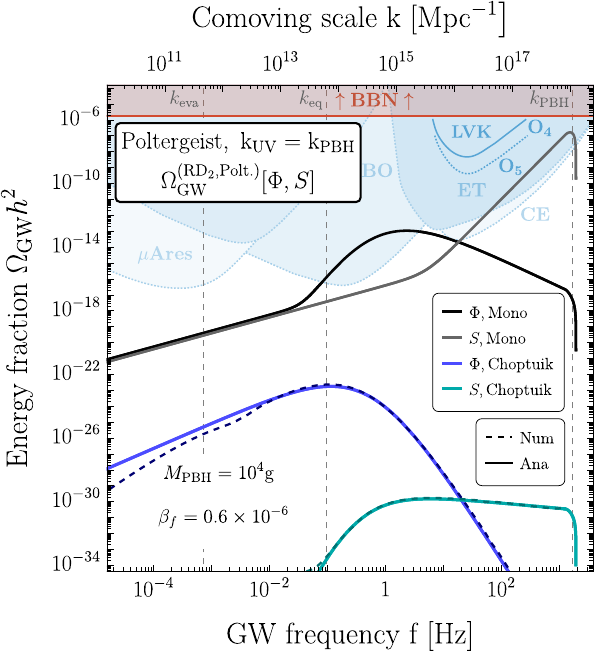}
\includegraphics[width=0.48\textwidth]{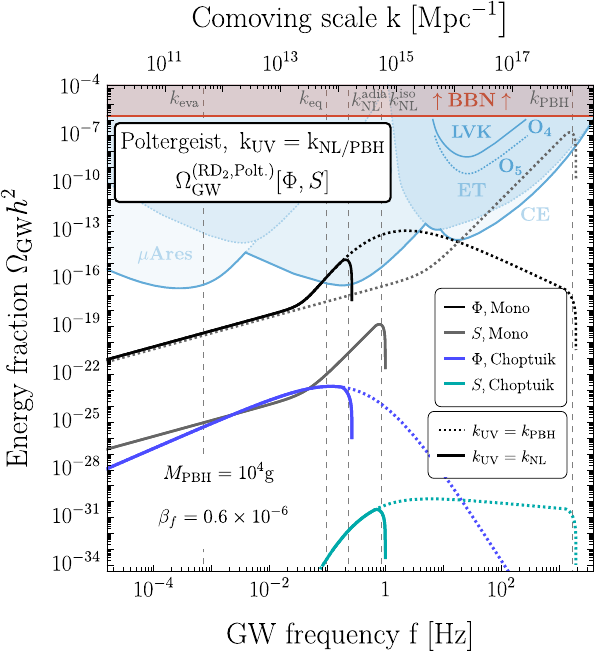}
\caption{%
\label{fig:GWSig_UV}
Comparison of the Poltergeist SIGW signal for adiabatic $\Phi$ and isocurvature $S$ initial conditions. We compare monochromatic (gray) and an extended
Choptuik-scaling (blue) PBH mass distribution. Numerical (dashed) and analytical (solid) results agree
well and demonstrate significant suppression in the extended case.
We fix $\beta_{\rm f}=0.6\times 10^{-6}$ and $M_{\rm \mathsmaller{PBH}}=10^4 \rm g$ for comparability with prior
works.  \textbf{Left:} Non-linear cutoff $k_{\rm \mathsmaller{UV}}=k_{\rm \mathsmaller{NL}}$. \textbf{Right:} Fluid cutoff $k_{\rm \mathsmaller{UV}}=k_{\rm \mathsmaller{PBH}}$. 
\textbf{Bottom:} Solid and dashed lines correspond to the fluid
cut-off $k_{\rm \mathsmaller{UV}}=k_{\rm \mathsmaller{PBH}}$ and the non-linear cut-off
$k_{\rm \mathsmaller{UV}}=k_{\rm \mathsmaller{NL}}$, respectively.}
\end{figure}

\clearpage

\section{SIGWs during PBH domination}
\label{chap:eMD}

In this section we compute the SIGW signal emitted during the PBH-dominated, early matter-dominated (eMD) era, between $\eta_{\rm eq}$ and the onset of reheating. Throughout this phase the Newtonian potential $\Phi$ does not decay on sub-horizon scales, so the quadratic source $f_X$ in Eq.~\eqref{eq:source_term} remains active and sources GWs continuously. The total signal,
\begin{equation}
\label{eq:Omega_GW_eMD_def}
    \Omega_{\rm GW}^{\rm (eMD,\,PBH)}
    =
    \Omega_{\rm GW}^{\rm (eMD,\,PBH)}[\Phi]
    +
    \Omega_{\rm GW}^{\rm (eMD,\,PBH)}[S],
\end{equation}
combines two contributions: a continuous build-up of the GW amplitude, proportional to the scale factor $a$, fed by the constant sub-horizon source deep inside the eMD era, and an overall transfer factor controlled by the way the eMD phase terminates at $\eta_{\rm eva}$. Here $[\Phi]$ and $[S]$ label the adiabatic and isocurvature seeds (cf.~Eqs.~\eqref{eq:P_Phi_main} and~\eqref{eq:P_S_main}).

\paragraph{Notation.} Inside this section, the suffix $(\rm eMD,\,PBH)$ distinguishes PBH-driven reheating from the instantaneous and gradual reference cases, labelled $(\rm eMD,\,inst.)$ and $(\rm eMD,\,grad.)$. Elsewhere in the paper only PBH reheating is considered, and we drop the qualifier, writing
\begin{equation}
\label{eq:eMD_notation_short}
    \Omega_{\rm GW}^{\rm (eMD)}\equiv \Omega_{\rm GW}^{\rm (eMD,\,PBH)}.
\end{equation}
We proceed in three steps. We first isolate the linear-in-$a$ growth in the idealised limit of a pure matter era and clarify its physical origin. We then include the finite eMD-to-$\rm RD_2$ transition and specialise to PBH-driven reheating, for which the Choptuik mass function produces a characteristic $4/3$ cusp in the decay of $\Phi$. Finally, we compare the resulting spectrum with the instantaneous and gradual reference cases and fix the ultraviolet cutoff. We show the representative benchmark spectra in Fig.~\ref{fig:eMD}.
The technical derivations are collected in App.~\ref{app:eMD}.

\subsection{The linear growth $\Omega_{\rm GW}\propto a $ during pure matter domination}
\label{sec:eMD_pureMD}

Deep inside the eMD era ($x_{\rm eq}\ll x\ll x_{\rm eva}$, with $x\equiv k\eta$
and $x_{\rm eva}\equiv k\eta_{\rm eva}$ marking reheating), the potential is
frozen, $\Phi'\simeq 0$, and the source $f_X$ becomes time-independent. The
GW kernel (Eq.~\eqref{eq:kernel_standard}) can then be evaluated in closed
form (see App.~\ref{app:eMD}) and reads
\begin{equation}
\label{eq:I_pure_eMD}
I_X^{\rm (pure~MD)}(x,u,v,x_{\rm eva}\gg x)
= \frac{5\alpha_X}{3}\,\frac{x^3 + 3x\cos x - 3\sin x}{x^3},
\end{equation}
where $X\in\{\Phi,S\}$ labels the adiabatic and isocurvature channels and
$\alpha_X\equiv T_X^{\rm (eMD)}(0)^2$ are the squared large-scale limits of
the scalar transfer functions $T_X^{\rm (eMD)}$ defined in
Eq.~\eqref{eq:transfer_adia_iso_main}, with $\alpha_\Phi=(9/10)^2$ and
$\alpha_S=(1/5)^2$. Crucially, $I_X^{\rm (pure~MD)}$ is independent of the
internal momenta $u,v$, so the scalar convolution factorizes cleanly from
the time-dependent kernel.
Splitting the GW spectrum into adiabatic and isocurvature pieces,
\begin{equation}
\label{eq:Omega_GW_MD_app}
\Omega^{\rm (pure~MD)}_{\rm GW}
=
\Omega_{\rm GW}^{({\rm pure~MD})}[\Phi]
+
\Omega_{\rm GW}^{\rm (pure~MD)}[S],
\end{equation}
the adiabatic contribution, sourced by the CMB-normalized curvature power
spectrum with amplitude $\mathcal{A}_{\mathcal R}$ (cf.\
Eq.~\eqref{eq:P_Phi_main}) and spectral index $n_s$, takes the compact form
\begin{equation}
\label{eq:Omega_GW_pure_MD_Phi}
\Omega_{\rm GW}^{({\rm pure~MD})}[\Phi]
=
\frac{3}{25}\,\mathcal{A}_{\mathcal R}^2
\left(\frac{k}{\mathcal H}\right)^{\!2}
\mathcal{J}_\Phi(n_s),
\end{equation}
with $\tilde k\equiv k/k_{\rm \mathsmaller{UV}}$ and
\begin{equation}
\label{eq:JPhi_main}
\mathcal{J}_\Phi(n_s = 1)=
\begin{cases}
\displaystyle
\frac{32}{15}\,\tilde{k}^{-1}-3+\frac{32}{35}\,\tilde{k}+\frac{1}{8}\,\tilde{k}^2,
& 0<\tilde{k}\le 1, \\[0.8em]
\displaystyle
\frac{1}{840}\!\left(1-\frac{2}{\tilde{k}}\right)^{\!4}
\!\left(105\tilde{k}^2+72\tilde{k}+16-\frac{32}{\tilde{k}}-\frac{16}{\tilde{k}^2}\right),
& 1<\tilde{k}\le 2, \\[0.8em]
0, & \tilde{k}>2.
\end{cases}
\end{equation}
Here, we assumed the power spectrum to be exactly scale invariant for simplicity and refer to App.~\ref{app:eMD} for the $n_s\neq1$ case. The shape function $\mathcal{J}_\Phi$ vanishes for $k>2k_{\rm \mathsmaller{UV}}$
by momentum conservation in the scalar convolution. In the deep infrared
$k\ll k_{\rm \mathsmaller{UV}}$, the leading term $\mathcal{J}_\Phi(1)\to(32/15)\,k_{\rm \mathsmaller{UV}}/k$ gives
\begin{equation}
\label{eq:Omega_pure_MD_Phi_IR}
\Omega_{\rm GW}^{({\rm pure~MD})}[\Phi]
\simeq
\frac{32}{125}\,\mathcal{A}_{\mathcal R}^2
\left(\frac{k}{\mathcal H}\right)^{\!2}
\frac{k_{\rm \mathsmaller{UV}}}{k}~\propto~a,
\end{equation}
reflecting the linear-in-$a$ growth of modes fed by a non-decaying source.

The isocurvature channel, driven by the Poisson PBH number-density
fluctuations of Eq.~\eqref{eq:P_S_main}, gives for $0<\tilde k\le 2$
\begin{equation}
\label{eq:Omega_GW_pure_MD_S}
\Omega_{\rm GW}^{({\rm pure~MD})}[S]
=
\frac{1}{10206000\pi^2}
\left(\frac{k}{\mathcal H}\right)^{\!2}
\!\left(\frac{k_{\rm \mathsmaller{UV}}}{k_{\rm \mathsmaller{PBH}}}\right)^{\!6}
\frac{(2-\tilde{k})^4}{\tilde{k}}
\left(128+116\tilde{k}+40\tilde{k}^2+5\tilde{k}^3\right),
\end{equation}
and $\Omega_{\rm GW}^{({\rm pure~MD})}[S]=0$ for $\tilde k>2$. In the same deep-infrared limit,
\begin{equation}
\label{eq:GW_eMD_pure_MD_IR}
\Omega_{\rm GW}^{({\rm pure~MD})}[S]
\simeq
\frac{128}{637875\pi^2}
\left(\frac{k}{\mathcal H}\right)^{\!2}
\!\left(\frac{k_{\rm \mathsmaller{UV}}}{k_{\rm \mathsmaller{PBH}}}\right)^{\!6}
\frac{k_{\rm \mathsmaller{UV}}}{k}~\propto~a,
\end{equation}
which displays the same $\Omega_{\rm GW}\propto a$ linear growth as the adiabatic channel of
Eq.~\eqref{eq:Omega_pure_MD_Phi_IR}. Details of the derivation are given in
App.~\ref{app:eMD}.

\label{sec:eMD_physical_origin}

We now give more details on the origin of the IR growth $\Omega_{\rm GW}\propto a$ in Eqs.~\eqref{eq:Omega_pure_MD_Phi_IR} and \eqref{eq:GW_eMD_pure_MD_IR}. During an eMD phase the scalar potential is frozen on all relevant scales,
$\Phi'\simeq 0$, so that the source in Eq.~\eqref{eq:source_f_X} becomes
effectively time-independent,
\begin{equation}
S_{\lambda}(\mathbf{k},\eta)\simeq \mathcal{S}_{\lambda}(\mathbf{k})=\text{const.}
\end{equation}
With this constant source, the tensor equation of motion
\begin{equation}
h_\lambda''(\mathbf{k},\eta)+\frac{4}{(1+3\omega)\eta}h_\lambda'(\mathbf{k},\eta)+k^2 h_\lambda(\mathbf{k},\eta)=\mathcal S_\lambda(\mathbf{k})
\end{equation}
admits, for $\omega=0$,
the closed-form solution~\cite{Assadullahi:2009nf}
\begin{equation}
\label{eq:h_eMD_full_improved}
h_{\lambda}(\mathbf{k},\eta)=
\frac{\mathcal{S}_{\lambda}(\mathbf{k})}{k^{2}}
\left[1+3\,\frac{k\eta\cos(k\eta)-\sin(k\eta)}{(k\eta)^{3}}\right],
\end{equation}
with super-horizon and sub-horizon limits
\begin{equation}
\label{eq:h_eMD_limits_improved}
h_{\lambda}(\mathbf{k},\eta)\simeq \frac{\mathcal{S}_{\lambda}(\mathbf{k})}{k^2}
\begin{cases}
(k\eta)^2/10, & k\eta\ll 1,\\[0.3em]
1, & k\eta\gg 1.
\end{cases}
\end{equation}
This shows that, once inside the horizon, the induced tensor mode settles to a constant particular solution rather than oscillating with a decaying envelope. From Eqs.~\eqref{eq:SIGW_basic_formula} and \eqref{eq:P_h_def}, we have
\begin{equation}
\label{eq:Omega_GW_hk_sqr_0}
\Omega_{\rm GW}(k)=\frac{k^5}{24\pi^2\mathcal{H}^2}
\sum_{\lambda} \left\langle \left|h_{\lambda}(\mathbf{k},\eta)\right|^2 \right\rangle_{\rm r} .
\end{equation}
Using the eMD scaling $k^2/\mathcal{H}^2=(k\eta/2)^2\propto a$, one obtains
\begin{equation}
\label{eq:Omega_GW_hk_sqr}
\Omega_{\rm GW}^{\rm (eMD,\,PBH)}(k,\eta)\;\propto\;\mathcal P_{\Phi}^{2}(k)
\begin{cases}
k^{5}\,\eta^{6}, & k\eta\ll 1,\\[0.3em]
k\,\eta^{2}\;\propto\;k\,a(\eta), & k\eta\gg 1 ,
\end{cases}
\end{equation}
namely a linear growth with the scale factor at sub-horizon scales, in agreement with Eqs.~\eqref{eq:Omega_GW_pure_MD_Phi}--\eqref{eq:Omega_GW_pure_MD_S}.
The situation during radiation domination is qualitatively different. From
Eq.~\eqref{eq:tensor_from_greens_function} with $\omega=1/3$,
\begin{equation}
\label{eq:h_RD_green_improved}
h_{\lambda}(\mathbf k,\eta)=\frac{1}{k\eta}
\int_{0}^{\eta}\!d\bar\eta\;\bar\eta\,\sin\!\big[k(\eta-\bar\eta)\big]\,
S_{\lambda}(\mathbf{k},\bar\eta),
\end{equation}
with a source that now depends explicitly on time through the
oscillating RD transfer function
\begin{equation}
\label{eq:Phi_RD_transfer_improved}
T_\Phi(x)=\frac{9}{x^{2}}
\left[\frac{\sin(x/\sqrt{3})}{x/\sqrt{3}}-\cos\!\left(\tfrac{x}{\sqrt{3}}\right)\right].
\end{equation}
Here, $x=k\eta$ and substituting into Eq.~\eqref{eq:source_f_X} yields
\begin{equation}
\mathcal{S}_\lambda(\mathbf{k},\eta)\simeq
\begin{cases}
\displaystyle 6\!\int\!\frac{d^3q}{(2\pi)^3}\,
e_\lambda^{ij}(k)\,q_iq_j\,\Phi_{\mathbf{q}}\Phi_{\mathbf{k}-\mathbf{q}},
& x\ll 1,\\[0.35cm]
\displaystyle\frac{54}{x^{2}}\!\int\!\frac{d^3q}{(2\pi)^3}\,
e_\lambda^{ij}(k)\,q_iq_j\,\Phi_{\mathbf{q}}\Phi_{\mathbf{k}-\mathbf{q}}\,
\frac{\cos\!\big(\tfrac{(v-u)x}{\sqrt{3}}\big)
-\cos\!\big(\tfrac{(v+u)x}{\sqrt{3}}\big)}{u^{2}v^{2}},
& x\gg 1,
\end{cases}
\end{equation}
so that, in sharp contrast to eMD, the source {decays} as
$S_\lambda\propto x^{-2}$ after horizon entry. Inserting this into
Eq.~\eqref{eq:h_RD_green_improved} gives
\begin{equation}
\label{eq:h_RD_ana}
h_\lambda(\mathbf{k},\eta)\simeq
\begin{cases}
\displaystyle \frac{\eta^{2}}{6}\,\mathcal{S}_\lambda^{(0)}(\mathbf{k}),
& x\ll 1,\\[0.3cm]
\displaystyle \frac{1}{k\eta}\Big[\mathcal{A}_\lambda(\mathbf{k})\sin(k\eta)
+\mathcal{B}_\lambda(\mathbf{k})\cos(k\eta)\Big],
& x\gg 1,
\end{cases}
\end{equation}
where $\mathcal{A}_\lambda,\mathcal{B}_\lambda\propto \mathcal P_\Phi^{1/2}(k)$
are fixed around horizon crossing. Sub-horizon RD tensor modes therefore
propagate as free gravitational waves whose amplitude redshifts as
$1/(k\eta)\propto 1/a$.
Since $k^2/\mathcal{H}^2=(k\eta)^2\propto a^2$ during RD, the redshifting of the tensor amplitude exactly compensates the background scaling, leading to
\begin{equation}
\Omega_{\rm GW}^{\rm (RD)}(k,\eta)\propto k^3\,\mathcal{P}_\Phi^2(k)\times \mathrm{const.},
\end{equation}
that is, the usual frozen sub-horizon SIGW spectrum with its causal infrared scaling $\Omega_{\rm GW}\propto k^3$.
The physical difference can thus be summarized as
\begin{equation}
\frac{\Omega_{\rm GW}^{\rm (eMD,\,PBH)}(k,\eta)}{\Omega_{\rm GW}^{\rm (RD)}(k,\eta)}
\propto \frac{a(\eta)}{k^2}.
\end{equation}
In RD, GW production is localized around horizon re-entry and is followed by free propagation. In eMD, instead, the source remains active well inside the horizon and continuously feeds the tensor mode, yielding a constant induced strain and a secular enhancement of $\Omega_{\rm GW}$. This sustained sub-horizon sourcing is the origin of the amplification factor $\propto a(\eta)$ appearing in Eq.~\eqref{eq:GW_eMD_pure_MD_IR}.

\subsection{PBH-driven reheating}
\label{sec:eMD_to_RD}

The eMD era ends at $\eta_{\rm eva}$, after which the Universe returns to a
radiation-dominated phase, denoted by $\mathrm{RD}_2$. Following the decomposition
introduced in Eq.~\eqref{eq:Phi_Polt_main}, we write the Newtonian potential as
\begin{equation}
\label{eq:Phi_eMD_main}
\Phi(k,\eta)
= \sum_{X \in \{\Phi,\,S\}}
T_X(k,\eta)\,X(k,0)\,
\Theta\!\left(k_{\rm \mathsmaller{UV}}^{\mathsmaller{(X)}}(\eta)-k\right),
\end{equation}
where the transfer function is defined piecewise as
\begin{equation}
\label{eq:transfer_fc_reh_eMD}
T_X(k,\eta)=
\begin{cases}
T_X^{\rm (eMD)}(k)\,F_\Phi(\eta),
& \eta \leq \eta_{\rm osc}(k), \\[4pt]
T_X^{\rm (eMD)}(k)\,\mathcal{S}_\Phi(k)\,T_\Phi^{\rm (RD_2)}(k,\eta),
& \eta > \eta_{\rm osc}(k).
\end{cases}
\end{equation}
Here $T_X^{\rm (eMD)}(k)$ is the eMD transfer function given in
Eq.~\eqref{eq:transfer_adia_iso_main}. In the analytical estimates below, we
will use its form in the limit relevant for modes entering the horizon after
the onset of eMD, namely $k\ll k_{\rm eq}$. The function $F_\Phi(\eta)$
describes the model-dependent evolution of $\Phi$ during reheating and will be
specified below for PBH reheating in Eq.~\eqref{eq:Phi_fit_main}. The factor
$\mathcal{S}_\Phi(k)$ is the suppression factor defined in
Eq.~\eqref{eq:suppression_factor_final_main}, while $T_\Phi^{\rm (RD_2)}(k,\eta)$ is
the standard radiation-era transfer function given in
Eq.~\eqref{eq:Phi_RD_transfer_improved}. Finally, the dimensionless time
$x_{\rm osc}\equiv k\eta_{\rm osc}$ marks the onset of oscillations of $\Phi$ on
the scale $k$, as defined in
Eq.~\eqref{eq:dec_limit_first_derivative_main}, and satisfies
$x_{\rm osc}\gtrsim x_{\rm eva}$.
Inserting Eq.~\eqref{eq:transfer_fc_reh_eMD} into Eq.~\eqref{eq:source_term}
with $\omega=0$, the eMD source takes the form
\begin{equation}
\label{eq:f_X_eMD_evap2}
f_X^{\rm (eMD,\,PBH)}(\bar{x},u,v)
= \alpha_X
  \left[
    F_\Phi^2
    + \frac{2}{3}
      \!\left(F_\Phi + \frac{F_\Phi'}{\mathcal{H}}\right)^{\!2}
  \right],
\qquad
\alpha_X \equiv
\begin{cases}
\left(\tfrac{9}{10}\right)^2, & X = \Phi, \\[4pt]
\left(\tfrac{1}{5}\right)^2,  & X = S,
\end{cases}
\end{equation}
which depends on time only through $F_\Phi$ and remains independent of
$u,v$. For $x>x_{\rm osc}$, the kernel splits into pre- and
post-oscillation contributions,
\begin{equation}
\label{eq:kernel_split_main}
I_{X}^{\rm (eMD \to RD_2)}(x,u,v)
= I^{\rm (eMD,\,PBH)}_{X}(x,u,v) + I^{\rm (RD_2)}_{X}(x,u,v),
\end{equation}
with
\begin{align}
\label{eq:I_X_eMD_main}
I^{\rm (eMD,\,PBH)}_{X}(x,u,v)
&= \int_0^{x_{\rm osc}}\! \mathrm{d}\bar{x}\;
   G^{\rm (eMD \to RD_2)}_h(x,\bar{x})\,
   f_X^{\rm (eMD,\,PBH)}(\bar{x},u,v), \\[4pt]
\label{eq:I_X_RD2_main}
I^{\rm (RD_2)}_{X}(x,u,v)
&= \int_{x_{\rm osc}}^{x}\! \mathrm{d}\bar{x}\;
   G^{\rm (RD_2)}_h(x,\bar{x})\,
   f_X^{\rm (RD_2)}(\bar{x},u,v),
\end{align}
where $G^{\rm (eMD\to RD_2)}_h$ (Eq.~\eqref{eq:G_mixed_Bessel}) is the mixed
tensor Green's function for GWs sourced during eMD and propagating into
$\mathrm{RD}_2$, and $G^{\rm (RD_2)}_h$ (Eq.~\eqref{eq:G_RD2}) is the
standard radiation-era Green's function. The second piece
$I^{\rm (RD_2)}_{X}$ coincides with the Poltergeist kernel of
Sec.~\ref{chap:poltergeist}. We now focus on $I^{\rm (eMD,\,PBH)}_{X}$, evaluated
in App.~\ref{app:eMD}.

\label{sec:eMD_PBH_reheating}

The physical case of interest is reheating driven by the
evaporation of a PBH population with a Choptuik mass function. The finite
width of the mass distribution smears the evaporation time and causes
$\Phi$ to decay in a characteristic, non-exponential way. We find that the
envelope $F_\Phi$ entering Eqs.~\eqref{eq:transfer_fc_reh_eMD}
and~\eqref{eq:f_X_eMD_evap2} is well described by the fit (Eq.~\eqref{eq:Phi_fit})
\begin{equation}
\label{eq:Phi_fit_main}
F_\Phi(\eta)\simeq
\begin{cases}
\exp\!\left[-0.84\left(\dfrac{\eta}{\eta_{\rm eva}}\right)^{\!3.43}\right],
& \eta<\eta_{\rm eva},
\\[8pt]
3.5\left(\dfrac{\eta_{\rm cut}-\eta}{\eta_{\rm cut}}\right)^{\!4/3},
& \eta_{\rm eva}\le \eta\le \eta_{\rm cut},
\end{cases}
\end{equation}
where $\eta_{\rm eva}$ is the mean evaporation time and
$z_{\rm cut}\equiv \eta_{\rm cut}/\eta_{\rm eva}\simeq 1.26$ is set by the
heaviest PBHs surviving in the tail of the mass distribution. The key
qualitative feature of Eq.~\eqref{eq:Phi_fit_main} is that $F_\Phi$ vanishes
at $\eta_{\rm cut}$ as a $4/3$ cusp, which sits between the hard step
of instantaneous reheating and the exponential softening characteristic of
perturbative particle decay (see Sec.~\ref{sec:eMD_gradual} and Fig.~\ref{fig:GWSig_RD_comp_MD}).

Inserting Eq.~\eqref{eq:Phi_fit_main} into Eq.~\eqref{eq:f_X_eMD_evap2} and
performing the time integral in Eq.~\eqref{eq:I_X_eMD_main}
(App.~\ref{app:eMD}), the eMD kernel admits compact asymptotic forms. In
the infrared, $x_{\rm eva}\ll 1$,
\begin{equation}
\label{eq:kernel_small_main}
I^{\rm (eMD,\,PBH)}_{X}(x,u,v)
\simeq
0.222\,\alpha_X\,
\frac{\sin x}{x}\,x_{\rm eva}^{2},
\qquad x_{\rm eva}\ll 1,
\end{equation}
while in the ultraviolet, $x\gg x_{\rm eva}\gg 1$,
\begin{equation}
\label{eq:kernel_large_main}
I^{\rm (eMD,\,PBH)}_{X}(x,u,v)
\simeq
2.23\,\alpha_X\,
\frac{x_{\rm eva}^{1/3}}{x}\,
\sin\!\left(x-z_{\rm cut}\,x_{\rm eva}+\frac{5\pi}{6}\right).
\end{equation}
The fractional UV scaling $I_X\propto x_{\rm eva}^{1/3}$ is a distinctive
signature of PBH reheating, milder than the $x_{\rm eva}^{1}$ scaling of
instantaneous reheating and stronger than the $x_{\rm eva}^{-1}$ suppression
of perturbative (gradual) reheating. Physically, it reflects the $4/3$ cusp
at $z=z_{\rm cut}$ in Eq.~\eqref{eq:Phi_fit_main}, which softens, but does
not erase, the endpoint of the time integral. The full intermediate
expression and the comparison with the instantaneous and gradual cases are
given in Sec.~\ref{sec:eMD_gradual} and App.~\ref{app:eMD}.

Because $I^{\rm (eMD,\,PBH)}_{X}$ in both asymptotic regimes
\eqref{eq:kernel_small_main}--\eqref{eq:kernel_large_main} remains
independent of the internal momenta $u,v$, as did $I_X^{\rm (pure~MD)}$ in
Eq.~\eqref{eq:I_pure_eMD}, the scalar convolution factorizes out of the
full GW spectrum. The SIGW spectrum sourced during a PBH-dominated eMD era
and observed after reheating then inherits the same spectral shape as
$\Omega_{\rm GW}^{\rm (pure~MD)}$, up to an overall rescaling,
\begin{equation}
\label{eq:Omega_GW_MD_PBH_main}
\Omega^{\rm (eMD,\,PBH)}_{\rm GW}[X]
=
R^{\rm (eMD,\,PBH)}\times\Omega_{\rm GW}^{\rm (pure~MD)}[X],
\end{equation}
which is the quantity denoted $\Omega^{\rm (eMD)}_{\rm GW}[X]$ in the rest of the paper, in line with the convention introduced in Eq.~\eqref{eq:eMD_notation_short}.
$\Omega_{\rm GW}^{\rm (pure~MD)}[X]$ is given by
Eqs.~\eqref{eq:Omega_GW_MD_app}, \eqref{eq:Omega_GW_pure_MD_Phi},
\eqref{eq:JPhi_main} and~\eqref{eq:Omega_GW_pure_MD_S}, and the transfer
factor defined as the ratio of oscillation-averaged squared kernels,
\begin{equation}
\label{eq:R_PBH_main}
R^{\rm (eMD,\,PBH)}\,\equiv\,
\overline{\bigl(I^{\rm (eMD,\,PBH)}_{X}\bigr)^2}\Big/
\overline{\bigl(I_X^{\rm (pure~MD)}\bigr)^2},
\end{equation}
where $I_X^{\rm (pure~MD)}$ is given by Eq.~\eqref{eq:I_pure_eMD}, and $I^{\rm (eMD,\,PBH)}_{X}$ is given by
Eqs.~\eqref{eq:kernel_small_main}--\eqref{eq:kernel_large_main} for the analytical calculation, and by Eq.~\eqref{eq:I_X_eMD_main} for the numerical calculation. Eq.~\eqref{eq:Omega_GW_MD_PBH_main} makes the physics transparent: the overall
shape is controlled by the pure-MD convolution encoded in
Eqs.~\eqref{eq:Omega_GW_pure_MD_Phi}--\eqref{eq:Omega_GW_pure_MD_S}, while
the entire dependence on the reheating mechanism is given by the factor $R^{\rm (eMD,\,PBH)}$. The analogous factorisation for instantaneous and gradual reheating, which we summarise next, is derived in detail in App.~\ref{app:eMD}.

\subsection{Comparison with instantaneous and gradual reheating}
\label{sec:eMD_gradual}

To assess what is specific to PBH reheating, we compare it with two reference scenarios in which the transition from matter to radiation domination proceeds at extremely different rates, namely instantaneous reheating, in which the source is switched off abruptly~\cite{Inomata:2019ivs}, and gradual reheating driven by perturbative particle decay at a constant rate $\Gamma\sim H$, for which the source is suppressed smoothly~\cite{Ananda:2006af,Baumann:2007zm,Kohri:2018awv,Espinosa:2018eve,Inomata:2019zqy,Pearce:2023kxp,Pearce:2025ywc}. The three reheating histories are characterised by the envelope of the Newtonian potential,
\begin{equation}
\label{eq:F_Phi_three_cases_main}
F_\Phi(\eta)=
\begin{cases}
\Theta(\eta_{\rm eva}-\eta), 
& \text{instantaneous reheating}, \\[6pt]
\left(\dfrac{\eta_{\rm cut}-\eta}{\eta_{\rm cut}}\right)^{\!4/3} \;\;\text{near $\eta_{\rm cut}$}, & \text{PBH reheating}, \\[6pt]
\exp\!\left[-\dfrac{2}{3}\left(\dfrac{\eta}{\eta_{\rm eva}}\right)^{\!3}\right], 
& \text{gradual reheating},
\end{cases}
\end{equation}
which differ in the way the source is switched off. The instantaneous case is a hard step at $\eta=\eta_{\rm eva}$, the gradual case is a smooth exponential, and PBH reheating is intermediate between the two limits: the Choptuik mass function produces a continuous but non-analytic $4/3$ cusp at $\eta=\eta_{\rm cut}$ (cf.~Eq.~\eqref{eq:Phi_fit_main}).

This distinction matters because the ultraviolet behaviour of the GW kernel is controlled entirely by how the source switches off, that is, by the behaviour of $F_\Phi$ near the endpoint of the time integral in Eq.~\eqref{eq:I_X_eMD_main}. The three reheating histories therefore generate three qualitatively different ultraviolet kernels. In the regime $x\gg x_{\rm eva}\gg 1$, one finds
\begin{align}
\label{eq:UV_kernel_inst_main}
I^{\rm (eMD,\,inst.)}_{X}(x_{\rm eva},u,v)
&\simeq
\frac{5\alpha_X}{6}\,\cos(x-x_{\rm eva})
&&\propto\,x_{\rm eva}^{0},\\[4pt]
\label{eq:UV_kernel_PBH_main}
I^{\rm (eMD,\,PBH)}_{X}(x_{\rm eva},u,v)
&\simeq
2.23\,\alpha_X\,x_{\rm eva}^{-2/3}\sin\!\left(x-z_{\rm cut}x_{\rm eva}+\tfrac{5\pi}{6}\right)
&&\propto\,x_{\rm eva}^{-2/3},\\[4pt]
\label{eq:UV_kernel_grad_main}
I^{\rm (eMD,\,grad.)}_{X}(x_{\rm eva},u,v)
&\simeq
\frac{5\alpha_X\cos x}{2\,x_{\rm eva}^2}
&&\propto\,x_{\rm eva}^{-2}.
\end{align}
A sharper endpoint of $F_\Phi$ yields a stronger ultraviolet kernel, with PBH reheating closer to the instantaneous case than to the gradual one, consistent with the rapid termination of the eMD phase by Hawking evaporation. The infrared regime, by contrast, is insensitive to the endpoint and gives a common scaling $I_X\propto x_{\rm eva}\sin x$ in all three cases.

Since none of the kernels in Eqs.~\eqref{eq:UV_kernel_inst_main}--\eqref{eq:UV_kernel_grad_main} depends on the internal momenta $u,v$, the scalar convolution factorises as in Eq.~\eqref{eq:Omega_GW_MD_PBH_main}, and the three GW spectra inherit the same spectral shape as $\Omega_{\rm GW}^{\rm (pure~MD)}[X]$, differing only by an overall transfer factor,
\begin{equation}
\label{eq:Omega_GW_comparison_main}
\Omega^{\rm (eMD,\,r)}_{\rm GW}[X]\,=\,R^{\rm (eMD,\,r)}\times\Omega_{\rm GW}^{\rm (pure~MD)}[X],
\qquad
R^{\rm (eMD,\,r)}\,\equiv\,\overline{\bigl(I^{\rm (eMD,\,r)}_{X}\bigr)^2}/\overline{\bigl(I_X^{\rm (pure~MD)}\bigr)^2},
\end{equation}
with $r\in\{\text{inst.},\,\text{PBH},\,\text{grad.}\}$. All the dependence on the reheating mechanism is repackaged into the scalar factor $R^{\rm (eMD,\,r)}$, whose parametric scaling follows immediately from Eqs.~\eqref{eq:UV_kernel_inst_main}--\eqref{eq:UV_kernel_grad_main},
\begin{equation}
\label{eq:R_hierarchy_main}
R^{\rm (eMD,\,inst.)}\,\propto\,x_{\rm eva}^{0}
\;\gtrsim\;
R^{\rm (eMD,\,PBH)}\,\propto\,x_{\rm eva}^{-4/3}
\;\gtrsim\;
R^{\rm (eMD,\,grad.)}\,\propto\,x_{\rm eva}^{-4}.
\end{equation}
Eq.~\eqref{eq:R_hierarchy_main} is the central result of this comparison: the GW amplitudes follow the same ordering as the kernels, with the PBH-reheating signal lying between the instantaneous and gradual predictions and carrying a parametric dependence on $x_{\rm eva}$ distinct from either. An explicit numerical comparison is shown in Fig.~\ref{fig:GWSig_RD_comp_MD}, and full derivations of the kernels and transfer factors, including the intermediate regime $x_{\rm eva}\sim 1$, are collected in App.~\ref{app:eMD}.

\subsection{The UV cutoff}
\label{sec:eMD_cutoffs}

For the GW signal $\Omega_{\rm GW}^{\rm (eMD)}$ generated during the eMD phase, we take the ultraviolet cutoff $k_{\rm \mathsmaller{UV}}$ to be set by the nonlinear scale $k_{\rm \mathsmaller{NL}}^{\rm adia/iso}$, defined in Eqs.~\eqref{eq:kNL_adi} and~\eqref{eq:kNL_iso}, beyond which linear perturbation theory ceases to be valid. This choice is motivated by numerical simulations of structure growth during an eMD era, which show that GW production is suppressed once nonlinear halo formation sets in~\cite{Fernandez:2023ddy}.\footnote{The reheating setup studied in Ref.~\cite{Fernandez:2023ddy} corresponds to the gradual-reheating scenario discussed in App.~\ref{app:eMD}. We assume that the same cutoff prescription can also be applied to the PBH-reheating scenarios considered here.}
In addition, modes with $k>k_{\rm eq}$ are already suppressed before the onset of eMD due to the behaviour of the transfer function $T_X^{\rm (eMD)}(k)$ in Eq.~\eqref{eq:transfer_adia_iso_main}. However, in our analytical estimate, in Eq.~\eqref{eq:transfer_fc_reh_eMD}, we have used the transfer functions $T_X^{\rm (eMD)}(k)$ in the regime relevant for modes entering the horizon after the beginning of eMD, namely $k\ll k_{\rm eq}$. To account conservatively for the suppression of earlier-entering modes, we therefore impose an additional cutoff at $k_{\rm eq}$ by hand~\cite{Baumann:2007zm,Kohri:2018awv}.
We thus adopt the prescription summarized in Tab.~\ref{tab:summary_cut-off},
\begin{equation}
\label{eq:cutoff_eMD}
k_{\rm \mathsmaller{UV}}^{\rm adia/iso}
=
\min\!\left[
k_{\rm \mathsmaller{NL}}^{\rm adia/iso},\,k_{\rm eq}
\right],
\end{equation}
where $k_{\rm \mathsmaller{NL}}^{\rm adia/iso}$ are given in Eqs.~\eqref{eq:kNL_adi} and~\eqref{eq:kNL_iso}.

\begin{figure}[t]
\centering
\includegraphics[width=0.65\linewidth]{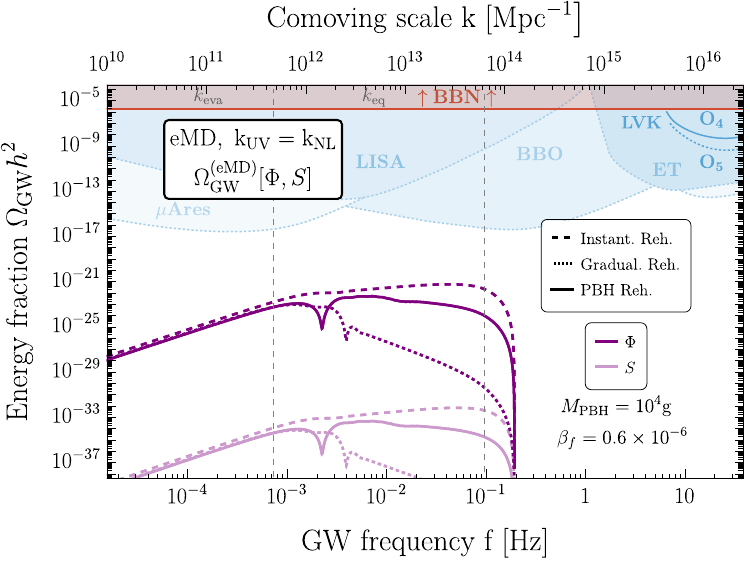}
\caption{\label{fig:GWSig_RD_comp_MD} SIGW signal generated during the PBH-induced eMD phase, $\Omega_{\rm GW}^{\rm (eMD,\,PBH)}$, compared with the instantaneous and gradual reheating reference cases of Sec.~\ref{sec:eMD_gradual}. In the eMD case the cutoff is unambiguous, since $k_{\rm eq}<k_{\rm \mathsmaller{NL,UV}}$ acts as the effective UV cutoff, see Eq.~\eqref{eq:cutoff_eMD}.}
\end{figure}

\clearpage

\section{SIGWs during the early radiation era}
\label{chap:eRD}

In this section we compute the SIGW signals sourced during a radiation-dominated era. Two channels are active during the first radiation era~$\rm RD_1$, between PBH formation at $\eta_f$ and matter--radiation equality at $\eta_{\rm eq}$, corresponding in the taxonomy of Sec.~\ref{sec:SIGW_sources} to the adiabatic (irreducible) and isocurvature (universal) seeds,
\begin{equation}
\label{eq:Omega_RD1_split}
\Omega_{\rm GW}^{\rm (RD_1)}
~=~\Omega_{\rm GW}^{\rm (RD_1)}[\Phi]
~+~\Omega_{\rm GW}^{\rm (RD_1)}[S]
~\equiv~\Omega_{\rm GW}^{\rm (RD_1,irred.)}
~+~\Omega_{\rm GW}^{\rm (RD_1,univ.)}.
\end{equation}
In addition, the post-evaporation radiation era~$\rm RD_2$ carries its own irreducible adiabatic contribution, already announced in Sec.~\ref{chap:poltergeist},
\begin{equation}
\label{eq:Omega_RD2_irred_split}
\Omega_{\rm GW}^{\rm (RD_2,irred.)}
~\equiv~
\Omega_{\rm GW}^{\rm (RD_2)}[\Phi].
\end{equation}
The adiabatic components $\Omega_{\rm GW}^{\rm (RD_1)}[\Phi]$ and $\Omega_{\rm GW}^{\rm (RD_2)}[\Phi]$ are sourced by the standard primordial curvature spectrum and are present in any cosmology with a radiation-dominated phase~\cite{Ananda:2006af,Baumann:2007zm,Kohri:2018awv}. They share the same RD-era formalism, differing only in the integration window of the GW kernel and in the post-emission redshift history, and are therefore derived jointly below. The isocurvature component $\Omega_{\rm GW}^{\rm (RD_1)}[S]$ is sourced instead by the Poisson statistics of the discrete PBH distribution, $\left<S(\mathbf{x})S(\mathbf{y})\right>\simeq \left<\delta_{\rm \mathsmaller{PBH}}(\mathbf{x})\delta_{\rm \mathsmaller{PBH}}(\mathbf{y})\right>$, which seeds the Newtonian potential at second order in perturbation theory and generates the universal contribution characteristic of any localized-object population~\cite{Domenech:2021and,Lozanov:2023aez,Lozanov:2023knf,Domenech:2023fuz,Domenech:2025ffb}. Crucially, none of these three contributions benefits from the Poltergeist enhancement of Sec.~\ref{chap:poltergeist}, since GW production takes place entirely in a radiation background. The two $\rm RD_1$ contributions are further diluted by the entropy release at PBH evaporation, while the $\rm RD_2$ contribution is not. As we show below, the universal channel inherits an additional radiation-era $\kappa^{-2}$ suppression of the isocurvature-to-curvature transfer, fixing the geometric hierarchy $\Omega_{\rm GW}^{\rm (RD_1)}[S]\propto (k_{\rm eq}/k_{\rm \mathsmaller{PBH}})^4$, while the two adiabatic channels are set by the small primordial scalar variance $A_\zeta^2\sim 10^{-18}$. The combined radiation-era signal therefore lies well below the projected sensitivity of all planned GW observatories. We treat the universal channel first, then the joint adiabatic derivation for both eras. We show the representative benchmark spectra in Fig.~\ref{fig:IsoeRD}.

\subsection{The universal PBH-isocurvature signal}
\label{sec:eRD_universal}

For modes deep inside the horizon, $\kappa\equiv k/k_{\rm eq}\gg 1$, the coupled scalar evolution Eqs.~\eqref{eq:Phi_eom_short}--\eqref{eq:S_eom_short} specialised to radiation domination ($\omega=c_s^2=1/3$, $\mathcal H=1/\eta$) with purely isocurvature initial conditions,
\begin{equation}
    \Phi(k,0)=0,\qquad S(k,0)\neq 0,
\end{equation}
admits the solution (see App.~\ref{subsec:RD_analytic} and Ref.~\cite{Domenech:2021and,Lozanov:2023aez}
\begin{align}
\label{eq:TS_eRD}
T_S^{\rm (RD_1)}(k,\eta)
\equiv\frac{\Phi(k,\eta)}{S(k,0)}
&=\frac{3}{2\sqrt{2}\,\kappa}\,\mathcal{F}(k\eta)
+\mathcal{O}(\kappa^{-2}),\qquad \textrm{for}\quad \eta<\eta_{\rm eq},
\end{align}
with the shape function
\begin{equation}
\label{eq:F_eRD}
\mathcal{F}(x)\equiv\frac{1}{x^3}\!\left[
6+x^2-2\sqrt{3}\,x\sin(c_s x)-6\cos(c_s x)\right].
\end{equation}
The prefactor $\kappa^{-1}$ in Eq.~\eqref{eq:TS_eRD} encodes the inefficient subhorizon transfer $S(k,0)\to \Phi(k,\eta)$ during radiation domination. We emphasize that $T_S^{\rm (RD_1)}(k,\eta)$ is a time-dependent sourcing function active throughout $\rm RD_1$, and is conceptually distinct from the late-time matter-era freeze-out value $T_S^{\rm (eMD)}(\kappa)\simeq \kappa^{-2}$ entering Eq.~\eqref{eq:transfer_adia_iso_main}. The two coincide only in their asymptotic $\kappa^{-2}$ scaling at $\eta\gtrsim\eta_{\rm eq}$. Since each factor of $T_S^{\rm (RD_1)}$ in the source contributes one power of $\kappa^{-1}$, one anticipates parametrically
\begin{equation}
\label{eq:OmegaGW_eRD_parametric}
\Omega_{\rm GW}^{\rm (RD_1)}[S]
~\propto~\big(T_S^{\rm (RD_1)}\big)^4
~\propto~\kappa^{-4},
\end{equation}
which is the origin of the smallness of the signal at its peak $k\sim k_{\rm \mathsmaller{PBH}}\gg k_{\rm eq}$.

We now substitute Eq.~\eqref{eq:TS_eRD} into the source function~\eqref{eq:source_term} with $\omega=1/3$. Since $\mathcal H=1/\bar\eta$ in RD, one has $T_S^{{\rm (RD_1)}\prime}/\mathcal H=z\,{\rm d}T_S^{\rm (RD_1)}/{\rm d}z$ with $z\equiv u\bar x$ or $v\bar x$. It is convenient to introduce the auxiliary shape function
\begin{equation}
\label{eq:U_eRD}
\mathcal{U}(z)\equiv \mathcal{F}(z)+z\,\mathcal{F}'(z).
\end{equation}
in terms of which the source takes the factorised form
\begin{equation}
\label{eq:fS_eRD}
f_S(\bar x,k,u,v)
=
\frac{9}{8\,\kappa^2\,u\,v}
\left[
\mathcal{F}(u\bar x)\,\mathcal{F}(v\bar x)
+\frac{1}{2}\,\mathcal{U}(u\bar x)\,\mathcal{U}(v\bar x)
\right].
\end{equation}
The overall factor $\kappa^{-2}$ makes the parametric suppression manifest at the level of the source. Combining $f_S$ with the radiation-era tensor Green function obtained from Eq.~\eqref{eq:hgreen} at $\omega=1/3$,
\begin{equation}
\label{eq:Gh_RD1_eRD}
G_h^{\rm (RD_1)}(x,\bar x)=\frac{\bar x}{x}\sin(x-\bar x)\,\Theta(x-\bar x),
\end{equation}
the isocurvature kernel of Eq.~\eqref{eq:kernel_standard} reads
\begin{equation}
\label{eq:I_S_eRD}
I_S^{\rm (RD_1)}(x_c,u,v)
\equiv\int_{x_f}^{x_{\rm eq}}\!{\rm d}\bar x\,
G_h^{\rm (RD_1)}(x_c,\bar x)\,f_S(\bar x,k,u,v),
\end{equation}
where $x_f\equiv k\eta_f$ marks the time at which the Poisson source is imprinted, and $x_{\rm eq}\equiv k\eta_{\rm eq}$ the time at which the radiation phase ends. For the modes that source the bulk of the signal, $k\sim k_{\rm \mathsmaller{PBH}}$, one finds $x_{\rm eq}\simeq k_{\rm \mathsmaller{PBH}}/k_{\rm eq}\simeq\beta_f^{-2/3}\gamma_{\rm H}^{-1/2}\gg 1$, so the kernel accumulates many oscillations within $[x_f,x_{\rm eq}]$ and the extension
\begin{equation}
\label{eq:eRD_kernel_limits}
x_f\to 0,\qquad x_{\rm eq}\to\infty,
\end{equation}
is an excellent approximation, providing moreover a conservative upper envelope of the signal~\cite{Domenech:2021and,Lozanov:2023aez}.\footnote{The hierarchy $x_{\rm eq}\gg 1$ degrades for $\beta_f\gtrsim 0.03$, but the resulting finite-time corrections only decrease the signal, leaving our analytic estimate as an upper bound.}

In the limit~\eqref{eq:eRD_kernel_limits}, introducing $w_\pm\equiv(u\pm v)/\sqrt{3}$, $w_u\equiv u/\sqrt{3}$, $w_v\equiv v/\sqrt{3}$ together with $\mathcal{L}(w)\equiv\ln|1-w^2|+i\pi\,\Theta(|w|-1)$, the oscillation-averaged kernel reads~\cite{Domenech:2021and}
\begin{equation}
\label{eq:Ibar_eRD}
\overline{I_S^{\rm (RD_1)\,2}}(u,v)
=\frac{1}{2}\!\left|\frac{9}{32\,u^4v^4\,\kappa^2}\,\mathcal{J}(u,v)\right|^{2},
\end{equation}
with
\begin{multline}
\label{eq:J_eRD}
\mathcal{J}(u,v)
=-3u^2v^2
+(u^2-3)(u^2+2v^2-3)\,\mathcal{L}(w_u)
+(v^2-3)(v^2+2u^2-3)\,\mathcal{L}(w_v)
\\
-\frac{1}{2}(u^2+v^2-3)^2\bigl[\mathcal{L}(w_+)+\mathcal{L}(w_-)\bigr].
\end{multline}
The logarithmic and step-function singularities at $w_u,w_v,w_\pm=1$ reflect the acoustic resonances of the radiation fluid: $u,v=\sqrt{3}$ corresponds to individual scalar modes resonating with the tensor mode, while $u\pm v=\sqrt{3}$ corresponds to resonances involving the sum or difference of the two scalar acoustic frequencies.

Inserting Eq.~\eqref{eq:Ibar_eRD} into the master formula~\eqref{eq:tensorpowerspectrum_general} yields the induced spectrum at equality,
\begin{equation}
\label{eq:Omega_GW_eRD_master}
\Omega_{\rm GW}^{\rm (RD_1)}[S](k,\eta_{\rm eq})
=\frac{1}{12}\!\left(\frac{k}{\mathcal H_{\rm eq}}\right)^{\!2}\!
\int_0^\infty\!{\rm d}v\!\int_{|1-v|}^{1+v}\!{\rm d}u\,
\mathcal{K}(u,v)\,
\overline{I_S^{\rm (RD_1)\,2}}\,
\mathcal{P}_S(uk)\,\mathcal{P}_S(vk),
\end{equation}
together with the causal infrared cutoff on the source spectrum,
\begin{equation}
\label{eq:PS_IRcut}
\mathcal{P}_S(k)\;\longrightarrow\;\mathcal{P}_S(k)\,\Theta(k-k_{\rm eq}),
\end{equation}
which removes modes that remain super-horizon throughout $\rm RD_1$ and therefore cannot source GWs during that epoch. The spectrum dependence on the IR-cutoff is shown in Fig.~\ref{fig:IsoeRD}-top. Combining $k/\mathcal H_{\rm eq}\sim\kappa$, $\overline{I_S^{\rm (RD_1)\,2}}\propto \kappa^{-4}$, and the white-noise scaling $\mathcal P_S\propto k^3$ from Eq.~\eqref{eq:P_S_main}, one finds the intermediate-regime amplitude
\begin{equation}
\label{eq:Omega_GW_eRD_scaling}
\Omega_{\rm GW}^{\rm (RD_1)}[S](k,\eta_{\rm eq})
\;\simeq\;
\mathcal{A}_S^{2}\,
\frac{k^{2}\,k_{\rm eq}^{4}}{k_{\rm \mathsmaller{PBH}}^{6}}\,
\mathcal{G}(k),
\qquad k_{\rm eq}\lesssim k\lesssim k_{\rm \mathsmaller{PBH}},
\end{equation}
where $\mathcal{A}_S\equiv 2/(3\pi)$ is the dimensionless Poisson amplitude defined by Eq.~\eqref{eq:P_S_main}, and $\mathcal{G}(k)$ collects the residual cutoff dependence of the double momentum integral. This dependence is mild, with $\mathcal{G}(k)\to \mathcal{G}_\infty\simeq 2.38$ as $k_{\rm IR}\to 0$ and $k_{\rm UV}\to\infty$. The overall prefactor $(k_{\rm eq}/k_{\rm \mathsmaller{PBH}})^4$ makes explicit the $\kappa^{-4}$ suppression anticipated in Eq.~\eqref{eq:OmegaGW_eRD_parametric}.

The convolution in Eq.~\eqref{eq:Omega_GW_eRD_master} can be evaluated in closed form using the analytic kernel of Eqs.~\eqref{eq:Ibar_eRD}--\eqref{eq:J_eRD}. The result is well reproduced by the broken power-law fit
\begin{equation}
\label{eq:OmegaGW_eRD_fit}
\Omega_{\rm GW}^{\rm (RD_1)}[S](k,\eta_{\rm eq})
\simeq
\mathcal{A}_S^2\,\mathcal{G}_\infty\,
\left(\frac{k_{\rm eq}}{k_{\rm \mathsmaller{PBH}}}\right)^{\!4}\!
\left[
10^{3/c}
+\left(\frac{k}{k_{\rm \mathsmaller{PBH}}}\right)^{\!-2/c}
+\left(\frac{100\,k^3}{k_{\rm eq}\,k_{\rm \mathsmaller{PBH}}^2}\right)^{\!-1/c}
\right]^{-c}
\Theta(k_{\rm \mathsmaller{PBH}}-k),
\end{equation}
with $c=2$ and where the explicit $\Theta(k_{\rm \mathsmaller{PBH}}-k)$ enforces the ultraviolet cutoff at the inverse PBH separation, beyond which the fluid description of the PBH gas breaks down (see Sec.~\ref{sec:UV_cut_off}). The three terms inside the bracket correspond, respectively, to a flat plateau at the largest momenta (first term), a Poisson intermediate regime inherited from $\mathcal{P}_S\propto k^3$ (second term), and a causal infrared suppression enforced by Eq.~\eqref{eq:PS_IRcut} (third term). The corresponding asymptotic scalings read
\begin{equation}
\label{eq:OmegaGW_eRD_asymptotic}
\Omega_{\rm GW}^{\rm (RD_1)}[S](k,\eta_{\rm eq})
\simeq \mathcal{A}_S^2\,\mathcal{G}_\infty
\begin{cases}
\displaystyle
\frac{10^2\,k^3\,k_{\rm eq}^3}{k_{\rm \mathsmaller{PBH}}^6},
& k\lesssim 10^{-2}\,k_{\rm eq}, \\[1.2ex]
\displaystyle
\frac{k^2\,k_{\rm eq}^4}{k_{\rm \mathsmaller{PBH}}^6},
& k_{\rm eq}\lesssim k\lesssim k_{\rm \mathsmaller{PBH}}, \\[1.2ex]
\displaystyle
10^{-3}\left(\frac{k_{\rm eq}}{k_{\rm \mathsmaller{PBH}}}\right)^{\!4},
& k\lesssim k_{\rm \mathsmaller{PBH}}\ \text{(plateau)},
\end{cases}
\end{equation}
all controlled by the same $(k_{\rm eq}/k_{\rm \mathsmaller{PBH}})^{4}$ hierarchy. The fit and the asymptotic limits are shown in the left panel of Fig.~\ref{fig:IsoeRD} for several choices of $k_{\rm IR}$.

Because $\Omega_{\rm GW}^{\rm (RD_1)}[S]$ is produced before PBH domination, it is subsequently diluted by the entropy release at PBH evaporation in exactly the same way as any pre-existing radiation-like relic. Using the dilution factor $D$ of Eq.~\eqref{eq:dilution_factor_final}, the present-day spectrum reads
\begin{equation}
\label{eq:OmegaGW_eRD_today}
\Omega_{\rm GW,0}^{\rm (RD_1)}[S](k)
=\Omega_{\rm r,0}\,D^{-4/3}\,
\Omega_{\rm GW}^{\rm (RD_1)}[S](k,\eta_{\rm eq}),
\end{equation}
with $\Omega_{\rm r,0}\simeq 4.18\times 10^{-5}$. The right panel of Fig.~\ref{fig:IsoeRD} shows the diluted spectrum for representative PBH parameters: at fixed $\beta_f$, heavier $M_{\rm \mathsmaller{PBH}}$ extends the eMD duration $N_{\rm MD}$, increasing $D$ and thereby suppressing the present-day signal further.

\subsection{The irreducible inflationary signal}
\label{App:lowerbound}

We derive here the irreducible adiabatic SIGW signal sourced during a radiation-dominated era, applicable to both $\rm RD_1$ and $\rm RD_2$. The source is the standard near-scale-invariant primordial curvature spectrum
\begin{equation}
\label{eq:Pzeta_def}
\mathcal{P}_\zeta(k)
=
A_\zeta\left(\frac{k}{k_{\rm CMB}}\right)^{n_s-1},
\qquad
A_\zeta\simeq 2.1\times 10^{-9},\quad n_s\simeq 0.96,
\end{equation}
which generates a SIGW background in any cosmology with a radiation-dominated phase~\cite{Ananda:2006af,Baumann:2007zm,Kohri:2018awv}. The two emission epochs share the same RD-era integrand of the kernel~\eqref{eq:kernel_standard} --- the source function $f_\Phi$ of Eq.~\eqref{eq:source_term} evaluated at $\omega=1/3$, convolved against the radiation-era tensor Green function~\eqref{eq:Gh_RD1_eRD} --- and differ {only} in the conformal-time window $[\bar\eta_1,\bar\eta_2]$ over which that kernel is integrated:
\begin{equation}
\label{eq:RD12_windows}
[\bar\eta_1,\bar\eta_2]=
\begin{cases}
[\eta_f,\eta_{\rm eq}],
& \rm RD_1, \\[2pt]
[\eta_{\rm osc},\eta],
& \rm RD_2,
\end{cases}
\end{equation}
i.e.~from PBH formation to matter--radiation equality for $\rm RD_1$, and from the onset of post-evaporation oscillations to the time of observation for $\rm RD_2$.

It is convenient to first compute the result obtained when the kernel is integrated over the unrestricted radiation-era window $\bar\eta\in[0,\infty]$ --- the radiation-era envelope --- and then re-impose the finite windows~\eqref{eq:RD12_windows} as effective momentum cutoffs. In the unrestricted limit, the oscillation-averaged kernel becomes time-independent up to an overall $(k\eta)^{-2}$ scaling~\cite{Kohri:2018awv},
\begin{equation}
\label{eq:Ibar_RD_def}
\overline{I_{\rm RD}^2(u,v;x\to\infty)}
=
\frac{9}{16\,x^2}\!
\left[\frac{3(u^2+v^2-3)}{4u^3 v^3}\right]^{\!2}\!
\mathcal{T}(u,v),
\end{equation}
with
\begin{equation}
\label{eq:T_RD_def}
\mathcal{T}(u,v)
\equiv
\Bigg[
-4uv+(u^2+v^2-3)\,
\log\!\left|\frac{3-(u+v)^2}{3-(u-v)^2}\right|
\Bigg]^{\!2}
+\pi^2(u^2+v^2-3)^2\,\Theta(u+v-\sqrt{3}).
\end{equation}
Substituting Eq.~\eqref{eq:Ibar_RD_def} together with the power-law spectrum~\eqref{eq:Pzeta_def} into the master formula~\eqref{eq:tensorpowerspectrum_general} yields the radiation-era envelope
\begin{align}
\label{eq:OmegaGW_RD_powerlaw}
\Omega_{\rm GW}^{\rm (RD)}[\Phi](k)
~&\equiv~
Q(n_s)\,A_\zeta^2\,
\left(\frac{k}{k_{\rm CMB}}\right)^{\!2(n_s-1)}\\
Q(n_s)
~&\equiv~
\frac{1}{27}\!\int_0^\infty\!{\rm d}v\!\int_{|1-v|}^{1+v}\!{\rm d}u\,
\mathcal{K}(u,v)\,
\overline{I_{\rm RD}^2(u,v)}\,
u^{n_s-1}v^{n_s-1},
\end{align}
with $Q(n_s)\simeq 0.81$ for $n_s\simeq 0.96$~\cite{Kohri:2018awv}. We use the bare superscript ``$(\rm RD)$'' in Eq.~\eqref{eq:OmegaGW_RD_powerlaw} to denote the unrestricted-window envelope. It is identical for both radiation-dominated phases and serves as a master result from which the era-specific amplitudes are extracted.

The era-specific amplitudes follow by re-imposing the windows~\eqref{eq:RD12_windows} on the kernel of Eq.~\eqref{eq:kernel_standard}. Since SIGW production during radiation domination is concentrated around horizon entry $\eta\sim k^{-1}$, restricting the kernel to $\bar\eta\in[\bar\eta_1,\bar\eta_2]$ selects modes whose horizon entry lies inside that interval, $\bar\eta_2^{-1}\lesssim k\lesssim \bar\eta_1^{-1}$. Applying this to Eq.~\eqref{eq:RD12_windows} yields the causal momentum cutoffs
\begin{align}
\label{eq:OmegaGW_RD1_eq}
\Omega_{\rm GW}^{\rm (RD_1)}[\Phi](k,\eta_{\rm eq})
&\simeq~
\Omega_{\rm GW}^{\rm (RD)}[\Phi](k)\,\Theta(k-k_{\rm eq}),\\
\label{eq:OmegaGW_RD2_eva}
\Omega_{\rm GW}^{\rm (RD_2)}[\Phi](k,\eta_{\rm eva})
&\simeq~
\Omega_{\rm GW}^{\rm (RD)}[\Phi](k)\,\Theta(k_{\rm eva}-k).
\end{align}
The Heaviside factors are sharp idealisations of the smooth transitions at horizon entry. Treating them as such provides a conservative estimate of the irreducible lower bound on the SIGW signal.
The two epochs differ further in their post-emission redshift history: the $\rm RD_1$ contribution is diluted during the intervening eMD phase, while the $\rm RD_2$ contribution is not. Using the dilution factor $D$ of Eq.~\eqref{eq:dilution_factor_final} and the redshift factor $\mathcal{D}(T_{\rm eva},T_0)$ of Eq.~\eqref{eq:def_mathcalD}, the present-day spectra read
\begin{align}
\label{eq:OmegaGW_RD1_today}
\Omega_{\rm GW,0}^{\rm (RD_1)}[\Phi](k)
&\simeq~
D^{-4/3}\,\mathcal{D}(T_{\rm eva},T_0)\,
\Omega_{\rm GW}^{\rm (RD_1)}[\Phi](k,\eta_{\rm eq}),\\
\label{eq:OmegaGW_RD2_today}
\Omega_{\rm GW,0}^{\rm (RD_2)}[\Phi](k)
&\simeq~
\mathcal{D}(T_{\rm eva},T_0)\,
\Omega_{\rm GW}^{\rm (RD_2)}[\Phi](k,\eta_{\rm eva}).
\end{align}
In both cases the amplitude is set entirely by the small primordial scalar variance $A_\zeta^2\sim 10^{-18}$, providing only an irreducible {lower bound} on the GW background that lies many orders of magnitude below current observational sensitivity.

\begin{figure}[h!]
    \centering
    \includegraphics[width=0.65\linewidth]{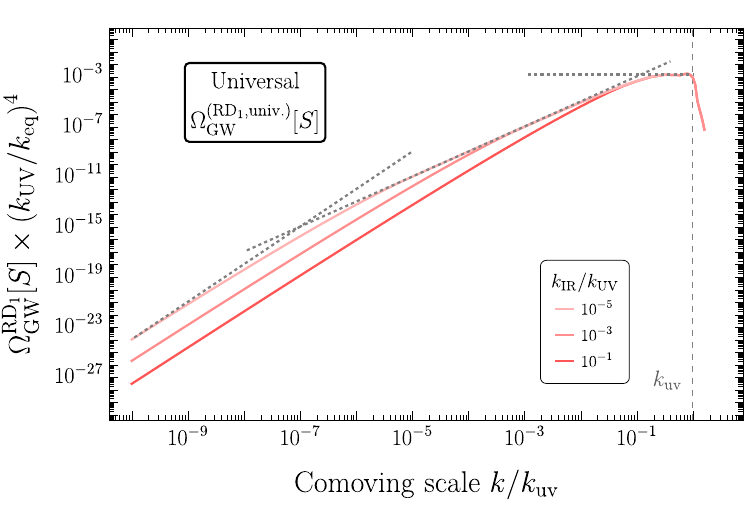}
    \includegraphics[width=0.65\linewidth]{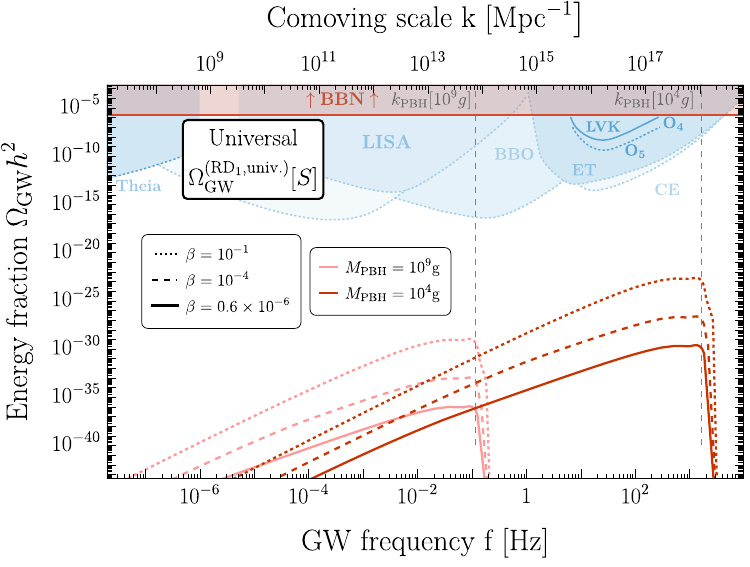}
    \caption{{\textbf{Top:}} Cutoff dependence of SIGWs from PBH Isocurvature during $RD_1$. We show the induced spectra for different IR cutoffs (red curves) and compare with our analytic broken PL approximations (dashed gray lines). In reality $k_{\rm IR}$ is chosen as $k_{\rm eq}$. \textbf{Bottom:} Resulting SIGW signal including dilution during early and late time MD adjusted to the PBH scenario. }
    \label{fig:IsoeRD} 
\end{figure}

\clearpage
\section{SIGWs from PBH formation}
\label{chap:SIGW_formation}

We now turn to the SIGW component sourced at PBH formation, $\Omega_{\rm GW}^{\rm (form)}[\Phi]$.
PBH formation requires a primordial curvature power spectrum
$\mathcal{P}_{\mathcal{R}}(k)$ that is strongly enhanced around the
formation scale $k_f$ defined in Eq.~\eqref{eq:kf_kUV}. We model this
enhancement by the log-normal profile of Eq.~\eqref{eq:log_normal},
characterized by an amplitude $\mathcal{A}_{\mathcal{R}}$, a width
$\Delta$, and a peak position $k_f$. In the narrow-width approximation
(cf.~Eq.~\eqref{eq:beta_f_asymptotic_final}), the initial PBH energy
fraction is exponentially sensitive to the smoothed density variance and
can be written as
\begin{equation}
\label{eq:beta_f_form}
\beta_f
\simeq
\frac{\sigma_0(M)}{\sqrt{2\pi}\,\delta_c}
\exp\!\left[
-\frac{\delta_c^2}{2\sigma_0^2(M)}
\right],
\end{equation}
where $\delta_c$ is the collapse threshold and $\sigma_0(M)$ denotes the
rms density contrast smoothed on the mass scale $M$, as defined in
Eq.~\eqref{eq:sigma_j_moments}. The threshold depends on the shape of the
collapsing perturbation, and typically lies in the range
$\delta_c\simeq 0.40$--$0.67$~\cite{Musco:2018rwt,Escriva:2019phb}. For
definiteness, we fix $\delta_c\simeq 0.5$ in the numerical analysis. In the plots, however, we do not rely on the asymptotic expression in
Eq.~\eqref{eq:beta_f_form}. Instead, we use the full numerical
Press--Schechter prediction for $\beta_f$ derived in
App.~\ref{app:PBH_mass_distrib}, see Eqs.~\eqref{eq:beta_k_MPBH} and
\eqref{eq:Prob_PS_Gauss}. The same enhanced curvature spectrum that
triggers PBH formation also sources, at second order, an adiabatic
scalar-induced gravitational-wave background
$\Omega_{\rm GW}^{\rm (form)}[\Phi]$ at the horizon re-entry of the scale
$k_f$. We show the representative benchmark spectra in Fig.~\ref{fig:formation}.

\subsection{Spectrum at formation}
\label{sec:form_spectrum}

Inserting the log-normal curvature spectrum into the master formula~\eqref{eq:tensorpowerspectrum_general} gives analytic approximations for the SIGW spectrum at formation~\cite{Pi:2020otn}. We use the dimensionless wavenumber
\begin{equation}
\label{eq:kappa_f_def}
\kappa_f \equiv \frac{k}{k_f}.
\end{equation}
For broad spectra, $\Delta\gtrsim 0.1$, the result can be written as
{\small
\begin{align}
\frac{\Omega_{{\rm GW}}^{\rm (form)}[\Phi]}{\mathcal{A}_{\mathcal{R}}^2}
&\simeq
\frac{4\kappa_f^3}{5\sqrt{\pi}}\,
\frac{e^{9\Delta^2/4}}{\Delta}
\Bigg[
\left[
\left(\ln\kappa_f+\frac{3\Delta^2}{2}\right)^2
+\frac{\Delta^2}{2}
\right]
{\rm erfc}\!\left[
\frac{\ln\!\left(\sqrt{3/2}\,\kappa_f\right)+3\Delta^2/2}{\Delta}
\right]
\notag\\
&\hspace{1.7cm}
-\frac{\Delta}{\sqrt{\pi}}\,
\exp\!\left[
-\frac{\left(\ln\!\left(\sqrt{3/2}\,\kappa_f\right)+3\Delta^2/2\right)^2}{\Delta^2}
\right]
\left[
\ln\!\left(\sqrt{2/3}\,\kappa_f\right)+\frac{3\Delta^2}{2}
\right]
\Bigg]
\\
&\hspace{-2cm}
+\frac{0.0659}{\Delta^2}\,
\kappa_f^2 e^{\Delta^2}
\exp\!\left[
-\frac{\left(\Delta^2+\ln\!\left(\sqrt{3/4}\,\kappa_f\right)\right)^2}{\Delta^2}
\right]
+\frac{1}{3}\sqrt{\frac{2}{\pi}}\,
\kappa_f^{-4}\,
\frac{e^{8\Delta^2}}{\Delta}
\exp\!\left[
-\frac{\ln^2\kappa_f}{2\Delta^2}
\right]
{\rm erfc}\!\left[
\frac{4\Delta^2-\ln(\kappa_f/4)}{\sqrt{2}\Delta}
\right].\notag
\label{eq:OmegaGW_form_broad}
\end{align}}
For narrow spectra, $\Delta\lesssim 0.1$, it is useful to define
\begin{equation}
\label{eq:chi_f_def}
\chi_f \equiv \kappa_f e^{\Delta^2}.
\end{equation}
The spectrum is then approximated by
{\small
\begin{align}
\frac{\Omega_{{\rm GW}}^{\rm (form)}[\Phi]}{\mathcal{A}_{\mathcal{R}}^2}
&\simeq
3\kappa_f^2 e^{\Delta^2}
\Bigg[
{\rm erf}\!\left(
\frac{1}{\Delta}{\rm arcsinh}\frac{\chi_f}{2}
\right)
-
{\rm erf}\!\left(
\frac{1}{\Delta}{\rm Re}\,{\rm arccosh}\frac{\chi_f}{2}
\right)
\Bigg]
\left(1-\frac{\chi_f^2}{4}\right)^2
\left(1-\frac{3\chi_f^2}{2}\right)^2
\notag\\
&\quad\times
\Bigg\{
\left[
\frac{1}{2}
\left(1-\frac{3\chi_f^2}{2}\right)
\ln\!\left|1-\frac{4}{3\chi_f^2}\right|
-1
\right]^2
+
\frac{\pi^2}{4}
\left(1-\frac{3\chi_f^2}{2}\right)^2
\Theta\!\left(2-\sqrt{3}\chi_f\right)
\Bigg\}.
\label{eq:OmegaGW_form_narrow}
\end{align}}

\subsection{Present-day amplitude}
\label{sec:form_today}

The SIGW signal $\Omega_{\rm GW}^{\rm (form)}[\Phi]$ is generated at PBH formation, deep inside the first radiation era $\rm RD_1$. It is therefore subsequently diluted by the entropy release at PBH evaporation in exactly the same way as the $\rm RD_1$ contributions of Sec.~\ref{chap:eRD}. Combining the dilution factor $D$ of Eq.~\eqref{eq:dilution_factor_final} with the redshift factor $\mathcal{D}(T_f,T_0)$ of Eq.~\eqref{eq:mathcal_D_def}, the present-day amplitude reads
\begin{equation}
\label{eq:OmegaGW_form_today}
\Omega_{\rm GW,0}^{\rm (form)}[\Phi](k)
\simeq
D^{-4/3}\,\mathcal{D}(T_f,T_0)\,
\Omega_{\rm GW}^{\rm (form)}[\Phi](k,\eta_f),
\end{equation}
and the corresponding frequency redshift is
\begin{align}
\label{eq:freq_redshift_form}
\frac{f_0}{f_f}=\frac{a(T_f)}{a(T_0)}&=
D^{-1}\left(\frac{g_{\star,s}(T_0)}{g_{\star,s}(T_f)}\right)^{\!1/3}\!\frac{T_0}{T_f}\\
&\simeq
2.5\times 10^{-25}\,
D^{-1}\,
\gamma_{\rm H}^{-1/2}
\left(\frac{M_{{\rm PBH},f}}{10^9~{\rm g}}\right)^{1/2}
\left(\frac{g_{\star}(T_f)}{106.75}\right)^{1/4}
\left(\frac{106.75}{g_{\star,s}(T_f)}\right)^{1/3},
\end{align}
where in the second line we have used the expression of $T_f$ in Eq.~\eqref{eq:T_f_def}. The SIGW frequency at production $f_f$ is related to the comoving scale at formation $k_{f}=2\pi f_f$, which we can derive from Eqs.~\eqref{eq:kPBH_D} and \eqref{eq:kf_kUV_D}
\begin{align}
k_f
&\simeq
\frac{5.4\times 10^{18}~{\rm Mpc}^{-1}}{D^{1/3}}\,
\frac{\gamma_{\rm H}^{1/2}}{h_{\star}^{1/3}(T_f)}
\left(\frac{106.75}{g_{\star}(T_f)}\right)^{1/12}
\left(\frac{10^9~{\rm g}}{M_{\rm PBH}}\right)^{1/2}.
\end{align}
We deduce
\begin{equation}
\label{eq:f0_form}
f_0
\simeq
\frac{8.2~ {\rm kHz}}{D^{1/3}}\,
\frac{\gamma_{\rm H}^{1/2}}{h_{\star}^{1/3}(T_f)}
\left(\frac{106.75}{g_{\star}(T_f)}\right)^{1/12}
\left(\frac{10^9~{\rm g}}{M_{\rm PBH}}\right)^{1/2}.
\end{equation}

\begin{figure}[h!]
    \centering
    \includegraphics[width=0.65\linewidth]{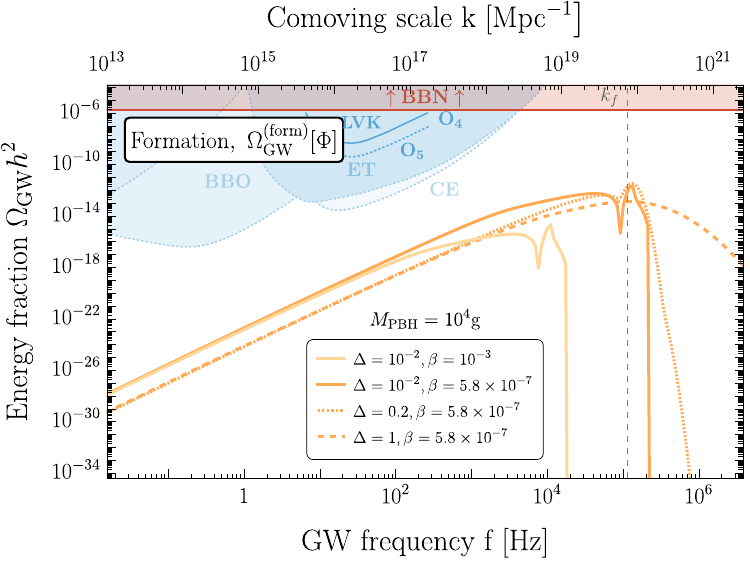}
    \caption{Present-day SIGW signal $\Omega_{\rm GW,0}^{\rm (form)}[\Phi]$ from PBH formation, Eq.~\eqref{eq:OmegaGW_form_today}, in the $(M_{\rm \mathsmaller{PBH}},\beta_f)$ plane. The width $\Delta$ of the curvature peak is kept as a free parameter.}
    \label{fig:formation}
\end{figure}

\clearpage
\section{High-frequency GWs: PBH mergers and Hawking emission}
\label{sec:HF_GW}
In addition to SIGWs, a gas of PBHs can also generate gravitational waves through mergers and direct graviton emission. We briefly summarize the relevant production channels here and refer to the companion paper~\cite{YannNicoPedro} for a detailed treatment:

\begin{enumerate}
    \item {\bf Isolated Mergers.}  
    Because PBHs are randomly distributed in space, they naturally exhibit local over- and underdensities. When two or three PBHs are sufficiently close, their mutual gravitational attraction can overcome the Hubble expansion, leading to the formation of bound systems that decouple from the Hubble flow. These systems can subsequently inspiral and merge—potentially mediated by the presence of a third body—thereby emitting gravitational waves.

    \item {\bf Clusters.}  
    Beyond isolated few-body systems, PBHs can form clusters of various sizes throughout the eMD era. Within these clusters, PBHs virialize, and dynamical interactions—such as hyperbolic encounters—can lead to the formation of bound binaries or three-body systems. These systems then merge and emit gravitational waves.

    \item {\bf Hawking evaporation.}  
    PBHs emit Hawking radiation with an (approximately) blackbody spectrum (see Sec.~\ref{sec:PBHgas}). This emission includes gravitons, with characteristic frequencies set by the initial PBH temperature.
\end{enumerate}
All of these mechanisms produce gravitational waves at very high frequencies, rendering them inaccessible to current and planned interferometers. In principle, they could contribute to the effective number of relativistic degrees of freedom, $N_{\rm eff}$. In~\cite{YannNicoPedro}, these contributions and their associated spectra are computed in detail, and conservative upper bounds are derived. It is shown that the high-frequency contributions remain negligible for $N_{\rm eff}$ and therefore do not modify the newly obtained constraints.

\clearpage
\section{Conclusion}
\label{conclusion}

The main result of this work is that the scalar-induced gravitational-wave (SIGW) signal from a realistic gas of evaporating ultra-light primordial black holes (PBHs) is orders of magnitude weaker than the monochromatic Poltergeist amplitudes commonly quoted in the recent literature. The physical origin of this suppression is the irreducible width of the PBH mass function. Spherical gravitational collapse in general relativity, through Choptuik's critical scaling, imposes the universal infrared behaviour (see Fig.~\ref{fig:psi_PBH})
\[
\psi_f\propto M_{\rm \mathsmaller{PBH}}^{1+1/\gamma_{\rm M}}\simeq M_{\rm \mathsmaller{PBH}}^{3.78}.
\]
This tail cannot be removed by tuning the primordial curvature spectrum. Once this realistic mass spread is included, PBHs evaporate over a finite time interval rather than simultaneously. The matter-to-radiation transition is therefore smoothed, and the post-evaporation suppression of the Newtonian potential steepens from the monochromatic scaling $\mathcal{S}_\Phi\propto k^{-1/3}$ to $\mathcal{S}_\Phi\propto k^{-4/3}$, see Figs.~\ref{fig:pbhepoch_longpaper} and \ref{fig:alpha_vs_gamma_M}. As a result, the SIGW spectrum is no longer dominated by the parametrically enhanced Poltergeist signal. Instead, it becomes qualitatively similar to the spectrum produced in a generic early matter-dominated era with gradual reheating, such as one driven by the decay of a heavy long-lived particle. This suppression relaxes the BBN bound on $\Delta N_{\rm eff}$ from PBH-induced SIGWs and reopens regions of the $(M_{\rm \mathsmaller{PBH}},\beta_f)$ plane that were previously thought to be excluded or within reach of future GW searches, see Fig.~\ref{fig:beta_vs_MPBH_standard}.

We have also provided a complete inventory of the SIGW signal in PBH reheating scenarios. In total, we computed eight distinct contributions to $\Omega_{\rm GW}$, summarized in Tab.~\ref{tab:GW_sources}. These contributions are generated throughout the cosmological history
\[
{\rm form}\to{\rm RD}_1\to{\rm eMD}\to{\rm RD}_2,
\]
where PBH formation is followed by an initial radiation era $\rm (RD_1)$, a PBH-driven early matter era $\rm (eMD)$, and a final radiation era $\rm (RD_2)$ after PBH evaporation. This decomposition shows that, once the Poltergeist enhancement is suppressed by the realistic mass distribution, the remaining SIGW channels become comparable to it and can even dominate in parts of parameter space, although they remain below the projected reach of future GW observatories. For the benchmark of Fig.~\ref{fig:GW_all_sources},the dominant effect of the realistic mass spread is confined to the post-evaporation Poltergeist component, $\Omega_{\rm GW}^{\rm (RD_2,Polt.)}$, which is highly sensitive to the duration of reheating. By contrast, the eMD contribution changes only mildly, while the other SIGW channels in Tab.~\ref{tab:GW_sources} are essentially insensitive to the details of the PBH-to-radiation transition. We also revisited earlier results for the SIGW signal from a monochromatic PBH gas and from a generic eMD phase with a constant decay rate. We confirm these results and provide new analytic expressions in Apps.~\ref{app:eMD} and~\ref{app:SIGW_mono_PBH}. 

The study of the high-frequency contributions from Hawking gravitons and binary mergers, summarised in Sec.~\ref{sec:HF_GW}, are reported in the companion paper~\cite{YannNicoPedro} where we show that they do not impact the constraints in Fig.~\ref{fig:beta_vs_MPBH_standard}.

Since our result differs from the usual expectation~\cite{Inomata:2020lmk,Papanikolaou:2020qtd,Domenech:2020ssp,Domenech:2021wkk,Bhaumik:2022pil,Gross:2024wkl,Gross:2025hia,Balaji:2024hpu,Domenech:2024cjn,Domenech:2025ffb,Inomata:2025wiv,Papanikolaou:2025ddc,Domenech:2023fuz,Domenech:2024kmh}, we now discuss its robustness. Three independent arguments support this conclusion.

\begin{itemize}
    \item 
\textbf{The IR slope of the PBH mass function is universal.} The exponent $1+1/\gamma_{\rm M}\simeq 3.78$ induced by Choptuik's critical scaling is fixed by spherical gravitational collapse in general relativity. It is therefore independent of the PBH formation channel, whether PBHs originate from inflationary peaks, first-order phase transitions, topological defects, scalar-field fragmentation, or other mechanisms. Naively, departure from sphericity is expected to broaden the mass function further and can only strengthen the suppression. While the critical exponent $\gamma_M$ depends on the equation of state $\omega$ at horizon crossing~\cite{Maison:1995cc}, the high-$k$ scaling of the suppression factor, $\mathcal{S}_\Phi\propto k^{-4/3}$, does not. Only the overall prefactor of $\mathcal{S}_\Phi$ depends on $\gamma_M$, making the suppression weakly sensitive to the equation of state. Most of this dependence is instead expected to enter through the transfer functions of $S$ and $\Phi$, as studied in~\cite{Domenech:2024wao}.
\item
\textbf{Our prescription is conservative.} We have made choices that maximize, rather than minimize, the SIGW signal. First, we impose a sharp ultraviolet cutoff on $\psi(M_{\rm \mathsmaller{PBH}})$. This removes the high-mass tail of the PBH distribution. In a realistic mass function, this tail would make evaporation last longer than $\eta_{\rm cut}\simeq 1.26\,\langle\eta_{\rm eva}\rangle$, smooth the $4/3$ cusp of the envelope $F_\Phi$, and further suppress the Poltergeist signal. Second, for both adiabatic and isocurvature seeds, we consider the two cutoff prescriptions summarized in Tab.~\ref{tab:summary_cut-off}. Cutting the GW convolution at the non-linear scale $k_{\rm \mathsmaller{NL}}$ gives a lower estimate, while extending it up to the fluid scale $k_{\rm \mathsmaller{PBH}}$ gives an upper envelope. Our most aggressive prescription therefore combines the narrowest mass distribution allowed by general relativity with the largest physically motivated UV cutoff. It should be interpreted as an upper bound on the SIGW signal. The middle-left panel of Fig.~\ref{fig:beta_vs_MPBH_standard} shows that even with this upper envelope choice, the SIGW bounds from $\Delta N_{\rm eff}$ and the projected reach of GW interferometers are relaxed wherever the pre-existing BBN bound from energy injection does not already dominate (green region).
\item
\textbf{Agreement with the monochromatic limit.} As shown in Fig.~\ref{fig:GWSigMono_longversion}, in the limit $\gamma_{\rm M}\to 0$ the analytic and numerical pipelines of this work reproduce the monochromatic Poltergeist amplitudes of Refs.~\cite{Inomata:2020lmk,Domenech:2024wao}, confirming that the orders-of-magnitude reduction we report is driven entirely by the universal IR slope.
\end{itemize}
A major remaining uncertainty is the contribution from genuinely non-linear halo dynamics during eMD. Adiabatic lattice simulations~\cite{Fernandez:2023ddy}, whose results are summarised in App.~\ref{app:NL_eMD}, find that the GW signal extends beyond the $k_{\rm \mathsmaller{NL}}$ scale, until the shell-crossing scale $k_{\mathsmaller{NL}}\,\simeq\,8\,k_{\rm \mathsmaller{NL}}]$, while the isocurvature analogue remains unexplored. This makes the latter a natural target for future numerical work.

\section*{Acknowledgements}

We thank Amirah Aljazaeri, Marco Calzà, Guillem Domènech, Gabriele Franciolini, Keisuke Inomata, Antonio Iovino, Riccardo Maule, Gabriele Perna, Jan Tränkle, Ville Vaskonen, and Hardi Veermäe for useful discussions. 
\\We acknowledge support by the Cluster of Excellence ``PRISMA$^{++}$'' funded by the German Research Foundation (DFG) within the German Excellence Strategy (Project No. 390831469). N.L. is grateful to the
German Academic Scholarship Foundation for the award of a PhD fellowship.

\newpage
\appendix

\section{PBH Dominated Era}
\label{App:PBHgas}
In this section, we provide the full derivation of all relevant quantities of the PBH era and provide the exact expression in terms of $g_\star(T)$ and $g_{\star,s}(T)$.
For that we assume a monochromatic mass function centered at $\langle M_{\rm \mathsmaller{PBH}} \rangle$ given by Eq.~\eqref{eq:M_f_long}, and we model the PBH-dominated era with the smooth scale factor of Eq.~\eqref{eq:scale_factor_full}, for which $\eta_\star=\eta_{\rm eq}/(\sqrt{2}-1)$ and the horizon-crossing values are
\begin{equation}
    k_{\rm eq}\eta_{\rm eq}=4-2\sqrt{2},\qquad\textrm{and}\qquad  k_{\rm eva}\left<\eta_{\rm eva}\right>\to2,
\end{equation}
in the deep matter-dominated regime $\left<\eta_{\rm eva}\right>/\eta_{\rm eq}\gg1$.

\subsection{PBH formation}
\label{sec:lightest_PBH}
In this work, we focus on the so-called {ultra-light} PBH window evaporating before the onset of Big-Bang Nucleosynthesis (BBN), $t_{\rm eva}\lesssim 1~\rm s$, which upon using Eq.~\eqref{eq:t_eva_def} imply that they are lighter than about $M_{{\rm PBH},f}\lesssim 10^{9}~\rm g$.  From inverting Eq.~\eqref{eq:M_f_long}, they would have formed at the temperature
\begin{equation}
\label{eq:T_f_def}
T_f \simeq 3.1\times 10^{11}~{\rm GeV}~\gamma_{\rm H}^{1/2}\left(\frac{106.75}{g_{\star}(T_f)}\right)^{1/4}\left(\frac{10^9~\rm g}{M_{{\rm PBH},f}}\right)^{1/2}.
\end{equation}
The quantity $M_{{\rm PBH},f}$ refers to the initial average PBH mass at formation 
\begin{equation}
\label{eq:M_f_long}
M_{{\rm PBH},f} \equiv \gamma_{\rm H} M_{{\rm H}, f} = \gamma_{\rm H} \frac{4\pi M_{\rm pl}^2}{H_{\rm f}} = 12\sqrt{\frac{10}{g_\star{ (T_f)}}} \gamma_{\rm H}\frac{M_{\rm pl}^3}{T_f^2},
\end{equation}
where $M_{{\rm H}, f} $ is the mass within the horizon when the Hubble expansion rate is $H_f$ and the temperature is $T_f$.
The horizon mass fraction falling into the PBH is $\gamma_{\rm H}=\mathcal{O}(1)$ for superhorizon collapse, see Sec.~\ref{sec:PBH_mass_distrib} for more details.

The PBH formation temperature $T_{\rm f}$ cannot be higher than the maximal temperature $T_{\rm max}$ of the universe given by
\begin{equation}
\label{eq:Tmax_uni}
    T_{\rm max}\simeq \left(\frac{3M_{\rm pl}^2H_{\rm end}^2}{\pi^2g_{\star}/30}\right)^{1/4} \; ,
\end{equation}
where $H_{\rm end}$ is the Hubble scale at the end of inflation.
The non-observation of primordial B modes in CMB anisotropies by BICEP/Keck~\cite{BICEP:2021xfz} provides the most stringent upper bound on the tensor-to-scalar ratio 
\begin{equation}
\label{eq:tensor_to_scalar_ratio}
    r\equiv \frac{\Delta^2_t(k_{\rm CMB})}{\Delta^2_s(k_{\rm CMB})}\lesssim 0.036,
\end{equation}
where $\mathcal{A}_{\mathcal{R}}(k_{\rm CMB})\simeq 2.099\times 10^{-9}$ is the curvature power spectrum measured by Planck at the CMB scale $k_{\rm CMB}=0.05~\rm Mpc^{-1}$~\cite{Aghanim:2018eyx}. In the slow-roll approximation, the tensor power spectrum reads~\cite{Baumann:2022mni}
\begin{equation}
\label{eq:tensor_power_spectrum}
    \Delta_t^2(k) = \frac{2}{\pi^2}\frac{H_k^2}{M_{\rm pl}^2} \; .
\end{equation}
Using Eq.~\eqref{eq:tensor_power_spectrum}, we can translate the BICEP/Keck bound in Eq.~\eqref{eq:tensor_to_scalar_ratio} to an upper bound on the maximal reheating temperature of the universe in Eq.~\eqref{eq:Tmax_uni}
\begin{equation}
\label{eq:T_max}
    T_{\rm max} \simeq 5.8\times 10^{15}~{\rm GeV}\left(\frac{r}{0.036} \right)^{1/4}\left(\frac{106.75}{g_{\star}} \right)^{1/4}\left(\frac{H_{\rm end}}{H_{\rm CMB}} \right)^{1/2},
\end{equation}
where $H_{\rm CMB}$ is the Hubble rate when $k_{\rm CMB}$ exits the horizon. The final ratio in Eq.~\eqref{eq:T_max} depends on the running of the tensor power spectrum $d\Delta^2_t(k)/d\ln(k)$. Assuming the power-law $\Delta_t^2(k)= A_t\left( k/k_{\rm CMB}\right)^{n_t}$, we obtain $\left({H_{\rm end}/H_{\rm CMB}} \right)^{\!1/2}= \left({k_{\rm end}}/{k_{\rm CMB}} \right)^{\!n_t/2}\leq 1$ leading to $H_{\rm end}/ H_{\rm CMB} \to 1$ for $n_t\to 0$. Plugging Eqs.~\eqref{eq:M_f_long} and \eqref{eq:T_max} into the condition $T_f<T_{\rm max}$ leads to the following lower bound on the PBH mass
\begin{equation}
\label{eq:M_min_app}
    M_{{\rm PBH},f} > M_{\rm min}\simeq 2.8~{g}~\gamma_{\rm H}\left(\frac{0.036}{r} \right)^{1/2}\left(\frac{H_{\rm CMB}}{H_{\rm end}} \right),
\end{equation}
where $1~{\rm g}\simeq 5.0\times 10^{-34}~M_{\odot}$. This explain the gray regions in Figs.~\ref{fig:beta_vs_MPBH_standard}.

\subsection{PBH domination}

Using $H=1/2t$ in Eq.~\eqref{eq:t_eva_def}, we obtain the temperature of the universe when PBHs evaporate
\begin{equation}
\label{eq:T_eva_def}
T_{\rm eva} =\left(\frac{405}{2\pi^2}\right)^{\!1/4}
\frac{A^{1/2}M_{\rm pl}^{5/2}}{g_{\star}^{1/4}(T_{\rm eva})\,M_{{\rm PBH},f}^{3/2}} \simeq 0.76~{\rm MeV}\,\left( \frac{g_{{\rm H} \star}}{108}\right)^{1/2}\left( \frac{106.75}{g_{\star}(T_{\rm eva})}\right)^{1/4}\left(\frac{10^9~\rm g}{M_{{\rm PBH},f}}\right)^{3/2}.
\end{equation}
where $g_\star$ is the number of relativistic degrees of freedom and  $A\equiv \pi\,\mathcal{G}\,g_{{\rm H}\star}/480\simeq 2.7$ (see Eq.~\eqref{eq:Gamma_def}). This would also be the reheating temperature of the universe if PBHs evaporate while dominating the universe.
The universe becomes PBH-dominated if the reheating temperature $T_{\rm eva}$ is lower than the temperature $T_{\rm eq}$ when PBHs start dominating the universe
\begin{equation}
\label{eq:T_eq_def}
    T_{\rm eq} =\frac{h_{\star}(T_{\rm eq})}{h_{\star}(T_f)} \, T_f \,\beta_f = 9.7~{\rm TeV}~\gamma_{\rm H}^{1/2}\left(\frac{h_{\star}(T_{\rm eq})}{h_{\star}(T_f)}\right)\left(\frac{106.75}{g_{\star}(T_f)}\right)^{\!1/4}\!\left(\frac{10^8~\rm g}{M_{\rm \mathsmaller{PBH}}}\right)^{\!1/2}\!\left(\frac{\beta_f}{10^{-8}} \right).
\end{equation}
where we have introduced
\begin{equation}
\label{eq:h_star_def}
h_{\star}(T) \equiv \frac{g_{\star,s}(T)}{g_{\star}(T)},
\end{equation} 
which equals $h_{\star}(T)=1$ if $g_{\star}(T)=g_{\star,s}(T)$. 
PBHs dominate the energy density before they
evaporate if their initial abundance $\beta_f$, defined as the fraction of the total energy density at formation, 
\begin{equation}
\beta_f\equiv\Omega_{\mathrm{PBH}, \mathrm{f}}=\frac{\rho_{\mathrm{PBH}, \mathrm{f}}}{3 H_{\mathrm{f}}^2 M_{\mathrm{pl}}^2}.
\end{equation}
exceeds a critical threshold $\beta_f>\beta_c$.
Plugging Eqs.~\eqref{eq:T_eva_def} and \eqref{eq:T_eq_def} into the condition for PBH domination $T_{\rm eq}>T_{\rm eva}$, we find
\begin{equation}
\label{eq:beta_min_long}
\beta_{\rm c} = \frac{h_{\star}(T_f)} {h_{\star}(T_{\rm eq})}\frac{T_{\rm eva}}{T_f}= 2.5\times 10^{-14} ~\gamma_{\rm H}^{-1/2}\left(\frac{g_{\mathrm{H} \star}}{108}\right)^{\!1 / 2}\left(\frac{h_{\star}(T_f)} {h_{\star}(T_{\rm eq})}\right)\left(\frac{g_{\star}(T_f)}{g_{\star}(T_{\rm eva})}\right)^{\!1 / 4}\!\left(\frac{10^8\, {\rm g}}{M_{{\rm PBH},f}}\right)
.
\end{equation}
From Eqs.~\eqref{eq:T_eq_def} and \eqref{eq:beta_min_long}, notice also the relationship
\begin{equation}
    \beta_c ~=~ \beta_f  
    \frac{T_{\rm eva}}{T_{\rm eq}}.
\end{equation}
 PBH evaporation following domination changes the comoving entropy density of the universe, $S \propto g_{\star,s} T^3 a^3$, by a factor
\begin{equation}
\label{eq:Dil_def_S}
    D \equiv \frac{S(T_{\rm eva})}{S(T_{\rm eva}^{-})} = \frac{S(T_{\rm eva})}{S(T_{\rm eq})} = \frac{g_{\star,s}(T_{\rm eva})}{g_{\star,s}(T_{\rm eq})} \left(\frac{T_{\rm eva}\,a(T_{\rm eva})}{T_{\rm eq}\,a(T_{\rm eq})}\right)^{\!3},
\end{equation}
where $T_{\rm eva}^-$ and $T_{\rm eva}$ denote the temperatures of the universe immediately before and after evaporation, respectively. In the second equality of Eq.~\eqref{eq:Dil_def_S}, we have used conservation of the comoving entropy from $T_{\rm eq}$ to $T_{\rm eva}^{-}$, that is $S(T_{\rm eva}^{-})=S(T_{\rm eq})$, while the third equality is the definition $S\propto g_{\star,s}T^3a^3$.
It can be useful to rewrite $D$ as
\begin{equation}
\label{eq:Dil_def}
    D = \frac{g_{\star,s}(T_{\rm eva})}{g_{\star,s}(T_{\rm eq})}\left(\frac{g_{\star}(T_{\rm eq})}{g_{\star}(T_{\rm eva})}\right)^{\!3/4}\left(\frac{a_{\rm eva}}{a_{\rm eq}}\right)^{\!3/4}=\frac{h_{\star}(T_{\rm eva})}{h_{\star}(T_{\rm eq})} \frac{T_{\rm eq}}{T_{\rm eva}} =\frac{h_{\star}(T_{\rm eva})}{h_{\star}(T_{\rm eq})} \frac{\beta_f}{\beta_c},
\end{equation}
where $h_\star(T)$ is defined in Eq.~\eqref{eq:h_star_def}. In the first equality of Eq.~\eqref{eq:Dil_def}, we have substituted the temperature ratio $T_{\rm eva}/T_{\rm eq}$ that follows from $a^3(T_{\rm eva})/a^3(T_{\rm eq})=\rho_{\rm m}(T_{\rm eq})/\rho_{\rm m}(T_{\rm eva})$ together with the radiation-matter equality conditions $\rho_{\rm m}(T_{\rm eq})=\rho_{\rm r}(T_{\rm eq})$ and $\rho_{\rm m}(T_{\rm eva})=\rho_{\rm r}(T_{\rm eva})$. Using $\rho_{\rm r}=(\pi^2/30)\,g_{\star}T^4$, these conditions give $T_{\rm eva}/T_{\rm eq}=\left(g_{\star}(T_{\rm eq})/g_{\star}(T_{\rm eva})\right)^{1/4}\left(a_{\rm eq}/a_{\rm eva}\right)^{3/4}$. The evaporation is assumed to occur instantaneously, i.e. $a(T_{\rm eva}^-)\simeq a(T_{\rm eva})$. The second equality of Eq.~\eqref{eq:Dil_def} follows from inserting the same relation once more to trade $\left(a_{\rm eva}/a_{\rm eq}\right)^{3/4}$ for $T_{\rm eq}/T_{\rm eva}$ and regrouping the statistical weights into $h_{\star}=g_{\star,s}/g_{\star}$. In the last equality, we have used Eqs.~\eqref{eq:T_eq_def} and \eqref{eq:beta_min_long} to express the dilution factor $D$ in terms of the ratio of the PBH energy fraction at formation $\beta_f$ divided by the threshold value $\beta_c$ for generating a PBH-dominated era. 
Using Eqs.~\eqref{eq:beta_min_long}, we can express $D$ as a function of PBH parameters
\begin{equation}
\label{eq:dilution_factor_final}
     D  \simeq 4.1 \times 10^{6}~\gamma_{\rm H}^{1/2}\left( \frac{108}{g_{\star \rm H}} \right)^{\!1/2}\left( \frac{h_\star(T_{\rm eva})}{h_\star(T_f)} \right)\left( \frac{g_\star(T_{\rm eva})}{g_\star(T_f)} \right)^{\!1/4}\!\left( \frac{M_{{\rm PBH},f}}{10^{9}~\rm g} \right)\left( \frac{\beta_f}{10^{-8}} \right).
\end{equation}
PBH evaporation following a PBH-dominated era injects entropy $D>1$. The case of an adiabatic universe expansion, e.g. when PBHs never dominate the energy density of the universe, can be recovered for $D=1$. 
In order to ensure that $D\geq 1$, we shift $D\to 1+D$ in our numerical treatment.

\subsection{Hubble expansion during matter era}
From the last equality of Eq.~\eqref{eq:Dil_def},
\begin{equation}
\label{eq:beta_ratio_a}
\frac{\beta_f}{\beta_c}=\left(\frac{g_{\star}(T_{\rm eva})}{g_{\star}(T_{\rm eq})}\right)^{\!1/4}\left(\frac{a_{\rm eva}}{a_{\rm eq}}\right)^{\!3/4},
\end{equation}
which requires the expansion factor $a_{\rm eva}/a_{\rm eq}$ across the PBH-dominated era. Evaluating the first branch of the smooth scale factor Eq.~\eqref{eq:scale_factor_full} at $\eta=\left<\eta_{\rm eva}\right>$, in the deep matter-dominated regime $\left<\eta_{\rm eva}\right>/\eta_{\rm eq}\gg1$,
\begin{equation}
\label{eq:aeva_aeq_x}
    \frac{a_{\rm eva}}{a_{\rm eq}}=\left(\frac{\left<\eta_{\rm eva}\right>}{\eta_\star}\right)^{\!2}+2\,\frac{\left<\eta_{\rm eva}\right>}{\eta_\star}\simeq\left(\frac{\left<\eta_{\rm eva}\right>}{\eta_\star}\right)^{\!2}.
\end{equation}
The horizon-crossing definitions $k_{\rm eq}=(4-2\sqrt{2})/\eta_{\rm eq}$ and $k_{\rm eva}\to2/\left<\eta_{\rm eva}\right>$, with $\eta_\star=\eta_{\rm eq}/(\sqrt{2}-1)$, give
\begin{equation}
\label{eq:keq_keva_x}
    \frac{k_{\rm eq}}{k_{\rm eva}}=(2-\sqrt{2})\,\frac{\left<\eta_{\rm eva}\right>}{\eta_{\rm eq}}
    =\frac{2-\sqrt{2}}{\sqrt{2}-1}\,\frac{\left<\eta_{\rm eva}\right>}{\eta_\star}=\sqrt{2}\,\frac{\left<\eta_{\rm eva}\right>}{\eta_\star},
\end{equation}
using $\eta_{\rm eq}=(\sqrt{2}-1)\eta_\star$ and $(2-\sqrt{2})/(\sqrt{2}-1)=\sqrt{2}$. Squaring Eq.~\eqref{eq:keq_keva_x} and inserting Eq.~\eqref{eq:aeva_aeq_x} gives the geometric link\footnote{Neglecting the order-unity matching at matter-radiation equality, i.e. extrapolating the deep matter-dominated relation $\mathcal{H}=2/\eta$ down to $\eta_{\rm eq}$ so that $k_{\rm eq}\eta_{\rm eq}\to2$, gives the commonly quoted $a_{\rm eva}/a_{\rm eq}=(k_{\rm eq}/k_{\rm eva})^2$, hence $D=c_2(k_{\rm eq}/k_{\rm eva})^{3/2}$ and $N_{\rm MD}=2\ln(k_{\rm eq}/k_{\rm eva})$ as in Ref.~\cite{Inomata:2020lmk}. Using instead the equality value $k_{\rm eq}\eta_{\rm eq}=4-2\sqrt{2}$ of the smooth background Eq.~\eqref{eq:scale_factor_full} introduces the constant factor $1/2$ in $a_{\rm eva}/a_{\rm eq}$, equivalently $2^{-3/4}\simeq0.60$ in $D$ and a $-\ln2$ shift in $N_{\rm MD}$.}
\begin{equation}
\label{eq:aeva_aeq_smooth}
    \frac{a_{\rm eva}}{a_{\rm eq}}\simeq\frac{1}{2}\left(\frac{k_{\rm eq}}{k_{\rm eva}}\right)^{\!2}.
\end{equation}

\subsection{Comoving scales}
An important scale is the comoving PBH interspacing above which the PBH gas description fails~\cite{Papanikolaou:2020qtd}
\begin{equation}
\label{eq:kUV_exp_0}
    k_{\rm \mathsmaller{PBH}} \equiv a(T_f)\left(\frac{4\pi}{3}\overline{n}_{\rm \mathsmaller{PBH}}(t_f)\right)^{1/3} = a(T_f)\left(\frac{4\pi}{3}\frac{\beta_f \rho_{\rm r}(T_f)}{M_{{\rm PBH},f}}\right)^{1/3} .
\end{equation}
Using $a(T_f)/a(T_0)=(T_0/T_f)(g_{\star,s}(T_0)/g_{\star,s}(T_f))^{1/3}/D^{1/3}$, the comoving wavenumber associated with the mean PBH interspacing reads
\begin{align}
\label{eq:kPBH_D}
    k_{\rm PBH}&=\left(\frac{2\pi^3}{45}\right)^{1/3}
\left(\frac{g_{\star,s}(T_0)}{h_{\star}(T_f)}\right)^{1/3}
\left(\frac{\beta_f T_f}{D\,M_{\rm \mathsmaller{PBH}}}\right)^{1/3}
T_0 \\
&\simeq\frac{3.6\times 10^{16}~{\rm Mpc^{-1}}}{D^{1/3}}\frac{\gamma_{H}^{1/6}}{h^{1/3}_{\star}(T_f)}\left(\frac{106.75}{g_\star(T_f)}\right)^{1/12}\left(\frac{\beta_f}{10^{-8}}\right)^{1/3}\left(\frac{10^8~\rm g}{M_{\rm PBH}}\right)^{1/2},
\end{align}
where we have set $a(T_0)=1$, $T_{0}\simeq 2.73~\rm K$, and $g_{\star,s}(T_0) \simeq 3.94$. For $D>1$, using Eqs.~\eqref{eq:Dil_def} and \eqref{eq:dilution_factor_final},
\begin{align}
\label{eq:kUV_exp_0b}
    k_{\rm \mathsmaller{PBH}}(D>1) &= \left(\frac{2\pi^3}{45}\right)^{1/3}\left(\frac{g_{\star,s}(T_0)}{h_{\star}(T_{\rm eva})}\right)^{1/3}\left(\frac{T_{\rm eva}}{M_{\rm \mathsmaller{PBH}}}\right)^{1/3}T_{\rm 0}\\
    &\simeq \frac{4.9 \times 10^{14}~{\rm Mpc^{-1}}}{h_{\star}^{1/3}(T_{\rm eva})} \left( \frac{106.75}{g_{\star}(T_{\rm eva})}\right)^{1/12}\left( \frac{g_{H\star}}{108} \right)^{1/6}\left( \frac{10^8~\rm g}{M_{{\rm PBH},f}} \right)^{5/6} .
\end{align}
The universe temperature $T_H$ when the mode $k_H$ enters the horizon, $aH=k_H$, is
\begin{equation}
\label{eq:TH_vs_kH}
T_H = 5.8 \times 10^{12}~{\rm GeV}~ h_{\star}(T_H)^{1/3}D^{1/3}\left(\frac{106.75}{g_{\star}(T_H)}\right)^{1/6}\left(\frac{k_H}{10^{20}~\rm Mpc^{-1}}\right),
\end{equation}
where $D\equiv S(T_0)/S(T_H)$ is the today to early-time comoving entropy density ratio with $S \propto g_{\star,s}T^3a^3$.\footnote{We have also used $g_\star(T_0)\equiv 2+(7/8)\cdot 2\cdot N_{\rm eff}\cdot (4/11)^{4/3} \simeq 3.38$, and $g_{\star,s}(T_0)\equiv 2+(7/8)\cdot 2\cdot N_{\rm eff}\cdot (4/11) \simeq 3.94$ which assume $N_{\rm eff}\simeq 3.043$ \cite{Cielo:2023bqp}.}
Plugging Eqs.~\eqref{eq:T_f_def} and \eqref{eq:T_eq_def} into Eq.~\eqref{eq:TH_vs_kH}, and dividing by Eq.~\eqref{eq:kPBH_D}, we obtain
\begin{align}
&\frac{k_{\rm eva}}{k_{\rm \mathsmaller{PBH}}} ~=~ \left(\frac{3A}{8\pi}\right)^{\!1/3}\!\left(\frac{M_{\rm pl}}{M_{{\rm PBH},f}}\right)^{\!2/3}\!\left(\frac{h_{\star}(T_f)}{h_{\star}(T_{\rm eva})}\,\frac{D\,T_{\rm eva}}{\beta_f T_f}\right)^{\!1/3} \label{eq:keva_kUV_D}\\
    &\frac{k_{\rm eq}}{k_{\rm \mathsmaller{PBH}}} ~=~ \sqrt{2}\,c_2^{-2/3}\left(\frac{3A}{8\pi}\right)^{\!1/3}\!\left(\frac{M_{\rm pl}}{M_{{\rm PBH},f}}\right)^{\!2/3}\!D^{2/3}, \label{eq:keq_kUV_D}\\[4pt]
    &\frac{k_{\rm f}}{k_{\rm \mathsmaller{PBH}}} ~=~ \left(\frac{\gamma_{\rm H}}{\beta_{\rm c}}\frac{h_{\star}(T_{\rm eva})}{h_{\star}(T_{\rm eq})}\right)^{\!1/3}\!D^{-1/3}, \label{eq:kf_kUV_D}
\end{align}
with $c_2\equiv \left( {g_{\star}(T_{\rm eva})}/{g_{\star}(T_{\rm eq})} \right)^{1/4}\left( {h_{\star}(T_{\rm eva})}/{h_{\star}(T_{\rm eq})} \right)$. Equations~\eqref{eq:keva_kUV_D} and \eqref{eq:kf_kUV_D} are exact identities, since $k_{\rm eva}$ and $k_f$ cross the horizon deep in matter and radiation domination: the bracket in Eq.~\eqref{eq:keva_kUV_D} equals unity by Eqs.~\eqref{eq:Dil_def} and \eqref{eq:beta_min_long}, and Eq.~\eqref{eq:kf_kUV_D} collapses to $(\gamma_{\rm H}/\beta_f)^{1/3}$. The factor $\sqrt{2}$ in Eq.~\eqref{eq:keq_kUV_D} is that of Eq.~\eqref{eq:keq_keva_x}, reflecting that $k_{\rm eq}$ crosses at matter-radiation equality. Using Eq.~\eqref{eq:dilution_factor_final} for $D$, these become
\begin{align}
&\frac{k_{\rm eva}}{k_{\rm \mathsmaller{PBH}}}(D>1) ~=~ \left(\frac{3A}{8\pi}\right)^{1/3}\!\left(\frac{M_{\rm pl}}{M_{{\rm PBH},f}}\right)^{\!2/3} ~\simeq~ 8.5\times 10^{-10}\left(\frac{g_{\rm H\star}}{108} \right)^{\!1/3}\left(\frac{10^{8}~\rm g}{M_{{\rm PBH},f}}\right)^{\!2/3},\label{eq:keva_kUV}\\
    &\frac{k_{\rm eq}}{k_{\rm \mathsmaller{PBH}}}(D>1) ~=~ \sqrt{2}\,c_1\gamma_{\rm H}^{1/3}\beta_f^{2/3} ~\simeq~ 6.5\times 10^{-6}\,c_1\,\gamma_{\rm H}^{1/3}\left( \frac{\beta_f}{10^{-8}} \right)^{\!2/3},  \label{eq:keq_kUV}\\
    &\frac{k_{\rm f}}{k_{\rm \mathsmaller{PBH}}}(D>1) ~=~ \left(\frac{\gamma_{\rm H}}{\beta_f}\right)^{1/3} ~\simeq ~ 4.6\times 10^2~\gamma_{\rm H}^{1/3}\left(\frac{10^{-8}}{\beta_f}\right)^{1/3}, \label{eq:kf_kUV}
\end{align}
where $c_1\equiv \left( g_{\star}(T_{\rm eq})h_{\star}^4(T_{\rm eq}) /g_{\star}(T_{f})/ h_{\star}^4(T_{f})\right)^{1/6}$. Finally, inserting Eq.~\eqref{eq:aeva_aeq_smooth} into Eq.~\eqref{eq:beta_ratio_a} and into the entropy relation Eq.~\eqref{eq:Dil_def}, the dilution factor and the number of e-foldings of the PBH-dominated era read
\begin{equation}
\label{eq:D_NMD}
    D \simeq 2^{-3/4}c_2\left(\frac{k_{\rm eq}}{k_{\rm eva}}\right)^{\!3/2},
    \qquad
    N_{\rm MD}\equiv\ln\!\left(\frac{a(T_{\rm eva})}{a(T_{\rm eq})}\right)=\frac{4}{3}\ln\!\left(c_2^{-1}D\right)\simeq 2\ln\!\left(\frac{k_{\rm eq}}{k_{\rm eva}}\right)-\ln2\,,
\end{equation}
with $2^{-3/4}\simeq0.60$, the constant $2^{-3/4}$ being the cube-root-free form of the $\sqrt{2}$ in Eq.~\eqref{eq:keq_keva_x}.

\newpage

\section{PBH mass distribution following Choptuik's law}
\label{app:PBH_mass_distrib}
This appendix derives the PBH mass distribution used in Sec.~\ref{sec:PBH_mass_distrib}. We first specify the formation prescription and recall Choptuik scaling. We then introduce the compaction scale and the geometric factor $\kappa$, derive the mass distribution in the Press--Schechter and peak-theory approaches, summarize the characteristic mass scales in Tab.~\ref{tab:kappa_gammaH}, and finally discuss the narrow-peak limit.
\subsection{PBH formation theory}
We consider PBHs formed during radiation domination from the collapse of large superhorizon density perturbations. After horizon re-entry, a perturbation collapses if the comoving-gauge density contrast smoothed over a horizon volume, $\delta_m$, exceeds a critical threshold $\delta_c$~\cite{Carr:1975qj}. The fraction of the total energy density that collapses into PBHs of mass $M_{\rm \mathsmaller{PBH}}$ at the epoch when the horizon mass is $M_H$ can be written as~\cite{Press:1973iz,Bond:1990iw,Gow:2020bzo,Karam:2022nym}
\begin{equation}
\label{eq:beta_k_MPBH}
\frac{d\beta(M_{\rm \mathsmaller{PBH}},M_H)}{d\ln M_{\rm \mathsmaller{PBH}}}
=
\int_{\delta_{\ell,c}}^\infty d\delta_{\ell}\,
\frac{M_{\rm \mathsmaller{PBH}}}{M_H}\,
P(\delta_{\ell},M_H)\,
\delta_D\!\left[
\ln\frac{M_{\rm \mathsmaller{PBH}}}{M(\delta_m(\delta_\ell))}
\right].
\end{equation}
Here $P(\delta_\ell,M_H)$ is the probability distribution of the linear density contrast smoothed on the Hubble scale at fixed $M_H$, and $\delta_{\ell,c}$ is the value corresponding to the nonlinear threshold $\delta_c$. During radiation domination, the nonlinear and linear density contrasts are related by~\cite{DeLuca:2019qsy,Young:2019yug}
\begin{equation}
\label{eq:delta_m_delta_l}
\delta_m
=
\delta_\ell
-
\frac{3}{8}\delta_\ell^2 .
\end{equation}
The branch relevant for PBH formation is therefore
\begin{equation}
\label{eq:delta_l_inversion}
\delta_\ell
=
\frac{4}{3}
\left(
1-\sqrt{\Lambda}
\right),
\qquad
\Lambda
\equiv
1-\frac{3}{2}\delta_m .
\end{equation}
Close to the collapse threshold, the PBH mass is governed by critical collapse. For a one-parameter family of initial data labeled by an amplitude $p$, there is a critical value $p_c$ separating dispersion from black-hole formation. For $p>p_c$ and sufficiently close to threshold, the solution first approaches a universal self-similar critical solution. Since this critical solution has only one unstable mode, the final black-hole mass is controlled by the distance to threshold and follows the Choptuik scaling law~\cite{Choptuik:1992jv,Abrahams:1993wa,Evans:1994pj,Maison:1995cc,Neilsen:1998qc,Koike:1999eg,Gundlach:2007gc,Niemeyer:1997mt,Niemeyer:1999ak,Yokoyama:1998xd,Green:1999xm,Ianniccari:2024ltb}
\begin{equation}
\label{eq:critical_scaling}
M(\delta_m)
=
\mathcal{K}\,M_H\,
(\delta_m-\delta_c)^{\gamma_{\rm M}} .
\end{equation}
The coefficient $\mathcal{K}$ depends on the perturbation profile and is typically of order a few, with $\mathcal{K}\simeq4$ often used as a reference value~\cite{Niemeyer:1997mt,Niemeyer:1999ak}. The exponent $\gamma_{\rm M}$ is universal for a fixed matter content, and for a radiation fluid one finds $\gamma_{\rm M}\simeq0.36$~\cite{Niemeyer:1997mt,Niemeyer:1999ak,Yokoyama:1998xd}. The threshold $\delta_c$ is profile dependent, with numerical studies giving the range $0.40\lesssim\delta_c\lesssim0.67$~\cite{Musco:2018rwt,Escriva:2019phb}. The scaling in Eq.~\eqref{eq:critical_scaling} assumes approximate spherical symmetry, near-threshold collapse, and a fixed equation of state during the collapse.
The horizon mass at temperature $T_H$ is
\begin{equation}
\label{eq:m_horizon}
M_H(T_H)
=
\frac{4\pi\rho}{3H^3}
=
0.95\times10^6~{\rm g}
\left(
\frac{106.75}{g_\star(T_H)}
\right)^{1/2}
\left(
\frac{10^{13}~{\rm GeV}}{T_H}
\right)^2 ,
\end{equation}
where $10^{7}~{\rm g}=5\times10^{-27}~M_{\odot}$ and $g_\star$ is the number of relativistic degrees of freedom.
Substituting Eq.~\eqref{eq:critical_scaling} into Eq.~\eqref{eq:beta_k_MPBH} and using $\delta_D(g(x))=\sum_i\delta_D(x-x_i)/|g'(x_i)|$ gives
\begin{equation}
\label{eq:beta_k_MPBH_2}
\frac{d\beta(M_{\rm \mathsmaller{PBH}},M_H)}
{d\ln M_{\rm \mathsmaller{PBH}}}
=
\frac{2\mathcal{K}}{\gamma_{\rm M}}\,
\frac{q^{1+1/\gamma_{\rm M}}}{\sqrt{\Lambda}}\,
P\!\left(
\delta_\ell(M_{\rm \mathsmaller{PBH}}),M_H
\right),
\end{equation}
with
\begin{equation}
\label{eq:q_Lambda_def}
q
\equiv
\frac{M_{\rm \mathsmaller{PBH}}}{\mathcal{K}\,M_H},
\qquad
\delta_m(M_{\rm \mathsmaller{PBH}})
=
\delta_c+q^{1/\gamma_{\rm M}},
\qquad
\Lambda
=
1-\frac{3}{2}
\left(
\delta_c+q^{1/\gamma_{\rm M}}
\right).
\end{equation}
The condition that the linear density contrast remains on the physical branch of Eq.~\eqref{eq:delta_l_inversion} implies
\begin{equation}
\label{eq:M_H_min}
M_H
\geq
M_H^{\rm min}(M_{\rm \mathsmaller{PBH}})
=
\frac{M_{\rm \mathsmaller{PBH}}}
{\mathcal{K}\,(2/3-\delta_c)^{\gamma_{\rm M}}}.
\end{equation}
Equivalently, at fixed $M_H$ the PBH mass satisfies
\begin{equation}
M_{\rm \mathsmaller{PBH}}
\leq
M_{\rm \mathsmaller{PBH}}^{\rm max}
=
\mathcal{K}(2/3-\delta_c)^{\gamma_{\rm M}}M_H .
\end{equation}
The remaining ingredient is the probability density $P(\delta_\ell,M_H)$, which we model using either Press--Schechter or peak theory.
\subsection{From the peak scale $k_f$ to the collapse scale $r_m$}
The Fourier scale $k_f$ at which the curvature power spectrum peaks is not, in general, the real-space scale that controls collapse. The relevant scale is defined by the compaction function, which measures the excess mass inside a spherical region normalized by its areal radius~\cite{Shibata:1999zs,Harada:2015yda,Musco:2012au,Musco:2018rwt,Young:2019yug,Escriva:2019nsa,Escriva:2021aeh,Musco:2020jjb}
\begin{equation}
\label{eq:compaction_def}
\mathcal{C}(r,t)
\equiv
2\,\frac{\delta M(r,t)}{R_{\rm ar}(r,t)}
=
2\,\frac{M_{\rm MS}(r,t)-M_b(r,t)}{R_{\rm ar}(r,t)} .
\end{equation}
Here $R_{\rm ar}$ is the areal radius, $M_{\rm MS}$ is the Misner--Sharp mass, and $M_b=4\pi\rho_b R_{\rm ar}^3/3$ is the background FRW mass enclosed in the same areal radius. The characteristic radius of the overdensity is the radius $r_m$ at which the compaction function is maximal,
\begin{equation}
\label{eq:rm_def}
\mathcal{C}'(r_m)=0 .
\end{equation}
For a curvature power spectrum peaked at $k_f$, we define the geometric factor
\begin{equation}
\label{eq:kappa_def}
\kappa
\equiv
k_f r_m .
\end{equation}
For the log-normal spectrum of Eq.~\eqref{eq:log_normal_app}, $\kappa$ depends only on the width $\Delta$ and is well fitted by~\cite{Musco:2020jjb,Gouttenoire:2025jxe}
\begin{equation}
\label{eq:kappa_lognormal_fit_app}
\kappa(\Delta)
\simeq
1.55+
\frac{1.19}{1+\left(\Delta/0.45\right)^{2.06}},
\qquad
0\leq\Delta\leq3 .
\end{equation}
This fit reproduces the tabulated values with a maximum absolute error below $5\times10^{-3}$. Representative values are given in Tab.~\ref{tab:kappa_gammaH}. Since
\begin{equation}
\label{eq:Mstar_MH_scaling}
M_f \equiv M_H(\mathcal{H}=k_f),
\qquad
M_H \propto \mathcal{H}^{-2}
\quad\text{(radiation domination)},
\end{equation}
the horizon mass associated with the compaction scale is
\begin{equation}
\label{eq:MH_Mf_kappa}
M_H(r_m)
\equiv
M_H(\mathcal{H}=1/r_m)
=
M_H(\mathcal{H}=k_f/\kappa)
=
\kappa^2 M_f .
\end{equation}
Thus the same physical distribution is shifted by a factor $\kappa^2$ depending on whether masses are normalized to $M_f$ or to $M_H(r_m)$. With this convention, the critical-collapse relation becomes
\begin{equation}
\label{eq:choptuik_Mf}
M_{\rm \mathsmaller{PBH}}
=
\mathcal{K}\,\kappa^2 M_f\,
(\delta_m-\delta_c)^{\gamma_{\rm M}} .
\end{equation}
\subsection{PBH mass distribution}
The probability density $P(\delta_\ell,M_H)$ entering Eq.~\eqref{eq:beta_k_MPBH} can be modeled in two standard ways, the Press--Schechter formalism and peak theory, which differ in how collapsing regions are counted but share the same Choptuik induced infrared slope. Combining either prescription with the critical-collapse relation of Eq.~\eqref{eq:choptuik_Mf} and integrating over horizon masses yields the formation-time mass distribution $\psi_f(M_{\rm \mathsmaller{PBH}})$ used throughout this work.

\subsubsection*{Press--Schechter formalism}
In the Press--Schechter approach, the smoothed linear density contrast is evaluated at a random spatial point and modeled as a Gaussian field~\cite{Press:1973iz,Bond:1990iw},
\begin{equation}
\label{eq:Prob_PS_Gauss}
P(\delta_{\ell},M_H)
=
\frac{1}{\sqrt{2\pi}\,\sigma_0(r_H)}
\exp\!\left[
-\frac{\delta_{\ell}^2}{2\sigma_0^2(r_H)}
\right],
\end{equation}
where $r_H=k_H^{-1}$ is the comoving Hubble radius when the horizon mass is $M_H$. The spectral moments of the smoothed density contrast are
\begin{equation}
\label{eq:sigma_j_moments}
\sigma_j^2(r_H)
=
\frac{16}{81}
\int\frac{dk}{k}\,
(k r_H)^{4+2j}
\tilde{W}^2(k,r_H)\,
T^2(k,r_H)\,
\mathcal{P}_{\mathcal{R}}(k).
\end{equation}
Here $\tilde{W}(k,R)=3j_1(kR)/(kR)$ is a real-space top-hat window function~\cite{Young:2019osy}, and $T(k,R)=3j_1(kR/\sqrt{3})/(kR/\sqrt{3})$ is the radiation-era transfer function in comoving gauge~\cite{Josan:2009qn}.
\footnote{Peak theory is often written in terms of the BBKS moments~\cite{Bardeen:1985tr}, defined by $\sigma_{j,\rm BBKS}^2(R)=\int d^3k/(2\pi)^3\,k^{2j}P_\delta(k)W^2(kR)$ when peaks of the density field are counted. Our moments are written in terms of the dimensionless combination $kr_H$ and are related by $\sigma_j^2(r_H)=r_H^{2j}\sigma_{j,\rm BBKS}^2(R=r_H)$. If peaks of $\mathcal{R}$ are counted instead, $P_\delta$ should be replaced by $P_{\mathcal R}$~\cite{Ferrante:2022mui}.}
\subsubsection*{Peak theory}
Peak theory counts local maxima of the smoothed density contrast rather than random spatial points. Let $\delta_\ell(\mathbf{x})$ be the density contrast smoothed on the Hubble scale $r_H$, and define its Hessian as
\begin{equation}
H_{ij}(\mathbf{x})
\equiv
\partial_i\partial_j\delta_\ell(\mathbf{x}) .
\end{equation}
The differential number density of peaks of height $\nu=\delta_\ell/\sigma_0$ is~\cite{Bardeen:1985tr}
\begin{equation}
n_{\rm pk}(\nu)
\equiv
\left\langle
\delta_D^{(3)}\!\big(\nabla\delta_\ell(\mathbf{x})\big)\,
|\det H(\mathbf{x})|\,
\Theta_{\rm pk}\!\big(H(\mathbf{x})\big)\,
\delta_D\!\left(
\nu-\frac{\delta_\ell(\mathbf{x})}{\sigma_0}
\right)
\right\rangle ,
\end{equation}
where $\Theta_{\rm pk}(H)=\prod_{a=1}^3\Theta(-\lambda_a)$, and $\lambda_a$ are the eigenvalues of $H_{ij}$. In the high-peak limit $\nu\gg1$, the number density of peaks with height in $[\delta_\ell,\delta_\ell+d\delta_\ell]$ is~\cite{Bardeen:1985tr}
\begin{equation}
\label{eq:npk_highnu_PBH}
n_{\rm pk}(\delta_\ell,M_H)\,d\delta_\ell
\simeq
\frac{1}{(2\pi)^2 r_H^3}
\left(
\frac{\sigma_1}{\sqrt{3}\,\sigma_0}
\right)^3
\left(
\frac{\delta_\ell}{\sigma_0}
\right)^3
\exp\!\left[
-\frac{\delta_\ell^2}{2\sigma_0^2}
\right]
\frac{d\delta_\ell}{\sigma_0}.
\end{equation}
To use this result in Eq.~\eqref{eq:beta_k_MPBH}, the Press--Schechter probability density is replaced by the expected number of peaks inside one Hubble volume $V_H=4\pi r_H^3/3$,
\begin{equation}
\label{eq:Prob_PS_Peak}
P(\delta_{\ell},M_H)
\simeq
\frac{1}{3\pi}
\left(
\frac{\sigma_1}{\sqrt{3}\,\sigma_0}
\right)^3
\left(
\frac{\delta_{\ell}}{\sigma_0}
\right)^3
\exp\!\left[
-\frac{\delta_{\ell}^2}{2\sigma_0^2}
\right].
\end{equation}
The main effect of peak theory is the extra high-peak weight proportional to $\nu^3$. It enhances rare overdensities and gives a heavier high-mass tail than Press--Schechter, while the low-mass scaling induced by Choptuik critical collapse remains unchanged.

\subsubsection*{Mass distribution}
The differential energy fraction at temperature $T$ is obtained by integrating over horizon masses,
\begin{equation}
\frac{d\beta(M_{\rm \mathsmaller{PBH}},T)}
{d\ln M_{\rm \mathsmaller{PBH}}}
\equiv
\int d\ln M_H\,
\frac{d\beta(M_{\rm \mathsmaller{PBH}},M_H)}
{d\ln M_{\rm \mathsmaller{PBH}}}\,
\frac{\rho(T_H)}{\rho(T)}\,
\frac{s(T)}{s(T_H)} .
\end{equation}
At the reference temperature $T_f$, defined by $k_f=\mathcal{H}$, we write
\begin{equation}
\label{eq:beta_f_def}
\beta_f
\equiv
\int d\ln M_{\rm \mathsmaller{PBH}}\,
\frac{d\beta(M_{\rm \mathsmaller{PBH}},T_f)}
{d\ln M_{\rm \mathsmaller{PBH}}},
\qquad
\frac{d\beta(M_{\rm \mathsmaller{PBH}},T_f)}
{d\ln M_{\rm \mathsmaller{PBH}}}
\simeq
\int d\ln M_H\,
\frac{d\beta(M_{\rm \mathsmaller{PBH}},M_H)}
{d\ln M_{\rm \mathsmaller{PBH}}}.
\end{equation}
The last expression neglects the small redshifting correction across the narrow range of horizon masses that dominates the integral. The normalized mass distribution at formation is
\begin{equation}
\label{eq:psi_f_app_def}
\psi_f(M_{\rm \mathsmaller{PBH}})
\equiv
\frac{1}{\beta_f}
\frac{d\beta(M_{\rm \mathsmaller{PBH}},T_f)}
{d\ln M_{\rm \mathsmaller{PBH}}}.
\end{equation}
In this work we use the log-normal curvature spectrum
\begin{equation}
\label{eq:log_normal_app}
\mathcal{P}_{\mathcal{R}}(k)
=
\frac{\mathcal{A}_{\mathcal{R}}}{\sqrt{2\pi}\,\Delta}
\exp\!\left[
-\frac{\ln^2(k/k_f)}{2\Delta^2}
\right],
\end{equation}
where $k_f$ is the peak scale, $\Delta$ is the width, and $\mathcal{A}_{\mathcal{R}}$ is the amplitude. The resulting distributions are shown in Fig.~\ref{fig:psi_PBH}. Their infrared behavior is universal,
\begin{equation}
\psi_f(M_{\rm \mathsmaller{PBH}})
\propto
M_{\rm \mathsmaller{PBH}}^{1+1/\gamma_{\rm M}},
\qquad
1+\frac{1}{\gamma_{\rm M}}
\simeq
3.78 ,
\end{equation}
while the ultraviolet tail depends on the abundance prescription and on the shape of the curvature spectrum. Peak theory enhances rare peaks relative to Press--Schechter and therefore extends the distribution toward larger PBH masses, but it does not modify the universal low-mass power law.
The mean PBH mass is defined as the total PBH energy density divided by the total PBH number density. Since $\psi_f$ is normalized as an energy fraction, one has
\begin{equation}
\label{eq:mean_mass_app}
\langle M_{\rm \mathsmaller{PBH}}\rangle
=
\left[
\int d\ln M_{\rm \mathsmaller{PBH}}\,
\frac{\psi_f(M_{\rm \mathsmaller{PBH}})}
{M_{\rm \mathsmaller{PBH}}}
\right]^{-1}.
\end{equation}
We denote the mean mass in units of $M_f$ by
\begin{equation}
\gamma_{\rm H}^{(k_f)}
\equiv
\frac{\langle M_{\rm \mathsmaller{PBH}}\rangle}{M_f},
\end{equation}
and the same mean mass in units of the horizon mass at the compaction scale by
\begin{equation}
\gamma_{\rm H}^{(r_m)}
\equiv
\frac{\langle M_{\rm \mathsmaller{PBH}}\rangle}{M_H(r_m)}
=
\frac{\gamma_{\rm H}^{(k_f)}}{\kappa^2}.
\end{equation}
The numerical values used in this work are collected in Tab.~\ref{tab:kappa_gammaH}.
\begin{table}[t]
\centering
\small
\renewcommand{\arraystretch}{1.25}
\setlength{\tabcolsep}{4pt}
\begin{tabular}{|c|c|c|c|c|c|c|c|}
\hline
\multirow{4}{*}{\textbf{\makecell[c]{\rule[-1.0ex]{0pt}{4.2ex}Quantity}}}
& \multirow{4}{*}{\textbf{\makecell[c]{\rule[-1.0ex]{0pt}{4.2ex}PBH abundance\\calculation}}}
& \multicolumn{6}{c|}{\makecell[c]{\rule[-0.5ex]{0pt}{3.4ex}\textbf{Width} $\boldsymbol{\Delta}$}}\\
\cline{3-8}
& & \multicolumn{2}{c|}{$0.1$} & \multicolumn{2}{c|}{$0.5$} & \multicolumn{2}{c|}{$1$}\\
\cline{3-8}
& & \multicolumn{6}{c|}{\makecell[c]{\rule[-0.5ex]{0pt}{3.4ex}$\boldsymbol{\mathcal{A}_{\mathcal{R}}}$}}\\
\cline{3-8}
& & $0.01$ & $0.04$ & $0.01$ & $0.04$ & $0.01$ & $0.04$\\
\hline
$\kappa\equiv k_f r_m$ & $-$ & \multicolumn{2}{c|}{$2.69$} & \multicolumn{2}{c|}{$2.08$} & \multicolumn{2}{c|}{$1.75$}\\
\hline
\multirow{2}{*}{$\gamma_{\rm H}^{(k_f)}\equiv\dfrac{\langle M_{\rm \mathsmaller{PBH}}\rangle}{M_f}$}
& Press--Schechter & $1.19$ & $1.09$ & $1.89$ & $1.65$ & $2.06$ & $1.78$\\
\cline{2-8}
& peak theory & $2.68$ & $4.21$ & $2.17$ & $2.69$ & $1.85$ & $2.03$\\
\hline
\multirow{2}{*}{$\gamma_{\rm H}^{(r_m)}\equiv\dfrac{\langle M_{\rm \mathsmaller{PBH}}\rangle}{M_H(r_m)}$}
& Press--Schechter & $0.16$ & $0.15$ & $0.44$ & $0.38$ & $0.67$ & $0.58$\\
\cline{2-8}
& peak theory & $0.37$ & $0.58$ & $0.50$ & $0.62$ & $0.61$ & $0.67$\\
\hline
\end{tabular}
\caption{Values of the geometric factor $\kappa$, the mean PBH mass in units of $M_f$, $\gamma_{\rm H}^{(k_f)}=\langle M_{\rm \mathsmaller{PBH}}\rangle/M_f$, and the same mean mass in units of the horizon mass at the compaction scale, $\gamma_{\rm H}^{(r_m)}=\langle M_{\rm \mathsmaller{PBH}}\rangle/M_H(r_m)=\gamma_{\rm H}^{(k_f)}/\kappa^2$. The values are shown for the widths used in Fig.~\ref{fig:psi_PBH}, for the two amplitudes $\mathcal{A}_{\mathcal{R}}=0.01$ and $\mathcal{A}_{\mathcal{R}}=0.04$, and for both the Press--Schechter and peak-theory prescriptions. The commonly quoted estimate $\gamma_{\rm H}^{(r_m)}\sim0.2$ of Ref.~\cite{Carr:2020gox} is recovered in the narrow-peak limit $\Delta\simeq0.1$ with the Press--Schechter prescription.}
\label{tab:kappa_gammaH}
\end{table}
\subsection{Narrow-peak approximation}
When the curvature spectrum is sufficiently narrow, the integral over horizon masses is dominated by the horizon mass associated with the compaction scale,
\begin{equation}
\label{eq:MH_narrow_kappa}
M_H
\simeq
M_H(r_m)
=
\kappa^2 M_f .
\end{equation}
The total abundance then reduces to
\begin{equation}
\label{eq:beta_f_narrow_peak}
\beta_f
\simeq
\int d\ln M_{\rm \mathsmaller{PBH}}\,
\frac{d\beta(M_{\rm \mathsmaller{PBH}},\kappa^2 M_f)}
{d\ln M_{\rm \mathsmaller{PBH}}}
\simeq
\int_{\delta_{\ell,c}}^\infty d\delta_\ell\,
\frac{M(\delta_m(\delta_\ell))}{\kappa^2 M_f}\,
P(\delta_\ell,\kappa^2 M_f).
\end{equation}
The second equality follows from integrating over the Dirac distribution in Eq.~\eqref{eq:beta_k_MPBH}. For Press--Schechter statistics and rare events,
\begin{equation}
\label{eq:nu_def}
\nu
\equiv
\frac{\delta_{\ell,c}}{\sigma_0(r_m)}
\gg
1 ,
\end{equation}
the integral is exponentially dominated by values close to the threshold. For an order-of-magnitude estimate of the total abundance, the mild mass weighting can be neglected in the threshold-dominated region,
\begin{equation}
\label{eq:drop_mass_weighting}
\frac{M(\delta_m(\delta_\ell))}{\kappa^2 M_f}
\simeq
\mathcal{O}(1),
\qquad
\beta_f
\sim
\int_{\delta_{\ell,c}}^\infty d\delta_\ell\,
P(\delta_\ell,\kappa^2 M_f).
\end{equation}
The Gaussian tail gives
\begin{equation}
\label{eq:beta_erfc_again}
\int_{\delta_{\ell,c}}^\infty d\delta_\ell\,
P(\delta_\ell,\kappa^2 M_f)
=
\frac{1}{2}
\mathrm{erfc}\!\left(
\frac{\nu}{\sqrt{2}}
\right)
=
\frac{1}{\sqrt{2\pi}\,\nu}
\exp\!\left(
-\frac{\nu^2}{2}
\right)
\left[
1+\mathcal{O}(\nu^{-2})
\right],
\qquad
\nu\gg1 .
\end{equation}
Thus the standard rare-event estimate is
\begin{equation}
\label{eq:beta_f_asymptotic_final}
\beta_f
\simeq
\frac{\sigma_0(r_m)}
{\sqrt{2\pi}\,\delta_{\ell,c}}
\exp\!\left[
-\frac{\delta_{\ell,c}^2}{2\sigma_0^2(r_m)}
\right],
\qquad
\frac{\delta_{\ell,c}}{\sigma_0(r_m)}
\gg
1 .
\end{equation}
The linear threshold is obtained from Eq.~\eqref{eq:delta_m_delta_l},
\begin{equation}
\label{eq:deltaellc_from_deltac_again}
\delta_{\ell,c}
=
\frac{4}{3}
\left(
1-\sqrt{1-\frac{3}{2}\delta_c}
\right).
\end{equation}
The same narrow-peak approximation gives the differential distribution by evaluating Eq.~\eqref{eq:beta_k_MPBH_2} at $M_H\simeq\kappa^2M_f$,
\begin{equation}
\label{eq:beta_k_MPBH_3}
\frac{d\beta(M_{\rm \mathsmaller{PBH}},T_f)}
{d\ln M_{\rm \mathsmaller{PBH}}}
\simeq
\frac{2\mathcal{K}}{\gamma_{\rm M}}\,
\frac{\tilde q_f^{1+1/\gamma_{\rm M}}}{\sqrt{\tilde\Lambda_f}}\,
P\!\left(
\delta_\ell(M_{\rm \mathsmaller{PBH}}),
\kappa^2M_f
\right),
\end{equation}
where
\begin{equation}
\label{eq:qtilde_def}
\tilde q_f
\equiv
\frac{M_{\rm \mathsmaller{PBH}}}{\mathcal{K}\kappa^2M_f},
\qquad
\tilde\Lambda_f
\equiv
1-\frac{3}{2}
\left(
\delta_c+\tilde q_f^{1/\gamma_{\rm M}}
\right),
\qquad
\delta_\ell(M_{\rm \mathsmaller{PBH}})
=
\frac{4}{3}
\left(
1-\sqrt{\tilde\Lambda_f}
\right).
\end{equation}
For $\tilde q_f^{1/\gamma_{\rm M}}\ll1$,
\begin{equation}
\delta_\ell(M_{\rm \mathsmaller{PBH}})
\simeq
\delta_{\ell,c}
+
\frac{\tilde q_f^{1/\gamma_{\rm M}}}
{\sqrt{1-\frac{3}{2}\delta_c}}
+
\mathcal{O}\!\left(
\tilde q_f^{2/\gamma_{\rm M}}
\right).
\end{equation}
Combining the Gaussian Press--Schechter probability with Eqs.~\eqref{eq:beta_f_asymptotic_final} and \eqref{eq:beta_k_MPBH_3}, the normalized PBH mass distribution becomes
\begin{equation}
\label{eq:psi_asymptotic_choptuik}
\psi_f(M_{\rm \mathsmaller{PBH}})
\simeq
A\,
\left(
\frac{M_{\rm \mathsmaller{PBH}}}{\mathcal{K}\kappa^2M_f}
\right)^{1+1/\gamma_{\rm M}}
\exp\!\left[
-
B\,
\frac{
\left(
M_{\rm \mathsmaller{PBH}}/(\mathcal{K}\kappa^2M_f)
\right)^{1/\gamma_{\rm M}}
}
{\sigma_0^2(r_m)}
\right],
\end{equation}
with
\begin{equation}
\label{eq:AB_asymptotic}
A
\equiv
\frac{2\mathcal{K}}{\gamma_{\rm M}}
\frac{\delta_{\ell,c}}
{\sigma_0^2(r_m)\sqrt{1-\frac{3}{2}\delta_c}},
\qquad
B
\equiv
\frac{4}{3}
\frac{1-\sqrt{1-\frac{3}{2}\delta_c}}
{\sqrt{1-\frac{3}{2}\delta_c}} .
\end{equation}
This expression recovers the universal infrared scaling and the exponential ultraviolet cutoff, as introduced in the introduction
\begin{equation}
\psi_f(M_{\rm \mathsmaller{PBH}})
\propto
M_{\rm \mathsmaller{PBH}}^{1+1/\gamma_{\rm M}}\exp\left[-c_1\left(\frac{M_{\rm \mathsmaller{PBH}}}{\langle M_{\rm \mathsmaller{PBH}} \rangle}\right)^{c_2}\right].
\end{equation}
 Earlier derivations can be found in Refs.~\cite{Niemeyer:1997mt,Yokoyama:1998xd,Karam:2022nym} and in Ref.~\cite[Eq.~(3.12)]{Carr:2018rid}. In the limit $\Delta\to0$, with smoothing at the compaction scale $r_H\simeq r_m\simeq\kappa/k_f$, one finds $\sigma_0\simeq4\mathcal{A}_{\mathcal{R}}^{1/2}/9$ up to window-function factors. Therefore the relaxation of Eq.~\eqref{eq:psi_asymptotic_choptuik} toward a sharp cutoff requires both $\Delta\to0$ and $\mathcal{A}_{\mathcal{R}}\to0$. This is consistent with our numerical check that the distribution for $\Delta\ll0.1$ remains close to the $\Delta=0.1$ case shown in orange in Fig.~\ref{fig:psi_PBH}.

\newpage
\section{Evolution of $\Phi$ in a radiation-matter universe}
\label{app:Transfer_function}
In this appendix we derive the evolution of the Newtonian potential $\Phi$ in a universe filled with matter and radiation, and extract the transfer functions $T_\Phi(\kappa)$ and $T_S(\kappa)$ that map adiabatic and isocurvature initial conditions onto the late-time matter-era amplitude used in the main text. We first set up the coupled two-fluid perturbation equations in the Newtonian gauge, casting them as a closed system for $\Phi$ and the PBH-radiation isocurvature mode $S$. We then solve this system, numerically and through analytic interpolation formulae, and compare our transfer functions with those of Ref.~\cite{Domenech:2024wao}.

 \subsection{Two-fluid perturbation equations}
 \label{app:two_fluid_perturbations}
From the Einstein equations for a homogeneous and isotropic background one recovers the standard Friedmann and continuity relations,  
\begin{align}
\label{eq:friedmann_eq_cosmic_time}
H^2 = \frac{8\pi G}{3}\,\overline{\rho}, \qquad 
\dot{H} = -\tfrac{3}{2}(1+\omega)H^2,\qquad 
\dot{\overline{\rho}} + 3(1+\omega)H\,\overline{\rho} = 0,
\end{align}
where $H\equiv \dot{a}/a$ is the Hubble parameter and $\omega \equiv \overline{p}/\overline{\rho}$ denotes the equation-of-state parameter.  
Expressed in conformal time $\eta$, these relations take the form  
\begin{align}
\label{eq:Hubble_prime_app}
\mathcal{H}^2 = \frac{8\pi G}{3}\overline{\rho}a^2, \qquad
\mathcal{H}' = -\frac{1+3\omega}{2}\mathcal{H}^2,\qquad
\overline{\rho}' + 3(1+\omega)\mathcal{H}\overline{\rho} = 0,
\end{align}
where $\mathcal{H}=a'/a$ and $dt=a\,d\eta$. The two Hubble parameters satisfy $H=\mathcal{H}/a$ and $\dot{H}=(\mathcal{H}'-\mathcal{H}^2)/a^2$.  
For a universe filled with a perfect fluid of constant $\omega$, the background evolution is given by 
\begin{equation}
   \overline{\rho}\propto a^{-3(1+\omega)}, 
   \qquad a\propto t^{\frac{2}{3(1+\omega)}}\propto \eta^{\tfrac{2}{1+3\omega}}, 
   \qquad \mathcal{H}=\frac{2}{(1+3\omega)\eta}.
\end{equation}
We now work with scalar metric perturbations in the Newton gauge,
\begin{align}
\label{eq:space-time_metric_app}
ds^2 = a^2(\eta)\Big[(1+2\Psi)\,d\eta^2 - (1-2\Phi)\,\delta_{ij}\,dx^i dx^j \Big].
\end{align}
In the Newtonian  gauge, the stress–energy tensor can be written as
\begin{equation}
T^\mu_{\ \nu} \;=\;
\begin{pmatrix}
\bar\rho+\delta\rho
& -(\bar\rho+\bar p)\,v_j
\\[6pt]
(\bar\rho+\bar p)\,v^i
& -(\bar p+\delta p)\,\delta^i_{\ j}
\end{pmatrix},
\end{equation}
with the velocity convention \(v_i=\delta_{ij}v^j\). 
Setting the anisotropic stress to zero (so the spatial block is \(\propto \delta^i_{\ j}\)) enforces the equality of the Newtonian potentials, \(\Phi=\Psi\).
The linearized Einstein equations in position space read \cite{Peter:2013avv,Baumann:2022mni}
\begin{align}
\nabla^2 \Phi - 3\mathcal H\big(\Phi' + \mathcal H \Phi\big) &= \frac{3}{2}\mathcal{H}^2\,\delta, \label{eq:E00}\\
\!\left(\Phi' + \mathcal H \Phi\right) &= -\frac{3}{2}\mathcal{H}^2(1+\omega)\, v, \label{eq:E0i}\\
\Phi'' + 3\mathcal H \Phi' -3\omega\mathcal{H}^2\Phi &= \frac{3}{2}\mathcal{H}^2\,\frac{\delta p}{\overline{\rho}}. \label{eq:Eii}
\end{align}
where primes denote $d/d\eta$, and $\mathcal H \equiv a'/a$. We have introduced the velocity potential $\partial_i v \equiv v_i$, the density contrast $\delta \equiv \delta\rho/\bar\rho$, the equation of state $w \equiv \bar p/\bar\rho$, and the sound speed 
\begin{equation}
\label{eq:SoS}
c_s^2 \equiv \left.\frac{\delta p}{\delta \rho}\right|_{\delta p_{\rm nad}=0}
= \frac{\bar p'}{\bar\rho'} = \omega - \frac{\omega'}{3(1+\omega)\mathcal{H}}.
\end{equation}
Setting $\delta p_{\rm nad}=0$ restricts to adiabatic modes, meaning pressure perturbations are entirely induced by density perturbations. Consequently the local perturbative relation follows the background equation of state, $\delta p/\delta\rho = d\bar p/d\bar\rho$ \cite{Bardeen:1983qw}.
Energy–momentum conservation implies \cite{Baumann:2022mni}
\begin{align}
\delta' - (1+w)\,(-\Delta v + 3\Phi') + 3\mathcal H\,(c_s^2 - w)\,\delta &= -3\mathcal{H}\delta_{\rm p, nad}, \label{eq:cont-local}\\
 v' + \mathcal H  v
+ \frac{c_s^2}{1+w}\,\delta^C + \Phi &= -\frac{\delta_{\rm p, nad}}{1+\omega}, \label{eq:euler-local}
\end{align}
where $\delta^C\equiv\delta-3(1+\omega)\mathcal{H}v$ is the comoving density contrast and $\delta_{p,\rm nad}$ is the non-adiabatic pressure perturbation
\begin{equation}
\label{eq:delta_pnad_def}
    \delta_{p,\rm nad} \equiv \frac{\delta p_{\rm nad}}{\overline{\rho}},\qquad \textrm{where}\quad \delta p_{\rm nad} \equiv \delta p - c_s^2 \delta \rho.
\end{equation}
Combining Eqs.~\eqref{eq:E00} and \eqref{eq:Eii} leads to 
\begin{equation}
\label{eq:Phi_equation_of_motion_0}
    \Phi''+3\mathcal{H}(1+c_s^2)\Phi'+\left[3(c_s^2-\omega)\mathcal{H}^2-c_s^2\Delta\right]\Phi = \frac{3}{2} \mathcal{H}^2 \delta_{p,\rm nad}.
\end{equation}
We now suppose that the stress-energy tensor of the universe is a linear sum of two fluids such that we can write $(a=\{1,2\})$ 
\begin{equation}
    \rho =\sum_a\rho_a,\qquad p=\sum_a p_a,\qquad v=\sum_a\frac{1+\omega_a}{1+\omega}\Omega_av_a.
\end{equation}
It is straightforward to show that 
\begin{equation}
    \omega = \sum_a\Omega_a\omega_a,\qquad  c_s^2=\sum_a \frac{1+\omega_a}{1+\omega}\Omega_a c_a^2,\label{eq:omega_cs_multifluid}
\end{equation}
where $c_a$ is the speed of sound of fluid $a$, and $\Omega_a=\rho_a/\rho_a$ with $\sum_a\Omega_a=1$. Assuming that the perfect fluids are not coupled to each other, continuity and Euler equations hold separately
\begin{align}
\delta_a' - (1+w_a)\,(-\Delta v_a + 3\Phi') + 3\mathcal H\,(c_a^2 - \omega_a)\,\delta_a &= -3\mathcal{H}\delta_{\rm p, nad,a}, \label{eq:cont-local_a}\\
 v_a' +\mathcal H  v_a
+ \frac{c_a^2}{1+\omega_a}\,\delta_a^C + \Phi &= -\frac{\delta_{\rm p, nad,a}}{1+\omega}, \label{eq:euler-local_a}
\end{align}
where $\delta_a^C\equiv\delta_a-3(1+\omega_a)\mathcal{H}v_a$.
Using Eq.~\eqref{eq:delta_pnad_def}, the non-adiabatic pressure perturbation of the mixture can be expressed as\footnote{Note the minus sign difference with respect to  \cite[Eq.~5.191]{Peter:2013avv} which has a typo.} 
\begin{equation}
 \delta_{p,\rm nad}=\sum_a \Omega_a\delta_{p,\rm nad,a}-\sum_{a}(c_s^2-c_a^2)\Omega_a \delta_a,\qquad \textrm{where}\quad \delta_{p,\rm nad,a}\equiv\delta p_a -c_a^2\delta\rho_a.\label{eq:delta_p_multifluid}
\end{equation}
Plugging Eq.~\eqref{eq:omega_cs_multifluid} into Eq.~\eqref{eq:delta_p_multifluid},  we obtain 
\begin{align}
    \delta_{p,\rm nad} &=\sum_a \Omega_a\delta_{p,\rm nad,a}+c_{ab}S_{ab},\qquad c_{ab}\equiv \frac{\Omega_a\Omega_b(1+\omega_a)(1+\omega_b)}{1+\omega}(c_a^2-c_b^2)\label{eq:delta_p_multifluid_Sab}
\end{align}
where $c_{ab}$ is a parameter and $S_{ab}$ is the so-called {isocurvature perturbation} between fluids $a$ and $b$
\begin{equation}
    S_{ab} \equiv \frac{\delta_a}{(1+\omega_a)}-\frac{\delta_b}{(1+\omega_b)}.
\end{equation}
Using Eq.~\eqref{eq:cont-local_a}, we calculate
\begin{align}
    S_{ab}' &=  -\Delta v_{ab} -3\mathcal{H}\left(\frac{\delta_{\rm p,nad,a}}{1+\omega_a}-\frac{\delta_{{\rm p,nad,b}}}{1+\omega_b} \right),\qquad v_{ab}\equiv v_a-v_b ,
\end{align}
where $v_{\rm ab}$ is the relative fluid velocity between fluids $a$ and $b$.
We now assume the fluids are barotropic
\begin{equation}
{p_a=\omega_a(\rho_a)\rho_a}\qquad \implies \qquad \delta_{p,\rm nad,a}=\delta p_a -c_a^2\delta \rho_a=0,
\end{equation}
and for $b$ as well.
Using Eq.~\eqref{eq:euler-local_a}, we calculate
\begin{align}
\label{eq:Sab_eq_delta_C}
    S_{ab}'' +\mathcal{H}S_{ab}' - \frac{\Omega_a(1+\omega_a)c_b^2+\Omega_b(1+\omega_b)c_a^2}{1+\omega}\Delta S_{ab}=(c_a^2-c_b^2)\frac{\Delta \delta^C}{1+\omega},
\end{align}
where $\delta^{(C)}\equiv \delta-3(1+\omega)\mathcal{H}v$ is the comoving density contrast. Using Eqs.~\eqref{eq:E00} and \eqref{eq:E0i}, we  obtain Poisson's equation
\begin{equation}
\label{eq:Poisson_eq_C}
  \nabla^2 \Phi = \frac{3}{2}\mathcal{H}^2\,\delta^{(C)}.  
\end{equation}
Using Eqs.~\eqref{eq:Phi_equation_of_motion_0} and \eqref{eq:delta_p_multifluid_Sab} and Eqs.~\eqref{eq:Sab_eq_delta_C} and \eqref{eq:Poisson_eq_C} we get
\begin{align}
\label{eq:Phi_equation_of_motion_2}
    &\Phi''+3\mathcal{H}(1+c_s^2)\Phi'+\left[3(c_s^2-\omega)\mathcal{H}^2-c_s^2\Delta\right]\Phi =\frac{3}{2}\mathcal{H}^2c_{ab}S_{ab}, \\
\label{eq:Sab_equation_of_motion_2}
   & S_{ab}'' +\mathcal{H}S_{ab}' - \frac{\Omega_a(1+\omega_a)c_b^2+\Omega_b(1+\omega_b)c_a^2}{1+\omega}\Delta S_{ab} =2(c_a^2-c_b^2)\frac{\Delta^2\Phi}{3\mathcal{H}^2(1+\omega)}.
\end{align}
We can use the chain rules:
\begin{equation}
\label{eq:chain_rules_y_eta}
X'(\eta) = y'\dot{X}(y)=y\mathcal{H}\dot{X}(y),\qquad 
    X''(\eta)=(y')^2\ddot{X}(y)+y''\dot{X}(y)=y^2\mathcal{H}^2\ddot{X}(y)+\frac{1-3\omega}{2}y\mathcal{H}^2\dot{X}(y),
\end{equation}
where we used $y''=y(\mathcal{H}'+\mathcal{H}^2)=y(1-3\omega)\mathcal{H}^2/2$, cf. Eq.~\eqref{eq:Hubble_prime}.  Then Eqs.~\eqref{eq:Phi_equation_of_motion_2} and \eqref{eq:Sab_equation_of_motion_2} become, in Fourier space,
\begin{align}
\label{eq:Phi_equation_of_motion_3}
    &y^2\ddot{\Phi}_\mathbf{k}(y)+\frac{y}{2}\left(7+6c_s^2-3\omega\right)\dot{\Phi}_\mathbf{k}+\left[3(c_s^2-\omega)+c_s^2\left(\frac{k}{\mathcal{H}}\right)^2\right]\Phi =\frac{3}{2}c_{ab}S_{ab,\mathbf{k}}, \\
\label{eq:Sab_equation_of_motion_3}
   & y^2\ddot{S}_{ab,\mathbf{k}} + \frac{3}{2}y(1-\omega)\dot{S}_{ab,\mathbf{k}} + \frac{\Omega_a(1+\omega_a)c_b^2+\Omega_b(1+\omega_b)c_a^2}{1+\omega}\left(\frac{k}{\mathcal{H}}\right)^2 S_{ab,\mathbf{k}} =\frac{2(c_a^2-c_b^2)}{3(1+\omega)}\left(\frac{k}{\mathcal{H}}\right)^4\Phi_\mathbf{k}.
\end{align}
\begin{figure}[t]
\centering
\includegraphics[width=0.48\textwidth]{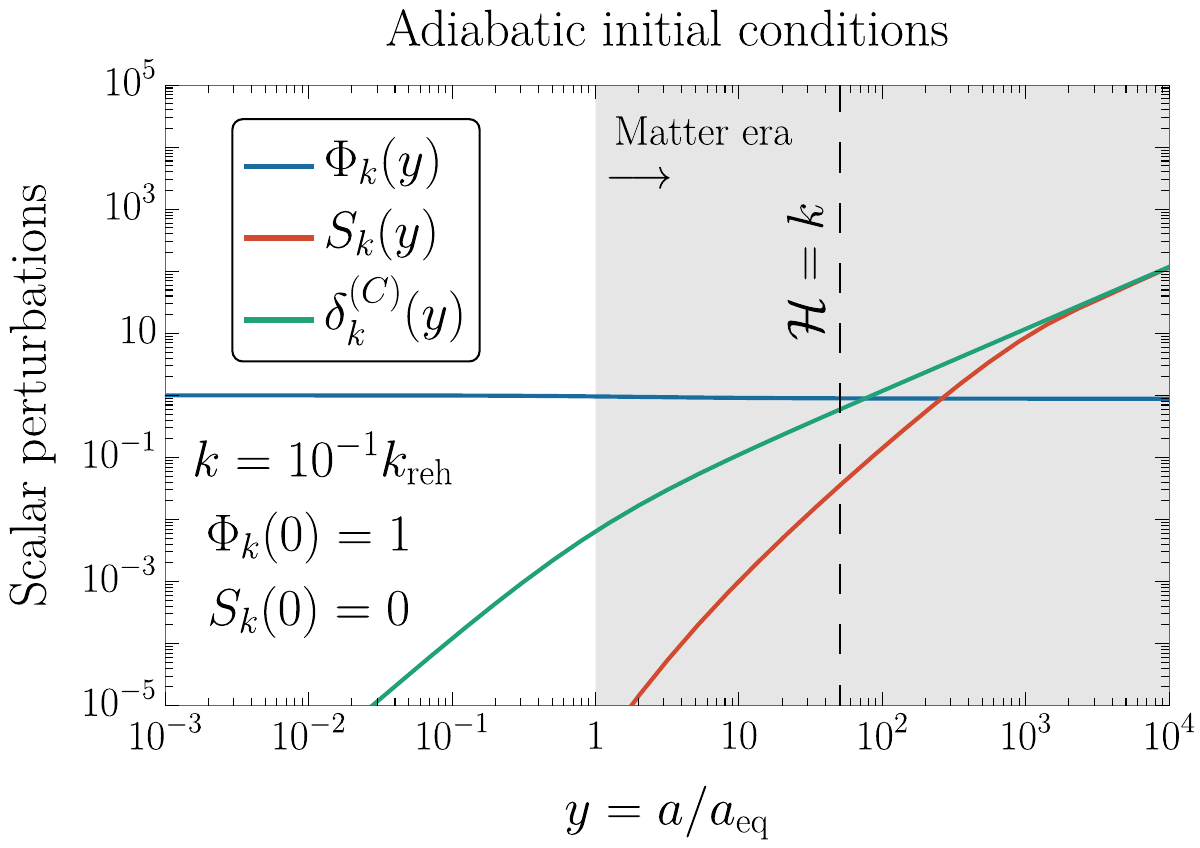}
\includegraphics[width=0.48\textwidth]{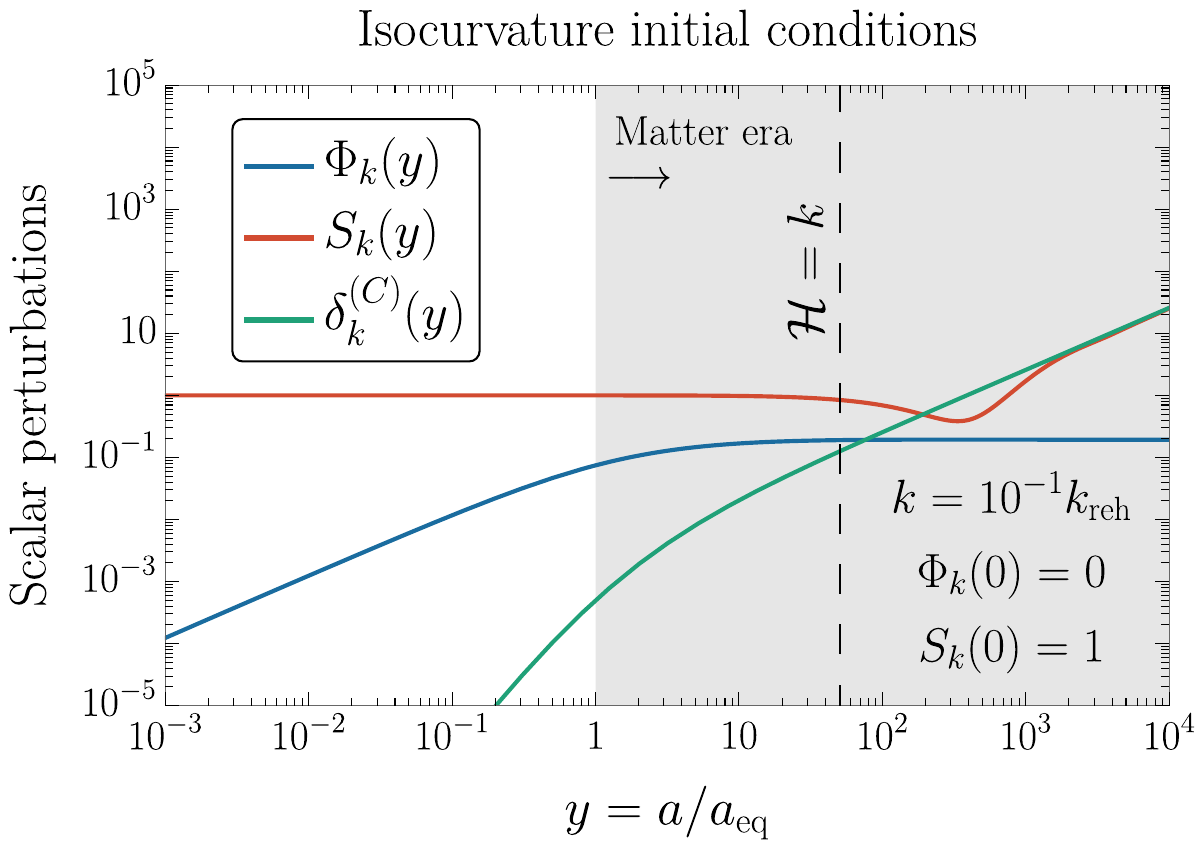}
\includegraphics[width=0.48\textwidth]{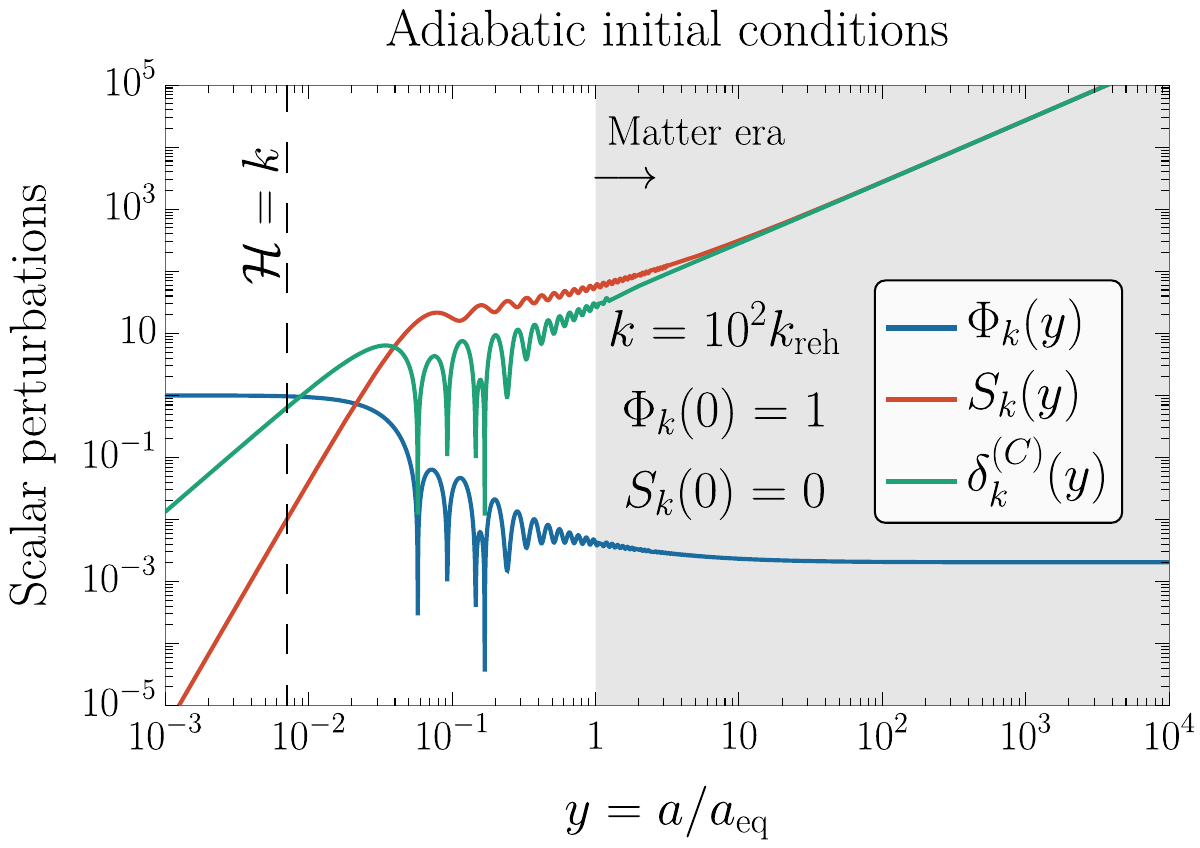}
\includegraphics[width=0.48\textwidth]{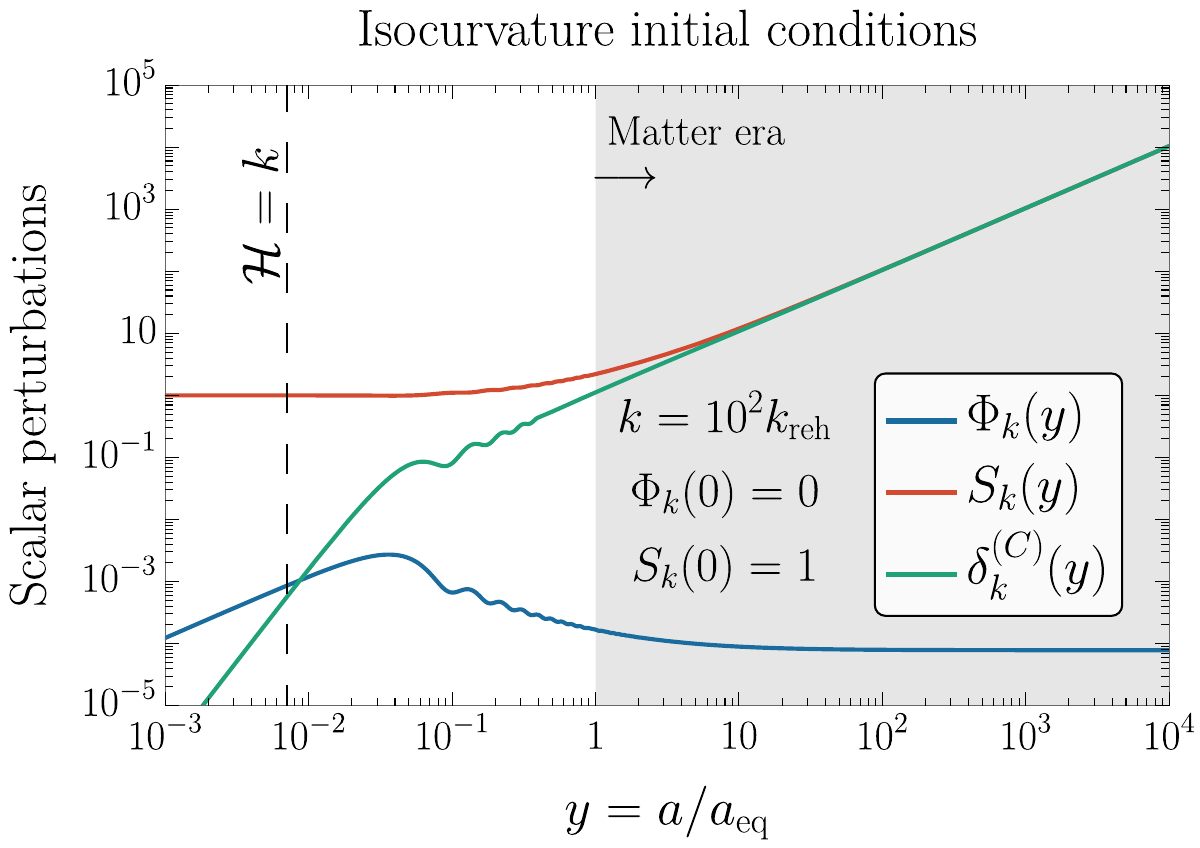}
\caption{\label{fig:Scalar_pert} Cosmological evolution of the Newtonian perturbation $\Phi$ and isentropic perturbation $S$ assuming adiabatic (\textbf{left}) and isocurvature  (\textbf{right}) initial conditions, for $\kappa=k/k_{\rm reh}=10^{-1}$ (\textbf{top}) and $10^2$ (\textbf{bottom}). The universe is dominated by radiation for $y<1$ and by matter for $y>1$. In all panels we plot the absolute value of the perturbations.}
\end{figure}
\begin{figure}[t]
\centering
\includegraphics[width=0.7\textwidth]{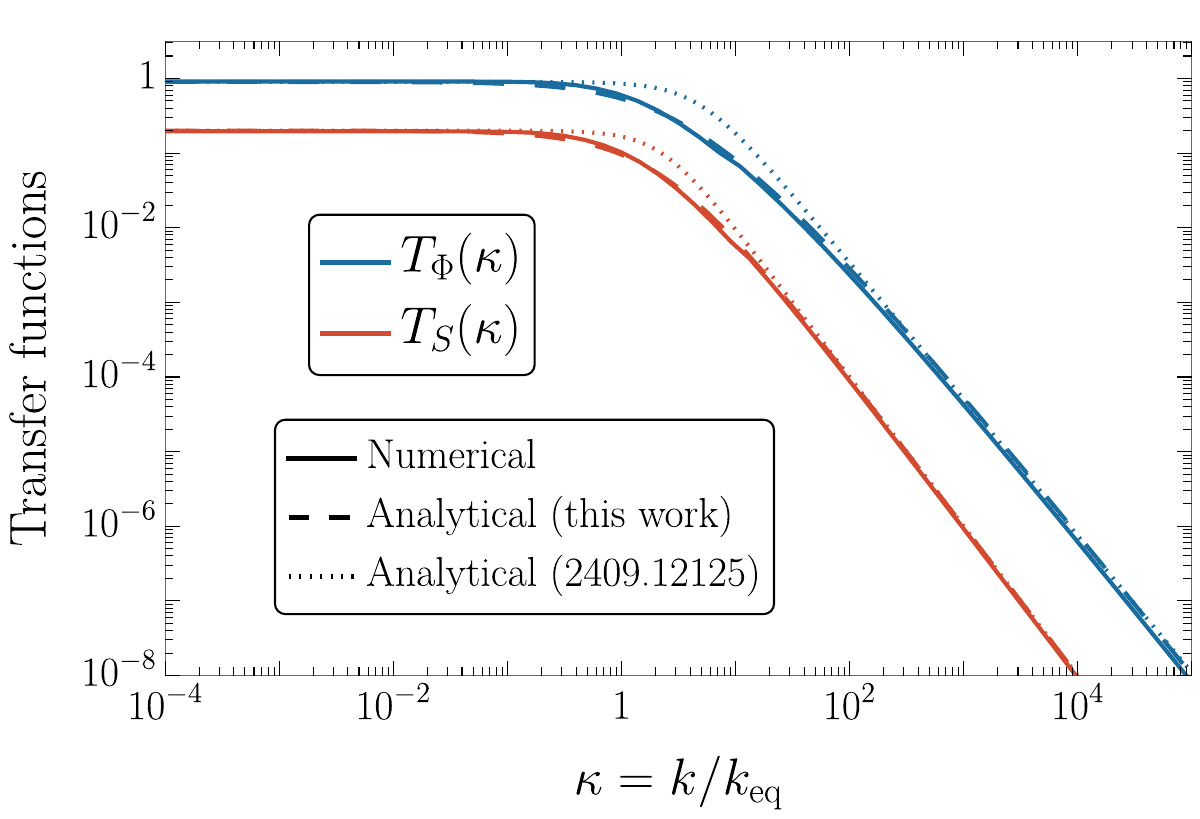}
\caption{\label{fig:Scalar_pert_Transfer_fct} Transfer function of the Newtonian potential $\Phi$ assuming adiabatic (\textbf{blue}) and isocurvature  (\textbf{red}) initial conditions. The solid lines are obtained from numerically integrating Eqs.~\eqref{eq:Phi_equation_of_motion_5} and \eqref{eq:Sab_equation_of_motion_5}, the dashed lines correspond to 
Eqs.~\eqref{eq:transfer_adia_app} and \eqref{eq:transfer_iso_app}, while the dotted lines correspond to Eq.~\eqref{eq:transfer_adia_iso}. }
\end{figure} 
Long before evaporation $\eta\ll \eta_{\rm eva}$, the universe containing PBH can be described as a mixture of a matter and radiation background, $\rho=\rho_m+\rho_r$ with $\rho_m\equiv \rho_{\rm \mathsmaller{PBH}}$. We have
\begin{equation}
 \Omega_m = \frac{y}{1+y},\qquad \Omega_r= \frac{1}{1+y},\qquad \mathcal{H}=k_{\rm eq}\sqrt{\frac{1+y}{2y^2}},
\end{equation}
where $y\equiv a/a_{\rm eq}$ which upon solving $a'(\eta)=\mathcal{H}a$ leads to
\begin{equation}
\label{eq:y_eta_matter_radiation}
 y(\eta) = \left(\frac{\eta}{\eta_\star}\right)^2 + 2\left(\frac{\eta}{\eta_\star} \right) ,\qquad \textrm{with}\quad \eta_{\star}\equiv \frac{\eta_{\rm eq}}{\sqrt{2}-1}=\frac{2\sqrt{2}}{k_{\rm eq}}.
\end{equation}
Using Eq.~\eqref{eq:omega_cs_multifluid}, we also have
\begin{equation}
   \omega_m=c_m^2=0,\qquad \omega_r=c_r^2=\frac{1}{3},\qquad\implies \qquad \omega= \frac{1}{3(1+y)},\qquad c_s^2 = \frac{1}{3(1+\frac{3}{4}y)}.
\end{equation} 
We denote by $S$ the isocurvature perturbation between PBHs and radiation
\begin{equation}
    S \equiv \delta_{\rm \mathsmaller{PBH}}- \frac{3}{4}\delta_{\rm r}.
\end{equation}
The system of equations \eqref{eq:Phi_equation_of_motion_2} and \eqref{eq:Sab_equation_of_motion_2} become (with $\kappa \equiv k/k_{\rm eq}$ and $x\equiv k\eta$)
\begin{align}
\label{eq:Phi_equation_of_motion_4}
    &\kappa^{2}\Phi_\mathbf{k}''(x)+\kappa\sqrt{\frac{1+y}{2y^2}}\frac{16+9y}{4+3y}\Phi_\mathbf{k}'(x)+ \frac{3+8\kappa^{2}y}{24y+18y^2}\Phi_\mathbf{k}(x) = -\frac{S_\mathbf{k}(x)}{4+3y}, \\
   & \kappa^{2}S_\mathbf{k}''(x) + \kappa\sqrt{\frac{1+y}{2y^2}}S_\mathbf{k}'(x) +\frac{y\kappa^{2}}{4+3y} S_\mathbf{k}(x) = -\frac{4y^2\kappa^{4}}{3(4+3y)}\Phi_\mathbf{k}(x).
   \label{eq:Sab_equation_of_motion_4}
\end{align}
where $y(x)$ follows from Eq.~\eqref{eq:y_eta_matter_radiation}.
Using the chain rules in Eq.~\eqref{eq:chain_rules_y_eta}, the system of equation becomes
\begin{align}
\label{eq:Phi_equation_of_motion_5}
    &\ddot{\Phi}_\mathbf{k}(y)+\left(7-\frac{1}{1+y}+\frac{8}{4+3y}\right)\frac{\dot{\Phi}_\mathbf{k}(y)}{2y}+\frac{(3+8\kappa^2y)}{3(1+y)(4+3y)}\frac{\Phi_\mathbf{k}(y)}{y} =- \frac{2S_\mathbf{k}(y)}{y(1+y)(4+3y)}, \\
\label{eq:Sab_equation_of_motion_5}
   & \ddot{S}_\mathbf{k}(y)  + \frac{2+3y}{2y(1+y)}\dot{S}_\mathbf{k}(y) + \frac{2y}{(1+y)(4+3y)}\kappa^2 S_\mathbf{k}(y) =-\frac{8y^2}{3(1+y)(4+3y)}\kappa^4\Phi_\mathbf{k}(y).
\end{align}
For more details, we refer the reader to pioneering works~\cite{Kodama:1986fg,Kodama:1986ud}, a textbook~\cite{Peter:2013avv}, and more recent Refs.~\cite{Domenech:2024wao,Zeng:2025tno}.

\subsection{Transfer functions}

The solution of the system of Eqs.~\eqref{eq:Phi_equation_of_motion_5} and \eqref{eq:Sab_equation_of_motion_5} are displayed in Fig.~\ref{fig:Scalar_pert}. The scalar metric perturbation approaches a constant at late times in the matter era, which we write as\footnote{With the metric convention of Eq.~\eqref{eq:space-time_metric_app}, the transfer function \(T_S\) differs by an overall minus sign from the convention commonly used in the literature, e.g.~\cite{Kodama:1986fg,Domenech:2024wao}. This sign originates from the source terms in Eqs.~\eqref{eq:Sab_equation_of_motion_4} and \eqref{eq:Sab_equation_of_motion_5}. Since only the scaling and magnitude of \(T_S\) are relevant, we omit this overall sign in this paper.\label{foontnote:TS_sign}}
\begin{equation}
\Phi_\mathbf{k}(y\gg 1)=T_\Phi(\kappa)\Phi_\mathbf{k}(0)+T_{S}(\kappa)S_\mathbf{k}(0).
\end{equation}
where $\Phi_\mathbf{k}(0)$, $S_\mathbf{k}(0)$ are the initial condition, and $T_\Phi(\kappa)$, $T_S(\kappa)$ are the transfer functions.  
For adiabatic perturbations and  $\kappa\equiv k/k_{\rm reh}\ll 1$, we can simply use conservation of the comoving curvature perturbation $\mathcal{R}\simeq \left( 5+3\omega\right)\Phi/\left( 3+3\omega\right)$ on superhorizon scales
\begin{equation}
\kappa \ll 1:\qquad T_\Phi(\kappa) \simeq \left( \frac{3+3\omega}{5+3\omega}\right)_{\omega = 0}\left( \frac{5+3\omega}{3+3\omega}\right)_{\omega=1/3}= \frac{9}{10},
\end{equation}
where the factor $\left( \frac{3+3\omega}{5+3\omega}\right)_{\omega = 0}$ arises because $\Phi(0)$ is defined in the pre-PBH radiation era, see Eq.~\eqref{eq:P_Phi_main}.
Ref.~\cite{Hu:1995en} calculated the transfer function for $\delta^{(C)}$ which upon using Poisson's equation in Eq.~\eqref{eq:Poisson_eq_C} can be converted in a transfer function for $\Phi$
\begin{equation}
\label{eq:delta_R}
\kappa\gg 1:\qquad T_\Phi(\kappa) \simeq 10.8  \frac{\textrm{ln}\left(0.13\kappa\right)}{\kappa^2}.
\end{equation}
For any given $k$, we find that a very good interpolation formulae is given by
\begin{equation}
\label{eq:transfer_adia_app}
T_\Phi(\kappa)\simeq \left(\frac{10}{9}+\frac{0.09\kappa^2}{\ln\left(1+0.13\kappa\right)} \right)^{-1}.
\end{equation}
For isocurvature perturbations, Ref.~\cite{Kodama:1986fg} found
\begin{equation}
\label{eq:T_iso_Kodama}
T_S(\kappa) \simeq
\begin{cases}
\dfrac{1}{5}, & \kappa \ll 1, \\[6pt]
\dfrac{C}{\kappa^2}, & \kappa \gg 1 .
\end{cases}
\end{equation}
where $C=3/4$. Instead, we find that the numerical solution prefers $C\simeq 1$, in agreement with Ref.~\cite{Domenech:2024wao}. The first line of Eq.~\eqref{eq:T_iso_Kodama} can be determined using conservation of $\zeta =-\Phi+\delta_{\rm tot}/3(1+\omega)$. We find that a very good interpolation function is
\begin{equation}
\label{eq:transfer_iso_app}
 T_S(\kappa) \simeq \left(5^{1/c}+\kappa^{2/c}\right)^{-c},\qquad c=2.
\end{equation} 
Ref.~\cite{Domenech:2024wao} calculates the transfer functions to be:
\begin{equation}
\label{eq:transfer_adia_iso}
 T_\Phi(\kappa)\simeq \left(10/9+0.06\kappa^{1.83} \right)^{-1},\qquad \textrm{and} \qquad  T_S(\kappa) \simeq \left(5+\kappa^{2}\right)^{-1}.
\end{equation}
The power-law (instead of a log) makes $T_\Phi(\kappa)$ more appropriate to derive an analytical function.
In Fig.~\ref{fig:Scalar_pert_Transfer_fct}, we compare the analytical formulae of Ref.~\cite{Domenech:2024wao}, with the more precise analytical formulae in Eqs.~\eqref{eq:transfer_adia_app} and \eqref{eq:transfer_iso_app} and the numerical solution obtained from solving the system of Eqs.~\eqref{eq:Phi_equation_of_motion_5} and \eqref{eq:Sab_equation_of_motion_5}. We conclude that the transfer function in Eq.~\eqref{eq:transfer_adia_iso} derived in Ref.~\cite{Domenech:2024wao} are good enough to approximate the numerical solution and hence we rely on them in the main text.

\subsection{Analytical solution in RD}
\label{subsec:RD_analytic}
 
For modes entering the horizon well before matter-radiation equality ($\kappa\gg 1$), the system \eqref{eq:Phi_equation_of_motion_4}--\eqref{eq:Sab_equation_of_motion_4} admits a closed-form solution throughout the radiation-dominated (RD) regime $y\ll 1$, equivalently $x/\kappa\ll 1$. Both limits are controlled by the same small parameter since, from Eq.~\eqref{eq:y_eta_matter_radiation} and $\eta_\star=2\sqrt{2}/k_{\rm eq}$,
\begin{equation}
\label{eq:y_expansion}
y=\frac{x}{\sqrt{2}\,\kappa}+\frac{x^{2}}{8\kappa^{2}}+\mathcal{O}(\kappa^{-3}),
\qquad
\frac{1}{y}=\frac{\sqrt{2}\,\kappa}{x}-\frac{1}{4}+\mathcal{O}(\kappa^{-1}).
\end{equation}
Expanding every coefficient of Eqs.~\eqref{eq:Phi_equation_of_motion_4}--\eqref{eq:Sab_equation_of_motion_4} to next-to-leading order in $\kappa^{-1}$ and dividing by $\kappa^{2}$ yields the reduced system\footnote{Our sign convention for the metric perturbation differs from that of Ref.~\cite{Domenech:2021ztg}: we use $ds^2=a^2[(1+2\Psi)d\eta^2-(1-2\Phi)\delta_{ij}dx^idx^j]$, while Ref.~\cite{Domenech:2021ztg} uses $ds^2=a^2[-(1+2\Psi)d\eta^2+(1+2\Phi_{\rm D})\delta_{ij}dx^idx^j]$, so that $\Phi_{\rm D}=-\Phi$. Eqs.~\eqref{eq:Phi_RD}--\eqref{eq:S_RD} match Eqs.~(2.8)--(2.9) of Ref.~\cite{Domenech:2021ztg} after this identification, which flips the sign of the $S$--$\Phi$ coupling terms.}
\begin{align}
\label{eq:Phi_RD}
\Phi_\mathbf{k}''+\frac{4}{x}\Phi_\mathbf{k}'+\frac{1}{3}\Phi_\mathbf{k}+\frac{1}{4\sqrt{2}\,\kappa x}\!\left[x\Phi_\mathbf{k}'+(1-x^2)\Phi_\mathbf{k}+2S_\mathbf{k}\right] &\simeq 0,\\
\label{eq:S_RD}
S_\mathbf{k}''+\frac{1}{x}S_\mathbf{k}'+\frac{x^2}{6}\Phi_\mathbf{k}-\frac{1}{2\sqrt{2}\,\kappa}\!\left[S_\mathbf{k}'-\frac{x}{2}S_\mathbf{k}+\frac{x^3}{12}\Phi_\mathbf{k}\right] &\simeq 0.
\end{align}
The hierarchy between the leading and bracketed terms motivates the perturbative ansatz $\Phi_\mathbf{k}=\kappa^{-1}\Phi_1+\mathcal{O}(\kappa^{-2})$ and $S_\mathbf{k}=S_\mathbf{k}(0)+\kappa^{-1}S_1+\mathcal{O}(\kappa^{-2})$, assuming pure isocurvature initial conditions $\Phi_\mathbf{k}(0)=0$, $S_\mathbf{k}(0)={\rm const}$. At leading order the bracketed terms decouple and we recover $\Phi_0=0$ and $S_0=S_\mathbf{k}(0)$. At order $\kappa^{-1}$, only $S_0$ sources $\Phi_1$, so that
\begin{equation}
\label{eq:Phi1_ODE}
\Phi_1''+\frac{4}{x}\Phi_1'+\frac{1}{3}\Phi_1=-\frac{S_\mathbf{k}(0)}{2\sqrt{2}\,x},
\end{equation}
whose regular-at-origin solution combines a particular piece $\propto (6/x^3+1/x)$ with a homogeneous contribution built from the spherical Bessel functions $j_1(c_sx)/x^2$ and $y_1(c_sx)/x^2$, where $c_s=1/\sqrt{3}$, yielding
\begin{equation}
\label{eq:Phi1_sol}
\Phi_1(x)=-\frac{3 S_\mathbf{k}(0)}{2\sqrt{2}\,x^3}\left[6+x^2-2\sqrt{3}\,x\sin(c_sx)-6\cos(c_sx)\right].
\end{equation}
Substituting \eqref{eq:Phi1_sol} into the order-$\kappa^{-1}$ part of \eqref{eq:S_RD}, the linear-in-$x$ pieces of the source cancel exactly, leaving $(xS_1')'=(S_\mathbf{k}(0)/4\sqrt{2})[6-2\sqrt{3}x\sin(c_sx)-6\cos(c_sx)]$, which integrates with $S_1(0)=0$ to
\begin{equation}
\label{eq:S1_sol}
S_1(x)=\frac{3 S_\mathbf{k}(0)}{2\sqrt{2}}\left[x+\sqrt{3}\sin(c_sx)-2\sqrt{3}\,{\rm Si}(c_sx)\right],
\end{equation}
where ${\rm Si}$ denotes the sine-integral function. Collecting \eqref{eq:Phi1_sol}--\eqref{eq:S1_sol}, the finite-$x$ RD transfer functions read
\begin{align}
\label{eq:TS_Phi_finite_x}
\frac{\Phi_\mathbf{k}(x)}{S_\mathbf{k}(0)} &= -\frac{3}{2\sqrt{2}\,\kappa\,x^3}\left[6+x^2-2\sqrt{3}\,x\sin(c_sx)-6\cos(c_sx)\right],\\
\label{eq:TS_S_finite_x}
\frac{S_\mathbf{k}(x)}{S_\mathbf{k}(0)} &= 1+\frac{3}{2\sqrt{2}\,\kappa}\left[x+\sqrt{3}\sin(c_sx)-2\sqrt{3}\,{\rm Si}(c_sx)\right],
\end{align}
valid throughout the RD regime $x/\kappa\ll 1$ at leading order in $\kappa^{-1}$, and in agreement with Eqs.~(2.11)--(2.12) of Ref.~\cite{Domenech:2021ztg}. The late-time transfer function $T_S(\kappa)$ appearing in \eqref{eq:T_iso_Kodama} follows by evaluating $\Phi_\mathbf{k}$ at the epoch where the RD solution matches the matter-era plateau. For $\kappa\gg 1$ the mode is deeply subhorizon at equality, so Poisson's equation \eqref{eq:Poisson_eq_C} together with $\mathcal{H}^2\delta_m^{C}\propto y^{-2}\cdot y=\text{const}$ during MD freezes $\Phi_\mathbf{k}$ at its value at equality, reached at $x_{\rm eq}\equiv k\eta_{\rm eq}=2(\sqrt{2}-1)\kappa$. Plugging $x_{\rm eq}$ into \eqref{eq:TS_Phi_finite_x} gives the closed-form expression, which after sign flip to match the literature convention, see footnote~\ref{foontnote:TS_sign}, reads
\begin{equation}
\label{eq:TS_finite_x}
T_S(\kappa)=\frac{3}{2\sqrt{2}\,\kappa\,x_{\rm eq}^3}\left[6+x_{\rm eq}^2-2\sqrt{3}\,x_{\rm eq}\sin(c_sx_{\rm eq})-6\cos(c_sx_{\rm eq})\right],
\end{equation}
valid for any $\kappa\gg 1$. In the deep-subhorizon regime the $x_{\rm eq}^2$ term dominates, the remaining contributions being suppressed by $1/x_{\rm eq}^2\sim\kappa^{-2}$, which yields
\begin{equation}
\label{eq:TS_leading}
T_S(\kappa)\simeq \frac{3(2+\sqrt{2})}{8\,\kappa^{2}}\approx \frac{1.28}{\kappa^{2}}.
\end{equation}
This reproduces the Kodama--Sasaki scaling in \eqref{eq:T_iso_Kodama} with an order-unity coefficient consistent with the numerical value $C\simeq 1$. A precise determination of $C$ would require matching \eqref{eq:TS_Phi_finite_x} onto the MD solution across $y\sim 1$, which lies beyond our leading-order RD treatment.

\newpage
\section{Evolution of $\Phi$ during PBH evaporation}
\label{app:Phi_PBH_evaporation}

Building on the two-fluid framework of App.~\ref{app:Transfer_function}, here we account for the energy exchange that accompanies PBH decay into radiation, which modifies the perturbation equations relative to the conserved-fluid case. Working in the synchronous gauge comoving with the PBH component, we follow the Newtonian potential $\Phi$ through PBH domination, evaporation, and reheating, and obtain an analytic estimate of the suppression factor $\mathcal{S}_\Phi(k)$ entering Eq.~\eqref{eq:Phi_osc_main}. We show that $\Phi$ stays on a plateau during PBH domination and decays into oscillations once PBHs evaporate, and compute $\mathcal{S}_\Phi(k)$ for the Choptuik mass distribution, recovering the monochromatic scaling of Ref.~\cite{Inomata:2020lmk} as a limiting case.

\subsection{Perturbation equations}
When PBH decays into dark radiation, energy and momentum are exchanged between the two fluids.  
At linear order, this exchange modifies the usual conservation equations.  
Splitting the time and spatial components of $\nabla_\mu T^{\mu\nu}$, one obtains~\cite{Audren:2014bca}
\begin{align}
\nabla_\mu T^{\mu 0}_{\rm \mathsmaller{PBH}} = -C, \qquad &
\nabla_\mu T^{\mu 0}_{\rm r} = +C, \label{eq:contC_rewrite} \\
\partial_i\!\left(\nabla_\mu T^{\mu i}_{\rm \mathsmaller{PBH}}\right) = -D, \qquad &
\partial_i\!\left(\nabla_\mu T^{\mu i}_{\rm r}\right) = +D, \label{eq:EulerD_rewrite}
\end{align}
where \(C\) characterises the rate of energy transfer and \(D\) the corresponding momentum transfer between the fluids.
A major simplification occurs in the synchronous gauge comoving with the PBH component defined by the metric~\cite{Audren:2014bca,Poulin:2016nat,Inomata:2020lmk}
\begin{equation}
    {\rm d}s^2 = a^2\left({\rm d}\eta^2 - H_{ij} {\rm d}x^i {\rm d}x^j\right),
\end{equation}
where $H_{ij} = \delta_{ij} + \hat{k}_i \hat{k}_j h + \left(\hat{k}_i \hat{k}_j - \frac{1}{3} \delta_{ij}\right) 6\varepsilon+\frac{1}{2}h_{ij}$, with $\hat{k}_i \equiv k_i/|k|$. The quantities $h$ and $\varepsilon$ are the trace and traceless component of the scalar mode in the synchronous gauge.\footnote{$h$ and $\varepsilon$ corresponds $h$ and $\eta$ in~\cite{Ma:1995ey,Audren:2014bca,Poulin:2016nat} and  to $\gamma$ and $\varepsilon$ in~\cite{Inomata:2020lmk}.} For simplicity, we neglect sources of anisotropic stress. One additional gauge degree of freedom must be fixed, and we impose that the PBH fluid has a vanishing velocity~\cite{Audren:2014bca,Poulin:2016nat}
\begin{equation}
     \theta_{\rm \mathsmaller{PBH}}=0.
\end{equation}
Since we have $\delta g_{00}=\delta g_{0i}=\theta_{\rm \mathsmaller{PBH}}=0$, then the decay is isotropic  so the only non-zero component of the interaction term is the energy transfer.  
Its perturbed expression follows directly from the local decay rate:
\begin{equation}
C^{(s)} = a\,\Gamma_{\rm \mathsmaller{PBH}}\,\rho_{\rm \mathsmaller{PBH}}\,\big(1+\delta_{\rm \mathsmaller{PBH}}^{(s)}\big),
\end{equation}
where the superscript $(s)$ indicates that the quantities are expressed in the synchronous gauge.
Because the decaying particles have no peculiar motion in this gauge and the decay products carry no net momentum, the spatial part of the interaction vector vanishes:
\begin{equation}
D^{(s)} = 0.
\end{equation}
With \(D^{(s)}=0\), the Euler equation for PBH reduces to $\theta_{\rm \mathsmaller{PBH}}^{(s)\prime} = -\mathcal H\,\theta_{\rm \mathsmaller{PBH}}^{(s)} = 0$,
so the velocity divergence stays zero at all times.  The continuity and Euler equations \eqref{eq:contC_rewrite} and \eqref{eq:EulerD_rewrite} in the synchronous gauge read~\cite{Ma:1995ey,Audren:2014bca,Poulin:2016nat,Inomata:2020lmk}:
\begin{align}
\label{eq:syncrhonous_gauge_eq_first_app}
    \delta_{\mathrm{m}}^{\prime} &= -\frac{h^{\prime}}{2} \\
    \delta_{\mathrm{r}}^{\prime} &= -\frac{4}{3} \left(\theta_{\mathrm{r}} + \frac{h^{\prime}}{2}\right) +  \frac{\left<\Gamma\rho_{\mathrm{m}}\right>}{\left<\rho_{\mathrm{r}}\right>} 
    \label{eq:syncrhonous_gauge_eq_second_app}
    \left(\delta_{\mathrm{m}} - \delta_{\mathrm{r}}\right), \\
    \theta_{\mathrm{r}}^{\prime} &= \frac{k^2}{4} \delta_{\mathrm{r}} - \frac{\left<\Gamma\rho_{\mathrm{m}}\right>}{\left<\rho_{\mathrm{r}}\right>} \theta_{\mathrm{r}},    \label{eq:syncrhonous_gauge_eq_third_app}\\
    k^2 \varepsilon-\frac{1}{2} \frac{a^{\prime}}{a} h^{\prime} & =-\frac{3}{2} \mathcal{H}^2\left(\frac{\left<\rho_{\mathrm{m}}\right>}{\left<\rho_{\mathrm{tot}}\right>} \delta_{\mathrm{m}}+\frac{\left<\rho_{\mathrm{r}}\right>}{\left<\rho_{\mathrm{tot}}\right>} \delta_{\mathrm{r}}\right),    \label{eq:syncrhonous_gauge_eq_four_app} \\
    \label{eq:syncrhonous_gauge_eq_last_app}
k^2 \varepsilon^{\prime} & =2 \mathcal{H}^2 \frac{\left<\rho_{\mathrm{r}}\right>}{\left<\rho_{\mathrm{tot}}\right>} \theta_{\mathrm{r}},
\end{align}
where $\theta \equiv \partial_i v^i$ is the velocity divergence. The background quantities $\left<\rho_{\mathrm{r}}\right>$, $\left<\rho_{\mathrm{m}}\right>$, and $\left<\Gamma\rho_{\mathrm{m}}\right>$ are given by Eqs.~\eqref{eq:rho_r_bkg}, \eqref{eq:rho_m_bkg}, \eqref{eq:Gamma_rho_m_bkg}.
The initial conditions for the perturbations  are~\cite{Inomata:2020lmk}
\begin{align}
& \delta_{\mathrm{r}}=-\frac{2}{3} C(k \eta)^2,\quad \delta_{\mathrm{m}}=\frac{3}{4} \delta_{\mathrm{r}}, \quad \theta_{\mathrm{r}}=-\frac{1}{18} C\left(k^4 \eta^3\right), \\
& h=C(k \eta)^2, \quad \varepsilon=2 C-\frac{1}{18} C(k \eta)^2,
\end{align}
where $C= \mathcal{R}/2$.
These initial conditions hold for radiation domination, which is why we start the numerical evolution long before the onset of PBH domination. To translate back to Newtonian gauge, where the metric in the absence of anisotropies is
\begin{equation}
ds^2 = a^2(\tau)\left[-(1+2\Psi)d\tau^2 + (1-2\Phi)\delta_{ij}dx^i dx^j\right],
\end{equation}
we use~\cite{Ma:1995ey,Poulin:2016nat,Inomata:2020lmk}: 
\begin{equation}
    \Phi = \varepsilon - \mathcal{H} \frac{(6\varepsilon + h)'}{2k^2}.
\end{equation}

Alternatively, we can perform the scalar gauge transformation~\cite{Ma:1995ey}
\begin{align}
\Psi &= \mathcal H\alpha + \alpha', \qquad
\Phi = \eta - \mathcal H\alpha, \\
\delta^{(n)} &= \delta^{(s)} + \frac{\bar\rho'}{\bar\rho}\alpha, \qquad
\theta^{(n)} = \theta^{(s)} + k^2\alpha, \qquad
\alpha = \frac{(6\eta + h)'}{2k^2},
\end{align}
where $\Psi=\Phi$ in the absence of anisotropic stress. One obtains~\cite{Poulin:2016nat}
\begin{align}
\delta'_{\rm m} &= -\theta_{\rm m} + 3\Phi' - \Gamma\,\Phi, \\
\theta'_{\rm m} &= -\mathcal H\,\theta_{\rm m} + k^2\Phi, \\[2mm]
\delta'_{\rm r} &= -\frac{4}{3}\left(\theta_{\rm r} - 3\Phi'\right)
        + \frac{\left<\Gamma\rho_{\mathrm{m}}\right>}{\left<\rho_{\mathrm{r}}\right>}
          \left(\delta_{\rm m} - \delta_{\rm r} + \Phi\right),\\
\theta'_{\rm r} &= \frac{k^2}{4}\delta_{\rm r} + k^2\Phi
        - \frac{3\left<\Gamma\rho_{\mathrm{m}}\right>}{4\left<\rho_{\mathrm{r}}\right>}
          \left(\frac{4}{3}\theta_{\rm r} - \theta_{\rm m}\right),
\end{align}
 The Newtonian potential obeys Poisson's relativistic equation~\cite{Mukhanov:2005sc}
\begin{equation}
\Phi' = -\frac{k^2\Phi + 3\mathcal H^2\Phi 
      + \frac{3}{2}\mathcal H^2
        \left( \frac{\left<\rho_{\mathrm{m}}\right>}{\left<\rho_{\mathrm{tot}}\right>}\delta_{\rm m}
             + \frac{\left<\rho_{\mathrm{r}}\right>}{\left<\rho_{\mathrm{tor}}\right>}\delta_{\rm r}\right)}
      {3\mathcal H},
\qquad \rho_{\rm tot}=\rho_{\rm m}+\rho_{\rm r},
\end{equation}
and the superhorizon adiabatic initial conditions used for numerical integration are
\begin{equation}
\delta_{{\rm m},{\rm ini}} = -2\Phi_{\rm ini}, \qquad
\delta_{{\rm r},{\rm ini}} = \frac{4}{3}\delta_{{\rm m},{\rm ini}}, \qquad
\theta_{{\rm m},{\rm ini}} = \theta_{{\rm r},{\rm ini}}
  = \frac{k^2\eta_{\rm ini}}{3}\Phi_{\rm ini}.
\end{equation}

\subsection{Suppression factor of the Newtonian potential}
\label{app:suppression_fac}
From the previous equations, we can calculate the Newtonian potential during PBH domination, evaporation and reheating. The amplitude right before it starts oscillating, during reheating, can be parametrized as (see Sec.~\ref{sec:pert_equation_synchronous})
\begin{equation}
\label{eq:Phi_osc_main}
\Phi_{\rm osc}(k)
=
\bigl[T_\Phi(\kappa)\,\Phi_{\bf k}(0)
+ T_S(\kappa)\,S_{\bf k}(0)\bigr]\,
\mathcal{S}_\Phi(k),
\end{equation}
where $T_\Phi$ and $T_S$ are the usual transfer functions for modes entering the
horizon during the early radiation era, and $\mathcal{S}_\Phi(k)$ is a
{suppression factor} encoding the decay of $\Phi$ during PBH evaporation.
Its physical origin is transparent from the subhorizon Poisson equation
\begin{equation}
\label{eq:poisson_subhorizon_main}
k^2 \Phi \simeq \frac{3}{2}\,\mathcal{H}^2
\left(
\frac{\langle\rho_{\rm \mathsmaller{PBH}}\rangle}{\langle\rho_{\rm tot}\rangle}\,\delta_{\rm m}
+
\frac{\langle\rho_{\rm r}\rangle}{\langle\rho_{\rm tot}\rangle}\,\delta_{\rm r}
\right),
\end{equation}
which shows that $\Phi$ is maintained constant by PBH isocurvature during the early
matter–dominated phase, and is driven into oscillations once the PBH
fraction decays and the radiation contribution takes over.
We can also understand the suppression factor from the evolution equation for $\Phi$, cf. Eq.~\eqref{eq:Phi_equation_of_motion_2}, expressed directly in terms of the background fractions $\langle\Omega_{\rm m}\rangle$ and $\langle\Omega_{\rm r}\rangle$, where $\left<\cdot\right>$ denotes the averaging over the PBH mass distribution:
\begin{equation}
\label{eq:Phi_during_eva}
\Phi''+\left(\frac{9+7\left<\Omega_r\right>}{3+\left<\Omega_r\right>} \right)\mathcal{H}\Phi'+\left<\Omega_r\right>\left(\frac{3(1-\left<\Omega_r\right>)\mathcal{H}^2+4k^2}{3(3+\left<\Omega_r\right>)} \right)\Phi=-\frac{2\left<\Omega_m\right>\left<\Omega_r\right>}{3+\left<\Omega_r\right>}\mathcal{H}^2\delta_{\rm \mathsmaller{PBH}},
\end{equation}
where we neglected $\delta_{r}\ll \delta_{\rm \mathsmaller{PBH}}$.
During PBH domination, $\eta\ll \left<\eta_{\rm eva}\right>$, we have $\left<\Omega_r\right>\simeq 0$ and $\left<\Omega_m\right>\simeq 1$ such that Eq.~\eqref{eq:Phi_during_eva} becomes
\begin{equation}
    \eta\ll \left<\eta_{\rm eva}\right>:\qquad  \Phi''+3\mathcal{H}\Phi'\simeq 0,
\end{equation}
and $\Phi$ stays on a near-constant plateau, $\Phi'\simeq 0$.
During PBH evaporation, $\eta\sim \left<\eta_{\rm eva}\right>$, we have $\left<\Omega_m\right>\propto\exp \left(-\int^\eta_0\! {\rm d}\tilde{\eta}\, \Gamma\right) $, cf. Eq.~\eqref{eq:rhom_general}. Taking the sub-horizon limit $k\gg\mathcal{H}$, and approximating $\left<\Omega_r\right>\simeq 1$, in Eq.~\eqref{eq:Phi_during_eva} gives
\begin{equation}
\label{eq:Phi_EoM_eta_eva}
        \eta\sim \left<\eta_{\rm eva}\right>:\qquad  k^2\Phi=-\frac{3}{2}\mathcal{H}^2\left<\Omega_m\right>\delta_{\rm \mathsmaller{PBH}}-3\left(\Phi''+4\mathcal{H}\Phi'\right).
\end{equation}
We define $\eta_{\rm osc}$ the time when the two terms on the right hand side are equal to each other
\begin{equation}
\label{eq:eta_dec_def}
   -\frac{1}{2}\mathcal{H}^2\left<\Omega_m\right>\delta_{\rm \mathsmaller{PBH}}(\eta_{\rm osc})\equiv \Phi''(\eta_{\rm osc})+4\mathcal{H}\Phi'(\eta_{\rm osc}).
\end{equation}
For $\eta\lesssim\eta_{\rm osc}$, the first term in Eq.~\eqref{eq:Phi_EoM_eta_eva}  dominate leading to
\begin{equation}
\label{eq:phi_proto_rhom}
    \Phi \simeq -\frac{3}{2}\mathcal{H}^2\left<\Omega_m\right>\delta_{\rm \mathsmaller{PBH}} \propto \left<\Omega_m\right>\propto \exp \left[-\int^\eta_0\! {\rm d}\tilde{\eta}\, \Gamma\right].
\end{equation}
For $\eta\gtrsim\eta_{\rm osc}$, the second term in Eq.~\eqref{eq:Phi_EoM_eta_eva} dominates and we have 
\begin{equation}
\Phi''+4\mathcal{H}\Phi'+\frac{k^2}{3}\Phi \simeq 0,
\end{equation}
whose solution, $\Phi \propto a^{-2}\cos(k\eta/\sqrt{3})$, features acoustic oscillations with a decaying envelope. 
The suppression factor $\mathcal{S}_\Phi(k)$ can be estimated as 
\begin{equation}
\label{eq:suppression_factor_def}
    \mathcal{S}_\Phi(k) \simeq \frac{\Phi(\eta_{\rm osc})}{\Phi(\eta\ll \eta_{\rm osc}))},
\end{equation}
where $\Phi(\eta\ll \eta_{\rm osc})$ is the plateau value before the exponential decrease, taken as a constant, and $\Phi(\eta_{\rm osc})$ is the amplitude at the onset of the oscillation, estimated to occur at the time $\eta_{\rm osc}$. According to Eq.~\eqref{eq:phi_proto_rhom}, during PBH evaporation before it starts oscillating, the Newtonian potential is proportional to 
\begin{equation}
\label{eq:Phi_ana}
   \eta\lesssim\eta_{\rm osc}:\qquad   \Phi~\propto~\left<\Omega_{\rm m}\right>,
\end{equation}
were $\left<\Omega_{\rm m}\right>$ is the average of $\Omega_{\rm m}\equiv\rho_m/\rho_{\rm tot}$ in Eq.~\eqref{eq:rhom_general} over the Choptuik's mass distribution $\psi(M)$ in Eq.~\eqref{eq:psi_f_maintext}:
\begin{equation}
\label{eq:rho_avg_0}
    \left<\Omega_{\rm m}\right>(\eta)=\int\!\frac{dM}{M}\psi(M)\Omega_{\rm m}(\eta;M)=\frac{1+\gamma_{\rm M}}{\gamma_{\rm M}} \int^{M_{\rm cut}}_{M(\eta)} \!\frac{dM}{M}\!\!\left(\frac{M}{M_{\rm cut}}\right)^{\! \frac{1}{\gamma_{\rm M}}+1}\!\! \left(1-\frac{t(\eta)}{t_{\rm eva}(M)}\right)^{\!1/3}.
\end{equation}
 The lower bound of the integral prevents the piece $1-t(\eta)/t_{\rm eva}(M)$ to turn negative.
Given that $t_{\rm eva} \propto M^3$ and $t(\eta) \propto \eta^3$ during matter domination, we can change the integration variable from $M$ to $\eta_{\rm eva}$ using the relation $M \propto \eta_{\rm eva}$. Eq.~\eqref{eq:rho_avg_0} becomes
\begin{equation}
\label{eq:rho_avg_1}
    \left<\Omega_{\rm m}\right>(\eta)=\frac{1+\gamma_{\rm M}}{\gamma_{\rm M}} \int^{\eta_{\rm cut}}_{\eta} \frac{d\eta_{\rm eva}}{\eta_{\rm eva}}\left(\frac{\eta_{\rm eva}}{\eta_{\rm cut}}\right)^{\! 1+1/\gamma_{\rm M}} \left(1-\frac{\eta^3}{\eta^3_{\rm eva}}\right)^{\!1/3},
\end{equation}
where $\eta_{\rm cut}=(1+2\gamma_{\rm M})\left<\eta_{\rm eva}\right>/(1+\gamma_{\rm M})$ is determined from $\int d\log(\eta_{\rm eva})\, \psi(\eta_{\rm eva}) \,\eta_{\rm eva}=\left<\eta_{\rm eva}\right>$.
The integral has an exact analytical expression
\begin{equation}
\label{eq:fully_analytic_sol_rhom_app}
\langle \Omega_{\rm m} \rangle = -(-1)^{2/3}(1+\gamma_{\rm M})\left.\frac{\eta}{\eta_{\rm eva}}
\left(\frac{\eta_{\rm eva}}{\eta_{\rm cut}}\right)^{\! 1+\frac{1}{\gamma_{\rm M}}}
\!\!{}_{2}F_1\left( -\frac{1}{3}, \frac{1}{3\gamma_{\rm M}}; 1 + \frac{1}{3\gamma_{\rm M}};\frac{\eta_{\text{eva}}^3}{\eta^3} \right)
\right|_{\eta_{\rm eva}=\eta}^{\eta_{\rm eva}=\eta_{\rm cut}},
\end{equation}
where ${}_2F_1$ denotes the hypergeometric function. Using the approximation $\Phi\propto \left<\Omega_m\right>$ in Eq.~\eqref{eq:Phi_ana}, the suppression factor defined in Eq.~\eqref{eq:suppression_factor_def} can be calculated {analytically} as 
\begin{equation}
\label{eq:suppression_factor_ana}
    \mathcal{S}_\Phi(k) \simeq \left<\Omega_m(\eta_{\rm osc})\right>,
\end{equation}
where $\left<\Omega_m(\eta_{})\right>$ is given by Eq.~\eqref{eq:fully_analytic_sol_rhom} and where have used $\left<\Omega_m(\eta\ll \eta_{\rm osc})\right>=1$. The time $\eta_{\rm osc}$ when $\Phi$ starts oscillating can be determined analytically from plugging Eq.~\eqref{eq:eta_dec_def} into Eq.~\eqref{eq:Phi_EoM_eta_eva}, which leads to

\begin{figure}[t!]
\centering
\raisebox{0cm}
{\makebox{\includegraphics[width=0.7\textwidth, scale=1]{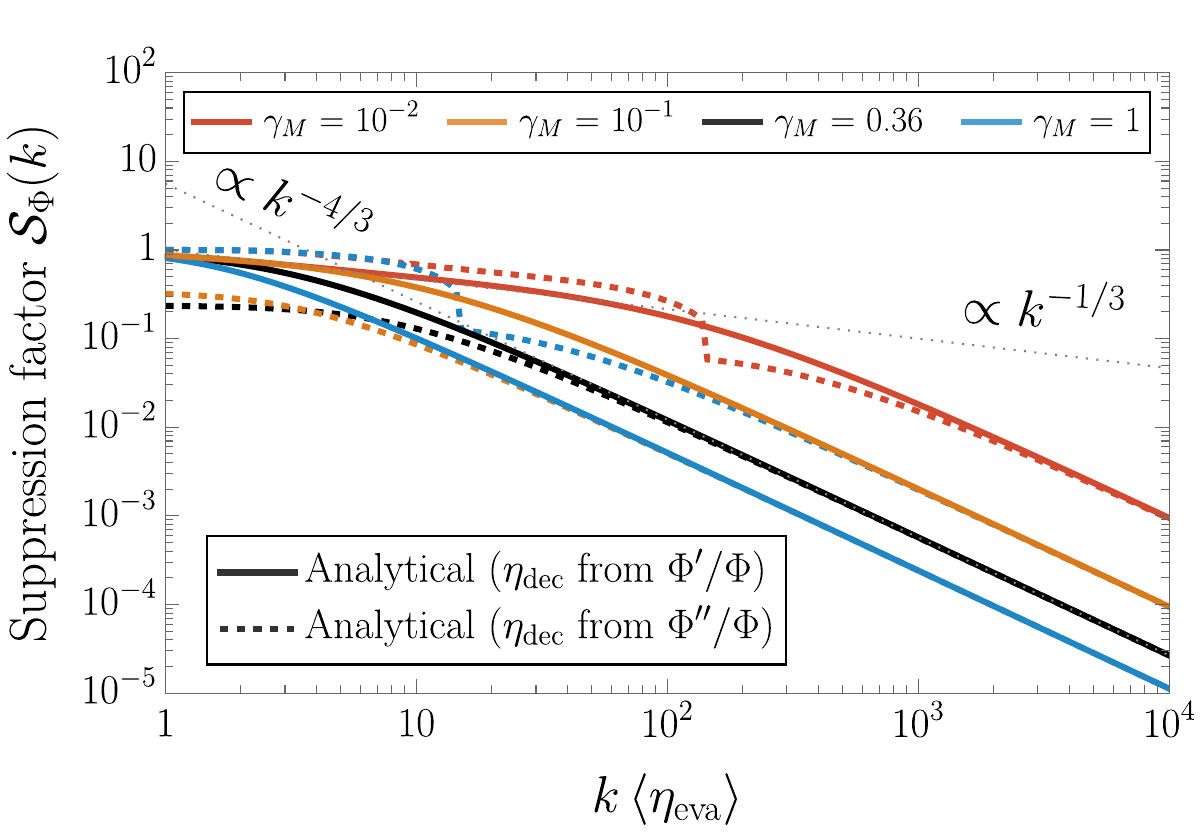}}}
\caption{Suppression factor $\mathcal{S}_\Phi(k)$ of the Newtonian potential $\Phi$ for different values of the 
critical exponent $\gamma_{\rm M}$ entering Choptuik's PBH mass distribution $\psi(M)$ in 
Eq.~\eqref{eq:psi_f_maintext}.  
We compare the analytical expression in Eq.~\eqref{eq:suppression_factor_ana} using 
$\eta_{\rm osc}$ determined from Eq.~\eqref{eq:dec_limit_first_derivative} (\textbf{solid lines}) and 
from Eq.~\eqref{eq:dec_limit_second_derivative} (\textbf{dashed lines}), which rely respectively on 
the first and second derivatives of the Newtonian potential, $\Phi'$ and $\Phi''$, with $\Phi$ given by Eq.~\eqref{eq:fully_analytic_sol_rhom}.  
The discontinuity in the dashed curves is caused by a sign change of $\Phi''$ occurring near 
$\eta_{\rm osc}\simeq \eta_{\rm cut}(1 - \gamma_{\rm M}/2)$.  
This sign flip produces an abrupt shift in the solution value for $\eta_{\rm osc}$ when using the 
second-derivative condition of Eq.~\eqref{eq:dec_limit_second_derivative}.  
In contrast, the first-derivative criterion in Eq.~\eqref{eq:dec_limit_first_derivative} (solid lines) 
eliminates this step-like behavior and yields a smooth suppression factor.
}
\label{fig:alpha_vs_gamma_M_app}
\end{figure}

\begin{equation}
\label{eq:dec_limit_second_derivative}
\left|\frac{\Phi^{\prime\prime}}{\Phi}\right| \simeq \frac{k^2}{6}\quad\text{at}\quad \eta=\eta_{\rm osc}(k).
\end{equation}
where we have neglected $4\mathcal{H}\Phi'/\Phi''\ll 1$ which is valid for $k\gg \mathcal{H}$. 
We note that the second derivative of $\Phi \propto \Omega_m$ changes sign at 
$\eta_{\rm osc}=\eta_{\rm zero}$, defined by $\langle\Omega_m\rangle''(\eta_{\rm zero})=0$, 
which occurs at $\eta_{\rm zero}\simeq \eta_{\rm cut}(1-\gamma_M/2)$.  
As shown by the dashed curves in Fig.~\ref{fig:alpha_vs_gamma_M_app}, this produces a step-like behavior in the suppression factor $\mathcal{S}_\Phi(k)$ when $k\langle\eta_{\rm eva}\rangle \sim \gamma_M^{-1}$.  
Since the physical origin of these discontinuities is unclear, we avoid them by replacing the 
second-derivative criterion with a first-derivative one, substituting 
Eq.~\eqref{eq:dec_limit_second_derivative} with
\begin{equation}
\label{eq:dec_limit_first_derivative}
\left|\frac{\Phi^{\prime}}{\Phi}\right| \simeq \sqrt{\frac{2}{3}}\,k
\qquad\text{at}\qquad \eta=\eta_{\rm osc}(k),
\end{equation}
where the factor $\sqrt{2/3}$ is chosen such that 
Eqs.~\eqref{eq:dec_limit_second_derivative} and \eqref{eq:dec_limit_first_derivative} return the same 
start-of-oscillation time $\eta_{\rm osc}$ in the sub-horizon limit $k\langle\eta_{\rm eva}\rangle\gg 1$.  
As illustrated by the solid curves in Fig.~\ref{fig:alpha_vs_gamma_M_app}, the suppression factor 
$\mathcal{S}_\Phi(k)$ from Eq.~\eqref{eq:suppression_factor_ana}, with $\eta_{\rm osc}$ determined using 
Eq.~\eqref{eq:dec_limit_first_derivative}, now exhibits a smooth behavior. 

\subsection{High-$k$ limit for the Choptuik mass function}
We next derive an analytic estimate for the suppression factor $\mathcal{S}_\Phi(k)$ in the limit 
$k\langle\eta_{\rm eva}\rangle \gg 1$, for which the start-of-oscillation time approaches the evaporation time 
of the heaviest PBHs, i.e.\ $\eta_{\rm osc} \to \eta_{\rm cut}^-$.  
Using Mathematica, we calculate the limit
\begin{equation}
\label{eq:rho_m_longversion}
\langle\Omega_{\rm m}\rangle 
~\xrightarrow[\eta\to \eta_{\rm cut}^-]{}~ 
\frac{3^{4/3}}{4}\,\frac{1+\gamma_{\rm M}}{\gamma_{\rm M}}
\left(1-\frac{\eta}{\eta_{\rm cut}}\right)^{4/3}.
\end{equation}
We therefore deduce that the Newtonian potential scales as
\begin{equation}
\label{eq:Phi_ana_2}
    \Phi \propto \langle\Omega_{\rm m}\rangle 
    \propto (\eta_{\rm cut}-\eta)^{4/3}.
\end{equation}
Substituting Eq.~\eqref{eq:Phi_ana_2} into either 
Eq.~\eqref{eq:dec_limit_second_derivative} or, equivalently, 
Eq.~\eqref{eq:dec_limit_first_derivative}, yields the time $\eta_{\rm osc}$ at which $\Phi$ starts oscillating
\begin{equation}
\label{eq:eta_dec_final_avg}
    \frac{4}{9}\,(\eta_{\rm cut}-\eta_{\rm osc})^{-2} 
    \simeq \frac{k^2}{6}
    \qquad\Longrightarrow\qquad
    \eta_{\rm cut}-\eta_{\rm osc}
    \;\simeq\; \sqrt{\frac{8}{3}}\,k^{-1}.
\end{equation}
Plugging Eq.~\eqref{eq:eta_dec_final_avg} into 
Eqs.~\eqref{eq:rho_m_longversion} and \eqref{eq:suppression_factor_ana} 
yields the high-$k$ limit of the suppression factor,
\begin{equation}
\label{eq:suppression_factor_final}
    \mathcal{S}_\Phi(k)
    \simeq 
    \kappa_0 \left(\frac{2}{k\langle\eta_{\rm eva}\rangle}\right)^{4/3}
    \simeq 
    \kappa_0 \left(\frac{k_{\rm eva}}{k}\right)^{4/3},
    \qquad
    \kappa_0 \equiv
    \frac{3^{2/3}(1+\gamma_M)^{7/3}}{2^{4/3}\gamma_M(1+2\gamma_M)^{4/3}}
    \simeq 2.3,
\end{equation}
where we have used $k_{\rm eva}\equiv\mathcal{H}_{\rm eva}
= 2/\langle\eta_{\rm eva}\rangle$, following Eq.~\eqref{eq:Hubble_full} for $\left<\eta_{\rm eva}\right>\gg \eta_{\rm eq}$,
together with $\eta_{\rm cut}/\langle\eta_{\rm eva}\rangle
= (1+2\gamma_{\rm M})/(1+\gamma_{\rm M})$ and $\gamma_{\rm M}\simeq 0.36$.
In Fig.~\ref{fig:alpha_vs_gamma_M} of the main text, we show that the analytic solution is validated against a full numerical integration of the
perturbation equations in the sub-horizon limit $k\left<\eta_{\rm eva}\right>\gg 1$.

\subsection{Monochromatic limit}
The expression in Eq.~\eqref{eq:fully_analytic_sol_rhom} can also be used to derive the suppression 
factor in the monochromatic limit ($\gamma_{\rm M}\rightarrow 0$), as done in 
Ref.~\cite{Inomata:2020lmk}.  
To this end, we expand the hypergeometric function ${}_2F_1$ in the limit where its second and third 
arguments become large:
\begin{equation}
\label{eq:hypergeometric_Taylor}
    {}_2 F_1\!\left(-\frac{1}{3},\frac{1}{3\gamma_{\rm M}};\,1+\frac{1}{3\gamma_{\rm M}},
    \frac{\eta_{\rm eva}^3}{\eta^3}\right)
    \xrightarrow[\gamma_{\rm M}\to 0]{} 
    \left(1-\frac{\eta_{\rm eva}^3}{\eta^3}\right)^{1/3},
\end{equation}
which, when combined with the prefactor in Eq.~\eqref{eq:fully_analytic_sol_rhom}, gives
\begin{equation}
\label{eq:rho_m_longversion_mono}
    \langle\Omega_{\rm m}\rangle 
    \xrightarrow[\gamma_{\rm M}\to 0]{} 
    (1+\gamma_{\rm M})
    \left(1-\frac{\eta^3}{\eta_{\rm cut}^3}\right)^{1/3}
    \;\propto\;
    \left(1-\frac{t}{t_{\rm cut}}\right)^{1/3},
\end{equation}
where we used $\eta a = 3t$ to convert from cosmic to conformal time.  
Substituting Eq.~\eqref{eq:rho_m_longversion_mono} into Eq.~\eqref{eq:dec_limit_second_derivative} with $\Phi\propto \langle\Omega_{\rm m}\rangle $, 
yields the time $\eta_{\rm osc}$ at which $\Phi$ starts oscillating
\begin{equation}
\label{eq:eta_dec_mono}
 \frac{\eta_{\rm cut}^3}{\eta_{\rm osc}^3}~\simeq~1+ \frac{\sqrt{12}}{k\eta_{\rm osc}},
\end{equation}
which after plugging nack into Eq.~\eqref{eq:rho_m_longversion_mono} gives the suppression factor
\begin{equation}
\label{eq:supp_mono}
    \mathcal{S}_\Phi(k)
    \xrightarrow[\gamma_{\rm M}\to 0]{}
    \left(\frac{\sqrt{12}}{k\langle\eta_{\rm eva}\rangle}\right)^{1/3},
\end{equation}
where we used $\eta_{\rm cut}\simeq\langle\eta_{\rm eva}\rangle$ in the monochromatic limit and $\eta_{\rm osc}\simeq \eta_{\rm cut}$ in the high $k$ limit.  
This reproduces the scaling obtained in Ref.~\cite{Inomata:2020lmk} for a monochromatic PBH mass 
function.
As shown in Fig.~\ref{fig:alpha_vs_gamma_M}, this scaling is reached only for unrealistically small 
values of $\gamma_{\rm M}$.  
For small but finite $\gamma_{\rm M}$, we observe that the monochromatic scaling 
$\mathcal{S}_\Phi(k)\propto k^{-1/3}$ in Eq.~\eqref{eq:supp_mono} transitions to the Choptuik-mass-function behavior 
$\mathcal{S}_\Phi(k)\propto k^{-4/3}$ in Eq.~\eqref{eq:suppression_factor_final} once
\begin{equation}
\label{eq:transition_mono_ext}
    k\langle\eta_{\rm eva}\rangle ~\gtrsim~ k_{\rm trans}\langle\eta_{\rm eva}\rangle ~\equiv ~\frac{\sqrt{3}}{2^{1/3}\gamma_{\rm M}}\sim \gamma_{\rm M}^{-1},
\end{equation}
where $k_{\rm trans}$ is defined as the value of $k$ for which Eqs.~\eqref{eq:suppression_factor_final} and \eqref{eq:supp_mono} intersect, expanded in the limit $\gamma_{\rm M}\ll 1$.

The transition scale $k_{\rm trans}$ in Eq.~\eqref{eq:transition_mono_ext} may also be identified as the wavenumber at which the next-to-leading–order correction to the monochromatic limit in Eq.~\eqref{eq:hypergeometric_Taylor} becomes non-negligible.
We need to expand
\begin{equation}
{}_2F_1\!\left(-\frac13,\frac{1}{3\gamma_M};1+\frac{1}{3\gamma_M};z\right),
\qquad 
z=\left(\frac{\eta_{\rm cut}}{\eta}\right)^3,
\end{equation}
for $\gamma_M\to0$. First we use
\begin{equation}
    _2F_1(a,b,c,z) = (1-z)^{-a}  _2F_1\left(a,c-b,c,\frac{z}{z-1}\right)
\end{equation}
to write 
\begin{equation}
\label{eq:hypergeo_rewritten}
    _2F_1\left(-\frac{1}{3},\frac{1}{3\gamma_M},1+\frac{1}{3\gamma_M},z\right)=  \left| (1-z)^{1/3} \ _2F_1\left(-\frac{1}{3},1,1+\frac{1}{3\gamma_M},\frac{z}{z-1}\right)\right|
\end{equation}
and move the singular $1/\gamma_M$ to the third argument only. We added such absolute values for the case of $z>1$. As we will show, the limit $\gamma_M\rightarrow0$ is purely real, such that we can drop the absolute value. Next, we employ 
\begin{equation}
_2F_1(\tilde{a},\tilde{b},\tilde{c}+\lambda,\tilde{z})\xrightarrow{\lambda\rightarrow\infty}\frac{\Gamma(\tilde{c} + \lambda)}{\Gamma(\tilde{c}-\tilde{b} +\lambda)}\sum_{s=0}^{\infty}q_s(\tilde{z})(\tilde{b})_s\lambda^{-s-\tilde{b}},
\end{equation}
where we identify $(\tilde{a},\tilde{b},\tilde{c},\lambda,\tilde{z})$ as $(-1/3,1,1,1/(3\gamma_M),z/(z-1))$. The Pochhammer symbols $(\tilde{b})_s=\tilde{b}(\tilde{b}+1)\dots(\tilde{b}+s-1)$ take the values $(\tilde{b})_0=1$ and $(\tilde{b})_1=1$, respectively. The coefficients $q_s(\tilde{z})$ are defined from the generating function order-by-order
\begin{equation}
    \left(\frac{\mathrm{e}^t-1}{t}\right)^{\tilde{b}-1} \mathrm{e}^{t(1-\tilde{c})}\left(1-z+z \mathrm{e}^{-t}\right)^{-\tilde{c}}=\sum_{s=0}^{\infty} q_s(\tilde{z}) t^s.
\end{equation}
At next-to-leading order we find 
\begin{equation}
    1 + \frac{1}{2} (1 + \tilde{b} - 2 \tilde{c} + 2 \tilde{a} \tilde{z}) t + \mathcal{O}(t^2)= q_0(\tilde{z}) + q_1(\tilde{z})t +\mathcal{O}(t^2).
\end{equation}
Comparing coefficient order-by-order yields $q_0(\tilde{z})=1$ and $q_1(\tilde{z})=-1/3\tilde{z}$. With this we find at next-to-leading order 
\begin{equation}
    \frac{\Gamma(\tilde{c} + \lambda)}{\Gamma(\tilde{c}-\tilde{b} +\lambda)}\sum_{s=0}^{\infty}q_s(\tilde{z})(\tilde{b})_s\lambda^{-s-\tilde{b}}\simeq\frac{1}{\lambda}(\lambda-\frac{1}{3}\tilde{z}) + \mathcal{O}(\lambda^{-2}), 
\end{equation}
which, after substituting into Eq.~\eqref{eq:hypergeo_rewritten} and replacing $(\tilde{a},\tilde{b},\tilde{c},\lambda,\tilde{z})$ with $(-1/3,1,1,1/(3\gamma_M),z/(z-1))$, yields
\begin{equation}
\label{eq:mono+approx}
    _2F_1(-\frac{1}{3},\frac{1}{3\gamma_M},1+\frac{1}{3\gamma_M},z)=(1-z)^{1/3}\left|1-\gamma_M \frac{z}{z-1}\right| +\mathcal{O}(\gamma_M^2),
\end{equation}
The monochromatic behaviour persists until these 
corrections become order unity.
Setting the NLO correction equal to a 
fraction $\Delta={\cal O}(1)$,
\begin{equation}
\gamma_M\frac{z}{z-1}\simeq\Delta,
\end{equation}
defines the transition time $\eta_{\rm trans}$. Using $z=\left({\eta_{\rm cut}}/{\eta_{\rm osc}}\right)^3$ with $\eta_{\rm osc}$ given by Eq.~\eqref{eq:eta_dec_mono} and $\eta_{\rm osc}\simeq \left<\eta_{\rm eva}\right>$ yields the transition scale
\begin{equation}
    k_{\rm trans}\,\langle\eta_{\rm eva}\rangle
    \simeq
    \frac{\sqrt{12}\,\Delta}{\gamma_M},
\end{equation}
which matches Eq.~\eqref{eq:transition_mono_ext} for $\Delta\simeq0.4$.

\newpage

\section{Green's function for tensor modes}
\label{app:Green_fct_tensor_mode}

We derive the retarded Green's function $G_h(x,\bar{x})$ for the tensor
equation of motion
\begin{equation}
G_h''(x,\bar{x})
+2\frac{\mathcal{H}}{k}\,G_h'(x,\bar{x})
+G_h(x,\bar{x})
=\delta(x-\bar{x}),
\end{equation}
cf.\ Eq.~\eqref{eq:green_function_diff_eq}, across the
matter-to-radiation transition induced by PBH evaporation at
$\eta_{\rm eva}\equiv\langle\eta_{\rm eva}\rangle$,
\begin{equation}
{\rm eMD}\;\longrightarrow\;{\rm RD}_2\,,
\end{equation}
see Eq.~\eqref{eq:different_eras}.
We further define $x_{\rm eva}\equiv k\eta_{\rm eva}$ for the
corresponding dimensionless time.
Primes denote derivatives with respect to $x\equiv k\eta$ and
$\bar{x}\equiv k\bar{\eta}$, and the retarded boundary condition
$G_h(x,\bar{x})=0$ for $x<\bar{x}$ is imposed throughout.

\subsection{Background}

In the limit of a prolonged matter-dominated era,
$\eta_{\rm eq}\ll\eta_{\rm eva}$
(equivalently $N_{\rm MD}\gg 1$), the scale factor in
Eq.~\eqref{eq:scale_factor_full} simplifies to the piecewise form
\begin{equation}
\frac{a(\eta)}{a_{\rm eq}}
\;\simeq\;
\begin{cases}
\left(\dfrac{\eta}{\eta_\star}\right)^2,
& \eta\leq\eta_{\rm eva},
\\[10pt]
\dfrac{(2\eta-\eta_{\rm eva})\,\eta_{\rm eva}}{\eta_\star^2},
& \eta>\eta_{\rm eva},
\end{cases}
\end{equation}
where $\eta_\star\equiv\eta_{\rm eq}/(\sqrt{2}-1)$.
In the same limit, the conformal Hubble parameter takes the form
\begin{equation}
\frac{\mathcal{H}}{k}
\;\simeq\;
\begin{cases}
\dfrac{2}{x},
& x\leq x_{\rm eva}
\quad(\text{eMD}),
\\[10pt]
\dfrac{1}{2x-x_{\rm eva}},
& x>x_{\rm eva}
\quad(\mathrm{RD}_2).
\end{cases}
\end{equation}

\subsection{Homogeneous solutions within each phase}

Within each constant-$\omega$ phase the tensor Green's function
is given by Eq.~\eqref{eq:hgreen},
\begin{equation}
\label{eq:hgreen_app}
G_h(x,\bar{x}) = \frac{\pi}{2}\,\frac{\bar{x}^{\,b+3/2}}{x^{\,b+1/2}}
\left[
J_{b+1/2}(\bar{x})\,Y_{b+1/2}(x)
-
J_{b+1/2}(x)\,Y_{b+1/2}(\bar{x})
\right]
\Theta(x-\bar{x}),
\end{equation}
where $b \equiv (1-3\omega)/(1+3\omega)$, and $J_\alpha$ and $Y_\alpha$
denote the Bessel functions of the first and second kind, respectively.
We now specialise to each phase in turn.

\begin{itemize}

\item \textit{$\rm RD_1$ phase} ($\omega=1/3$, $b=0$).
Inserting $b=0$ into Eq.~\eqref{eq:hgreen_app}
the Green's function reduces to
\begin{equation}
\label{eq:G_RD1}
G_h^{\mathrm{RD_1}}(x,\bar{x})
=\frac{\bar{x}}{x}\sin(x-\bar{x})
\Theta(x-\bar{x}).
\end{equation}

\item \textit{$\rm eMD$ phase} ($\omega=0$, $b=1$).
Inserting $b=1$ into Eq.~\eqref{eq:hgreen_app} and using the
identities $J_{3/2}(z)=\sqrt{2/(\pi z)}\,(\sin z/z-\cos z)$ and
$Y_{3/2}(z)=-\sqrt{2/(\pi z)}\,(\cos z/z+\sin z)$,
the Green's function reduces to
\begin{equation}
\label{eq:G_eMD}
G^{\rm (eMD)}_h(x,\bar{x})
=\frac{\bar{x}}{x^3}
\Bigl[
(1+x\bar{x})\sin(x-\bar{x})
+(\bar{x}-x)\cos(x-\bar{x})
\Bigr]
\Theta(x-\bar{x}).
\end{equation}

\item \textit{$\mathrm{RD}_2$ phase} ($\omega=1/3$, $b=0$).
After evaporation the scale factor reads $a\propto\eta-\eta_{\rm eva}/2$
(see footnote~\ref{footnote:etabar} and Eq.~\eqref{eq:scale_factor_full}),
so the conformal Hubble parameter is
$\mathcal{H}/k=(x-x_{\rm eva}/2)^{-1}$.
The two independent homogeneous solutions are
\begin{equation}
u_1(x)=\frac{\sin(x-x_{\rm eva}/2)}{x-x_{\rm eva}/2},
\qquad
u_2(x)=\frac{\cos(x-x_{\rm eva}/2)}{x-x_{\rm eva}/2},
\end{equation}
and imposing the retarded jump condition at $x=\bar{x}$ yields
\begin{equation}
\label{eq:G_RD2}
G^{\rm (RD_2)}_h(x,\bar{x})
=\frac{\bar{x}-x_{\rm eva}/2}{x-x_{\rm eva}/2}\,\sin(x-\bar{x})\,
\Theta(x-\bar{x}),
\qquad x>\bar{x}>x_{\rm eva}.
\end{equation}
 Compared to the standard RD result obtained from Eq.~\eqref{eq:hgreen_app} for $b=0$ in Eq.~\eqref{eq:G_RD1}, the effect of the prior $\rm eMD$ phase is a shift in conformal time,
\begin{equation}
\tilde{x} = x - \frac{x_{\mathrm{eva}}}{2},
\qquad
\tilde{\bar{x}} = \bar{x} - \frac{x_{\mathrm{eva}}}{2},
\end{equation}
such that
\begin{equation}
\label{eq:G_RD2b}
G_h^{\mathrm{RD_2}}(x,\bar{x})
\;\rightarrow\;
\frac{\tilde{\bar{x}}}{\tilde{x}}
\sin(\tilde{x}-\tilde{\bar{x}}).
\end{equation}

\end{itemize}
\subsection{Sourced in eMD, evaluated in
$\mathrm{RD}_2$}

For the eMD contribution to the SIGW spectrum (Sec.~\ref{chap:eMD}),
the relevant configuration has the scalar source acting during the
eMD phase and the tensor mode evaluated after evaporation,
i.e.\ $\bar{x}<x_{\rm eva}<x$. In this case $G_h$ must be constructed
by matching $G^{\rm (eMD)}_h$ to a linear combination of the $\mathrm{RD}_2$
homogeneous solutions $u_1$ and $u_2$ at the transition $x=x_{\rm eva}$,
\begin{equation}
\label{eq:G_mixed}
G^{\rm (eMD\to RD_2)}_h(x,\bar{x})
=\frac{
A(\bar{x})\cos(x-x_{\rm eva})+B(\bar{x})\sin(x-x_{\rm eva})
}{x-x_{\rm eva}/2},
\qquad \bar{x}<x_{\rm eva}<x.
\end{equation}
The matching coefficients $A(\bar{x})$ and $B(\bar{x})$ are fixed by
requiring continuity of $G_h$ and $G_h'$ at $x=x_{\rm eva}$,\footnote{References~\cite{Pearce:2023kxp,Pearce:2025ywc} pointed out that
the oscillatory behaviour in the GW kernel found in
Ref.~\cite{Inomata:2019zqy} and observed in this work
(see Figs.~\ref{fig:GWSig_RD_comp_MD}, \ref{fig:eMDGrad},
and~\ref{fig:EMD_Max_RD_NL}) is non-physical and stems from the
piecewise definition of the Green's function across $\eta_{\rm eva}$.
A more refined treatment replaces the sharp transition by a smooth
numerical interpolation, which does not alter the amplitude or spectral
shape qualitatively but averages over the oscillations. We adopt the
simpler piecewise approximation throughout, which suffices for the
qualitative estimates presented here.}
\begin{equation}
\left[G^{\rm (eMD\to RD_2)}_h(x,\bar{x})\right]_{x=x_{\rm eva}}
=g_0(\bar{x}),
\qquad
\left[\partial_x G^{\rm (eMD\to RD_2)}_h(x,\bar{x})\right]_{x=x_{\rm eva}}
=g_1(\bar{x}),
\end{equation}
where we have defined the boundary data
\begin{equation}
g_0(\bar{x})\equiv G^{\rm (eMD)}_h(x_{\rm eva},\bar{x}),
\qquad
g_1(\bar{x})\equiv
\partial_x G^{\rm (eMD)}_h(x,\bar{x})\big|_{x=x_{\rm eva}}.
\end{equation}
Solving for $A$ and $B$ gives
\begin{equation}
A(\bar{x})=\frac{x_{\rm eva}}{2}\,g_0(\bar{x}),
\qquad
B(\bar{x})=g_0(\bar{x})+\frac{x_{\rm eva}}{2}\,g_1(\bar{x}).
\end{equation}
Inserting the explicit eMD result Eq.~\eqref{eq:G_eMD}, the boundary
data evaluate to
\begin{align}
g_0(\bar{x})
&=\frac{\bar{x}}{x_{\rm eva}^3}
\Bigl[
(1+x_{\rm eva}\bar{x})\sin(x_{\rm eva}-\bar{x})
+(\bar{x}-x_{\rm eva})\cos(x_{\rm eva}-\bar{x})
\Bigr],
\\[4pt]
g_1(\bar{x})
&=\frac{\bar{x}}{x_{\rm eva}^4}
\Bigl[
(x_{\rm eva}^2-3x_{\rm eva}\bar{x}-3)\sin(x_{\rm eva}-\bar{x})
+(x_{\rm eva}^2\bar{x}-3\bar{x}+3x_{\rm eva})\cos(x_{\rm eva}-\bar{x})
\Bigr].
\end{align}
For use in the eMD kernel integrals of Sec.~\ref{chap:eMD}, it is
convenient to re-express $G^{\rm (eMD\to RD_2)}_h$ on the spherical
Bessel basis $j_1(\bar{x})$ and $y_1(\bar{x})$,
\begin{equation}
j_1(\bar{x})
=\frac{\sin\bar{x}}{\bar{x}^2}-\frac{\cos\bar{x}}{\bar{x}},
\qquad
y_1(\bar{x})
=-\frac{\cos\bar{x}}{\bar{x}^2}-\frac{\sin\bar{x}}{\bar{x}},
\end{equation}
since these diagonalise the $\sin\bar{x}$ and $\cos\bar{x}$ structure
of the boundary data. Writing
\begin{equation}
\label{eq:G_mixed_Bessel}
G^{\rm (eMD\to RD_2)}_h(x,\bar{x})
=\frac{1}{2x/x_{\rm eva}-1}\,\frac{\bar{x}^3}{x_{\rm eva}^2}
\Bigl[
C(x,x_{\rm eva})\,j_1(\bar{x})
+D(x,x_{\rm eva})\,y_1(\bar{x})
\Bigr],
\end{equation}
and matching to Eq.~\eqref{eq:G_mixed} expanded in $\sin\bar{x}$
and $\cos\bar{x}$, one reads off
\begin{align}
\label{eq:C_coeff}
C(x,x_{\rm eva})
&=\frac{
\sin x
-2x_{\rm eva}(\cos x+x_{\rm eva}\sin x)
+\sin(x-2x_{\rm eva})
}{2x_{\rm eva}^2},
\\[4pt]
\label{eq:D_coeff}
D(x,x_{\rm eva})
&=\frac{
(2x_{\rm eva}^2-1)\cos x
-2x_{\rm eva}\sin x
+\cos(x-2x_{\rm eva})
}{2x_{\rm eva}^2}.
\end{align}

\subsection{Summary}

Collecting all three cases, the retarded Green's function across
the eMD-to-$\mathrm{RD}_2$ transition reads
\begin{equation}
\label{eq:Gh_full_appendix}
G_h(x,\bar{x})
=
\begin{cases}
G^{\rm (eMD)}_h(x,\bar{x}),
& \bar{x}<x\leq x_{\rm eva},
\\[4pt]
G^{\rm (eMD\to RD_2)}_h(x,\bar{x}),
& \bar{x}\leq x_{\rm eva}<x,
\\[4pt]
G^{\rm (RD_2)}_h(x,\bar{x}),
& x_{\rm eva}<\bar{x}<x,
\end{cases}
\end{equation}
with each piece given in Eqs.~\eqref{eq:G_eMD},~\eqref{eq:G_mixed_Bessel}--\eqref{eq:D_coeff},
and~\eqref{eq:G_RD2}, respectively.
$G^{\rm (eMD\to RD_2)}_h$ enters the eMD kernel integrals of
Sec.~\ref{chap:eMD}, while $G^{\rm (RD_2)}_h$ governs the Poltergeist
contribution of Sec.~\ref{chap:poltergeist}. Our result agrees with Ref.~\cite{Inomata:2019zqy} once the difference in notation is taken into account, see footnote~\ref{footnote:green_fct_notation}.

\newpage
\section{More details on SIGWs following PBH evaporation}
\label{app:Poltergeist_formalism}

This appendix derives the Poltergeist contribution to the SIGW spectrum produced at PBH evaporation.  Our goal is to obtain an explicit expression for the oscillatory kernel and its time-averaged square entering the GW spectrum.

\subsection{The oscillatory kernel}

Immediately after PBH evaporation, for $\eta>\eta_{\rm osc}$, the Newtonian potential evolves as
\begin{equation}
\label{eq:transfer_fc_reh_app}
\Phi(k,\eta)
=
\sum_{X=\{\Phi,S\}}
T_X(k,\eta)\,X(k,0),
\qquad
T_X(k,\eta)
=
T_X^{\rm (eMD)}(k)\,\mathcal{S}_\Phi(k)\,T_\Phi^{\rm (RD_2)}(k,\eta),
\end{equation}
where $T_\Phi^{\rm (RD_2)}(k,\eta)$ is the oscillatory RD transfer function defined in Eq.~\eqref{eq:TRD2_def}. The eMD transfer functions $T_X^{\rm (eMD)}(k)$ and the evaporation-induced suppression factor $\mathcal{S}_\Phi(k)$ are given in Eqs.~\eqref{eq:transfer_adia_iso_main} and~\eqref{eq:suppression_factor_final_main}, respectively.

The transfer function describing oscillations in the RD era is \cite{Domenech:2020ssp}
\begin{equation}
\label{eq:PhiRD_app}
T_\Phi^{\rm (RD_2)}(k,\eta)
=\frac{1}{c_s k \bar{\eta}}
\left(C_1 j_1(c_s k \bar{\eta})+C_2 y_1(c_s k \bar{\eta})\right),
\qquad c_s=\frac{1}{\sqrt{3}},
\end{equation}
where we introduced the shifted conformal time $\bar{\eta}\equiv \eta-\eta_{\rm eva}/2$.
Matching $\Phi$ and $\Phi'$ across the transition fixes
\begin{equation}
\label{eq:C1C2_app_app}
C_1=-\frac{1}{8}(c_s k \eta_{\rm eva})^3\,y_2(c_s k \eta_{\rm eva}/2),
\qquad
C_2=\frac{1}{8}(c_s k \eta_{\rm eva})^3\,j_2(c_s k \eta_{\rm eva}/2),
\end{equation}
with $j_n$ and $y_n$ the spherical Bessel functions.

The Poltergeist contribution arises from the enhancement of the $\Phi'$-terms in the SIGW source. Keeping only the dominant piece, the kernel $I_X$ defined in Eq.~\eqref{eq:kernel_standard} is approximated by
\begin{align}
\label{eq:KernelIRDGen_app}
I_{X}(\bar{x},u,v)\simeq \frac{1}{2k^2}\int_{x_{\rm eva}/2}^{\bar{x}} d\tilde{\bar{x}}
~ \tilde{\bar{x}}^2 \, G_h^{\rm (RD_2)}(\bar{x},\tilde{\bar{x}})
T_X'\left(uk,\tilde{\eta}\right)
T_X'\left(vk,\tilde{\eta}\right),
\qquad X=\{\Phi,S\},
\end{align}
where $\bar{x}=k\bar{\eta}=x-x_{\rm eva}/2$, $x\equiv k\eta$, $x_{\rm eva}\equiv k\eta_{\rm eva}$, and $\tilde{\eta}=(\tilde{\bar{x}}+x_{\rm eva}/2)/k$. In the limit $x_{\rm eva}\to +\infty$, differentiating Eq.~\eqref{eq:transfer_fc_reh_app} gives \cite{Domenech:2020ssp}
\begin{align}
\label{eq:tprime_app}
T_X'\left(uk,\tilde{\eta}\right)\simeq
-\,\mathcal{S}_\Phi(uk)\,T_X^{\rm (eMD)}(uk)\,c_s u k
\left(\frac{x_{\rm eva}}{2\tilde{\bar{x}}}\right)^2
\sin\!\left[c_s u\left(\tilde{\bar x}-x_{\rm eva}/2\right)\right].
\end{align}
The tensor Green function in the post-evaporation phase is given by Eq.~\eqref{eq:G_RD2} which we report here for clarity
\begin{equation}
\label{eq:hgreen_app_RD_app}
G_h^{\rm (RD_2)}(x,\tilde{x})=\frac{\tilde x-x_{\rm eva}/2}{x-x_{\rm eva}/2} \sin(x-\tilde x)\,\Theta(x-\tilde{x}).
\end{equation}
Using Eqs.~\eqref{eq:tprime_app} and \eqref{eq:hgreen_app_RD_app} in Eq.~\eqref{eq:KernelIRDGen_app} yields
\begin{equation}
\label{eq:IX_RD_final_app_app}
I_{X}(\bar{x}, u, v)
\simeq
\frac{c_s^2 u v}{32 \bar{x}}\,x_{\mathrm{eva}}^4\,
\mathcal{S}_{\Phi}(u k)\mathcal{S}_{\Phi}(v k)\,
T_{X}^{\mathrm{eMD}}(u k)T_{X}^{\mathrm{eMD}}(v k)\,
\mathcal{I}_{\mathrm{osc}}(\bar{x}, u, v),
\end{equation}
where the oscillatory integral is
\begin{equation}
\label{eq:I_osc_app}
\mathcal{I}_{\mathrm{osc}}(\bar{x}\to\infty,u,v)=
\int_0^{\infty} \frac{{\rm d} x_2}{x_2+x_3}
\sin(x_1-x_2)\,\sin(c_s u x_2)\,\sin(c_s v x_2),
\end{equation}
with $x_1\equiv \bar{x}-x_{\rm eva}/2$, $x_2\equiv \tilde{\bar{x}}-x_{\rm eva}/2$, and $x_3\equiv x_{\rm eva}/2$.

The integral \eqref{eq:I_osc_app} can be performed exactly and results in a sum of eight terms \cite{Domenech:2020ssp}. Writing the primitive explicitly,
\begin{equation}\hspace{-0.3cm}
\label{eq:Iosc_exact_app}
\begin{aligned}
\mathcal{I}_{\mathrm{osc}}(\bar{x},u,v)=\frac{1}{4} \Bigg[
& - \mathrm{Ci}\!\left[ \left(-1 - \frac{u}{\sqrt{3}} + \frac{v}{\sqrt{3}} \right)(x_2 + x_3) \right]
    \sin\!\left( x_1 + \left(1 + \frac{u}{\sqrt{3}} - \frac{v}{\sqrt{3}} \right) x_3 \right) \\
& + \mathrm{Ci}\!\left[ \left(1 - \frac{u}{\sqrt{3}} + \frac{v}{\sqrt{3}} \right)(x_2 + x_3) \right]
    \sin\!\left( x_1 + \left(1 - \frac{u}{\sqrt{3}} + \frac{v}{\sqrt{3}} \right) x_3 \right) \\
& - \mathrm{Ci}\!\left[ \left(-1 + \frac{u}{\sqrt{3}} + \frac{v}{\sqrt{3}} \right)(x_2 + x_3) \right]
    \sin\!\left( x_1 + \left(1 - \frac{u}{\sqrt{3}} - \frac{v}{\sqrt{3}} \right) x_3 \right) \\
& - \mathrm{Ci}\!\left[ \left(1 + \frac{u}{\sqrt{3}} + \frac{v}{\sqrt{3}} \right)(x_2 + x_3) \right]
    \sin\!\left( x_1 + \left(1 + \frac{u}{\sqrt{3}} + \frac{v}{\sqrt{3}} \right) x_3 \right) \\
& - \cos\!\left( x_1 + \left(1 - \frac{u}{\sqrt{3}} + \frac{v}{\sqrt{3}} \right) x_3 \right)
    \mathrm{Si}\!\left[ \left(1 - \frac{u}{\sqrt{3}} + \frac{v}{\sqrt{3}} \right)(x_2 + x_3) \right] \\
& + \cos\!\left( x_1 + \left(1 + \frac{u}{\sqrt{3}} + \frac{v}{\sqrt{3}} \right) x_3 \right)
    \mathrm{Si}\!\left[ \left(1 + \frac{u}{\sqrt{3}} + \frac{v}{\sqrt{3}} \right)(x_2 + x_3) \right] \\
& + \cos\!\left( x_1 + \left(1 - \frac{u}{\sqrt{3}} - \frac{v}{\sqrt{3}} \right) x_3 \right)
    \mathrm{Si}\!\left[ \left(1 - \frac{u}{\sqrt{3}} - \frac{v}{\sqrt{3}} \right)(x_2 + x_3) \right] \\
& - \cos\!\left( x_1 + \left(1 + \frac{u}{\sqrt{3}} - \frac{v}{\sqrt{3}} \right) x_3 \right)
    \mathrm{Si}\!\left[ \left(1 + \frac{u}{\sqrt{3}} - \frac{v}{\sqrt{3}} \right)(x_2 + x_3) \right]
\Bigg]\Bigg|_{0}^\infty .
\end{aligned}
\end{equation}
The sine and cosine integral functions are defined as
\begin{equation}
\label{eq:CI_SI_app}
\mathrm{Ci}(x)\equiv -\int_x^\infty \frac{\cos t}{t}\,{\rm d}t,
\qquad
\mathrm{Si}(x)\equiv \int_0^x \frac{\sin t}{t}\,{\rm d}t,
\end{equation}
and arise from the standard integrals
\begin{equation}
\label{eq:CI_SI_id_app}
\int_x^{+\infty}\frac{\cos(\beta t)}{t}\,{\rm d}t=\mathrm{Ci}(|\beta|x),
\qquad
\int_0^{x}\frac{\sin(\beta t)}{t}\,{\rm d}t=\mathrm{Si}(\beta x).
\end{equation}
Time averaging over many oscillations removes mixed $\sin\times\cos$ contributions and yields a factor $1/2$ from purely $\sin^2$ and $\cos^2$ terms. The resulting averaged squared kernel can be written compactly as \cite{Domenech:2020ssp}
\begin{equation}
\label{eq:Iosc_avg_sq_app}
\begin{aligned}
\overline{\mathcal{I}_{\mathrm{osc}}(\bar{x}, u, v)}^2=\frac{1}{32}\Bigg( &-\text{Ci}\left[\left(1 + \frac{u}{\sqrt{3}} - \frac{v}{\sqrt{3}}\right) \frac{k}{k_{\rm eva}}\right]
- \text{Ci}\left[\left(1 - \frac{u}{\sqrt{3}} + \frac{v}{\sqrt{3}}\right) \frac{k}{k_{\rm eva}}\right] \\
&+ \text{Ci}\left[\left(1 + \frac{u}{\sqrt{3}} + \frac{v}{\sqrt{3}}\right) \frac{k}{k_{\rm eva}}\right]
+ \text{Ci}\left[\frac{k}{k_{\rm eva}} \left|-1 + \frac{u}{\sqrt{3}} + \frac{v}{\sqrt{3}}\right|\right] \Bigg)^2 \\
+ \frac{1}{32} \Bigg( &+\text{Si}\left[\left(1 - \frac{u}{\sqrt{3}} + \frac{v}{\sqrt{3}}\right) \frac{k}{k_{\rm eva}}\right]
- \text{Si}\left[\left(1 + \frac{u}{\sqrt{3}} + \frac{v}{\sqrt{3}}\right) \frac{k}{k_{\rm eva}}\right] \\
& - \text{Si}\left[\frac{k}{k_{\rm eva}} \left(1 - \frac{u}{\sqrt{3}} - \frac{v}{\sqrt{3}}\right)\right]
+ \text{Si}\left[\frac{k}{k_{\rm eva}} \left(1 + \frac{u}{\sqrt{3}} - \frac{v}{\sqrt{3}}\right)\right] \\
& + \frac{1}{2} \pi \left(-1 + \text{Sign}\left[3 - \sqrt{3} u - \sqrt{3} v\right]\right) \Bigg)^2 .
\end{aligned}
\end{equation}
The absolute value originates from the evenness of $\cos(\beta x)$ in Eq.~\eqref{eq:CI_SI_id_app}.

\subsection{Tensor spectrum}

Plugging Eq.~\eqref{eq:IX_RD_final_app} into the general SIGW expression \eqref{eq:tensorpowerspectrum_general} gives \cite{Domenech:2020ssp,Domenech:2024wao}
\begin{multline}
\label{eq:Ph_RD_uv_app}
\overline{\mathcal{P}_{h}}(k,\overline{x}\gg 1)
\simeq
\frac{c_s^4\, x_{\rm eva}^8}{2048\,\overline{x}^2}
\int_0^\infty {\rm d}v
\int_{|1-v|}^{1+v} {\rm d}u
\left[4v^2 - (1+v^2 - u^2)^2\right]^2
\,\overline{\mathcal{I}_{\rm osc}^2}(\overline{x},u,v)
\\
\times
\mathcal{S}_\Phi^{2}(uk)\,\mathcal{S}_\Phi^{2}(vk)
\sum_{X=\{\Phi,S\}}
T_{X}^{(\rm eMD)\,2}(uk)T_{X}^{(\rm eMD)\,2}(vk)\mathcal{P}_X(uk)\mathcal{P}_X(vk),
\end{multline}
where $\overline{x}=k\overline{\eta}\simeq k\eta$ at late times and $x_{\rm eva}=2k/k_{\rm eva}$ indicates whether a mode re-enters before or after reheating.

Finally, inserting Eq.~\eqref{eq:Ph_RD_uv_app} into Eq.~\eqref{eq:SIGW_basic_formula} yields the Poltergeist contribution at PBH evaporation,
\begin{align}
\label{eq:Omega_GW_0_Reh_app}
\Omega_{\rm GW}^{\rm (RD_2)}(k,\eta_{\rm eva})=\Omega_{\rm GW}^{\rm (RD_2)}\left[\Phi\right]+\Omega_{\rm GW}^{\rm (RD_2)}\left[S\right],
\end{align}
with
\begin{multline}
\label{eq:Omega_GW_reh_X_app}
\Omega_{\rm GW}^{\rm (RD_2)}[X]=
\frac{1}{864}\left(\frac{k}{k_{\rm eva}}\right)^{\!8}\!
\int_0^\infty {\rm d}v
\int_{|1-v|}^{1+v} {\rm d}u
\left[4v^2 - \left(1+v^2 - u^2\right)^2\right]^2
\,\overline{\mathcal{I}_{\rm osc}^2}(\overline{x},u,v)
\\
\times
\mathcal{S}_\Phi^{2}(uk)\,\mathcal{S}_\Phi^{2}(vk)
T_{X}^{(\rm eMD)\,2}(uk)T_{X}^{(\rm eMD)\,2}(vk)\mathcal{P}_X(uk)\mathcal{P}_X(vk),
\qquad X=\{\Phi,S\}.
\end{multline}
The present-day spectrum follows from multiplying Eq.~\eqref{eq:Omega_GW_0_Reh_app} by the redshift factor $\mathcal{D}(T_{\rm eva},T_0)$ defined in Eq.~\eqref{eq:def_mathcalD}.

\subsection{Analytic approximation: resonant contributions}

In the limit $k/k_{\rm eva}\gg 1$, the kernel $\overline{\mathcal{I}_{\mathrm{osc}}^2(\bar{x}, u, v)}$ in Eq.~\eqref{eq:Omega_GW_0_Reh_app} has two main contributions: a resonant one induced by the pole $u+v=\sqrt{3}$ in ${\rm Ci}(x)$ and a low-frequency one along the condition $u\sim v\gg1$
\begin{align}
    \overline{\mathcal{I}^2_{\rm osc,res}}(\overline{x},u,v)  &\simeq \frac{1}{32} {\rm Ci}(|1-c_s(u+v)k/k_{\rm eva}|)^2, \label{eq:I_osc_res}\\
    \overline{\mathcal{I}^2_{\rm osc,LF}}(\overline{x},u,v)  &\simeq {\frac{1}{8}\left(\mathrm{Ci}\left(k/k_{\rm eva}\right)^2+\left(\pi / 2-\mathrm{Si}\left(k/k_{\rm eva}\right)\right)^2\right)} 
    \simeq\frac{1}{8} \frac{k^2_{\rm eva}}{k^2} + {\mathcal{O}\left(\frac{k_{\rm eva}}{k}\right)^{\!3}}.\label{eq:I_osc_LF_2}
\end{align}
In the resonant case, we perform the change of variables $t = u + v - 1$ and $s = u - v$, which introduces a Jacobian factor of $1/2$ and Eq.~\eqref{eq:Omega_GW_reh_X_app} becomes 
\begin{multline}
\label{eq:Omega_GW_0_Reh_st}
    \Omega_{\rm GW}^{\rm (RD_2)}[X]=
\frac{1}{1728}\left(\frac{k}{k_{\rm eva}}\right)^{\!8}\! 
    \int_0^\infty {\rm d}t 
    \int_{-1}^{1} {\rm d}s\ 
    \left[t(t+2)(s^2-1)\right]^2
    \,\overline{\mathcal{I}_{\rm osc}^2}(\overline{x},u,v)
    \\
    \times
    \mathcal{S}_\Phi^{2}(uk)\,\mathcal{S}_\Phi^{2}(vk)
    T_{X}^{(\rm eMD)\,2}(uk)T_{X}^{(\rm eMD)\,2}(vk)\mathcal{P}_X(uk)\mathcal{P}_X(vk),
\end{multline}
with $\overline{\mathcal{I}_{\rm osc}^2}\simeq  \overline{\mathcal{I}^2_{\rm osc,res}}+ \overline{\mathcal{I}^2_{\rm osc,LF}}$.
The resonance occurs at $c_s^{-1} = u + v = t + 1$, where the term 
${\rm Ci}\!\left(|1 - c_s(u+v)k/k_{\rm eva}|\right)^2$ develops a logarithmic 
divergence. In practice, this sharp peak can be regularized and treated, in the 
distributional sense, as a Dirac delta. Near the resonance, one may approximate
\begin{equation}
 {\rm Ci}^2\!\left(|1 - c_s(u+v)k/k_{\rm eva}|\right)
 \;\longrightarrow\;
c_s^{-1}\,\frac{\pi k_{\rm eva}}{k}\,
 \delta(c_s^{-1}-1-t).
\end{equation}
The factor of $\pi$ originates from integrating the squared logarithmic 
divergence, while the prefactor $c_s^{-1}\,k_{\rm eva}/k$ follows from the 
identity $\delta(f(t)) = \delta(t-t_0)/|f'(t_0)|$, evaluated at the resonant 
point $t_0=c_s^{-1}-1$.
 Using this representation, the original double integral in Eq.~\eqref{eq:Omega_GW_0_Reh_st} can be reduced to a one-dimensional one
  \begin{multline}
    \Omega_{\rm GW,res}^{\rm (RD_2)}[X]\simeq
\frac{\pi c_s}{4608}\left(\frac{k}{k_{\rm eva}}\right)^{\!7}
    \int_{-1}^{1} {\rm d}s\ 
    (s^2-1)^2
    \mathcal{S}_\Phi^{2}(uk)\,\mathcal{S}_\Phi^{2}(vk)\\
    \times T_{X}^{(\rm eMD)\,2}(uk)\, T_{X}^{(\rm eMD)\,2}(vk) \mathcal{P}_X(uk)\,\mathcal{P}_X(vk)\,
    \label{eq:Ph_RD_ts_res}
\end{multline}
with $u\simeq (\sqrt{3}+s)/2$ and $v\simeq (\sqrt{3}-s)/2$.
In the low-frequency regime, the region where $u \sim v$ dominates the double integral. Approximating by setting $u = v$ thus reduces the double integral to a single one, providing a suitable simplification with good precision \cite{Domenech:2020ssp}
  \begin{multline}
    \Omega_{\rm GW,LF}^{\rm (RD_2)}[X]\simeq
\frac{1}{13824}\left(\frac{k}{k_{\rm eva}}\right)^{\!10}
\int_{0}^{\infty} {\rm d}t\ t^2(t+2)^2\left[
    \mathcal{S}_\Phi^{4}(uk)T_{X}^{(\rm eMD)\,4}(uk)\mathcal{P}^2_X(uk)\right]_{u=(1+t)/2}\,.
    \label{eq:Ph_RD_ts_LF}
\end{multline}
The Poltergeist component of the GW spectrum in Eq.~\eqref{eq:Omega_GW_0_Reh_st} can be approximated as the superposition of the resonant and low-frequency contributions for both adiabatic and isocurvature modes
\begin{align}
\label{eq:Omega_GW_0_Reh_ana}
\Omega_{\rm GW}^{\rm (RD_2)}(k,\eta_{\rm eva})\simeq \Omega_{\rm GW,res}^{\rm (RD_2)}[\Phi]+ \Omega_{\rm GW,LF}^{\rm (RD_2)}[\Phi]+\Omega_{\rm GW,res}^{\rm (RD_2)}[S]+ \Omega_{\rm GW,LF}^{\rm (RD_2)}[S].
\end{align}
The resonant contributions are enough to describe most of the GW spectrum $\Omega_{\rm GW,res}^{\rm (RD_2)}[X]$ and the low-frequency component  $\Omega_{\rm GW,LF}^{\rm (RD_2)}[X]$ only describe the IR tail.

We now proceed in deriving analytical approximations for Eq.~\eqref{eq:Omega_GW_0_Reh_ana} starting with the adiabatic contribution $X=\Phi$. Following 
Ref.~\cite{Domenech:2024wao}, the logarithmic behaviour of the transfer 
function $T_{\Phi}^{(\rm eMD)}$ in Eq.~\eqref{eq:transfer_adia_iso_main} at 
$k\gg k_{\rm eq}$ can be accurately captured by the power-law fit
\begin{equation}
\label{eq:transfer_adia_PL}
T_{\Phi}^{(\rm eMD)}(k)\simeq 
\begin{cases}
15.6\,(k/k_{\rm eq})^{n(\omega)}, & k \gtrsim k_{\rm eq}\,,\\[4pt]
9/10, & k \lesssim k_{\rm eq}\,,
\end{cases}
\end{equation}
with $n(1/3)\simeq -1.83$.
Using Eqs.~\eqref{eq:transfer_adia_PL}, \eqref{eq:P_Phi_main} and 
\eqref{eq:suppression_factor_final_main} for $T_\Phi^{\rm (eMD)}$, $\mathcal{P}_\Phi$ 
and $\mathcal{S}_\Phi$, respectively, the integral appearing in the resonant contribution 
Eq.~\eqref{eq:Ph_RD_ts_res} can be performed, which gives \footnote{ Contrary to~\cite{Domenech:2024wao}, we find that the peak region is accurately described by the resonant contribution together with the use of transfer functions. In~\cite{Domenech:2024wao}, the LF approximation was applied on the left side of the peak region. However, our results indicate that explicitly separating the resonant and LF domains yields better agreement with numerical computations.}
\begin{equation}
\label{eq:Omega_GW_adia_app}
\Omega_{\rm GW,\rm res}^{\rm (RD_2)}[\Phi]
\simeq 
\left(\tfrac{2}{3}\right)^{\!4}
\begin{cases}
1.2\times 10^5\,\mathcal{A}_{\mathcal{R}}^2 c_s^4\left(\frac{k}{k_{\rm eva}}\right)^{5/3}
\left(\frac{k}{k_{\rm eq}}\right)^{4n(\omega)}
\left(\frac{k}{k_{\rm CMB}}\right)^{2n_s-2}
&
k \gtrsim k_{\rm eq}\,,
\\[10pt]
0.35 \, c_s^4\,\mathcal{A}_{\mathcal{R}}^2 \left(\frac{k}{k_{\rm eva}}\right)^{5/3}
\left(\frac{k}{k_{\rm CMB}}\right)^{2n_s-2},
&
k \lesssim k_{\rm eq}\,,
\end{cases}
\end{equation}
where the numerical coefficients are evaluated for $n_s\simeq 0.96$. As noted 
also in Ref.~\cite{Domenech:2024wao}, a residual $1/2$ mismatch between 
numerical and analytical results persists; the expressions above include the 
appropriate correction.
To model the transition region, we adopt a simple interpolation whose 
coefficient is fitted to numerical results (see 
Fig.~\ref{fig:GWSig_UV} and validated for multiple parameter choices):
\begin{equation}
\Omega_{\rm GW,res}^{\rm (RD_2)}[\Phi]
\simeq 
\mathcal{A}_{\mathcal{R}}^2
\left(\frac{k}{k_{\rm eva}}\right)^{\!5/3}\left(\frac{k}{k_{\rm CMB}}\right)^{\!2 n_s-2}
\frac{
7.7\times 10^{-3}
}{
\bigl[1+ 0.24 \,(k/k_{\rm eq})^{-4n(\omega)/9}\bigr]^9
}.
\end{equation}
which is in good agreement with our numerical calculations.

We now derive the analytical expression for the isocurvature contribution 
($X=S$). Using Eqs.~\eqref{eq:transfer_adia_iso_main}, \eqref{eq:P_S_main} and 
\eqref{eq:suppression_factor_final_main} for $T_S^{\rm (eMD)}$, $\mathcal{P}_S$ 
and $\mathcal{S}_\Phi$, respectively, the integral for the resonant contribution in 
Eq.~\eqref{eq:Ph_RD_ts_res} can be performed analytically. This yields
\begin{equation}
\label{eq:Omega_GW_iso_app}
 \Omega_{\rm GW,res}^{\rm (RD_2)}[S]\simeq
    \begin{cases} 
0.034\,c_s^4 c_w^4
\left(\frac{k_{\rm eq}}{k_{\rm UV}}\right)^6
\left(\frac{k_{\rm eq}}{k_{\rm eva}}\right)^{5/3}
\left(\frac{k_{\rm eq}}{k}\right)^{1/3},
    & k \gtrsim k_{\rm eq},

    \\[10pt]
   1.4\times 10^{-5}\,c_s^4
\left(\frac{k}{k_{\rm \mathsmaller{PBH}}}\right)^6
\left(\frac{k}{k_{\rm eva}}\right)^{5/3},
    & k \lesssim k_{\rm eq},
    \end{cases}
\end{equation}
with $C \simeq 1$ (see Eq.~\eqref{eq:TS_leading}),which agrees well with the numerical results. The same factor-of-two 
correction identified in the adiabatic case has been applied.
In order to recover the transition regime, we propose the following expression\footnote{Note that the spectral slopes for $k \lesssim k_{\rm eq}$ in Eq.~\eqref{eq:Omega_GW_iso} can be recovered by setting $n_s = 4$ in Eq.~\eqref{eq:Omega_GW_adia}. Similarly, the prefactor in Eq.~\eqref{eq:Omega_GW_iso} arises from the general power-law expression by taking the limits $n_s \to 4$, $\mathcal{A}_{\mathcal{R}} \to 2/3\pi$ and $\mathcal{T}= 9/10 \to 1/5$.}
\begin{equation}
\Omega_{\rm GW,res}^{\rm (RD_2)}[S]\simeq
\left(\frac{k}{k_{\rm \mathsmaller{PBH}}}\right)^6
\left(\frac{k}{k_{\rm eva}}\right)^{5/3}
\,
\frac{
1.5\times 10^{-6}
}{
\left[
1+0.38\,\left(k/k_{\rm eq}\right)
\right]^8
}.
\end{equation}
We cross-checked the above formula for different parameter choices, and found equally close agreement between numerical and analytical results, see Fig.~\ref{fig:GWSig_UV}.
In contrast to the monochromatic case, the isocurvature-induced gravitational wave signal features a distinct but relatively flat peak for $k \gtrsim k_{\rm eq}$ due to the considerably stronger suppression factor. This makes the resulting spectrum considerably less sensitive to the ultraviolet cutoff scale $k_{\rm \mathsmaller{UV}}$.

\subsection{Validity of the low-frequency approximation}
Unlike the resonant contribution, the low-frequency component cannot be computed fully analytically in the case of extended PBH mass functions. To see this, we split Eq.~\eqref{eq:Ph_RD_ts_LF} into two integration domains,
\begin{equation}
\label{eq:ana_IRtal}
\Omega_{\rm GW,LF}^{\rm (RD_2)}[X]\propto
\int_{v_{\rm IR}}^{v_{\rm eq}}{\rm d}v\,v^{-10/3+2n_s+4n}
+
\int_{v_{\rm eq}}^{v_{\rm \mathsmaller{UV}}}{\rm d}v\,v^{-10/3+2n_s+4n}.
\end{equation}
For adiabatic perturbations, the exponent of the integrand is negative for both transfer-function branches, implying that both integrals are dominated by their lower boundary. However, this region lies outside the regime of validity of the LF approximation ($v\gg1$), rendering an analytic estimate unreliable. Consequently, the infrared tail must be evaluated numerically.

For isocurvature perturbations with spectral index $n_s=4$, the situation is partially improved. In the range $v\in[0,v_{\rm eq}]$, the integral is dominated by its upper boundary and remains within the LF regime, while for $v\in[v_{\rm eq},v_{\rm \mathsmaller{UV}}]$ it is dominated by the lower boundary and falls outside it. Nevertheless, as in the monochromatic case~\cite{Domenech:2024wao}, the latter contribution is subdominant.

In light of these considerations, we rely on numerical computations to determine the full low-frequency tail. Our results show that this contribution is subdominant compared to the resonant part, and it will therefore be neglected in the remainder of this work.

\clearpage

\section{More details on SIGWs during early matter domination}
\label{app:eMD}

In addition to GW production from scalar oscillations after reheating (the Poltergeist mechanism), the constant scalar potential $\Phi$ during eMD (or slowly decaying during reheating) also acts as a source (cf.~Eq.~\ref{eq:source_term}). Hence, we refer to this GW contribution as eMD induced, and denotes its spectrum by $\Omega_{\rm GW}^{\rm (eMD)}$. First, we review the case of instantaneous reheating, and then continue with the extended reheating phase induced by the evaporation of PBHs with the Choptuik mass distribution.

\subsection{Pure MD}

An early matter-dominated (eMD) phase, followed by instantaneous reheating at $x_{\rm eva}\equiv k\eta_{\rm eva}$, generates a characteristic SIGW signal, which can be derived from the general expression of $\overline{\mathcal{P}_h(k,\eta)}$ in Eq.~\eqref{eq:tensorpowerspectrum_general}. During matter domination, the Newtonian potential remains constant ($\Phi'=0$), which significantly simplifies the source term $f_X$ defined in Eq.~\eqref{eq:source_term}.
Deep in the matter era, $x_{\rm eq}\ll x \ll x_{\rm eva}$, the source term asymptotes to\footnote{Our expression, \(f_\Phi^{\rm (eMD)}\), is larger than the one of Ref.~\cite{Kohri:2018awv} by a factor $9/8$, see footnote~\ref{footnote:convention_KT}. We have verified that both conventions lead to consistent final results.}
\begin{align}
\label{eq:f_X_eMD}
f_X^{\rm (eMD)}(x,v,u,x_{\rm eva}\gg x)
=
\left(1+\frac{2}{3(1+\omega)}\right)\, T_X^{\rm (eMD)}(0)^2
=\frac{5}{3}\,\alpha_X
\end{align}
where 
\begin{align}
\label{eq:alpha_X_def}
\alpha_X\equiv T_X^{\rm (eMD)}(0)^2= 
\begin{cases}
\left(\tfrac{9}{10}\right)^2, & X=\Phi,\\[4pt]
\left(\tfrac{1}{5}\right)^2, & X=S.
\end{cases}
\end{align}
The transfer function $T_X^{\rm (eMD)}$, defined in Eq.~\eqref{eq:transfer_adia_iso_main}, is used in the limit relevant for modes that enter the horizon after the onset of the eMD era, namely $k\ll k_{\rm eq}$. The Green function deep inside the eMD ($x_{\rm eq}\ll x\ll x_{\rm eva}$) is given by Eq.~\eqref{eq:hgreen} for $\omega=0$
\begin{equation}
G_h(x,\bar{x})
=
\frac{\bar{x}}{x^{3}}
\left[
(x\bar{x}+1)\sin(x-\bar{x})-(x-\bar{x})\cos(x-\bar{x})
\right]\Theta(x-\bar{x}) \, .
\end{equation}
We obtain the contribution to the kernel function in Eq.~\eqref{eq:kernel_standard} corresponding to GW production and evaluation both occurring during the MD era
\begin{align}
\label{eq:I_pure_eMD_app}
I_X^{\rm (pure~MD)}(x,u,v) &=\frac{5}{3x^3}\,\alpha_X \int_0^x \mathrm{d} \overline{x} \, \overline{x}
\left[
(x\bar{x}+1)\sin(x-\bar{x})-(x-\bar{x})\cos(x-\bar{x})
\right] \notag\\
&= \frac{5\alpha_X}{3}\frac{x^3 + 3x\cos x - 3\sin x }{x^3}.
\end{align}
At late times, the kernel approaches a constant,
\begin{equation}
\label{eq:I_pure_MD}
I_X^{\rm (pure~MD)}(x,u,v)
\xrightarrow[1\ll x\ll x_{\rm eva}]{} \frac{5\alpha_X}{3}.
\end{equation}
Accordingly, its time-averaged square becomes
\begin{equation}
\label{eq:I_pure_MD_square}
\overline{\big(I_X^{\rm (pure~MD)}(x,u,v)\big)^2}
\xrightarrow[1\ll x\ll x_{\rm eva}]{}
\frac{1}{2}\cdot\frac{25}{9}\alpha_X^2.
\end{equation}
Following Ref.~\cite{Kohri:2018awv}, we have included the factor $1/2$ in Eq.~\eqref{eq:I_pure_MD_square} to account for the fact that, when GWs are both generated and observed during a matter-dominated era, only the gradient term contributes to the GW energy density, $\rho_{\rm GW}\sim h'h' + \partial_i h \partial_i h$. Indeed, in this regime the tensor perturbation is constant, $h^{\rm (pure~MD)}\sim I_X^{\rm (pure~MD)}$, as follows from Eq.~\eqref{eq:I_pure_MD}, so that $h'=0$. By contrast, during radiation domination and for subhorizon modes, $k\gg \mathcal{H}$, one has $h'\simeq kh$, and the kinetic and gradient contributions are equal.\footnote{
The physical interpretation of this constant mode has long been debated~\cite{Ali:2020sfw,Inomata:2019yww,Sipp:2022kmb,Domenech:2025ccu}. Ref.~\cite{Domenech:2025ccu} clarifies this issue by focusing on what GW detectors actually measure, namely a time-dependent strain. Since a constant strain produces no time variation, it yields no observable signal, whether in interferometers, pulsar timing arrays, or any other detector. In this sense, the constant mode is not a propagating GW and does not contribute to the observable GW energy density. Therefore, the quantities $\Omega_{\rm GW}^{({\rm pure~MD})}$ in Eqs.~\eqref{eq:Omega_GW_pure_MD_Phi} and \eqref{eq:Omega_GW_pure_MD_S}, as well as the corresponding results of Ref.~\cite{Kohri:2018awv}, are dominated by an unobservable constant contribution and hence substantially overestimate the physical GW background. The observable signal is instead controlled by the first time-dependent correction in Eq.~\eqref{eq:I_pure_eMD}, implying the replacement
\begin{equation}
\label{eq:correct_Omega_GW_eMD}
\Omega_{\rm GW}^{({\rm pure~MD})}
~\longrightarrow~
\frac{9}{x^4}\Omega_{\rm GW}^{({\rm pure~MD})},
\end{equation}
in Eqs.~\eqref{eq:Omega_GW_pure_MD_Phi} and \eqref{eq:Omega_GW_pure_MD_S} rather than the simple factor-of-$2$ suppression encoded in Eq.~\eqref{eq:I_pure_MD_square}.
This conclusion applies when GWs are both produced and observed during MD. The situation is different if the GWs are generated during MD but observed only after the Universe has re-entered a RD phase. In that case, the constant mode inherited from the matter era starts oscillating (cf. Eqs.~\eqref{eq:C_coeff} and \eqref{eq:D_coeff}) and contributes to the kinetic term in $\rho_{\rm GW}\sim h'h'+\partial_i h\partial_i h$, with $h'\simeq kh$. As a result, the factor $1/2$ in Eq.~\eqref{eq:I_pure_MD_square} does not affect the final expressions, since it cancels out in  $\Omega_{\rm GW}^{\rm eMD}(\eta>\eta_{\rm eva})$ in Eqs.~\eqref{eq:Omega_GW_MD_inst_app}, \eqref{eq:Omega_GW_MD_PBH_app}, and \eqref{eq:Omega_GW_MD_grad_app}. For this reason, we keep this factor only to facilitate comparison with Ref.~\cite{Kohri:2018awv}: although its interpretation is incorrect for $\Omega_{\rm GW}^{({\rm pure~MD})}$, it leads to the correct final result for $\Omega_{\rm GW}^{\rm eMD}(\eta>\eta_{\rm eva})$, as discussed in Sec.~\ref{app:eMD_to_RD_transition}.
}

The GW power spectrum is given by the convolution integral in Eq.~\eqref{eq:tensorpowerspectrum_general_s_t},
\begin{equation}
\label{eq:tensorpowerspectrum_general_s_t_app}
\Omega_{\rm GW}(k,\eta)
\equiv
\frac{k^2}{12\mathcal H^2}\,\overline{\mathcal P_h(k,\eta)}
=
\frac{k^2}{3\mathcal H^2}\int \mathrm{d}t\,\mathrm{d}s\,
\mathcal{K}(s,t)
\sum_{X=\{\Phi,S\}}
\overline{I_X^2}\,
\mathcal{P}_X(ku)\mathcal{P}_X(kv),
\end{equation}
where
\begin{equation}
\mathcal{K}(s,t)
=
\left[
\frac{t(2+t)(1-s^2)}{(1+t)^2-s^2}
\right]^2.
\end{equation}
We decompose
\begin{equation}
\label{eq:Omega_GW_MD_app_app}
\Omega^{\rm (pure~MD)}_{\rm GW}
=
\Omega_{\rm GW}^{({\rm pure~MD})}[\Phi]
+
\Omega_{\rm GW}^{\rm (pure~MD)}[S].
\end{equation}

\subsubsection*{Adiabatic contribution}
Using Eq.~\eqref{eq:P_Phi_main} and \(\overline{I_\Phi^2}=729/800\) in Eq.~\eqref{eq:I_pure_MD_square}, the GW energy density from adiabatic modes during MD, in Eq.~\eqref{eq:Omega_GW_MD_app}, reads
\begin{equation}
\Omega_{\rm GW}^{({\rm pure~MD})}[\Phi]
=
\frac{k^2}{12\mathcal H^2}\,
\frac{36}{50}\,\mathcal{A}_{\mathcal R}^2
\left(\frac{k}{k_{\rm CMB}}\right)^{2n_s-2}
\mathcal{J}_\Phi(n_s),
\end{equation}
with
\begin{equation}
\mathcal{J}_\Phi(n_s)
=
\int_{\rm domain} \mathrm{d}t\,\mathrm{d}s\,
\mathcal{K}(s,t)
\left(\frac{1+t+s}{2}\right)^{n_s-1}
\left(\frac{1+t-s}{2}\right)^{n_s-1}.
\end{equation}
The integration domain is determined by the cutoff
\(\Theta(k_{\rm \mathsmaller{UV}}-ku)\Theta(k_{\rm \mathsmaller{UV}}-kv)\), i.e.
\begin{equation}
|s|\le \frac{2}{\tilde{k}}-1-t,
\qquad
t \le \frac{2}{\tilde{k}}-1,
\qquad
\tilde{k}\equiv \frac{k}{k_{\rm \mathsmaller{UV}}},
\end{equation}
together with \(|s|\le 1\), which yields
\begin{equation}
\label{eq:domain}
\int_{\rm domain} =
\begin{cases}
\displaystyle
\int_0^{2/\tilde{k}-2}\!\!\mathrm{d}t\!\int_{-1}^{1}\!\mathrm{d}s
+
\int_{2/\tilde{k}-2}^{2/\tilde{k}-1}\!\!\mathrm{d}t
\!\int_{-(2/\tilde{k}-1-t)}^{2/\tilde{k}-1-t}\!\mathrm{d}s,
& 0<\tilde{k}\le 1, \\[1em]
\displaystyle
\int_0^{2/\tilde{k}-1}\!\!\mathrm{d}t
\!\int_{-(2/\tilde{k}-1-t)}^{2/\tilde{k}-1-t}\!\mathrm{d}s,
& 1<\tilde{k}\le 2, \\[1em]
0, & \tilde{k}>2.
\end{cases}
\end{equation}
For \(n_s=1\), one finds
\begin{equation}
\mathcal{J}_\Phi(1)=
\begin{cases}
\displaystyle
\frac{32}{15}\,\tilde{k}^{-1}-3+\frac{32}{35}\,\tilde{k}+\frac{1}{8}\,\tilde{k}^2,
& 0<\tilde{k}\le 1, \\[1em]
\displaystyle
\frac{1}{840}\left(1-\frac{2}{\tilde{k}}\right)^4
\left(105\tilde{k}^2+72\tilde{k}+16-\frac{32}{\tilde{k}}-\frac{16}{\tilde{k}^2}\right),
& 1<\tilde{k}\le 2, \\[1em]
0, & \tilde{k}>2.
\end{cases}
\end{equation}
Thus,
\begin{equation}
\label{eq:Omega_GW_pure_MD_Phi_app}
\Omega_{\rm GW}^{({\rm pure~MD})}[\Phi]
=
\frac{3}{50}\,\mathcal{A}_{\mathcal R}^2
\left(\frac{k}{\mathcal H}\right)^2
\mathcal{J}_\Phi(1),
\end{equation}
and in the infrared limit \(k\ll k_{\rm \mathsmaller{UV}}\),
\begin{equation}
\Omega_{\rm GW}^{({\rm pure~MD})}[\Phi]
\simeq
\frac{16}{125}\,\mathcal{A}_{\mathcal R}^2
\left(\frac{k}{\mathcal H}\right)^2
\frac{k_{\rm \mathsmaller{UV}}}{k}~\propto~a.
\end{equation}

\subsubsection*{Isocurvature contribution}
Using Eq.~\eqref{eq:P_S_main} and \(\overline{I_S^2}=1/450\) in Eq.~\eqref{eq:I_pure_MD_square}, the GW energy density from isocurvature modes during MD, in Eq.~\eqref{eq:Omega_GW_MD_app}, reads
\begin{equation}
\Omega_{\rm GW}^{({\rm pure~MD})}[S]
=
\frac{k^2}{12\mathcal H^2}\,
\frac{1}{16200\pi^2}
\left(\frac{k}{k_{\rm \mathsmaller{PBH}}}\right)^6
\int_{\rm domain}\mathrm{d}t\,\mathrm{d}s\,
t^2(2+t)^2(1-s^2)^2
\big[(1+t)^2-s^2\big],
\end{equation}
with the same domain as in Eq.~\eqref{eq:domain}.
Performing the integral for \(0<\tilde{k}\le 2\),
\begin{equation}
\label{eq:Omega_GW_pure_MD_S_app}
\Omega_{\rm GW}^{({\rm pure~MD})}[S]
=
\frac{1}{20412000\pi^2}
\left(\frac{k}{\mathcal H}\right)^2\left(\frac{k_{\rm \mathsmaller{UV}}}{k_{\rm \mathsmaller{PBH}}}\right)^6
\frac{(2-\tilde{k})^4}{\tilde{k}}
\left(128+116\tilde{k}+40\tilde{k}^2+5\tilde{k}^3\right),
\end{equation}
and \(\Omega_{\rm GW}^{(S)}=0\) for \(\tilde{k}>2\), with $\tilde{k}= k/k_{\rm \mathsmaller{UV}}$. In the infrared limit \(k\ll k_{\rm \mathsmaller{UV}}\), we get
\begin{equation}
\Omega_{\rm GW}^{({\rm pure~MD})}[S]
\simeq
\frac{64}{637875\pi^2}
\left(\frac{k}{\mathcal H}\right)^2
\frac{k_{\rm \mathsmaller{UV}}}{k}~\propto~a.
\end{equation}

\subsection{eMD to RD transition}
\label{app:eMD_to_RD_transition}

\subsubsection*{Gravitational-wave kernels}

During the early matter-dominated (eMD) era followed by the second
radiation-dominated phase ($\mathrm{RD}_2$), i.e.\ for $\eta > \eta_{\rm eq}$,
the Newtonian potential takes the form (cf.\ Eq.~\eqref{eq:Phi_Polt_main})
\begin{equation}
\label{eq:Phi_eMD_app}
\Phi(k,\eta)
= \sum_{X \in \{\Phi,\,S\}}
  T_X(k,\eta)\, X(k,0)\,
  \Theta\!\left(k_{\rm \mathsmaller{UV}}^{\mathsmaller{(X)}}(\eta) - k\right),
\end{equation}
where the transfer function $T_X$ is defined piecewise as
\begin{equation}
\label{eq:transfer_fc_reh_eMD_app}
T_X(k,\eta) =
\begin{cases}
T_X^{\rm (eMD)}(k)\,F_\Phi(\eta),
  & \eta \leq \eta_{\rm osc}(k), \\[4pt]
T_X^{\rm (eMD)}(k)\,\mathcal{S}_\Phi(k)\,T_\Phi^{\rm (RD_2)}(k,\eta),
  & \eta > \eta_{\rm osc}(k).
\end{cases}
\end{equation}
Here $x_{\rm osc} \equiv k\eta_{\rm osc}$ denotes the dimensionless conformal
time at which the Newtonian potential $\Phi$ first begins to oscillate
(cf.\ Eq.~\eqref{eq:dec_limit_first_derivative_main}). In practice,
$x_{\rm osc} \sim x_{\rm eva}$, with $x_{\rm osc} > x_{\rm eva}$, where
$x_{\rm eva} \equiv k\eta_{\rm eva}$ is the reheating time defined by the
condition $\rho_{\rm PBH} = \rho_{\rm rad}$ under the assumption of
instantaneous reheating (see Eq.~\eqref{eq:scale_factor_full}).
For observation times $x \equiv k\eta > x_{\rm osc}$, the GW kernel
defined in Eq.~\eqref{eq:kernel_standard} naturally splits into
contributions from sources active before and after the onset of
oscillations,
\begin{equation}
\label{eq:kernel_split_appendix_short}
I_{X}^{\rm (eMD \to RD_2)}(x,u,v)
= I^{\rm (eMD)}_{X}(x,u,v)
+ I^{\rm (RD_2)}_{X}(x,u,v),
\end{equation}
where
\begin{align}
\label{eq:I_X_eMD}
I^{\rm (eMD)}_{X}(x,u,v)
&= \int_0^{x_{\rm osc}} \mathrm{d}\bar{x}\;
   G^{\rm (eMD \to RD_2)}_h(x,\bar{x})\,
   f_X^{\rm (eMD)}(\bar{x},u,v), \\[4pt]
\label{eq:I_X_RD2}
I^{\rm (RD_2)}_{X}(x,u,v)
&= \int_{x_{\rm osc}}^{x} \mathrm{d}\bar{x}\;
   G^{\rm (RD_2)}_h(x,\bar{x})\,
   f_X^{\rm (RD_2)}(\bar{x},u,v).
\end{align}
In these expressions, $G^{\rm (eMD \to RD_2)}_h(x,\bar{x})$, given in
Eq.~\eqref{eq:G_mixed_Bessel}, is the mixed Green's function describing
GWs sourced during the eMD phase ($\bar{x} < x_{\rm eva}$) and subsequently
propagating through the $\mathrm{RD}_2$ era ($x > x_{\rm eva}$).
The function $G^{\rm (RD_2)}_h(x,\bar{x})$, given in Eq.~\eqref{eq:G_RD2},
is the standard radiation-era Green's function evaluated with the
shifted argument $\bar{x} - x_{\rm eva}/2$, appropriate to the
post-reheating scale factor $a \propto \eta - \eta_{\rm eva}/2$
(see footnote~\ref{footnote:etabar}).
The functions $f_X^{\rm (eMD)}$ and $f_X^{\rm (RD_2)}$ in
Eqs.~\eqref{eq:I_X_eMD}--\eqref{eq:I_X_RD2} denote the GW source
terms during the eMD and $\mathrm{RD}_2$ eras, respectively\footnote{In~\cite{Inomata:2019zqy}, the Green functions were piecewise matched at $x_{\rm eva}$, which leads to a spurious oscillatory pattern in the GW kernel $I_X^{\rm eMD}$ as shown in~ \cite{Pearce:2023kxp, Pearce:2025ywc}. A more refined approach includes a numeric transition region for the Green function, which, however, does not change the amplitude or the shape of the signal qualitatively but rather averages over oscillations. We find, that matching piecewise at $x_{\rm osc}$ gives very similar results (see also Fig.~\ref{fig:kernelsComp} for comparison) and allows for analytical approximations. Hence, we use adopt our matching procedure for all further computations.}

The $\mathrm{RD}_2$ contribution $I^{\rm (RD_2)}_{X}$ in
Eq.~\eqref{eq:I_X_RD2} coincides with the Poltergeist contribution
computed in Sec.~\ref{chap:poltergeist} and is not recomputed here.
Instead, we focus on the eMD contribution $I^{\rm (eMD)}_{X}$ in
Eq.~\eqref{eq:I_X_eMD}, which arises from GW production during the eMD
phase. Substituting the first branch of Eq.~\eqref{eq:transfer_fc_reh_eMD}
into Eq.~\eqref{eq:source_f_X} with $\omega = 0$ yields
\begin{equation}
\label{eq:f_X_eMD_evap2_app}
f_X^{\rm (eMD)}(\bar{x},u,v)
= \alpha_X
  \left[
    F_\Phi^2
    + \frac{2}{3}
      \!\left(F_\Phi + \frac{F_\Phi'}{\mathcal{H}}\right)^{\!2}
  \right],
\qquad
\alpha_X \equiv
\begin{cases}
\left(\tfrac{9}{10}\right)^2, & X = \Phi, \\[4pt]
\left(\tfrac{1}{5}\right)^2,  & X = S,
\end{cases}
\end{equation}
where the coefficients $\alpha_X$ are the $k \ll k_{\rm eq}$ limits of
the transfer functions in Eq.~\eqref{eq:transfer_adia_iso_main}.
Introducing the dimensionless time $z \equiv \eta/\eta_{\rm eva}$ and
using $\mathcal{H} = 2/\eta$ during matter domination, one finds
\begin{equation}
\frac{F_\Phi'}{\mathcal{H}} = \frac{z}{2}\,\frac{\mathrm{d}F_\Phi}{\mathrm{d}z}.
\end{equation}
It is then convenient to define the source weight function
\begin{equation}
\label{eq:Q_def}
\mathcal{Q}(z)
\equiv F_\Phi^2(z)
+ \frac{2}{3}
  \!\left[F_\Phi(z) + \frac{z}{2}\,\frac{\mathrm{d}F_\Phi}{\mathrm{d}z}
  \right]^{\!2},
\end{equation}
which encodes the full reheating history through $F_\Phi$.
Substituting the mixed Green's function of Eq.~\eqref{eq:G_mixed_Bessel}
into the eMD contribution of Eq.~\eqref{eq:I_X_eMD}, and changing
variables to
\begin{equation}
z \equiv \frac{\bar{x}}{x_{\rm eva}}, \qquad
z_{\rm osc} \equiv \frac{x_{\rm osc}}{x_{\rm eva}},
\end{equation}
yields the master expression
\begin{equation}
\label{eq:IeMD_master}
I_{X}^{\rm (eMD)}(x,u,v;\,x_{\rm eva})
= \frac{\alpha_X\, x_{\rm eva}^2}{2x/x_{\rm eva} - 1}
  \left[
    C(x,x_{\rm eva})\,\mathcal{J}_j(x_{\rm eva})
    + D(x,x_{\rm eva})\,\mathcal{J}_y(x_{\rm eva})
  \right],
\end{equation}
where the oscillatory coefficients are
\begin{align}
\label{eq:C_coeff_app}
C(x,x_{\rm eva})
&= \frac{
    \sin x
    - 2x_{\rm eva}(\cos x + x_{\rm eva}\sin x)
    + \sin(x - 2x_{\rm eva})
   }{2x_{\rm eva}^2}, \\[4pt]
\label{eq:D_coeff_app}
D(x,x_{\rm eva})
&= \frac{
    (2x_{\rm eva}^2 - 1)\cos x
    - 2x_{\rm eva}\sin x
    + \cos(x - 2x_{\rm eva})
   }{2x_{\rm eva}^2},
\end{align}
and the source integrals, which depend on the reheating model only
through $\mathcal{Q}$, are
\begin{equation}
\label{eq:calJ_y_j}
\mathcal{J}_j(x_{\rm eva})
\equiv \int_0^{z_{\rm osc}} \mathrm{d}z\; z^3\,\mathcal{Q}(z)\,j_1(x_{\rm eva}\,z),
\qquad
\mathcal{J}_y(x_{\rm eva})
\equiv \int_0^{z_{\rm osc}} \mathrm{d}z\; z^3\,\mathcal{Q}(z)\,y_1(x_{\rm eva}\,z).
\end{equation}
All information about the reheating history is thus encoded in
$\mathcal{Q}(z)$, while the GW propagation history enters only through
the oscillatory coefficients $C$ and $D$. 

\paragraph{Instantaneous reheating.}

In the sudden-transition limit, the envelope function is simply
$F_\Phi(\eta) = \Theta(\eta_{\rm eva} - \eta)$, i.e.\ $F_\Phi(z) = 1$
for $0 \leq z < 1$, with the oscillation onset coinciding with the
evaporation time, $z_{\rm osc} = 1$.
In this case, the source weight function in Eq.~\eqref{eq:Q_def}
reduces to the constant
\begin{equation}
\mathcal{Q}_{\rm inst}(z) = 1 + \frac{2}{3} = \frac{5}{3}.
\end{equation}
The source integrals $\mathcal{J}_{j,y}$ are then analytically
tractable and evaluate to
\begin{align}
\mathcal{J}_j^{\rm inst}(x_{\rm eva})
&= \frac{5}{3}\,
   \frac{(3 - x_{\rm eva}^2)\sin x_{\rm eva} - 3x_{\rm eva}\cos x_{\rm eva}}
        {x_{\rm eva}^4}, \\[4pt]
\mathcal{J}_y^{\rm inst}(x_{\rm eva})
&= \frac{5}{3}\,
   \frac{(x_{\rm eva}^2 - 3)\cos x_{\rm eva} - 3x_{\rm eva}\sin x_{\rm eva} + 3}
        {x_{\rm eva}^4}.
\end{align}
Substituting these into Eq.~\eqref{eq:IeMD_master} yields the
closed-form eMD kernel in the instantaneous-reheating limit,
\begin{multline}
\label{eq:I_eMD_inst}
I^{\rm (eMD, inst.)}_{X}
= \frac{5\alpha_X}{3\,x_{\rm eva}^3\,(2x - x_{\rm eva})}
  \Big[
    x_{\rm eva}^4\cos(x - x_{\rm eva})
    + 2x_{\rm eva}^3\sin(x - x_{\rm eva}) \\
    + 3x_{\rm eva}^2\cos x
    - 3x_{\rm eva}\sin x
    + 3\sin x_{\rm eva}\sin(x - x_{\rm eva})
  \Big].
\end{multline}
As a consistency check, setting $x = x_{\rm eva}$ recovers the pure
matter-dominated result of Eq.~\eqref{eq:I_pure_eMD}, corresponding to
the case where both GW production and observation occur within the MD
era, as expected.
In the limit $x_{\rm eva} \ll 1$, expanding the trigonometric functions
in powers of $x_{\rm eva}$ shows that the bracket in
Eq.~\eqref{eq:I_eMD_inst} vanishes to order $x_{\rm eva}^4$, with the
leading surviving contribution at order $x_{\rm eva}^5$ giving
\begin{equation}
\label{eq:kernel_inst_small}
I^{\rm (eMD, inst.)}_{X}(x,u,v;\,x_{\rm eva})
\simeq \frac{\alpha_X}{3}\,\frac{\sin x}{x}\,x_{\rm eva}^2,
\qquad x_{\rm eva} \ll 1.
\end{equation}
In the opposite limit $x_{\rm eva} \gg 1$, the leading terms in the
bracket are of order $x_{\rm eva}^4$, and one obtains
\begin{equation}
\label{eq:kernel_inst_large}
I^{\rm (eMD, inst.)}_{X}(x,u,v;\,x_{\rm eva})
\simeq \frac{5\alpha_X}{6x}\,x_{\rm eva}\cos(x - x_{\rm eva}),
\qquad x \gg x_{\rm eva} \gg 1.
\end{equation}
Because the kernel functions $I^{\rm (eMD,\,inst.)}_{X}$ in Eq.~\eqref{eq:kernel_inst_small} and $I_X^{\rm (pure~MD)}$ in Eq.~\eqref{eq:I_pure_eMD} are independent of $u$ and $v$, they can be factored out of the integral over $u$ and $v$ (equivalently, over $s$ and $t$) in Eq.~\eqref{eq:tensorpowerspectrum_general_s_t_app}. As a result, the SIGW spectrum $\Omega^{\rm (eMD,\,inst.)}_{\rm GW}$ sourced during an eMD era terminated by instantaneous reheating, and observed after reheating, takes the factorized form
\begin{equation}
\label{eq:Omega_GW_MD_inst_app}
\Omega^{\rm (eMD,\,inst.)}_{\rm GW}
=
R^{\rm (eMD,\,inst.)}\times \Omega_{\rm GW}^{\rm (pure~MD)},
\end{equation}
where $\Omega_{\rm GW}^{\rm (pure~MD)}$ denotes the SIGW spectrum sourced and observed during a pure MD era, given by Eqs.~\eqref{eq:Omega_GW_MD_app}, \eqref{eq:Omega_GW_pure_MD_Phi}, and \eqref{eq:Omega_GW_pure_MD_S}. The prefactor $R^{\rm (eMD,\,inst.)}$ is the ratio of the averaged squared kernels,
\begin{equation}
    R^{\rm (eMD,\,inst.)}\,\equiv\,
    \overline{\bigl(I^{\rm (eMD,\,inst.)}_{X}\bigr)^2}/\overline{\bigl(I_X^{\rm (pure~MD)}\bigr)^2},
\end{equation}
with $I^{\rm (eMD,\,inst.)}_{X}$ and $I_X^{\rm (pure~MD)}$ defined in Eqs.~\eqref{eq:kernel_inst_small}, \eqref{eq:kernel_inst_large} and \eqref{eq:I_pure_eMD}, respectively.

\begin{figure}
    \centering
    \includegraphics[width=0.47\linewidth]{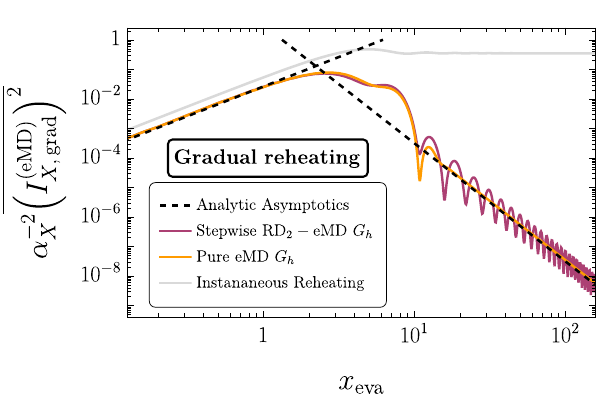}\hspace{2mm}\includegraphics[width=0.47\linewidth]{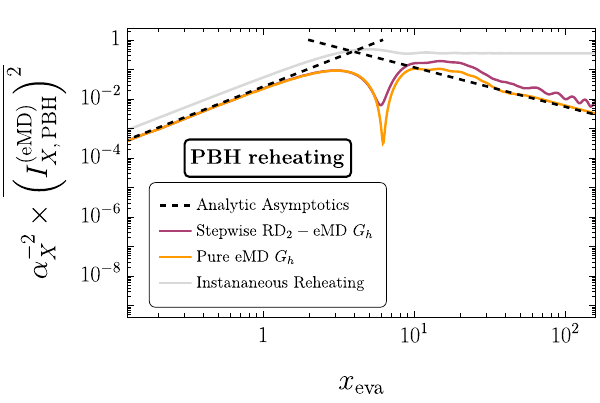}
    \caption{Oscillation avaraged squared GW kernels $\overline{I^2}$. Left: Kernels for a gradual reheating phase. Right: Kernels for a PBH reheating phase. We compare the results obtained using the stepwise definition of tensor Green functions $G_h$ during the $\rm eMD$ and $\rm RD_2$ matched at reheating, with the results obtained using only $G_h$ from the eMD. We also present our analytically derived asymptotics for the IR and UV tails, respectively. For comparison, we also show the oscillation-averaged kernel for instantaneous reheating.}
    \label{fig:kernelsComp}
\end{figure}

\paragraph{Gradual reheating.}

We now consider a smooth transition from eMD to $\mathrm{RD}_2$
driven by perturbative particle decay with rate $\Gamma \sim H$.
The envelope of the Newtonian potential in this case is~\cite{Inomata:2019zqy}
\begin{equation}
\label{eq:F_Phi_particle_decay}
F_\Phi(\eta)
= \exp\!\left(-\int^\eta \mathrm{d}\bar\eta\;a(\bar\eta)\Gamma\right)
=
\begin{cases}
\exp\!\left[-\dfrac{2}{3}\!\left(\dfrac{\eta}{\eta_{\rm eva}}\right)^{\!3}\right],
  & \eta < \eta_{\rm eva}, \\[10pt]
\exp\!\left[-2\!\left(\left(\dfrac{\eta}{\eta_{\rm eva}}\right)^{\!2}
  - \dfrac{\eta}{\eta_{\rm eva}} + \dfrac{1}{3}\right)\right],
  & \eta \geq \eta_{\rm eva},
\end{cases}
\end{equation}
which is approximately independent of $u$ and $v$ prior to the onset
of oscillations. We retain the notation $\eta_{\rm eva}$ for consistency
with the rest of the paper, although Eq.~\eqref{eq:F_Phi_particle_decay}
pertains to reheating by particle decay rather than PBH evaporation.
For analytical convenience, we extend the $\eta < \eta_{\rm eva}$ branch
of Eq.~\eqref{eq:F_Phi_particle_decay} to all times, which gives
\begin{equation}
F_\Phi(z) = e^{-2z^3/3}, \qquad
\frac{\mathrm{d}F_\Phi}{\mathrm{d}z} = -2z^2 F_\Phi, \qquad
\frac{F_\Phi'}{\mathcal{H}} = -z^3 F_\Phi.
\end{equation}
The corresponding source weight function is
\begin{equation}
\mathcal{Q}_{\rm grad}(z)
= e^{-4z^3/3}
  \left[1 + \frac{2}{3}(1 - z^3)^2\right].
\end{equation}
Taking $z_{\rm osc} \to +\infty$ in Eq.~\eqref{eq:calJ_y_j}, the
asymptotic behaviour of the source integrals is
\begin{align}
\mathcal{J}_j(x_{\rm eva})
&\simeq
\begin{cases}
\displaystyle
\frac{5}{54}\!\left(\frac{3}{4}\right)^{\!2/3}\!
\Gamma\!\left(\tfrac{2}{3}\right) x_{\rm eva}, & x_{\rm eva} \ll 1, \\[8pt]
\displaystyle -\frac{512}{x_{\rm eva}^7}, & x_{\rm eva} \gg 1,
\end{cases}
\\[6pt]
\mathcal{J}_y(x_{\rm eva})
&\simeq
\begin{cases}
\displaystyle
-\frac{17}{36}\!\left(\frac{3}{4}\right)^{\!2/3}\!
\Gamma\!\left(\tfrac{2}{3}\right)\frac{1}{x_{\rm eva}^2}, & x_{\rm eva} \ll 1, \\[8pt]
\displaystyle \frac{5}{x_{\rm eva}^4}, & x_{\rm eva} \gg 1,
\end{cases}
\end{align}
up to higher-order corrections.
Substituting these expressions, together with the large-$x$ asymptotics
of $C$ and $D$, into Eq.~\eqref{eq:IeMD_master}, the eMD kernel
simplifies to
\begin{equation}
\label{eq:IeMD_grad_small}
I^{\rm (eMD, grad.)}_{X}(x , u, v)
\simeq \frac{11}{54}\!\left(\frac{3}{4}\right)^{\!2/3}\!
\Gamma\!\left(\frac{2}{3}\right)
\frac{\alpha_X\, x_{\rm eva}^2\, \sin x}{x},
\qquad x_{\rm eva} \ll 1,
\end{equation}
and
\begin{equation}
\label{eq:IeMD_grad_large}
I^{\rm (eMD, grad.)}_{X}(x , u, v)
= \frac{5\alpha_X\cos x}{2x\, x_{\rm eva}},
\qquad x \gg x_{\rm eva} \gg 1.
\end{equation}
As shown in Fig.~\ref{fig:kernelsComp}-left, the analytical formulas in Eqs.~\eqref{eq:IeMD_grad_small}, \eqref{eq:IeMD_grad_large} show very good agreement with the numerical calculation, obtained from integrating Eq.~\eqref{eq:IeMD_master}.
Analogously to Eq.~\eqref{eq:Omega_GW_MD_inst_app}, the SIGW spectrum $\Omega^{\rm (eMD,\,grad.)}_{\rm GW}$ sourced during an eMD era terminated by gradual reheating (controlled by a constant decay rate), and observed after reheating, takes the factorized form
\begin{equation}
\label{eq:Omega_GW_MD_grad_app}
\Omega^{\rm (eMD,\,grad.)}_{\rm GW}
=
R^{\rm (eMD,\,grad.)}\times\Omega_{\rm GW}^{\rm (pure~MD)},
\end{equation}
where $\Omega_{\rm GW}^{\rm (pure~MD)}$ denotes the SIGW spectrum sourced and observed during a pure MD era, given by Eqs.~\eqref{eq:Omega_GW_MD_app}, \eqref{eq:Omega_GW_pure_MD_Phi}, and \eqref{eq:Omega_GW_pure_MD_S}. The prefactor $R^{\rm (eMD,\,grad.)}$ is the ratio of the averaged squared kernels,
\begin{equation}
    R^{\rm (eMD,\,grad.)}\,\equiv\,
  \overline{\bigl(I^{\rm (eMD,\,grad.)}_{X}\bigr)^2}/\overline{\bigl(I_X^{\rm (pure~MD)}\bigr)^2},
\end{equation}
with $I^{\rm (eMD,\,grad.)}_{X}$ and $I_X^{\rm (pure~MD)}$ defined in Eqs.~\eqref{eq:IeMD_grad_small}, \eqref{eq:IeMD_grad_large}, and \eqref{eq:I_pure_eMD}, respectively.

\paragraph{PBH reheating.}
We consider reheating driven by the evaporation of a PBH population
with an extended mass function obeying Choptuik scaling and a sharp UV cutoff.
We model the envelope of the Newtonian potential by the fit
\begin{equation}
\label{eq:Phi_fit_app_clean}
F_\Phi(\eta)\simeq
\begin{cases}
\exp\!\left[-0.84\left(\dfrac{\eta}{\eta_{\rm eva}}\right)^{3.43}\right],
& \eta<\eta_{\rm eva},
\\[6pt]
3.5\left(\dfrac{\eta_{\rm cut}-\eta}{\eta_{\rm cut}}\right)^{4/3},
& \eta_{\rm eva}\le \eta\le \eta_{\rm cut},
\end{cases}
\end{equation}
where $\eta_{\rm eva}$ denotes the evaporation time averaged over the PBH mass distribution.
Introducing the dimensionless variable $z\equiv \eta/\eta_{\rm eva}$ and
$z_{\rm cut}\equiv \eta_{\rm cut}/\eta_{\rm eva}\simeq 1.26$,
the source weight entering Eq.~\eqref{eq:Q_def} is defined by
\begin{equation}
\label{eq:Q_def_clean}
\mathcal Q(z)
\equiv
F_\Phi^2(z)
+\frac{2}{3}
\left[
F_\Phi(z)+\frac{z}{2}\frac{{\rm d}F_\Phi}{{\rm d}z}
\right]^2,
\end{equation}
and satisfies $\mathcal Q(z_{\rm cut})=0$.
The source integrals are then
\begin{equation}
\label{eq:J_def_clean}
\mathcal J_j(x_{\rm eva})
=
\int_0^{z_{\rm cut}}{\rm d}z\,z^3\mathcal Q(z)\,j_1(x_{\rm eva}z),
\qquad
\mathcal J_y(x_{\rm eva})
=
\int_0^{z_{\rm cut}}{\rm d}z\,z^3\mathcal Q(z)\,y_1(x_{\rm eva}z).
\end{equation}
In the infrared regime, $x_{\rm eva}\ll 1$, one finds
\begin{equation}
\label{eq:J_small_clean}
\mathcal J_j(x_{\rm eva})
\simeq
\frac{M_4}{3}\,x_{\rm eva},
\qquad
\mathcal J_y(x_{\rm eva})
\simeq
-\frac{M_1}{x_{\rm eva}^2},
\end{equation}
with truncated moments
\begin{equation}
M_n\equiv \int_0^{z_{\rm cut}}{\rm d}z\,z^n\mathcal Q(z).
\end{equation}
For $z_{\rm cut}=1.26$, we obtain numerically $M_1\simeq 0.533$ and $M_4\simeq 0.265$, yielding
\begin{equation}
\label{eq:kernel_small_clean}
I^{\rm (eMD,\,PBH)}_{X}(x,u,v)
\simeq
0.222\,\alpha_X\,
\frac{\sin x}{x}\,x_{\rm eva}^2,
\qquad x_{\rm eva}\ll 1.
\end{equation}
The infrared scaling is therefore universal, $I_X^{\rm (eMD)}\propto (\sin x/x)\,x_{\rm eva}^2$. 
In the ultraviolet regime \(x\gg x_{\rm eva}\gg 1\), the integral is dominated by the endpoint
\(z=z_{\rm cut}\), where \(\mathcal Q(z_{\rm cut})=0\). Expanding around the cusp,
\begin{equation}
\mathcal Q(z)\simeq2(z_{\rm cut}-z)^{2/3}z_{\rm cut}^2,
\end{equation}
and the kernel reduces to
\begin{equation}
\mathcal{J}_{j,y}^{\rm PBH}(x_{\rm eva}\gg1)\simeq1.96\frac{z_{\rm cut}^4}{x_{\rm eva}^{8/3}}
\int_0^{x_{\rm eva}z_{\rm cut}} {\rm d}u\,u^{2/3}\begin{cases}
    \sin\big(u-z_{\rm cut}x_{\rm eva}\big),\\
    \cos\big(u-z_{\rm cut}x_{\rm eva}\big),
\end{cases}
\end{equation}
with \(u=x_{\rm eva}(z_{\rm cut}-z)\). Setting the upper integration boundary to infinity and using the regularized limit
\begin{equation}
\lim_{\epsilon\rightarrow\infty}\int_0^\infty {\rm d}u\,u^{2/3} e^{-iu-\epsilon}
= e^{-i5\pi/6}\ \Gamma(5/3),
\end{equation}
one obtains the phase shift \(+5\pi/6\) and a numerical prefactor
\(1.96z_{\rm cut}^4\,\Gamma(5/3)\simeq 2.23\). Therefore,
\begin{equation}
\label{eq:kernel_large_clean}
I^{\rm (eMD, PBH)}_{X}(x,u,v)
\simeq
2.23\,\alpha_X\,
\frac{x_{\rm eva}^{1/3}}{x}
\sin\!\left(x-z_{\rm cut}\,x_{\rm eva}+\frac{5\pi}{6}\right),
\qquad x\gg x_{\rm eva}\gg 1.
\end{equation}
The ultraviolet scaling is thus $I_X^{\rm (eMD,\,PBH)}\propto x_{\rm eva}^{1/3}$, which lies between the instantaneous-reheating scaling $I_X^{\rm (eMD,\,inst.)}\propto x_{\rm eva}$ (Eq.~\eqref{eq:kernel_inst_large}) and the gradual-reheating scaling $I_X^{\rm (eMD,\,grad.)}\propto x_{\rm eva}^{-1}$ (Eq.~\eqref{eq:IeMD_grad_large}). This hierarchy can be traced to the behaviour of the source near the upper endpoint of the time integral. In the instantaneous case, the source is discontinuous at $z=1$, producing a hard boundary term linear in $x_{\rm eva}$. In the gradual case, the exponential decay of $F_\Phi$ effectively removes any endpoint contribution, so the kernel is dominated by the smooth bulk and is suppressed as $x_{\rm eva}^{-1}$. The PBH case is intermediate: the source vanishes continuously at $z=z_{\rm cut}$ through the cusp $\mathcal{Q}(z)\propto (z_{\rm cut}-z)^{2/3}$, which softens the step-like endpoint of the instantaneous case without erasing it as in the gradual case, resulting in the fractional scaling $x_{\rm eva}^{1/3}$. As shown in the right panel of Fig.~\ref{fig:kernelsComp}, the analytical formulas in Eqs.~\eqref{eq:kernel_small_clean} and \eqref{eq:kernel_large_clean} are in very good agreement with the numerical result obtained by integrating Eq.~\eqref{eq:IeMD_master}.

Analogously to Eq.~\eqref{eq:Omega_GW_MD_inst_app}, the SIGW spectrum $\Omega^{\rm (eMD,\,PBH)}_{\rm GW}$ sourced during a PBH-dominated era and observed after PBH evaporation takes the factorized form
\begin{equation}
\label{eq:Omega_GW_MD_PBH_app}
\Omega^{\rm (eMD,\,PBH)}_{\rm GW}
=
R^{\rm (eMD,\,PBH)}\times\Omega_{\rm GW}^{\rm (pure~MD)},
\end{equation}
where $\Omega_{\rm GW}^{\rm (pure~MD)}$ denotes the SIGW spectrum sourced and observed during a pure MD era, given by Eqs.~\eqref{eq:Omega_GW_MD_app}, \eqref{eq:Omega_GW_pure_MD_Phi}, and \eqref{eq:Omega_GW_pure_MD_S}. The prefactor $R^{\rm (eMD,\,PBH)}$ is the ratio of the averaged squared kernels,
\begin{equation}
    R^{\rm (eMD,\,PBH)}\,\equiv\,
\overline{\bigl(I^{\rm (eMD,\,PBH)}_{X}\bigr)^2}/\overline{\bigl(I_X^{\rm (pure~MD)}\bigr)^2},
\end{equation}
with $I^{\rm (eMD,\,PBH)}_{X}$ and $I_X^{\rm (pure~MD)}$ defined in Eqs.~\eqref{eq:kernel_small_clean}, \eqref{eq:kernel_large_clean}, and \eqref{eq:I_pure_eMD}, respectively.

In the main text, we restrict attention to reheating driven by PBH evaporation. To streamline the notation, we therefore drop the explicit ``PBH'' label and use
\begin{equation}
I_X^{\rm (eMD,\,PBH)} \;\to\; I_X^{\rm (eMD)},
\qquad
\Omega_{\rm GW}^{\rm (eMD,\,PBH)} \;\to\; \Omega_{\rm GW}^{\rm (eMD)}.
\end{equation}
Unless stated otherwise, all eMD quantities appearing in the main text thus refer to the PBH-reheating scenario discussed above.

\subsection{Time-dependent Cutoff}
\label{sec:Time-dependent_Cutoff}
In this section, we examine the time dependence of the UV cutoff introduced 
in Eq.~\eqref{eq:Phi_Polt_main}. During eMD, structures grow linearly with 
the scale factor, $\delta_{\rm \mathsmaller{PBH}}\sim a$, so the nonlinear 
scale $k_{\rm \mathsmaller{NL}}^{\mathsmaller{(X)}}(\eta)$ is itself 
time-dependent: at earlier times, structures have had less time to grow, 
and more modes remain in the linear regime (see also 
Eq.~\eqref{eq:cutoffs_final}). We parametrize this running as
\begin{equation}
\label{eq:runnning_kNL}
    k_{\rm \mathsmaller{NL}}^{\mathsmaller{(X)}}(\eta)=k_{\rm \mathsmaller{NL}}^{\mathsmaller{(X)}}(\eta_{\rm eva})\left(\frac{x_{\rm eva}}{x}\right)^{\beta},
\end{equation}
where $
k_{\rm \mathsmaller{NL}}^{\mathsmaller{(X)}}(\eta_{\rm eva})$ is defined in 
Eq.~\eqref{eq:Poisson_NL}, and $\beta>0$ is a source-dependent exponent. 
Because of this time dependence, the UV cutoff cannot be factored out of the 
time integral in the kernel, and must be retained inside the integration.
To handle this, we rewrite the product of cutoff step functions appearing 
inside the double integral of Eq.~\eqref{eq:tensorpowerspectrum_general} as
\begin{equation}
    \Theta\!\left( k_{\rm \mathsmaller{NL}}^{\mathsmaller{(X)}}(\eta)-u k\right)
    \Theta\!\left( k_{\rm \mathsmaller{NL}}^{\mathsmaller{(X)}}(\eta)-v k\right) 
    =\Theta\!\left(x^{\mathsmaller{(X)}}_{\rm cut}-x\right),
\end{equation}
where the critical conformal time $x^{\mathsmaller{(X)}}_{\rm cut}$ is 
determined by the more stringent of the two constraints on $u$ and $v$,
\begin{equation}
    x^{\mathsmaller{(X)}}_{\rm cut}= x_{\rm eva}\times 
    \min\left[\left(\frac{k_{\rm \mathsmaller{NL}}^{\mathsmaller{(X)}}(\eta_{\rm eva})}{u k}\right)^{1/\beta},
    \left(\frac{k_{\rm \mathsmaller{NL}}^{\mathsmaller{(X)}}(\eta_{\rm eva})}{v k}\right)^{1/\beta}\right].
\end{equation}
The quantity \(x_{\rm cut}^{(X)}\) is the latest conformal time at which both scalar legs \(uk\) and \(vk\) sourcing a GW mode \(k\) still lie below the nonlinear scale \(k_{\rm NL}^{(X)}(\eta)\). It therefore marks the end of GW production by that momentum triangle in linear theory: for \(x<x_{\rm cut}^{(X)}\) the source is active, while for \(x\ge x_{\rm cut}^{(X)}\) only the subsequent tensor propagation remains. Retaining this time dependence in the kernel 
of Eq.~\eqref{eq:kernel_standard} yields
\begin{equation}
 I_X^{\rm (eMD)}(x, u, v,x^{\mathsmaller{(X)}}_{\rm cut}) =
\begin{cases}
\displaystyle
\frac{1}{x^3}\Big[
3x\cos x - 3\sin x
\\[6pt]
\quad
+\big(3x^{\mathsmaller{(X)}}_{\rm cut} - 3x + x\,\left(x^{\mathsmaller{(X)}}_{\rm cut}\right)^2\big)\cos(x-x^{\mathsmaller{(X)}}_{\rm cut}) &x \geq x^{\mathsmaller{(X)}}_{\rm cut}
\\[6pt]
\quad
+\big(3 + 3x\,x^{\mathsmaller{(X)}}_{\rm cut} - \left(x^{\mathsmaller{(X)}}_{\rm cut}\right)^2\big)\sin(x-x^{\mathsmaller{(X)}}_{\rm cut})
\Big],
\\[12pt]
\displaystyle
1 + \frac{3\cos x}{x^2} - \frac{3\sin x}{x^3},
& x < x^{\mathsmaller{(X)}}_{\rm cut}.
\end{cases}
\end{equation}
When GWs are sourced during eMD but evaluated during ${\rm RD}_2$, the kernel 
takes different forms depending on whether the cutoff has already been 
reached. For $x \geq x^{\mathsmaller{(X)}}_{\rm cut}$,
\begin{align}
I_X^{\rm (eMD\to RD_2)}(x, u, v,x^{\mathsmaller{(X)}}_{\rm cut}) =\;
&\frac{1}{2(2x-x_{\rm eva})\,x_{\rm eva}^3}
\Big[
\big((2x_{\rm eva}^2-1)\cos x
+\cos(x-2x_{\rm eva})
-2x_{\rm eva}\sin x
\big)\nonumber\\
&\times
\big(x^{\mathsmaller{(X)}}_{\rm cut}\cos x^{\mathsmaller{(X)}}_{\rm cut}-2\sin x^{\mathsmaller{(X)}}_{\rm cut}\big)\nonumber\\
+\;&\big(2-2\cos x^{\mathsmaller{(X)}}_{\rm cut} - x^{\mathsmaller{(X)}}_{\rm cut}\sin x^{\mathsmaller{(X)}}_{\rm cut}\big)\nonumber\\
&\times
\big({-}2x_{\rm eva}\cos x
+(1-2x_{\rm eva}^2)\sin x
+\sin(x-2x_{\rm eva})
\big)\Big],
\end{align}
and for $x < x^{\mathsmaller{(X)}}_{\rm cut}$, the source has not yet been 
cut off and the kernel reduces to
\begin{align}
I_X^{\rm (eMD\to RD_2)}(x, u, v,x^{\mathsmaller{(X)}}_{\rm cut}) =\;
&\frac{1}{2(2x-x_{\rm eva})\,x_{\rm eva}^3}
\Big[
-x+4x_{\rm eva}+2x\,x_{\rm eva}^2
-4x_{\rm eva}\cos x\nonumber\\
&+x\cos(2x-2x_{\rm eva})
+2(1-2x_{\rm eva}^2)\sin x\nonumber\\
&+2\sin(x-2x_{\rm eva})
-2\sin(2x-2x_{\rm eva})
\Big].
\end{align}
We find that, to a very good approximation after averaging over oscillations,
\begin{equation}
\overline{I_X(x,u,v,x^{\mathsmaller{(X)}}_{\rm cut})^2}~\simeq~
\Theta\!\left(x^{\mathsmaller{(X)}}_{\rm cut}-x\right)
\overline{I_X(x,u,v,x^{\mathsmaller{(X)}}_{\rm cut}\to +\infty)^2},
\end{equation}
which holds exactly for $x < x^{\mathsmaller{(X)}}_{\rm cut}$. This result 
has a transparent physical interpretation: for an observer at conformal time 
$x$, only modes with triangle legs satisfying $\ell k \leq 
k_{\rm \mathsmaller{NL}}^{\mathsmaller{(X)}}(\eta)$ contribute, and the 
effective UV cutoff in $\ell \in \{u, v\}$ is
\begin{equation}
\ell_{\rm \mathsmaller{UV}}^{\mathsmaller{(X)}}(x)=\frac{k_{\rm \mathsmaller{NL}}^{\mathsmaller{(X)}}(\eta_{\rm eva})}{k}\left(\frac{x_{\rm eva}}{x}\right)^{\beta}.
\end{equation}
For an observer deep inside the matter era, $x < x_{\rm eva}$, this cutoff 
exceeds $k_{\rm \mathsmaller{NL}}^{\mathsmaller{(X)}}(\eta_{\rm eva})/k$, consistently with 
the fact that modes which only become nonlinear after $x$ must still be 
included. At reheating, $x = x_{\rm eva}$, the running cutoff reduces to 
the fixed value $k_{\rm \mathsmaller{NL}}^{\mathsmaller{(X)}}$, defined in 
Eq.~\eqref{eq:Poisson_NL}, which is 
precisely the relevant limit for a finite eMD terminating at $x_{\rm eva}$, 
as reflected by the upper integration boundary in 
Eq.~\eqref{eq:kernel_split_appendix_short}.
\clearpage

\section{Non-linearities during eMD}
\label{app:NL_eMD}

In this appendix we discuss the limitations of the perturbative SIGW
calculation during an eMD era. In eMD, density perturbations
grow with the scale factor and can become non-linear before reheating. The
perturbative prediction must therefore be supplemented by a physical cutoff,
which we take to be the scale at which the density contrast reaches order
unity. We compare this prescription with the fully non-linear simulations of
Ref.~\cite{Fernandez:2023ddy}, which provide the best available benchmark for
the adiabatic case.

\subsection{Adiabatic sources}

The simulations show that, once perturbations collapse, the GW signal is no
longer described by the standard second-order SIGW source alone. Two effects
become relevant:
\begin{enumerate}
    \item \emph{Non-linear collapse of matter overdensities.}
    During eMD, density perturbations grow with the scale factor and can reach
    order unity before reheating. The subsequent collapse, shell crossing and
    halo formation source GWs through genuinely non-linear dynamics.

    \item \emph{Inhomogeneous reheating.}
    The eMD--to--RD transition  is spatially
    modulated by the non-linear density field. This provides an additional GW source, although simulations indicate that it is subdominant compared to halo formation in the non-linear regime.
\end{enumerate}
For adiabatic initial conditions, the resulting non-perturbative spectrum is
well fitted by~\cite{Fernandez:2023ddy}
\begin{equation}
\Omega_{\mathrm{GW}}(k)
\simeq
0.05\,A_{\mathrm{s}}^{7/4}
\left(\frac{k}{\mathcal{H}_{\mathrm{eq}}}\right)^{3/2},
\end{equation}
over the finite range
\begin{equation}
k_{\rm low}
\lesssim
k
\lesssim
k_{\rm cross},
\qquad
k_{\rm low}
\simeq
15\,\mathcal{H}_{\mathrm{eq}},
\qquad
k_{\rm cross}
\simeq
\frac{14\,\mathcal{H}_{\mathrm{eq}}}{A_{\mathrm{s}}^{1/4}} \simeq 8\,k_{\rm \mathsmaller{NL}},\quad k_{\rm \mathsmaller{NL}}\simeq \frac{1.7\,\mathcal{H}_{\rm eq}}{A_s^{1/4}}.
\end{equation}
The lower scale is set by the latest structures that become non-linear before
reheating, while the upper scale is controlled by the light-crossing time of the
largest virialized halos.

Fig.~\ref{fig:eMDGrad} compares the perturbative spectrum with the lattice fit. For gradual reheating, truncating the perturbative integral at $k_{\rm \mathsmaller{NL}}$ reproduces the infrared part of the simulated signal and connects smoothly to the non-perturbative branch, supporting our fiducial choice. Extending it to the microscopic UV scale instead overestimates $\Omega_{\rm GW}$ by including modes already in the non-linear regime.

\subsection{Isocurvature sources}

For isocurvature sources, fully non-linear simulations are not yet available.
We nevertheless apply the same criterion and cut off the perturbative integral
when the total density contrast becomes order unity. The corresponding
non-linear scale depends on the PBH mass and is largest for the lightest
allowed PBHs. The maximal spectra obtained in this way are shown in
Fig.~\ref{fig:EMD_Max_RD_NL}.

Throughout this work we therefore include only the perturbative eMD signal with
a non-linear cutoff. Additional GWs from halo formation may be present, as
suggested by the adiabatic simulations, but their amplitude is at most
marginally within the reach of upcoming experiments and no corresponding
calculation exists for PBH-isocurvature sources. We leave a dedicated treatment
of these non-perturbative contributions to future work.

\begin{figure}[t!]
    \centering
    \includegraphics[width=0.65\linewidth]{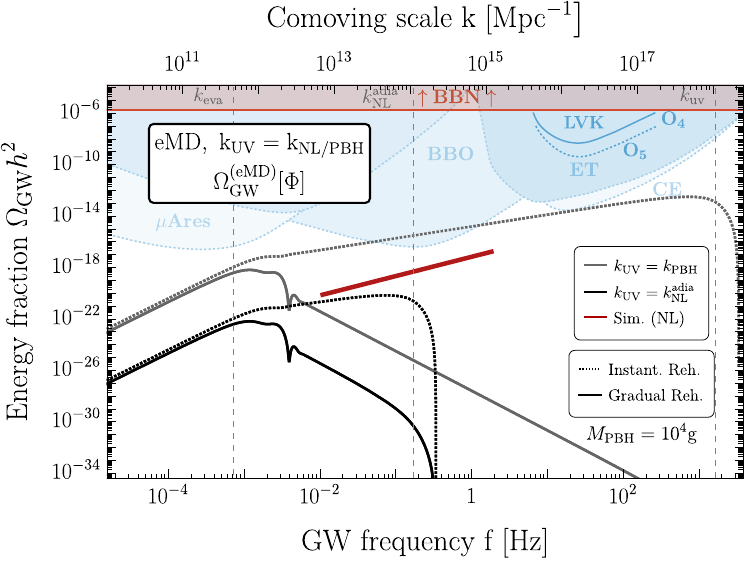}
    \caption{SIGWs produced during eMD for sudden and gradual reheating. The
    comparison with non-linear simulations favors the perturbative result with
    a non-linear cutoff, while extending the integral to the UV scale gives an
    unreliably large signal. We therefore use the non-linear cutoff for the eMD
    contribution in the parameter scans.}
    \label{fig:eMDGrad}
\end{figure}

\begin{figure}[h!]
    \centering
    \includegraphics[width=0.65\linewidth]{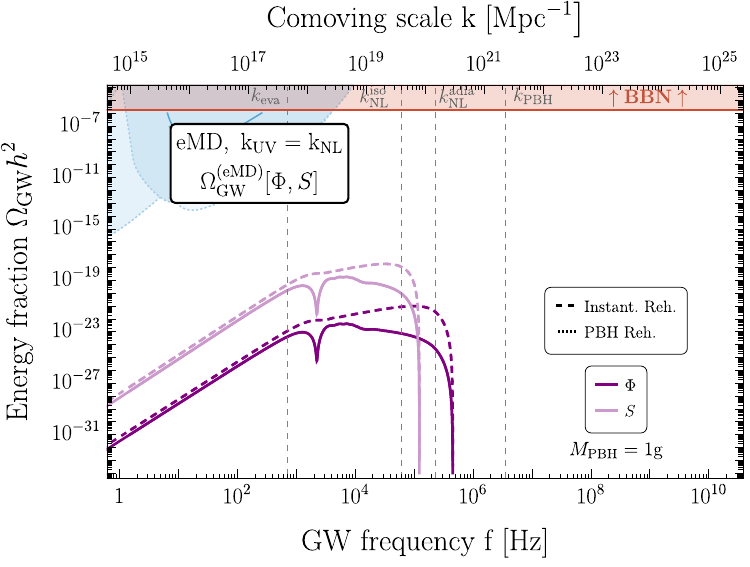}
    \caption{Maximal perturbative eMD signal from Choptuik-smeared PBHs using
    the non-linear cutoff. For comparison, the reheating times in the sudden and
    gradual scenarios are matched to $\eta_{\langle\rm eva\rangle}$ of PBH
    reheating. Monochromatic PBHs are approximated by sudden reheating (see
    Sec.~\ref{chap:eMD}).}
    \label{fig:EMD_Max_RD_NL}
\end{figure}

\clearpage
\section{SIGWs from monochromatic PBHs}
\label{app:SIGW_mono_PBH}

In this appendix we recover the SIGW spectrum for a monochromatic PBH mass
distribution. This limit provides a useful benchmark for the numerical
calculation and makes transparent why the Choptuik-smeared case differs
parametrically from the monochromatic result. Following
Ref.~\cite{Domenech:2024wao}, the finite duration of evaporation suppresses the
Newtonian potential after reheating according to~\cite{Inomata:2020lmk}
\begin{equation}
    \mathcal{S}_{\Phi, \mathrm{eva}}^{\rm mono}(k)
    =
    \frac{\Phi_{\mathrm{RD}}}{\Phi_{\mathrm{RD}}^{\rm instant}}
    \simeq
    \left(\sqrt{\frac{2}{3}}\frac{k}{k_{\mathrm{eva}}}\right)^{-1/3}.
\end{equation}
Since this suppression is weak, the post-evaporation radiation-era source gives
rise to the large monochromatic Poltergeist enhancement. We show the representative benchmark spectra in Fig.~\ref{fig:GWSigMono_longversion}.

\subsection{Adiabatic contribution}
Using the power-law approximation for the adiabatic transfer function,
Eq.~\eqref{eq:transfer_adia_PL}, the adiabatic contribution can be separated
into a resonant branch, an intermediate branch, and an infrared branch:
\begin{align}
\Omega_{\mathrm{GW}, \mathrm{res}}^{\mathrm{adia}}
&=
\mathcal{A}_{\mathcal{R}}^2 A_{\Phi}(w)^4
\frac{\pi}{2^{11}}
\left(\frac{2}{3}\right)^{1/3}
\frac{\left(c_s^2-1\right)^2}
{\left(2 c_s\right)^{1+2 n_{\mathrm{eff}}}}
\left(\frac{k_{\rm \mathsmaller{PBH}}}{k_{\mathrm{eq}}}\right)^{4 n(w)}
\left(\frac{k_{\rm \mathsmaller{PBH}}}{k_{\mathrm{eva}}}\right)^{17/3}
\left(\frac{k}{k_{\rm \mathsmaller{PBH}}}\right)^{2n_{\rm eff}+7}
\nonumber\\
&\quad\times
\left[
\frac{2}{5}s_0^5\,{}_2 F_1
\left(\frac{5}{2},-n_{\mathrm{eff}};\frac{7}{2};c_s^2 s_0^2\right)
-\frac{4}{3}s_0^3\,{}_2 F_1
\left(\frac{3}{2},-n_{\mathrm{eff}};\frac{5}{2};c_s^2 s_0^2\right)
\right.
\nonumber\\
&\hspace{4.2cm}
\left.
+2s_0\,{}_2 F_1
\left(\frac{1}{2},-n_{\mathrm{eff}};\frac{3}{2};c_s^2 s_0^2\right)
\right],
\\[0.5em]
\Omega_{\mathrm{GW}, \mathrm{mid}}^{\mathrm{adia}}(k)
&=
\frac{
\mathcal{A}_{\mathcal{R}}^2 A_{\Phi}(w)^4 c_s^4
}{
768\left(-1-2 n_{\mathrm{eff}}\right)
}
\left(\frac{3}{2}\right)^{2/3}
\xi_1^{1+2 n_{\mathrm{eff}}}
\left(\frac{k_{\mathrm{eq}}}{k_{\rm \mathsmaller{PBH}}}\right)^{2 n_s-7/3}
\left(\frac{k_{\rm \mathsmaller{PBH}}}{k_{\mathrm{eva}}}\right)^{14/3}
\left(\frac{k}{k_{\rm \mathsmaller{PBH}}}\right)^5 ,
\\[0.5em]
\Omega_{\mathrm{GW}, \mathrm{IR}}^{\mathrm{adia}}(k)
&=
\frac{
\mathcal{A}_{\mathcal{R}}^2 c_s^4(3 w+5)^4
}{
10^4\left(6 n_s+5\right)(w+1)^4
}
\left(\frac{3}{2}\right)^{2/3}
\left(\frac{\xi_2 k_{\mathrm{eq}}}{k_{\rm \mathsmaller{PBH}}}\right)^{5/3+2 n_s}
\left(\frac{k_{\rm \mathsmaller{PBH}}}{k_{\mathrm{eva}}}\right)^{14/3}
\left(\frac{k}{k_{\rm \mathsmaller{PBH}}}\right),
\end{align}
where
\begin{equation}
    n_{\mathrm{eff}}=-\frac{5}{3}+n_s+2n(w).
\end{equation}
The resonant branch is obtained from the pole contribution of the radiation-era
kernel, Eq.~\eqref{eq:I_osc_res}, and controls the peak region for
$k\lesssim k_{\rm eq}$. We reproduce the result of
Ref.~\cite{Domenech:2024wao}, up to minor differences in the scaling of some
branches, and generalize it to a tilted primordial spectrum by keeping the
explicit dependence on the CMB pivot scale. For $c_s=1/\sqrt{3}$ the resonant
contribution can be written as
\begin{align}
\Omega_{\mathrm{GW}, \mathrm{res}}^{\mathrm{adia}}
&=
\mathcal{A}_{\mathcal{R}}^2 A_{\Phi}(w)^4
\frac{\pi}{2^{11}}
\left(\frac{2}{3}\right)^{1/3}
\frac{\left(c_s^2-1\right)^2}
{\left(2 c_s\right)^{1+2 n_{\mathrm{eff}}}}
\left(\frac{k}{k_{\mathrm{eva}}}\right)^{17/3}
\left(\frac{k}{k_{\rm eq}}\right)^{4n(\omega)}
\left(\frac{k}{k_{\rm CMB}}\right)^{2n_s-2}
\nonumber\\
&\quad\times
\left[
\frac{2}{5}s_0^5\,{}_2 F_1
\left(\frac{5}{2},-n_{\mathrm{eff}};\frac{7}{2};c_s^2 s_0^2\right)
-\frac{4}{3}s_0^3\,{}_2 F_1
\left(\frac{3}{2},-n_{\mathrm{eff}};\frac{5}{2};c_s^2 s_0^2\right)
\right.
\nonumber\\
&\hspace{4.2cm}
\left.
+2s_0\,{}_2 F_1
\left(\frac{1}{2},-n_{\mathrm{eff}};\frac{3}{2};c_s^2 s_0^2\right)
\right].
\end{align}
The infrared branch follows from the low-frequency kernel of
Eq.~\eqref{eq:Ph_RD_ts_LF}. Its normalization depends on the matching parameter
$\xi_2$, which specifies the transition between the two branches of the transfer
function. For $\omega=1/3$ we find numerically
\begin{equation}
    \xi_2\simeq 0.65 ,
\end{equation}
for both the UV and non-linear cutoffs. This weak cutoff dependence reflects the
fact that the integral is dominated by modes near equality rather than by the
far ultraviolet tail. Including a general tilt gives
\begin{equation}
    \Omega_{\mathrm{GW}, \mathrm{IR}}^{\mathrm{adia}}(k)
    =
    \frac{
    \mathcal{A}_{\mathcal{R}}^2 c_s^4(3 w+5)^4
    }{
    10^4\left(6 n_s+5\right)(w+1)^4
    }
    \left(\frac{3}{2}\right)^{2/3}
    \xi_2^{5/3+2 n_s}
    \left(\frac{k_{\mathrm{eq}}}{k_{\mathrm{eva}}}\right)^{14/3}
    \left(\frac{k_{\rm CMB}}{k_{\mathrm{eq}}}\right)^{2-2n_s}
    \left(\frac{k}{k_{\mathrm{eq}}}\right),
\end{equation}
which is independent of the ultraviolet cutoff.

The intermediate branch requires more care. Combining the low-frequency kernel
with the power-law branch of the transfer function gives an integral over
$v\in[\xi_1 v_{\rm eq},v_{\rm cut}]$. Since the integrand falls with $v$, the
result is controlled by the lower endpoint and is again insensitive to the
ultraviolet cutoff:
\begin{equation}
    \Omega_{\mathrm{GW}, \mathrm{mid}}^{\mathrm{adia}}(k)
    =
    \frac{
    \mathcal{A}_{\mathcal{R}}^2 A_{\Phi}(w)^4 c_s^4
    }{
    768\left(-1-2 n_{\mathrm{eff}}\right)
    }
    \left(\frac{3}{2}\right)^{2/3}
    \xi_1^{1+3n_{\mathrm{eff}}}
    \left(\frac{k_{\mathrm{eq}}}{k_{\mathrm{eva}}}\right)^{14/3}
    \left(\frac{k_{\rm CMB}}{k_{\mathrm{eq}}}\right)^{2-2n_s}
    \left(\frac{k}{k_{\mathrm{eq}}}\right)^5 .
\end{equation}
This endpoint dominance is physically sensible. However, the derivation
formally extends the low-frequency approximation into the region
$k>k_{\rm eq}$, or equivalently $v>1$, where the expansion is no longer
controlled. For the rising part of the peak we therefore evaluate the resonant
kernel directly, using the constant branch of the transfer function. This gives
\begin{align}
\Omega_{\mathrm{GW}, \mathrm{mid}}^{\mathrm{adia}}
&=
\frac{
\mathcal{A}_{\mathcal{R}}^2 (5+3\omega)^4
c_s^{22/3-2n_s}
}{
3125\,3^{5/6}(1+\omega)^4
}
2^{-13/3+2n_s}
\pi s_0
\left(\frac{k}{k_{\rm eva}}\right)^{17/3}
\left(\frac{k}{k_{\rm CMB}}\right)^{2n_s-2}
\nonumber\\
&\quad\times
\Bigg[
15\,{}_2F_1
\left(\frac{1}{2},\frac{5}{3}-n_s;\frac{3}{2};c_s^2s_0^2\right)
-10s_0^2\,{}_2F_1
\left(\frac{3}{2},\frac{5}{3}-n_s;\frac{5}{2};c_s^2s_0^2\right)
\nonumber\\
&\hspace{4.2cm}
+3s_0^4\,{}_2F_1
\left(\frac{5}{2},\frac{5}{3}-n_s;\frac{7}{2};c_s^2s_0^2\right)
\Bigg].
\end{align}

Fig.~\ref{fig:GWSigMono_longversion} shows that the analytic expressions
capture the numerical spectra in the monochromatic limit. Applying the same
procedure to extended PBH mass functions gives equally good agreement. The peak
is generated by the resonant part of the double integral, more precisely by the
integrable pole structure of the radiation-era kernel. This treatment avoids
introducing an additional fitting parameter for the transition between
transfer-function branches. The remaining difference between the analytic and
numerical curves is mainly caused by deviations of the transfer function from a
pure power law near $k_{\rm eq}$.  Imposing the power-law approximation in the
numerical calculation removes this discrepancy.

\subsection{Isocurvature contribution}
The isocurvature contribution can be treated analogously. For a monochromatic
PBH mass distribution the result of Ref.~\cite{Domenech:2024wao}, confirmed by
our analytic and numerical calculation, is
\begin{align}
\Omega_{\mathrm{GW}, \mathrm{res}}^{\mathrm{iso}}
&=
C^4(w)
\frac{
c_s^{7/3}\left(c_s^2-1\right)^2
}{
576\times6^{1/3}\pi
}
\left(\frac{k_{\mathrm{eq}}}{k_{\rm \mathsmaller{PBH}}}\right)^8
\left(\frac{k_{\rm \mathsmaller{PBH}}}{k_{\mathrm{eva}}}\right)^{17/3}
\left(\frac{k}{k_{\rm \mathsmaller{PBH}}}\right)^{11/3},
\\
\Omega_{\mathrm{GW}, \mathrm{IR}}^{\rm iso}(k)
&=
\frac{1}{2}
C^4(w)
\frac{c_s^4}{120\pi^2}
\left(\frac{2}{3}\right)^{1/3}
\left(\frac{k_{\mathrm{eq}}}{k_{\rm \mathsmaller{PBH}}}\right)^8
\left(\frac{k_{\rm \mathsmaller{PBH}}}{k_{\mathrm{eva}}}\right)^{14/3}
\left(\frac{k}{k_{\rm \mathsmaller{PBH}}}\right).
\end{align}
The factor $1/2$ in the infrared branch reflects the improved transfer function,
which reduces the amplitude of $\Phi$ around equality. After reheating the
background is radiation dominated, so we set $\omega=1/3$ and
$c_s=1/\sqrt{3}$. The corresponding transfer-function coefficient is
$C(1/3)\simeq1$, see Sec.~\ref{subsec:RD_analytic}. Consistent with
Ref.~\cite{Domenech:2024kmh}, we find that for both adiabatic and isocurvature
initial conditions the numerical spectra are better described by
$\Omega_{\rm res,mid}(k)/2$.

\begin{figure}[t]
\centering
\includegraphics[width=0.47\textwidth]{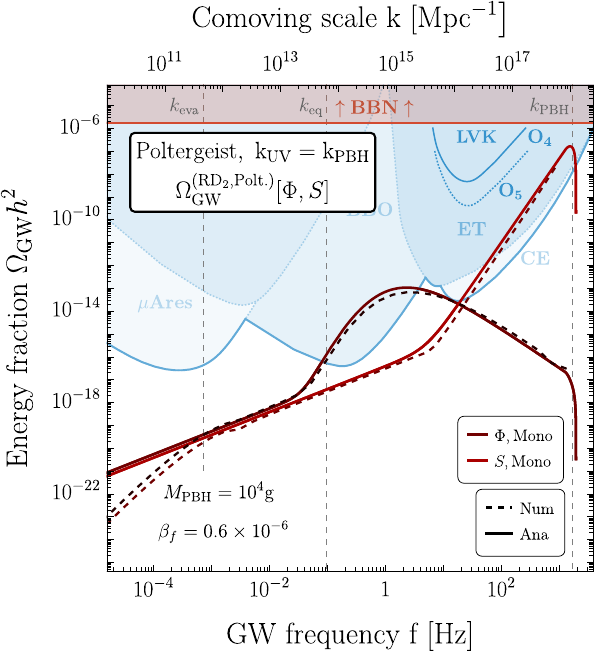}
\includegraphics[width=0.47\textwidth]{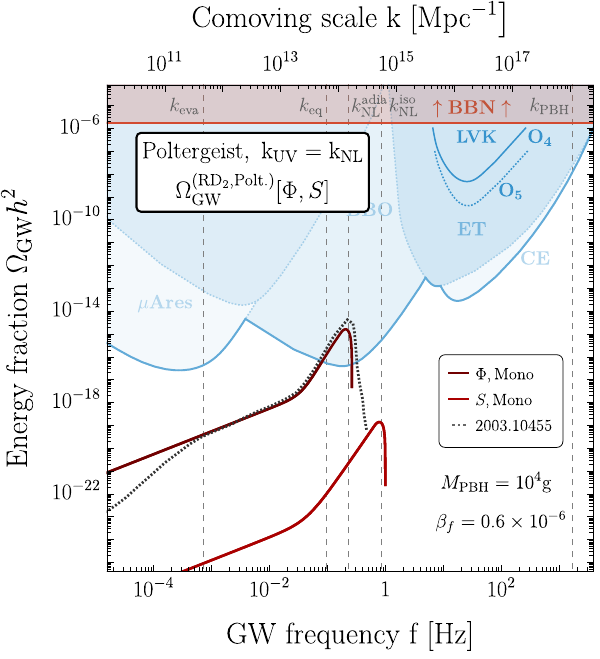}
\caption{\label{fig:GWSigMono_longversion}
SIGW spectrum for a monochromatic PBH mass distribution, including adiabatic
and isocurvature sources. Left: comparison with previous results for the
non-linear cutoff. Right: comparison between the numerical result (dashed) and
the analytic approximation (solid) for the UV cutoff. We choose $\beta_{\rm f}=0.6\times10^{-6}$ and $M_{\rm \mathsmaller{PBH}}=10^4\,{\rm g}$ 
to facilitate comparison with
Ref.~\cite{Inomata:2020lmk}.}
\end{figure}

\clearpage

\section{GW sensitivities}
\label{chap:sensitivities}

\subsection{GW observatories and BBN bound}
In Fig.~\ref{fig:GW_reach_longversion}, we show the regions of parameter space constrained by current Pulsar Timing Array observations (NG15)~\cite{NANOGrav:2023ctt} and by ground-based interferometers from the LIGO–Virgo–KAGRA (LVK) O3 observing run \cite{KAGRA:2021kbb}, together with projected sensitivities for the forthcoming LVK O5 run \cite{LIGOScientific:2014pky}. We further include sensitivity forecasts for future pulsar timing arrays such as SKA \cite{Janssen:2014dka}, the proposed lunar laser ranging experiment \cite{Blas:2021mqw,Blas:2021mpc}, atom-interferometer detectors including AION~km \cite{Badurina:2019hst,Abend:2023jxv} and AEDGE \cite{AEDGE:2019nxb}, as well as next-generation space- and ground-based interferometers. The latter comprise LISA \cite{Audley:2017drz,Robson:2018ifk}, the Einstein Telescope (ET) \cite{Punturo:2010zz,ET:2019dnz}, Cosmic Explorer (CE) \cite{Reitze:2019iox}, $\mu$-Ares \cite{Sesana:2019vho}, the Big-Bang Observer (BBO) \cite{Yagi:2011wg}, and THEIA \cite{Garcia-Bellido:2021zgu}.
The Big Bang Nucleosynthesis (BBN) constraint is obtained by imposing $\Delta N_{\rm eff}\lesssim 0.34$, following the procedure detailed in \cite{Gouttenoire:2019kij}.

\subsection{Detector sensitivities}
Sensitivity estimates are constructed using power-law integrated curves (PLICs), assuming signal-to-noise ratios and observation times of $(\mathrm{SNR}=5, T=15~\mathrm{yr})$ for NG15, $(\mathrm{SNR}=2, T=160~\mathrm{days})$ for LVK~O3, $(\mathrm{SNR}=10, T=1~\mathrm{yr})$ for LVK~O5, $(\mathrm{SNR}=2, T=15~\mathrm{yr})$ for lunar ranging, $(\mathrm{SNR}=10, T=5~\mathrm{yr})$ for AION~km, $(\mathrm{SNR}=10, T=7~\mathrm{yr})$ for $\mu$-Ares and $(\mathrm{SNR}=10, T=10~\mathrm{yr})$ for SKA, THEIA, LISA, BBO, AEDGE, ET, and CE. The blue violins show the GW signal detected by Pulsar Timing Arrays (NG15)~\cite{NANOGrav:2023gor,EPTA:2023fyk,Reardon:2023gzh,Xu:2023wog}.

\subsection{Astrophysical foregrounds}
Astrophysical foregrounds are also shown. These include the galactic white-dwarf binary background modeled following \cite{Robson:2018ifk} (see \cite{Lamberts:2019nyk,Boileau:2021gbr} for alternative treatments), the contribution from extragalactic supermassive black hole binaries taken from \cite{Rosado:2011kv}, and the background from extragalactic compact binaries (neutron stars and black holes), modeled as a power law $\Omega \propto f^{2/3}$ calibrated to LVK O3 observations \cite{KAGRA:2021kbb}.

\begin{figure}[th!]
\centering
\raisebox{0cm}{\makebox{\includegraphics[width=0.8\textwidth, scale=1]{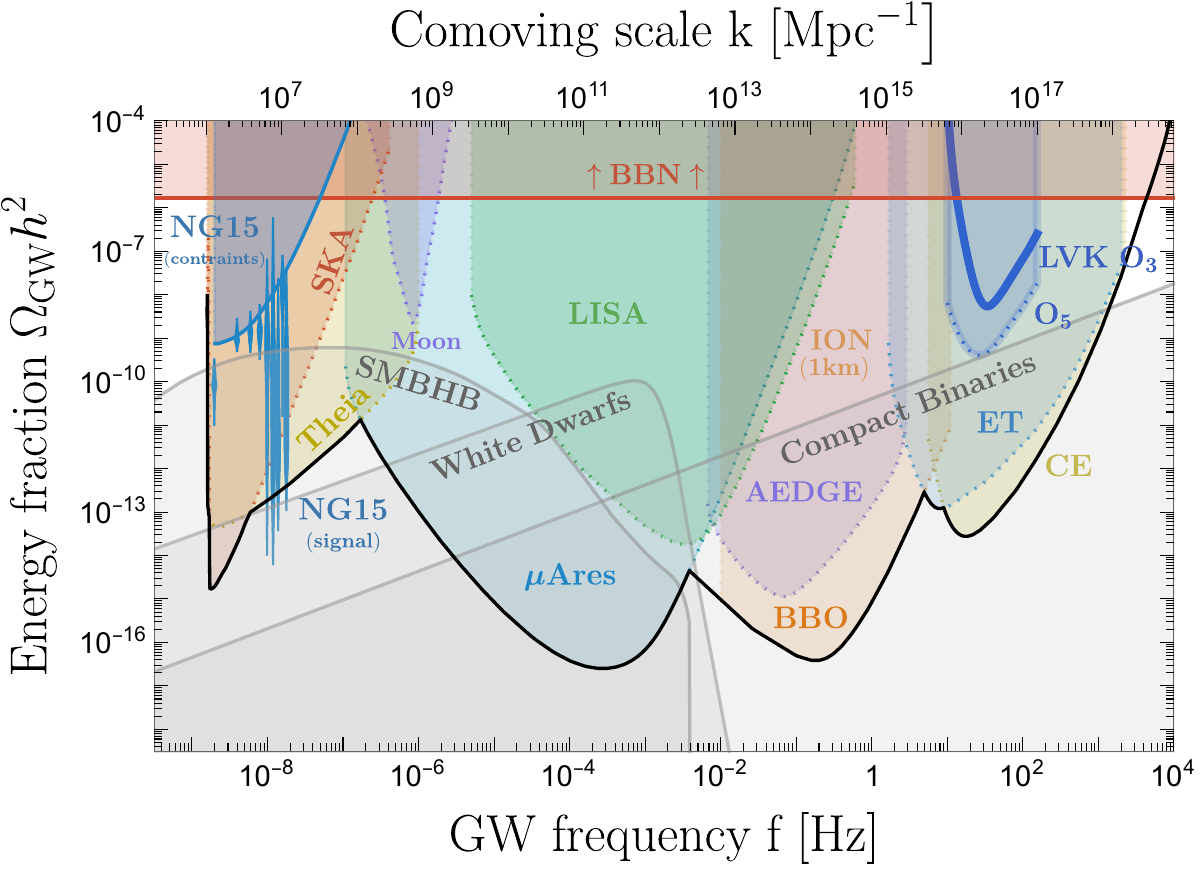}}}
\caption{ The black lines shows the most optimistic GW reach. We refer the reader to the text for the details. } 
\label{fig:GW_reach_longversion}
\end{figure}

\clearpage
\bibliographystyle{JHEP}
\bibliography{Bib}
\end{document}